\documentclass[times,sort&compress,3p]{elsarticle}
\journal{ }
\usepackage[labelfont=bf]{caption}

\usepackage{amsthm}
\usepackage{amsmath, amsfonts, amssymb, bbm, bm}

\usepackage[utf8]{inputenc} 
\usepackage[T1]{fontenc}
\usepackage{graphicx}
\usepackage{lmodern}
\usepackage{enumerate}
\usepackage{setspace}
\usepackage{xcolor}
\usepackage{soul}

\makeatletter
\def\ps@pprintTitle{%
  \let\@oddhead\@empty
  \let\@evenhead\@empty
  \let\@oddfoot\@empty
  \let\@evenfoot\@oddfoot
}
\makeatother

\usepackage[english]{babel}

\usepackage{hyperref}

\renewcommand{\title}[1]{ \noindent{\centering \Large \textbf{ #1 }\\} }

\theoremstyle{plain}
\newtheorem{theo}{Theorem}

\newtheorem{rem}{Remark}

\newtheorem{ass}{Assumption}
\newtheorem{algo}{Algorithm}

\theoremstyle{definition}

\newcommand{\R}{\mathbb{R}}
\newcommand{\T}{\mathcal{T}}
\newcommand{\E}{\mathrm{E}}

\renewcommand{\P}{\mathrm{P}}

\begin{document}

\pagenumbering{arabic}
\setcounter{page}{1}

\begin{sloppypar}

\begin{frontmatter}

\title{
General multiple tests 
for functional data}

\author[1]{Merle Munko} \corref{mycorrespondingauthor}
\author[1]{Marc Ditzhaus}
\author[2,3]{Markus Pauly}
\author[4]{{\L}ukasz Smaga}
\author[5]{Jin-Ting Zhang}

\address[1]{Faculty of Mathematics, Otto-von-Guericke University Magdeburg, Universitätsplatz 2, 39106 Magdeburg, Germany}
\address[2]{Faculty of Statistics, TU Dortmund University, August-Schmidt-Straße 1, 44227 Dortmund, Germany}
\address[3]{Research Center Trustworthy Data Science and Security, UA Ruhr, Joseph-von-Fraunhofer-Straße 25, 44227 Dortmund, Germany}
\address[4]{Faculty of Mathematics and Computer Science, Adam Mickiewicz University, Uniwersytetu Pozna\'nskiego 4, 61-614 Pozna\'{n}, Poland}
\address[5]{Faculty of Science, National University of Singapore, 6 Science Drive 2, Singapore 117546, Singapore}

\cortext[mycorrespondingauthor]{Corresponding author. Email address: \url{merle.munko@ovgu.de}}

\begin{abstract}
While there exists several inferential methods for analyzing functional data in factorial designs, there is a lack of statistical tests that are valid (i) in general designs, (ii) under non-restrictive assumptions on the data generating process and (iii) allow for coherent post-hoc analyses. In particular, most existing methods assume Gaussianity or equal covariance functions across groups (homoscedasticity) and are only applicable for specific study designs that do not allow for evaluation of interactions.
Moreover, all available 
strategies are only designed for testing global hypotheses 
and do not directly allow a more in-depth analysis of multiple local 
hypotheses. To address the first two problems (i)-(ii), we propose flexible integral-type test statistics that are applicable in general factorial designs under minimal assumptions on the data generating process. In particular, we neither postulate homoscedasticity nor Gaussianity. 
To approximate the statistics' null distribution, we adopt a resampling approach and validate it methodologically.
Finally, we 
use our flexible testing framework 
to (iii) infer several local null hypotheses simultaneously. To allow for powerful data analysis, we thereby take the complex dependencies of the different local tests statistics into account. 
In extensive simulations we confirm that the new methods are flexibly applicable. Two illustrate data analyses complete our study. 
The new testing procedures are implemented in the R package multiFANOVA, 
which will be available on CRAN soon.
\end{abstract}


\end{frontmatter}

\section{Introduction} 
Functional data are 
observations measured in time or space that can be represented by functions, curves, or surfaces. This 
allows to avoid 
problems of classical methods, such as missing data or 
the 
curse of dimensionality (i.e., a smaller number of observations in comparison to the number of time points). Due to the fast development of precise measurement instruments, functional data occur more frequently in several scientific 
fields \cite[p.~17]{Zhang2013}, especially when the data are collected over a continuum like time or space. This includes audiology, biology, ergonomy, meteorology, growth studies, and environmentology~\cite[p.~2]{Zhang2013}. 
Moreover, the analysis of stock prices in economy \cite{das2019} and of electroencephalography (EEG) curves in medicine \cite{scheffler2019, shangguan2020} may require suitable tools for functional data analysis (FDA). As a result, the statistical analysis of functional data is becoming more and more important. This is also documented by
the existence of several well-cited 
textbooks \cite{Ferraty_Vieu_2006,HorvathKokoszka_2012,kokoszka2017introduction,ramsay_silverman_2005,Zhang2013} and a large number of scientific publications in the last few years (e.g., a Google Scholar search with 'functional data' output more than 14,000 hits from January 2022 to June 2023). 

When functional data is collected in a factorial design, the question arises whether there are differences between the groups' mean functions or whether certain interactions exist.
These questions 
can be answered by 
a functional analysis of variance (FANOVA). For one-way designs, e.g., FANOVA tests the null hypothesis of equal mean functions across groups. 
Numerous FANOVA procedures 
have been developed: \citeauthor{ramsay_silverman_1997} \cite[p.~144]{ramsay_silverman_1997} proposed a point-wise F-test statistic for the one-way layout 
assuming 
that all groups share 
the same covariance structure 
(homoscedasticity). Integrating this statistic leads to the so-called GPF (globalizing point-wise F) test statistic proposed by \citeauthor{ZhangLiang2013}~\cite{ZhangLiang2013} which was shown to be asymptotically valid for homoscedastic non-Gaussian settings. 
Under the same assumptions, \citeauthor{zhangEtAl2013} \cite{zhangEtAl2013} proposed the Fmax FANOVA test. 
Here, the supremum over the point-wise F-test statistics is taken instead of integrating. 
The GPF- and Fmax-procedures were recently extended to a general linear hypothesis testing problem covering factorial FANOVA designs 
by \citeauthor{SmaZha} \cite{SmaZha2020, SmaZha}, still assuming homoscedasticity.

However, similar to (but worse than) the case of univariate metric observations, 
homoscedasticity is a very restrictive assumption which is almost never fulfilled in practical analyses, let alone verified.
Thus, using 
homoscedastic FANOVA methods 
might lead to misleading test decisions if the assumption is violated \cite[p.~307]{Zhang2013}.
Therefore, testing procedures for heteroscedastic functional data were developed: \citeauthor{cuevasEtal2004}~\cite{cuevasEtal2004} proposed an $L^2$-norm-based one-way FANOVA test under heteroscedasticity and introduced a parametric bootstrap approach. \citeauthor{febrero} \cite{febrero}, \citeauthor{jimi} \cite{jimi} and \citeauthor{linEtAl} \cite{linEtAl} developed further methods for heteroscedastic one-way FANOVA. 
In addition, \citeauthor{febrero} \cite{febrero} also considered the heteroscedastic two-way FANOVA. %
Moreover, three different testing procedures for a general linear hypothesis testing problem including factorial FANOVA designs under heteroscedasticity are presented in \cite[pp.~323-330]{Zhang2013}. This includes an $L^2$-norm-based, F-type, and non-parametric bootstrap test, where the F-type test requires Gaussian functional data. However, the asymptotic validity and finite sample properties of these three testing procedures have not been investigated so far. Moreover, the problem of multiple testing in the FANOVA context has not yet been addressed, to the best of our knowledge. If a global testing procedure leads to a significant result, it is of special interest which concrete individual hypotheses/group differences have caused the rejection. In fact, this is often the central research question.

We close this gap in the present paper. Thereby, 
we focus on general multiple tests which provide more information than global tests by offering local and global test decisions without necessarily losing power \cite{konietschke2013}. 
Specifically, we derive different procedures
for a general multiple testing problem for functional data covering multiple contrast tests without assuming homoscedasticity nor Gaussianity. Our solutions cover 
FANOVA and post hoc testing in general factorial designs. Special cases contain all-pairs or Tukey-type comparisons of mean functions in a one-way design or the joint testing of main and interaction effects in a two-way design. 
To derive our methods, we start with a point-wise Hotelling's $T^2$-type statistic, which takes into account the possible heteroscedasticity of the functional data. This test statistic is invariant under scale and orthogonal transformations of the data, which distinguishes it from other well known test statistics from the literature. 
Following the basic idea of aggregating point-wise objects in FDA, we integrate the point-wise Hotelling's $T^2$ statistic to obtain an 
overall test statistic. 
We then establish its asymptotic distribution using empirical process theory. As 
it is too complicated for practical applications
, we propose a parametric bootstrap approach to compute critical values. We prove that our approach leads to an asymptotically valid testing procedure. 
This globalized point-wise Hotelling's $T^2$-test is initially only a 
global testing procedures. 
However, it is the base to obtain a multiple testing method. Performing multiple comparisons in a stepwise procedure may result in incompatible test decisions. In detail, the global null hypothesis can be rejected although none of the individual null hypotheses is or vice versa \cite[p.~63]{konietschke2013}. Solving the multiple testing problem via an adjustment of the level of significance by using, for example, the Bonferroni correction may lead to power loss \cite[pp.~63-64]{konietschke2013}. That is why we propose a multiple testing procedure that 
takes into account the concrete dependencies of the separate test statistics.
The derived methods have the desirable property that the global null hypothesis is rejected whenever at least one of the local null hypotheses is rejected. We also prove its asymptotic validity and analyze its finite sample behavior 
in numerical experiments. 
The results provide a guideline for practical use of their implementation in the R package \texttt{multiFANOVA}, which will be available on CRAN soon \cite{Rcore}.

The remainder of this paper is organized as follows. 
In Section~\ref{sec:setup}, the general setup of the heteroscedastic linear hypothesis testing problem is presented. 
A suitable test statistic for this global testing problem is proposed and studied in Section~\ref{sec:testingProc}. 
In Section~\ref{sec:Bootstrap} explains the parametric bootstrap procedure and derives its asymptotic correctness. The new multiple tests are finally derived in Section~\ref{sec:multiple}.
Moreover, the finite sample performance of the proposed tests is analyzed in extensive simulation studies in terms of size control and power in Section~\ref{sec:Simu}.
In Section~\ref{sec:Data}, our proposed methodologies are applied to two real data examples. Finally, the results of this paper are discussed in Section~\ref{sec:Discussion}. All technical proofs and simulation results are outlined in the supplement.

\section{Heteroscedastic Linear Hypothesis Testing Problem}
\label{sec:setup}
In this section, we first present the general statistical model and hypotheses. Secondly, we formulate and discuss the technical assumptions, which are needed for proving the asymptotic properties of the test statistic.

\subsection{Model and Hypotheses}
Throughout this paper, let $\T \subset \R$ be a closed finite interval, e.g., $\T=[0,1]$, and $\mathcal{L}_2(\T)$ be the set of all squared integrable functions over $\T$, which forms a Hilbert space with inner product $$ \mathcal{L}_2(\T) \times \mathcal{L}_2(\T) \ni ({g}, {h}) \mapsto \int_{\T} g(t)h(t) \,\mathrm{ d }t .$$ 
A real-valued stochastic process $\left(x(t)\right)_{t\in\T}$ with mean function $\eta: \T \to \R $ and covariance function $\gamma: \T^2 \to \R$ is denoted by $x \sim \text{SP}(\eta,\gamma)$. 
If $\left(x(t)\right)_{t\in\T}$ is additionally a Gaussian process, i.e., all finite-dimensional distributions are Gaussian, we write $x \sim \text{GP}(\eta,\gamma)$.
Analogously, we write $\mathbf{x} \sim \text{SP}_r(\boldsymbol{\eta},\boldsymbol{\Gamma})$ or $\mathbf{x} \sim \text{GP}_r(\boldsymbol{\eta},\boldsymbol{\Gamma})$ for an $r$-dimensional stochastic or Gaussian process $\mathbf{x}$, respectively, with mean function $\boldsymbol{\eta}: \T \to \R^r$, covariance function $\boldsymbol{\Gamma}: \T^2 \to \R^{r \times r}$ and $r\in\mathbb{N}$. 
Moreover, we use the abbreviation 'i.i.d.' for independent and identically distributed random variables. 

As described in the introduction, we want to cover different factorial FANOVA designs in a unified way. To this end, we consider the following FANOVA model given by 
$k \in \mathbb{N}$ independent groups of random processes
\begin{align}
\label{eq:samples}
    x_{i1},\dots,x_{in_i} \sim \text{SP}(\eta_i,\gamma_i)\: \text{ i.i.d. \quad for each } i\in\{1,\dots,k\},
\end{align}
which take values in $\mathcal{L}_2(\T)$. Here, the first index ($i$) represents the group or sample and the second stands for the experimental unit on which functional measurements were taken. Moreover, $n = \sum_{i=1}^k n_i$ is the total sample size. 
Here, $\eta_i: \T \to \R$ denotes the mean function and $\gamma_i: \T^2 \to \R$ denotes the covariance function of the $i$th sample for all $i \in \{1,\dots,k\}$, which are assumed to be unknown. In contrast to \cite[p.~101]{SmaZha}, the covariance functions of the different groups may differ from each other, i.e., heteroscedasticity is explicitly allowed. As in classical (M)ANOVA settings \cite{KONIETSCHKE2015, paulyETAL2015}, a more general factorial structure can be incorporated easily by splitting up the indices (see below). 
Since we put the greater importance on the accurate description of the multiple testing procedures and related standard examples, we merely focus on the one-way case. However, crossed two-, three- and higher-way layouts will be studied in more detail in the future.


Let us now formulate the hypotheses 
of interest. Assume that $r\in\mathbb{N}$ and $\mathbf{H} \in \R^{r \times k}$ is a known 
matrix. This could, e.g. be a contrast matrix fulfilling $\mathbf{H}\mathbf{1}_k = \mathbf{0}_r$, where here and throughout $\mathbf{1}_{k}\in\R^{k}$ and $\mathbf{0}_r\in\R^{r}$ represent vectors of ones and zeros, respectively. Contrast matrix are the most common matrices to formulate linear hypotheses. However, we do not explicitly need the assumption of $\mathbf{H}$ being a contrast matrix. 
Let $\boldsymbol{\eta} := (\eta_1,\dots,\eta_k)^{\top}: \T \to \R^k$ be the vector of the mean functions and \mbox{$\mathbf{c} 
: \T\to\R^r$} be a fixed, measurable function, e.g., $\mathbf{c}(t) = \mathbf{0}_r$ for all $t\in\T$. 
We consider the following null and alternative hypotheses
\begin{align}
    \label{eq:global}
    \mathcal H_0: \mathbf{H}\boldsymbol{\eta}(t) = \mathbf{c}(t) \text{ for all } t\in \T \quad \text{vs.} \quad \mathcal H_1: \mathbf{H}\boldsymbol{\eta}(t) \neq \mathbf{c}(t) \text{ for some } t\in \T.
\end{align}
This general formulation 
 contains many special cases like the one-way FANOVA problem of equal mean functions across groups.  
To be concrete, choose $\mathbf{c}(t)=\mathbf{0}_{k}$ for all $t\in\T$ and $\mathbf{H} = \mathbf{P}_k:= \mathbf{I}_k - \mathbf{J}_k/k$, 
where $\mathbf{I}_k \in\R^{k\times k}$ denotes the unit matrix and \mbox{$\mathbf{J}_k := \mathbf{1}_k\mathbf{1}_k^{\top} \in\R^{k\times k}$} is the matrix of ones. Then we have 
$\mathcal H_0: \eta_1(t)=\dots=\eta_k(t)$ for all $t\in \mathcal{T}$. 
There exist other possible choices of 
$\mathbf{H}$ which lead to this global null hypothesis \cite[pp.~64-65]{konietschke2013}. For example, the \textit{Dunnett-type} contrast matrix \cite{dunnett_1955} $\mathbf{H} = [-\mathbf{1}_{k-1}, \mathbf{I}_{k-1}]$ for many-to-one comparisons can be considered 
 together with $\mathbf{c}(t) \equiv \mathbf{0}_{k-1}$. This (later) corresponds to the multiple testing problem, where 
 each mean functions $\eta_2, \dots, \eta_k$ is compared to the mean function $\eta_1$ of the first group.
another choice is the \textit{Tukey-type} contrast matrix \cite{tukey}
\begin{align*}
    \mathbf{H} = \begin{bmatrix}
-1 & 1 & 0 & 0 & \cdots & \cdots & 0 \\
-1 & 0 & 1 & 0 &\cdots & \cdots & 0 \\
\vdots  & \vdots &\vdots & \vdots & \ddots & \vdots & \vdots  \\
-1 & 0 & 0 & 0& \cdots & \cdots & 1\\
0 & -1 & 1 & 0& \cdots & \cdots & 0 \\
0 & -1 & 0 & 1& \cdots & \cdots & 0 \\
\vdots  & \vdots & \vdots  & \vdots & \ddots & \vdots & \vdots \\
0 & 0 & 0 & 0 & \cdots & -1 & 1
\end{bmatrix}  \in \R^{k(k-1)/2 \times k}
\end{align*} 
 together with $\mathbf{c}(t) \equiv \mathbf{0}_{k(k-1)/2}$. 
In the corresponding multiple testing problem all pairs of mean functions are compared. 
For an overview of different contrast tests we refer to \cite{bretz2001}.\\ 
To illustrate that we can also treat higher way layouts we consider 
a two-way design with factors A ($a$~levels) and B ($b$ levels). We set $k := a b$ and split up the group index $i$ in two subindices \mbox{$(i_1, i_2)\in\{1,\dots,a\}\times\{1,\dots,b\}$.} Then our framework also covers the following 
null hypotheses:
\begin{itemize}
    \item  $\mathcal{H}_0^A: \mathbf{H}_A \boldsymbol\eta(t) = \mathbf{0}_a $ for all $t\in\T$ with $\mathbf{H}_A := \mathbf{P}_a \otimes (\mathbf{1}_b^{\top}/b)$ (no main effect of A),
    \item $\mathcal{H}_0^B: \mathbf{H}_B \boldsymbol\eta(t) = \mathbf{0}_b $ for all $t\in\T$ with $\mathbf{H}_B := (\mathbf{1}_a^{\top}/a)  \otimes \mathbf{P}_b$ (no main effect of B),
    \item $\mathcal{H}_0^{AB}: \mathbf{H}_{AB} \boldsymbol\eta(t) = \mathbf{0}_{ab} $ for all $t\in\T$ with $\mathbf{H}_{AB} := \mathbf{P}_a  \otimes \mathbf{P}_b$ (no interaction effect).
\end{itemize}
Here, $\otimes$ denotes the Kronecker product. Higher-way designs can be incorporated similarly as in \cite[p.~467]{paulyETAL2015}. 


\subsection{Assumptions}
In the following we state the mathematical assumptions we postulate to construct valid tests for (\ref{eq:global}). These are followed by a discussion on their practical meaning.
\begin{ass}
\label{assumptions}
Consider the $k$ independent samples (\ref{eq:samples}). Let $v_{ij} := x_{ij} - \eta_i,$ \mbox{$i\in\{1,\dots,k\},$} $j\in\{1,\dots,n_i\},$ denote the \textit{subject-effect functions}.  Additionally, assume  for all $i\in\{1,\ldots,k\}$:
\begin{enumerate}
    \item[(A1)] The functional data $x_{i1},\dots,x_{in_i} \sim SP(\eta_i, \gamma_i)$ are i.i.d.~and take values in $\mathcal{L}_2(\T)$.
    \item[(A2)] The mean functions satisfy $\eta_i \in \mathcal{L}_2(\T)$. 
    \item[(A3)] The function $\T\ni t \mapsto \gamma_i(t,t) \in \R$ is continuous and positive.
    \item[(A4)] There exists $\tau_i > 0$ such that ${n_i}/{n} \to \tau_i$ as $n \to \infty$.
    \item[(A5)] It holds $\E\left[\left( \int\limits_{\T} v_{i1}^2(t)\,\mathrm{ d }t \right)^2\right] < \infty$.
    \item[(A6)] For all $j\in\{1,2\}$, it holds 
    $$  \E_{\mathrm{out}}\left[\sup\limits_{\substack{t,s\in\T, \:|t-s|<\delta}} |v_{i1}(t)v_{ij}(t)-v_{i1}(s)v_{ij}(s)|\right] \to 0 \quad \text{as $\delta\to 0$}, $$
    where here and throughout $\E_{\mathrm{out}}$ denotes the outer expectation as defined in \cite[p.~6]{vaartWellner1996}.
\end{enumerate}
\end{ass}


\begin{rem}[On the Assumptions] \label{AssumptBemerkungen}
\begin{enumerate}
\item Assumptions (A1)-(A4) ensure that the mean function estimators are asymptotically Gaussian. The uniform consistency of the covariance function estimators is guaranteed by (A1), (A3) and (A6).
Furthermore, (A5) is needed for the parametric bootstrap consistency, see Section~S2 in the supplement for details.
\item Note that (A3) implies $\int_{\T} \gamma_i(t,t) \,\mathrm{ d }t < \infty$ 
    for all $i\in\{1,\dots,k\}$ immediately since $\T\ni t \mapsto \gamma_i(t,t)$ attains its maximum as a continuous function over the compact set~$\T$. Moreover, it follows from the Cauchy-Schwarz inequality
    $$ \int_{\T^2} \gamma_i^2(t,s) \,\mathrm{ d }(t,s) \leq \int_{\T^2} \gamma_i(t,t)\gamma_i(s,s) \,\mathrm{ d }(t,s) = \left(\int_{\T} \gamma_i(t,t) \,\mathrm{ d }t\right)^2 < \infty  $$
    for all $i\in\{1,\dots,k\}$ under (A3).
    \item By the triangular inequality and (A1), we have
\begin{align*}
    &\E_{\mathrm{out}}\left[\sup\limits_{\substack{t,s\in\T, \:|t-s|<\delta}} |v_{i1}(t)v_{i2}(t)-v_{i1}(s)v_{i2}(s)|\right] 
    \\
    &\leq 2\E_{\mathrm{out}}\left[\sup\limits_{t\in\T}|v_{i1}(t)|\right]\E_{\mathrm{out}}\left[ \sup\limits_{\substack{t,s\in\T, \:|t-s|<\delta}} |v_{i1}(t)-v_{i1}(s)|\right]
\end{align*}
for all $i\in\{1,\dots,k\}$. Hence, (A6) is satisfied if 
\begin{align*}
    &\E_{\mathrm{out}}\left[\sup\limits_{t\in\T}|v_{i1}(t)|\right] < \infty, \qquad \E_{\mathrm{out}}\left[ \sup\limits_{\substack{t,s\in\T, \:|t-s|<\delta}} |v_{i1}(t)-v_{i1}(s)|\right] \to 0\\
    &\text{and}\quad \E_{\mathrm{out}}\left[\sup\limits_{\substack{t,s\in\T, \:|t-s|<\delta}} |v_{i1}^2(t)-v_{i1}^2(s)|\right] \to 0 \quad \text{as $\delta\to 0$}
\end{align*}
holds for all $i\in\{1,\dots,k\}$. This is particularly satisfied if $\E_{\mathrm{out}}\left[\sup_{t\in\T}|v_{i1}(t)|\right] < \infty$ holds and $v_{i1}$ takes values in an equicontinuous set of functions for all $i\in\{1,\dots,k\}$.
\item The supremum in assumption (A6) and, therefore, in Remark~\ref{AssumptBemerkungen}~(3) can be replaced by the essential supremum since we only need the $\mathcal L_{\infty}(\T)$-convergence of the covariance function estimators in the proofs, see Section~S2 in the supplement.
\item Assumption (A6) 
implies that $v_{i1}^2$ is sample continuous for all \mbox{$i\in\{1,\dots,k\}$}. 
In order to show this, we firstly note that convergence in mean implies convergence in probability. Moreover, $\sup\limits_{\substack{t,s\in\T, \:|t-s|<\delta}} |v_{i1}^2(t)-v_{i1}^2(s)|$ is monotone in $\delta$. Hence, it follows
$$ \P_{\mathrm{out}}\left( \lim\limits_{\delta\to 0}\sup\limits_{\substack{t,s\in\T, \:|t-s|<\delta}} |v_{i1}^2(t)-v_{i1}^2(s)| = 0 \right) = 1 $$ and, thus, $v_{i1}$ is sample continuous for all $i\in\{1,\dots,k\},$
where $\P_{\mathrm{out}}$ denotes the outer probability.
Therefore, it turns out that the outer expectation in (A6) can be replaced by the usual expectation since the supremum of a continuous function is measurable.
Since $\T$ is a compact interval, it follows that $v_{i1}$ is bounded almost surely for all $i\in\{1,\dots,k\}.$
Even the stronger result
\begin{align*}
    \E\left[\sup\limits_{t\in\T} v_{i1}^2(t) \right]
    < \infty
\end{align*}
follows easily from (A1) and (A6) for all $i\in\{1,\dots,k\}$. 
\end{enumerate}
\end{rem}

\section{Globalizing point-wise Hotelling's $T^2$-Test Statistic}\label{sec:testingProc}

Following the illustration of the testing problem and the assumptions on the functional data in Section~\ref{sec:setup}, a suitable test statistic for the heteroscedastic linear hypothesis testing problem is defined and its asymptotic behaviour is analyzed in this section.
First of all, we need estimators of all unknown quantities, i.e., the mean and covariance functions. An unbiased estimator for the mean function of the $i$th group at $t$ is given by
$$ \widehat{\eta}_i(t) := \frac{1}{n_i}\sum_{j=1}^{n_i} x_{ij}(t)$$
and for the covariance function of the $i$th group at $(t,s)$ by
\begin{align} \label{eq:gammaHat}
     \widehat{\gamma}_i(t,s) := \frac{1}{n_i-1}\sum_{j=1}^{n_i} \left(x_{ij}(t) - \widehat{\eta}_i(t)\right)\left(x_{ij}(s) - \widehat{\eta}_i(s)\right) 
\end{align}
for all $t,s\in\T, i\in\{1,\dots,k\}$ \cite[p.~310]{Zhang2013}. In comparison to \cite[p.~101]{SmaZha}, we have to estimate the covariance functions of the different groups separately due to a potential heteroscedasticity.

It follows from the weak law of large numbers, that the estimators for mean and covariance functions are point-wise consistent under assumption (A1).
In Lemma~S2 in the supplement, we even prove that 
the covariance function estimators are uniformly consistent over $\{(t,t)\in\T^2 \mid t\in\T\}$ if (A1), (A3), and (A6) hold. 
For $t,s \in \T$, let
\begin{align*}
	\widehat{\boldsymbol{\Sigma}}(t,s) &:= \mathrm{diag}\left( \frac{n}{n_1}\widehat{\gamma}_1(t,s),\ldots,\frac{n}{n_k}\widehat{\gamma}_k(t,s)\right)
\end{align*}
be an estimator for
\begin{align*}
	\boldsymbol{{\Sigma}}(t,s) &:= \mathrm{diag}\left( \frac{1}{\tau_1}{\gamma}_1(t,s),\ldots,\frac{1}{\tau_k}{\gamma}_k(t,s)\right) \in \R^{k\times k}.
\end{align*}
It should be noted that the matrix $\boldsymbol{{\Sigma}}(t,t)$ is invertible under (A3) and (A4) for all $t \in \T$. 
The point-wise consistency of the covariance function estimators 
immediately yield the point-wise consistency of $\widehat{\boldsymbol{\Sigma}}$ assuming (A1) and (A4).
Moreover, the uniform consistency over $\{(t,t)\in\T^2 \mid t\in\T\}$ follows from Lemma~S2 in the supplement under the assumptions (A1), (A3), (A4), and (A6).

Now, we are ready to formulate 
the \textit{point-wise Hotelling's $T^2$-test statistic}
\begin{align}
\label{eq:testfunc}
	\mathrm{TF}_{n,\mathbf{H},\mathbf{c}}(t) &:= n( \mathbf{H} \widehat{\boldsymbol\eta}(t) - \mathbf{c}(t) )^{\top} (\mathbf{H}\mathbf{\widehat \Sigma}(t,t) \mathbf{H}^{\top})^+ ( \mathbf{H} \widehat{\boldsymbol\eta}(t) - \mathbf{c}(t) )
\end{align}
for all $t \in\T$, where $\widehat{\boldsymbol\eta} := (\widehat\eta_1,\dots,\widehat\eta_k)^{\top}$ denotes the vector of all mean function estimators and $\mathbf{A}^+$ denotes the Moore-Penrose inverse of the matrix $\mathbf{A}$. Under the null hypothesis, we expect that the point-wise Hotelling's $T^2$-test statistic is small since $\mathbf{H} \widehat{\boldsymbol\eta}(t)$ is an estimator for $\mathbf{H} \boldsymbol{\eta}(t) = \mathbf{c}(t)$. The term $\mathbf{H}\mathbf{\widehat \Sigma}(t,t) \mathbf{H}^{\top}$ approximates the covariance matrix of $\sqrt{n} \left(\mathbf{H} \widehat{\boldsymbol\eta}(t) - \mathbf{c}(t) \right)$. In comparison to the point-wise test statistic proposed in \cite[p.~101]{SmaZha}, we take the estimated covariance functions of the different groups into account. 
integrating over the point-wise Hotelling's $T^2$-test statistic as in \cite[p.~101]{SmaZha}
we define the \textit{globalizing point-wise Hotelling's $T^2$-test} (GPH) \textit{statistic} by 
\begin{align*}
    T_{n}(\mathbf{H}, \mathbf{c}) := \int_{\T} \mathrm{TF}_{n,\mathbf{H},\mathbf{c}}(t) \,\mathrm{ d }t.
\end{align*}
We reject the null hypothesis~\eqref{eq:global} for large values of $T_{n}(\mathbf{H}, \mathbf{c})$. This test statistic has nice invariance properties as stated in the subsequent remarks. 

\begin{rem}[Invariance under Orthogonal Transformations]\label{CTI} 
Multiplying the left and right side of the null hypothesis in (\ref{eq:global}) with an orthogonal matrix $\mathbf{P}\in\R^{r\times r}$ leads to the same testing problem. 
By the properties of the Moore-Penrose inverse, the point-wise test statistic (\ref{eq:testfunc}) is invariant under orthogonal transformations, i.e.,
\begin{align*}	\mathrm{TF}_{n,\mathbf{P}\mathbf{H},\mathbf{P}\mathbf{c}}(t) &= n( \mathbf{H} \widehat{\boldsymbol\eta}(t) - \mathbf{c}(t) )^{\top}\mathbf{P}^{\top} (\mathbf{P}\mathbf{H}\mathbf{\widehat \Sigma}(t,t) \mathbf{H}^{\top}\mathbf{P}^{\top})^+ \mathbf{P}( \mathbf{H} \widehat{\boldsymbol\eta}(t) - \mathbf{c}(t) )\\&= n( \mathbf{H} \widehat{\boldsymbol\eta}(t) - \mathbf{c}(t) )^{\top} (\mathbf{H}\mathbf{\widehat \Sigma}(t,t) \mathbf{H'})^+ ( \mathbf{H} \widehat{\boldsymbol\eta}(t) - \mathbf{c}(t) ) \\&= \mathrm{TF}_{n,\mathbf{H},\mathbf{c}}(t)\end{align*} regarding any orthogonal $\mathbf{P}\in\R^{r\times r}$ and for all $t \in\T$.\\
\end{rem}

\begin{rem}[Scale-invariance]\label{invariance}
The point-wise Hotelling's $T^2$-test statistic (\ref{eq:testfunc}) is scale-invariant in the sense of \cite{GuoEtAl2019}, i.e., it does not change when the observed functional data $x_{ij},$ $i\in\{1,\dots,k\},$ \mbox{$j\in\{1,\dots,n_i\},$} are multiplied by any fixed function $h:\T\to\R$ with $h(t)\neq 0$ for all $t\in\T$. We denote the scaled functional data with a superscript $h$, i.e., $x_{ij}^h(t):=h(t)x_{ij}(t)$ for all $i\in\{1,\dots,k\},$ \mbox{$j\in\{1,\dots,n_i\}$} and $t\in\T$. 
Analogously, we denote the counterparts of the statistics defined above for the scaled functional data $x_{ij}^h,$ $i\in\{1,\dots,k\},$ \mbox{$j\in\{1,\dots,n_i\},$} with a superscript $h$.
The corresponding null hypothesis of the scaled functional data is $\mathcal{H}_0:\mathbf{H}\boldsymbol{\eta}^h(t)=\mathbf{c}^h(t)$ for all $t\in\T$, where $\boldsymbol{\eta}^h(t):=h(t)\boldsymbol{\eta}(t)$ and $\mathbf{c}^h(t):=h(t)\mathbf{c}(t)$.
Then, we have the scale-invariance of the point-wise Hotelling's $T^2$-test statistic
\begin{align*}
    \mathrm{TF}_{n,\mathbf{H},\mathbf{c}^h}^h(t) &= n( \mathbf{H} \widehat{\boldsymbol\eta}^h(t) - \mathbf{c}^h(t) )^{\top} (\mathbf{H}\mathbf{\widehat \Sigma}^h(t,t) \mathbf{H}^{\top})^+ ( \mathbf{H} \widehat{\boldsymbol\eta}^h(t) - \mathbf{c}^h(t) )\\&= n( \mathbf{H} \widehat{\boldsymbol\eta}(t) - \mathbf{c}(t) )^{\top} (\mathbf{H}\mathbf{\widehat \Sigma}(t,t) \mathbf{H}^{\top})^+ (\mathbf{H} \widehat{\boldsymbol\eta}(t) - \mathbf{c}(t) )\\
	&=\mathrm{TF}_{n,\mathbf{H},\mathbf{c}}(t)
\end{align*}
for all $t\in\T$.
Consequently, it follows immediately that the GPH statistic is also scale-invariant.
\end{rem}

In order to construct asymptotic tests based on the GPH statistic, we have to examine its asymptotic behaviour. The following theorem describes the limiting distribution of the GPH statistic under the null hypothesis. 



\begin{theo}
\label{asy_GPH}
Let $\mathbf{z}\sim GP_r(\mathbf{0}_r,\mathbf{H}\boldsymbol{\Sigma}\mathbf{H}^{\top})$.
Under (A1)-(A4), (A6) and the null hypothesis in (\ref{eq:global}), we have
\begin{align}
    &T_{n}(\mathbf{H}, \mathbf{c}) \xrightarrow{d} \int_{\T} \mathbf{z}^{\top}(t)(\mathbf{H}\boldsymbol{\Sigma}(t,t)\mathbf{H}^{\top})^+\mathbf{z}(t) \,\mathrm{ d }t
\end{align}
as $ n \to\infty$, where here and throughout $\xrightarrow{d}$ denotes convergence in distribution in the sense of \cite[p.~17]{vaartWellner1996}. 
\end{theo}

Since the different covariance functions are taken into account in (\ref{eq:testfunc}), the limiting distribution in Theorem~\ref{asy_GPH} is not as simple as in the homoscedastic case \cite[p.~101]{SmaZha}, and we are generally not able to specify it as known distributions. 
Nevertheless, we can still use the results of the above theorem to construct tests by approximating the limiting distribution via resampling strategies, as we will see in the next section. 

\section{Parametric Bootstrap}
\label{sec:Bootstrap}
Theorem~\ref{asy_GPH} provides that the GPH statistic constructed in Section~\ref{sec:testingProc} converges in distribution. However, the limiting distribution still depends on the unknown covariance functions $\gamma_1,\dots,\gamma_k$.
Thus, we cannot directly derive 
critical values in form of quantiles. 
One possibility would be to to estimate the limit distribution plugging in $\mathbf{\widehat \Sigma}$ and compute critical values via Monte-Carlo. 
This sampling from the estimated limit distribution was, e.g., suggested by 
\citeauthor{cuevasEtal2004}~\cite[pp.~115-116]{cuevasEtal2004} in the FANOVA context. 
This approach, however, may result in conservative or too liberal behavior, see the simulation study by \cite{GoreckiSmaga2015}). Therefore, we propose a different approach based upon parametric bootstrapping. 
To this end we estimate $\gamma_1,\dots,\gamma_k$ by $\widehat{\gamma}_1,\dots,\widehat{\gamma}_k$ and (conditioned on the functional data $(x_{i1},\dots,x_{in_i})_{i\in\{1,\dots,k\}}$) generate parametric bootstrap samples via
\begin{align*}
    x_{i1}^{\mathcal{P}},\dots,x_{in_i}^{\mathcal{P}} \sim GP(0,\widehat{\gamma}_i), \quad\text{for each }i\in\{1,\dots,k\}.
\end{align*} 
Here, using a Gaussian process for generating the parametric bootstrap sample seems natural regarding the limiting distribution in Theorem~\ref{asy_GPH} since the mean function estimators are asymptotically Gaussian anyway. Moreover, using the estimators of sample covariance functions, we mimic the covariance structure of the given functional data, see \cite{KONIETSCHKE2015} for a similar motivation in the MANOVA context. Note, that we still postulate our general model, i.e. we do not assume a parametric model for our functional data $(x_{i1},\dots,x_{in_i})_{i\in\{1,\dots,k\}}$.

In the following, we denote the parametric bootstrap counterparts of the estimators and statistics defined in Section~\ref{sec:testingProc}
 with a superscript $\mathcal{P}.$ Then, the \textit{parametric bootstrap point-wise Hotelling's $T^2$-test statistic} at $t$ is defined by
    \begin{align*}
        \mathrm{TF}^{\mathcal{P}}_{n,\mathbf{H}}(t) :=  n\left(\mathbf{H}\widehat{\boldsymbol\eta}^{\mathcal{P}}(t)\right)^{\top}\left(\mathbf{H}\widehat{\boldsymbol{\Sigma}}^{\mathcal{P}}(t,t)\mathbf{H}^{\top}\right)^+\mathbf{H}\widehat{\boldsymbol\eta}^{\mathcal{P}}(t).
    \end{align*} 
Integrating 
yields the \textit{parametric bootstrap globalizing point-wise Hotelling's $T^2$-test statistic}
    \begin{align}\label{eq:paraGPH}
        T_{n}^{\mathcal{P}}(\mathbf{H}) := \int_{\T} \mathrm{TF}^{\mathcal{P}}_{n,\mathbf{H}}(t) \,\mathrm{ d }t.
\end{align}

The following theorem ensures 
that the parametric bootstrap procedure is asymptotically valid for the heteroscedastic linear hypothesis testing problem.

\begin{theo}
\label{ParaBS}
Let $\mathbf{z}\sim GP_r(\mathbf{0}_r,\mathbf{H}\boldsymbol{\Sigma}\mathbf{H}^{\top})$.
Under (A1)-(A6), 
it holds that
\begin{align*}
    &T_{n}^{\mathcal{P}}(\mathbf{H}) \xrightarrow{d^*} \int_{\T} \mathbf{z}^{\top}(t)(\mathbf{H}\boldsymbol{\Sigma}(t,t)\mathbf{H}^{\top})^+\mathbf{z}(t) \,\mathrm{ d }t
\end{align*}
as $ n \to\infty$,
where here and throughout $\xrightarrow{d^*}$ denotes conditional convergence in distribution given the data $(x_{i1},x_{i2},...)_{i\in\{1,...,k\}}$. 
\end{theo}

The following algorithm can be used for the implementation of the parametric bootstrap approach.

\begin{algo} 
\label{ParaBSAlgo}
Let $\alpha \in (0,1)$ be the significance level and $B\in\mathbb{N}$ sufficiently large.
\begin{enumerate}
    \item Calculate $\widehat{\gamma}_1,...,\widehat{\gamma}_k$ in (\ref{eq:gammaHat}).
    \item Simulate $B$ times $n$ independent Gaussian processes $x_{i1;b}^{\mathcal{P}}, ..., x_{in_i;b}^{\mathcal{P}} \sim GP(0,\widehat{\gamma}_i),$ \mbox{$i\in\{1,...,k\},$} $b\in\{1,...,B\}$.
    \item Compute the parametric bootstrap GPH statistic $T_{n;b}^{\mathcal{P}}(\mathbf{H})$ as in (\ref{eq:paraGPH}) for all $b\in\{1,...,B\}$.
    \item Determine the empirical $(1-\alpha)$-quantile $Q^{\mathcal{P}}_{n,1-\alpha}$ of the computed values.
    \item Reject the null hypothesis in (\ref{eq:global}) if and only if $T_{n}(\mathbf{H}, \mathbf{c}) > Q^{\mathcal{P}}_{n,1-\alpha}.$
\end{enumerate}
\end{algo}

In practice, the Gaussian processes in Step~(2) of Algorithm~\ref{ParaBSAlgo} can only be evaluated on a finite grid of values $\{t_1,...,t_m\} \subset \T, m\in\mathbb{N}$, due to limited hardware accuracy. In addition, we have to use numerical integration techniques in Step~(3). However, by choosing $m$ sufficiently large, the resulting approximation errors can be neglected. The resulting finite sample properties will be studied in simulations in Sections~\ref{sec:Simu}-\ref{sec:Data}.

\section{General Multiple Tests}
\label{sec:multiple}
Using the techniques stated in the previous sections, we are now able to test global linear hypotheses of the mean functions of different groups in a general heteroscedastic functional setting.
However, in many applications one is not just interested in the existence of a significant result but also in which specific linear combinations are causing the significance. For example, in one-way layouts, hypotheses about pairwise comparisons often contain the main research questions. To also tackle such situations, we construct general multiple tests for heteroscedastic functional data covering multiple contrast tests in this section. Multiple tests provide coherent local and global test decisions. 
Thus, multiple tests offer more information than global tests. Furthermore, multiple tests do not have necessarily less power than global tests in ANOVA \cite{konietschke2013} settings as well as in FANOVA settings as we will see in Sections~\ref{sec:Simu}-\ref{sec:Data}.

Usually, a stepwise procedure as described in \cite[p.~63]{konietschke2013} can be used to detect which specific linear combinations cause the significance. However, this procedure may lead to incompatible test decisions. In detail, it would be possible that the global null hypothesis is rejected but none of the individual null hypotheses (\textit{non-consonant}) or, the other way round, the global null hypothesis is not rejected but at least one of the individual null hypotheses is rejected (\textit{incoherent}). We aim to reject the global null hypothesis whenever at least one of the individual null hypotheses is rejected. Thus, the resulting local and global test decisions would be consonant and coherent as defined in \cite[pp.~229-231]{gabriel_1969}. Another usual solution for the multiple testing problem is the adjustment of the significance level for example by using the Bonferroni correction. However, this procedure may yield to lower power \cite[pp.~63-64]{konietschke2013}. This will be also shown in our hypotheses testing framework (Sections~\ref{sec:Simu}-\ref{sec:Data}).

In the following, we interpret $\mathbf{H}$ as a block matrix, that is $\mathbf{H}=\left[\mathbf{H}_1^{\top},\dots,\mathbf{H}_R^{\top}\right]^{\top}$ for matrices  $\mathbf{H}_{\ell} \in \R^{ r_{\ell} \times k }$, and \mbox{$\mathbf{c} = (\mathbf{c}_{\ell})_{\ell\in\{1,\dots,R\}} : \T\to\R^r$} as function with $\mathbf{c}_{\ell}:\T \to \R^{r_{\ell}}$ for all $\ell\in\{1,\dots,R\}$, where $R, r_{\ell}\in\{1,\dots,r\}$ such that $\sum_{l=1}^R r_{\ell} = r$. 
The main idea of multiple tests is to split up the global null hypothesis in (\ref{eq:global}) with 
$\mathbf{H}= [\mathbf{H}_1^{\top}, \dots, \mathbf{H}_R^{\top}]^{\top}$ and function $\mathbf{c} = (\mathbf{c}_{\ell})_{\ell\in\{1,\dots,R\}} $ into $R$ single 
tests with 
hypothesis matrices $\mathbf{H}_1, \dots, \mathbf{H}_R$ and functions $\mathbf{c}_1,\dots,\mathbf{c}_R$, respectively. Typically, the 
matrix $\mathbf{H}$ is partitioned in $r$ 
row vectors $\mathbf{H}_1, \dots, \mathbf{H}_r \in\R^{1\times k}$ but we do not want to restrict to this case.
This leads to the multiple testing problem 
\begin{align}
\label{eq:multiple}
	\mathcal H_{0,{\ell}} : \; \mathbf{H}_{\ell} \boldsymbol{\eta}(t) = \mathbf{c}_{\ell}(t) \;\text{ for all }t\in\T, \quad 
	\quad 
	\text{for }\ell\in \{1,\ldots, R\}.
\end{align}
This general formulation of the multiple testing problem covers the post-hoc testing problem and includes for example many-to-one comparisons 
$\mathcal H_{0,{\ell}} : \: \eta_{1}(t) = \eta_{\ell +1}(t)$ for all $t\in\T$, for $\ell\in \{1,\ldots, k-1\},$ by choosing $\mathbf{H}_{\ell} = \left(-1,\mathbf{e}_{\ell+1}\right)$ for all $\ell\in\{1,\dots,k-1\}$. Here, $\mathbf{e}_{i}\in\R^{k-1}$ denotes the $i$th unit vector. 
In this case, every individual hypothesis in (\ref{eq:multiple}) corresponds to a single contrast of the Dunnett-type contrast matrix $\mathbf{H}$ \cite{dunnett_1955}. Pairwise comparisons can be considered by choosing $\mathbf{H}_{1},\dots,\mathbf{H}_{{k(k-1)}/{2}}$ as rows of the Tukey-type contrast matrix \cite{tukey}. However, the testing problem (\ref{eq:multiple}) is formulated more general since we do not restrict to the partitioning by rows. Choosing, e.g., $\mathbf{H}_1 = \mathbf{H}_A, \mathbf{H}_2 = \mathbf{H}_B$ and $\mathbf{H}_3 = \mathbf{H}_{AB}$ in the two-way design as introduced in Section~\ref{sec:setup} allows to test for a main effect of the factors A and B as well as on an interaction effect simultaneously.

To construct a flexible procedure that can handle the general multiple testing problem (\ref{eq:multiple}), we study the joint distribution of different GPH statistics in more detail. In fact, the following theorem states joint asymptotic null distribution of the GPH statistic vector.
\begin{theo}
\label{Multiasy_GPH}
Let $\mathbf{z} = (\mathbf{z}_{\ell})_{\ell\in\{1,\dots,R\}} \sim GP_r(\mathbf{0}_r,\mathbf{H}\boldsymbol{\Sigma}\mathbf{H}^{\top})$.
Under (A1)-(A4), (A6) and the null hypothesis in (\ref{eq:global}), we have
\begin{align*}
    \left(T_n(\mathbf{H}_{\ell}, \mathbf{c}_{\ell}) \right)_{\ell\in\{1,\dots,R\}} \xrightarrow{d} \left( \int_{\T} \mathbf{z}_{\ell}^{\top}(t) \left(\mathbf{H}_{\ell}  \boldsymbol\Sigma(t,t)\mathbf{H}_{\ell}^{\top} \right)^+ \mathbf{z}_{\ell}(t) \;\mathrm{d}t   \right)_{\ell\in\{1,\dots,R\}}
\end{align*} 
as $ n \to\infty$.
\end{theo}

For approximating the limiting distribution, we can again use the parametric bootstrap approach as described in Section~\ref{sec:Bootstrap}. The following theorem ensures the parametric bootstrap consistency for the multiple testing problem.

\begin{theo}
\label{MultiParaBS}
Let $\mathbf{z} = (\mathbf{z}_{\ell})_{\ell\in\{1,\dots,R\}} \sim GP_r(\mathbf{0}_r,\mathbf{H}\boldsymbol{\Sigma}\mathbf{H}^{\top})$.
Under (A1)-(A6), 
it holds
\begin{align*}
    \left(T_n^{\mathcal{P}}(\mathbf{H}_{\ell}) \right)_{\ell\in\{1,\dots,R\}} \xrightarrow{d^*} \left( \int_{\T} \mathbf{z}^{\top}_{\ell}(t) \left(\mathbf{H}_{\ell}  \boldsymbol\Sigma(t,t)\mathbf{H}_{\ell}^{\top} \right)^+ \mathbf{z}_{\ell}(t) \;\mathrm{d}t   \right)_{\ell\in\{1,\dots,R\}}
\end{align*}  
as $ n \to\infty$. 
\end{theo}

A first naive approach for constructing a test regarding the multiple testing problem (\ref{eq:multiple}) is to consider the maximum of the GPH statistics based on the matrices $\mathbf{H}_1,...,\mathbf{H}_R$, that is
$\max\limits_{\ell\in\{ 1, \ldots, R\}} T_{n}(\mathbf{H}_{\ell}, \mathbf{c}_{\ell})$. By Theorem~\ref{Multiasy_GPH}, the test statistics $T_{n}(\mathbf{H}_1, \mathbf{c}_1), \dots, T_{n}(\mathbf{H}_R, \mathbf{c}_R)$ do not have the same asymptotic null distribution in general. Hence, this approach does not treat every linear combination in the same way. A typical solution for this problem is the studentization of the test statistics as in \cite[p.~647]{bretz2001}. 
However, we do not have Gaussian limits in Theorem~\ref{Multiasy_GPH} and, thus, a simple studentization does not really solve the problem of different limiting distributions.

Therefore, we adopt the idea for the construction of simultaneous confidence bands proposed by \citeauthor{buhlmann1998sieve} \cite[pp.~60-61]{buhlmann1998sieve} for a completely different problem. 
To this end, let $T_{n}^{1,\mathcal{P}}(\mathbf{H}_{\ell}),\dots,T_{n}^{B,\mathcal{P}}(\mathbf{H}_{\ell})$ denote $B$ parametric bootstrap counterparts of the GPH statistic based on the matrix $\mathbf{H}_{\ell}$, see Section~\ref{sec:Bootstrap}, where $B\in\mathbb{N}$ is sufficiently large and $\ell\in\{1,\dots,R\}$.
For each $b\in\{1,\dots,B\}$, the same bootstrap sample is used for calculating the resampling counterparts $T_{n}^{b,\mathcal{P}}(\mathbf{H}_1),\ldots,T_{n}^{b,\mathcal{P}}(\mathbf{H}_R)$ for the different matrices. 
Let $q_{\ell,\beta}^{\mathcal{P}}$ denote the empirical $(1-\beta)$-quantile of $T_{n}^{1,\mathcal{P}}(\mathbf{H}_{\ell}),\ldots,T_{n}^{B,\mathcal{P}}(\mathbf{H}_{\ell})$ for all $\ell\in\{1,\ldots,R\}$ and $\beta\in [0,1)$.
Furthermore, let 
\begin{align*}
 	\widetilde{\beta} := \max\left\{ \beta \in \left\{ 0, \frac{1}{B}, \dots, \frac{B-1}{B} \right\} \:\mid\:	\frac{1}{B} \sum_{b=1}^B \mathbbm{1}\left\{\exists\: \ell\in\{1,\ldots,R \}: \; T_{n}^{b,\mathcal{P}}(\mathbf{H}_{\ell}) > q_{\ell,\beta}^{\mathcal{P}} \right\} \leq \alpha \right\}
 	\end{align*} 
be the largest $\beta \in \{ 0, {1}/{B}, \dots, {(B-1)}/{B} \}$ such that the approximated family-wise type I error rate (FWER) is bounded by the significance level $\alpha$, where 
$\mathbbm{1}$ denotes the indicator function. It should be noted that it is enough to maximize over the set $\{ 0, {1}/{B}, \dots, {(B-1)}/{B} \}$ since the empirical quantiles $ q_{\ell,\beta}^{\mathcal{P}}$ can only take $B$ different values  
for each $\ell\in\{1,\dots,R\}$.
In addition, we only need to search for the desired $\widetilde{\beta}$ within the interval $$\left[\frac{1}{B}\left\lfloor\frac{B\alpha}{R}\right\rfloor, \alpha\right],$$ where $\lfloor . \rfloor$ represents the floor function.
This leads to the following decision rules:
\begin{itemize}
	\item For each $\ell\in\{1,\dots,R\}$, we reject $\mathcal H_{0,{\ell}}$ in (\ref{eq:multiple}) if and only if $T_{n}(\mathbf{H}_{\ell}, \mathbf{c}_{\ell})>q_{\ell,\widetilde{\beta}}^{\mathcal{P}}$ or, equivalently, ${T_{n}(\mathbf{H}_{\ell}, \mathbf{c}_{\ell})}/{q_{\ell,\widetilde{\beta}}^{\mathcal{P}}}>1$. Here, we set ${0}/{0} := 1$.
	\item We reject the global null hypothesis $\mathcal H_0$ in (\ref{eq:global}) whenever at least one of the hypotheses $\mathcal H_{0,1},\dots,\mathcal H_{0,R}$ is rejected. Hence, we reject the global null hypothesis $\mathcal H_0$ in (\ref{eq:global}) if and only if
	\begin{align*}
		\max_{\ell\in\{ 1, \ldots, R\}} \frac{T_{n}(\mathbf{H}_{\ell}, \mathbf{c}_{\ell})}{q_{\ell,\widetilde{\beta}}^{\mathcal{P}}} > 1.
	\end{align*}
\end{itemize}
Each test statistic $T_{n}(\mathbf{H}_{\ell}, \mathbf{c}_{\ell}), \ell\in\{1,\dots,R\},$ is treated in the same way and has the same impact since we use the same $\widetilde{\beta}$ for each linear combination. Moreover, the definition of $\widetilde{\beta}$ ensures that the level of significance for the global test and the FWER for the multiple testing problem is controlled asymptotically. Nevertheless, this multiple testing procedure also performs well for finite samples, as we will see in the next sections.

    Analogously to the proposed multiple testing procedure, we can construct simultaneous asymptotic confidence regions for the linear combinations $\mathbf{H}_1 \boldsymbol{\eta},...,\mathbf{H}_L\boldsymbol{\eta}$, that are
    $$ CR_{\ell} := \left\{ \boldsymbol{\xi}\in\mathcal{L}_2(\T) \mid T_n(\mathbf{H}_{\ell}, \boldsymbol{\xi})  \leq q_{\ell,\widetilde{\beta}}^{\mathcal{P}}\right\}, \quad \ell \in\{1,...,L\}.$$

\section{Simulation Studies}
\label{sec:Simu}
The last sections gurantee the asymptotic validity of the proposed methods. In this section, the finite sample performance of the proposed methods is investigated in extensive Monte Carlo simulation studies. We take into account the type I error control and power of the new and competing statistical tests.

\subsection{Setup}
The simulation setup is based on the simulations in \cite{Simu}.
We simulated $k=4$ samples with sample sizes $(n_1,n_2,n_3,n_4) = K\cdot (15,20,25,30)$, where $K\in\{1,2,4\}$ under the null hypothesis settings and $K=1$ under the alternative hypothesis settings, by
\begin{align*}
    v_{ij}(t) = h(t)\lambda_i\sum\limits_{q=1}^{10} \left(\sqrt{\frac{2}{q}}\sin(\pi qt)Y_{ijq} + \sqrt{\frac{1}{q}} \cos(\pi qt)Z_{ijq}\right)
    , \; t\in \T = [0,1]
\end{align*}
for all $j\in\{1,\dots,n_i\}, i\in\{1,2,3,4\}$.
The simulated functions are evaluated at $J=50$ equidistant points.
Here, the random variables 
$Y_{ijq},Z_{ijq}, j\in\{1,\dots,n_i\}, i\in\{1,2,3,4\}, q\in\{1,\dots,10\}$ are generated independently using standardized normal, $t_5$- and $\chi^2_5$-distributed random variables.
The factors $\lambda_i$ are determined by 1 for the homoscedastic case, by $\lambda_i = 0.75 + 0.25i$ for a positive pairing (i.e., the variability increases with sample size) and by $\lambda_i = 2 -  0.25i$ for a negative pairing (i.s., the variability decreases with sample size) for all $i\in\{1,2,3,4\}$. Furthermore, a scaling function $h$ is multiplied, where $h(t)=1$ for all $t\in[0,1]$ for a scenario without scaling function and $h(t) = {1}/{(t+1/J)}$ for all $t\in[0,1]$ for a scenario with scaling function. Note that using a scaling function should not have an impact on our proposed methodologies since the GPH statistic is scale-invariant, see Remark~\ref{invariance}.

For constructing the hypotheses, we chose the Dunnett- and Tukey-type contrast matrices as $\mathbf{H}$ and the constant zero vector function as $\mathbf{c}$. Here, the rows of the contrast matrix are considered as single contrast matrices for the multiple testing problem. Under the null hypothesis, we specified $\eta_1(t)=\eta_2(t)=\eta_3(t)=\eta_4(t)=0$ for all $t\in[0,1]$, i.e., $x_{ij} = v_{ij}$ for all $j\in\{1,\dots,n_i\}, i\in\{1,2,3,4\}.$
Under the alternative hypothesis A1, we consider the mean functions 
$\mu_1(t) = \mu_2(t) = \mu_3(t) = 0$ and $\mu_4(t)=2{(t-1)}/{(J-1)}$
for all $t\in[0,1]$. More alternatives can be found in Section~S4 of the supplement.

The global level of significance is set to $\alpha = 5\%$. For the parametric bootstrap, we used $B=1000$ bootstrap samples. All in all, 2000 repetitions have been simulated for each setting. The experiments were conducted in the R environment \cite{Rcore}.

Since multiple tests for functional data are not tackled in the literature up until now, 
we compared our multiple test (mGPH) with the following global tests by using the Bonferroni correction:
\begin{itemize}
        \item the Fmax- and GPF-test by \citeauthor{SmaZha} \cite{SmaZha} with nonparametric bootstrap approach,
        \item the bootstrap $L^2$-norm-based (L2b) and F-type (Fb) test by \citeauthor{Zhang2013}~\cite{Zhang2013},
        \item the projection test (CAFB) by \citeauthor{febrero}~\cite{febrero},
        \item the parametric bootstrap globalizing point-wise Hotelling's $T^2$-test (GPH) for the global testing problem, see Section~\ref{sec:testingProc}.
\end{itemize}
Here, the Bonferroni correction is used since it is directly related to the construction of confidence regions in contrast to a stepwise procedure as, e.g., the Holm correction. Moreover, we want to emphasize that the Fmax- and GPF-tests are procedures for an homoscedastic FANOVA setting. Furthermore, the L2b-, Fb- and CAFB-test statistics are not scale-invariant in the sense of \cite{GuoEtAl2019}.
A detailed description of each of these methods can be found in Section~S5 in the supplement.

\subsection{Simulation Results}
The most important findings from simulation results are summarized in Figures~\ref{fig_1_fwer}-\ref{fig_6}, while all results are shown in the supplement. For comparisons regarding the type-I error rate of the tests, we focus on the empirical FWER. Under the alternatives, the empirical power is compared, i.e., the rejection rate of the false null hypothesis.


\begin{figure}[!ht]
\centering
\includegraphics[width=
0.95\textwidth,height=0.9\textheight]{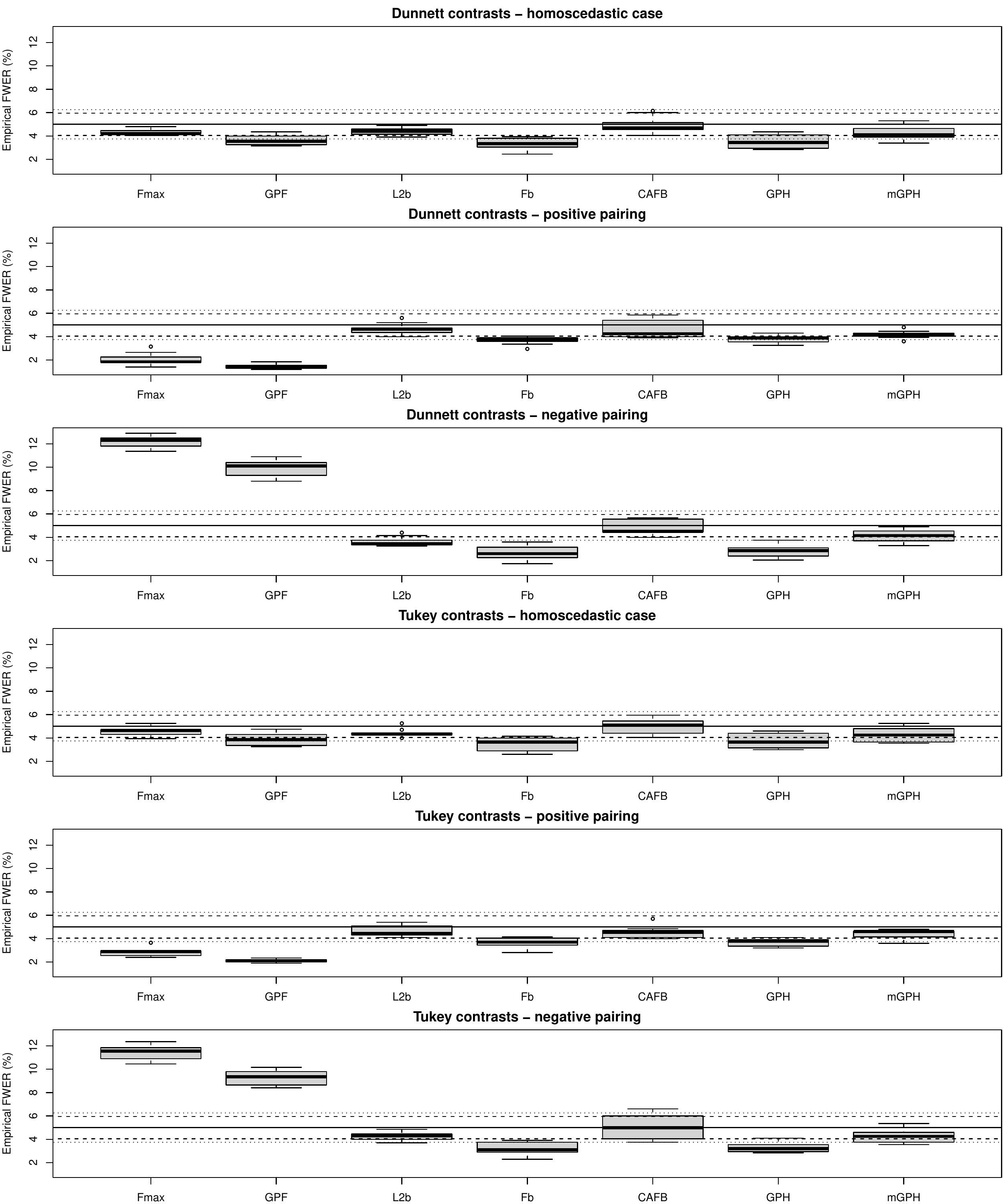}
\caption{Box-and-whisker plots for the empirical FWER (as percentages) of all tests obtained for the Dunnett and Tukey constrasts for homoscedastic and heteroscedastic cases}
\label{fig_1_fwer}
\end{figure}


\begin{figure}[!hb]
\centering
\includegraphics[width=
0.95\textwidth,height=0.6\textheight]{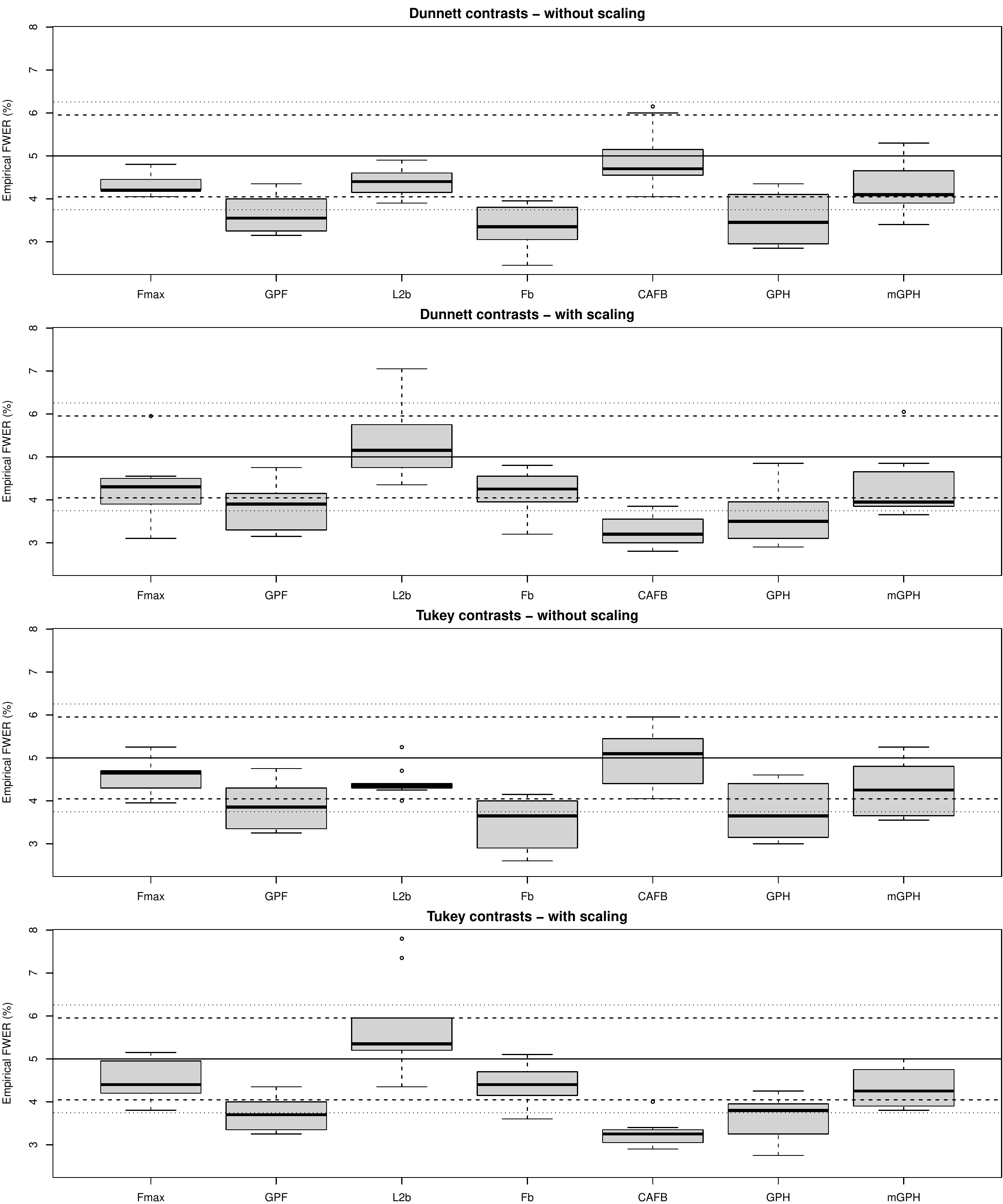}
\caption{Box-and-whisker plots for the empirical FWER (as percentages) of all tests obtained for the Dunnett and Tukey constrasts with and without scaling function and homoscedastic case} 
\label{fig_2_fwer}
\end{figure}



\begin{figure}[!ht]
\centering
\includegraphics[width=
0.95\textwidth,height=0.9\textheight]{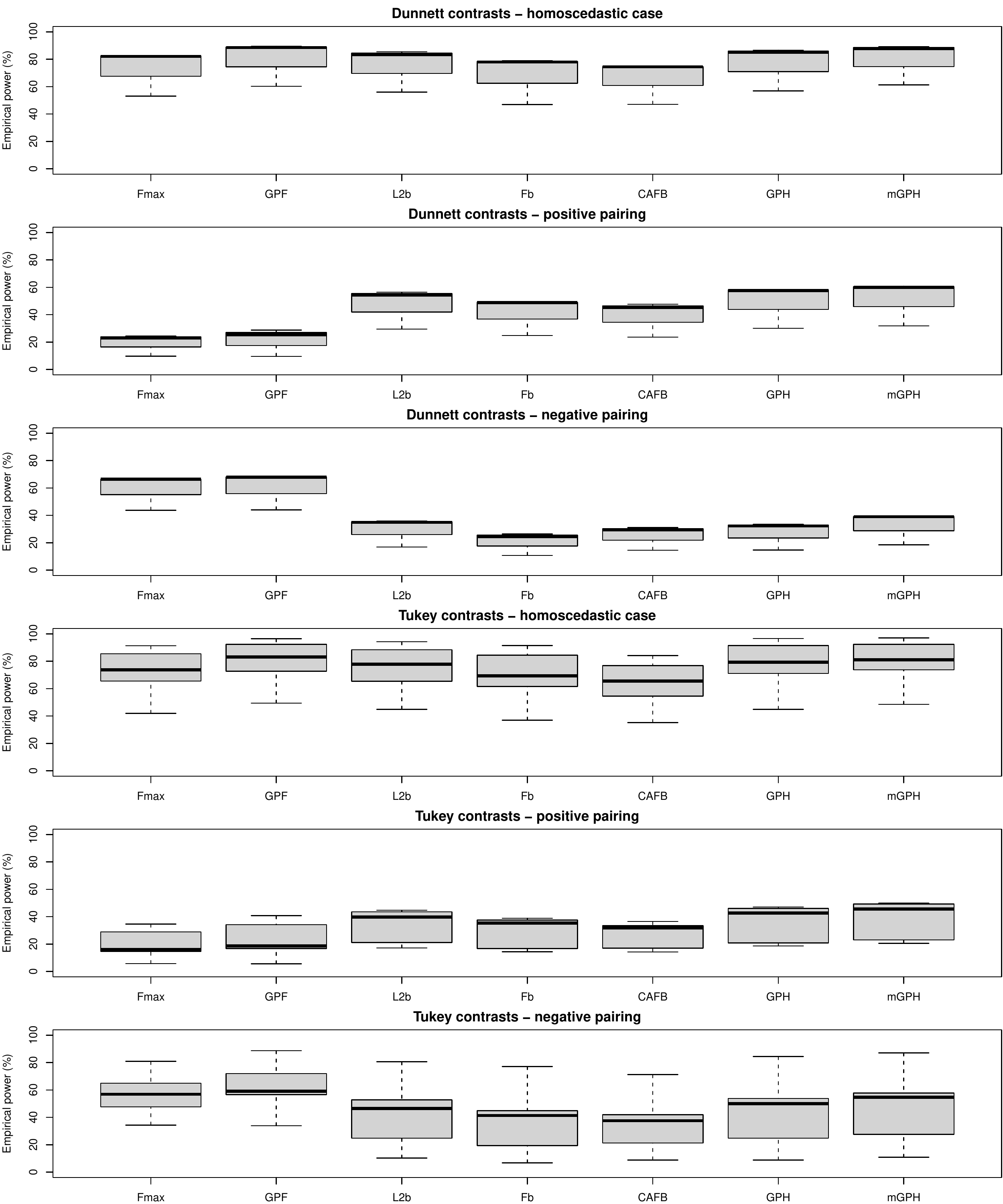}
\caption{Box-and-whisker plots for the empirical powers (as percentages) of all tests obtained under alternative A1 for the Dunnett and Tukey constrasts and homoscedastic and heteroscedastic cases}
\label{fig_5}
\end{figure}

\begin{figure}[!ht]
\centering
\includegraphics[width=
0.95\textwidth,height=0.9\textheight]{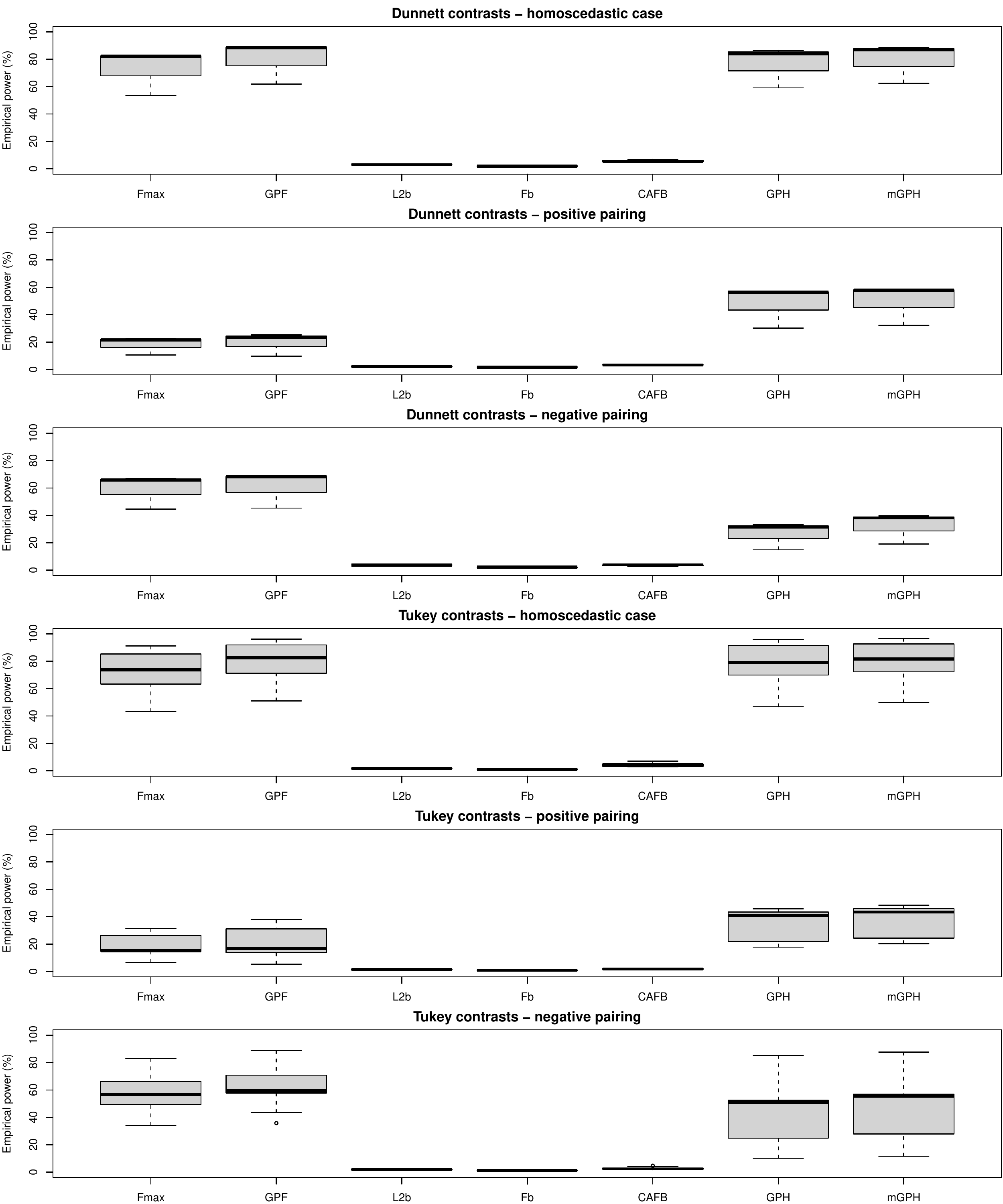}
\caption{Box-and-whisker plots for the empirical powers (as percentages) of all tests obtained under alternative A1 for the Dunnett and Tukey constrasts with and without scaling function and homoscedastic and heteroscedastic cases}
\label{fig_6}
\end{figure}

The simulation results given in the supplement suggest that the distribution of $Y_{ijq}, Z_{ijq}$, $j\in\{1,\dots,n_i\}$, $i\in\{1,2,3,4\}$, $q\in\{1,\dots,10\}$ does not significantly affect the properties of the tests. However, varying the covariance settings (homoscedastic, positive pairing, negative pairing) as well as the scaling of the data may have a notable impact. Thus, in the following, we describe the properties of the tests considered in detail, emphasizing their advantages and disadvantages. Finally, we present the recommendation for practical use.


\paragraph{Fmax- and GPF-tests} 
The Fmax-test and GPF-test perform slightly conservatively for homoscedastic functional data, as shown in Figure~\ref{fig_1_fwer}. In the case of positive pairing, both tests have a conservative character, which results in the loss of power. However, in the scenario of negative pairing, the empirical FWER exceeds the significant level of 5\%, indicating that the Fmax- and GPF-tests do not control the type-I error level in heteroscedastic scenarios. Thus, they are too liberal in this case, which is unacceptable. The scale-invariance in the sense of \cite{GuoEtAl2019} ensures that the simulation results are very similar for the Fmax- and GPF-tests in both with and without scaling function cases, see Figure~\ref{fig_2_fwer} and \ref{fig_6}. In Figure~\ref{fig_5} and \ref{fig_6}, it is observable that the empirical power of both tests is very similar under alternative A1. For the homoscedastic case, the empirical power of the Fmax- and GPF-tests is naturally quite high. In the case of negative pairing, this is also the case, but the power is not fair as the tests are too liberal. However, in the case of positive pairing, our methods outperform the Fmax- and GPF-tests in terms of empirical power, which are less powerful due to the conservative character of the latter. 


\paragraph{L2b- and Fb-tests}
The empirical FWER of the L2b- and Fb-tests are mainly less than 5\% across all covariance scenarios, as shown in Figure~\ref{fig_1_fwer}, particularly in the case of negative pairing. Thus, the L2b- and Fb-tests perform conservatively. Interestingly, the Fb-test appears to have even lower empirical FWER than the L2b-test. The L2b- and Fb-tests are not scale-invariant, as demonstrated in Figure~\ref{fig_2_fwer}, where scaling the data leads to a tendency towards larger empirical FWER values for the homoscedastic case. The effect of scaling is even more pronounced for the empirical power under alternative A1, as shown in Figure~\ref {fig_6}. Here, scaling the data results in an empirical power of less than 4\% for all scenarios. Thus, the L2b- and Fb-tests are unlikely to detect alternatives of type A1 when the data is scaled in a certain way. A loss of power after scaling the data can also be noted under other alternatives, as shown in the supplement.  

\paragraph{CAFB-test}
Regarding type-I error level control, the CAFB-test appears to be relatively accurate, as shown in Figure~\ref{fig_1_fwer}. The empirical FWER of the CAFB-test fluctuates close to 5\% under all covariance scenarios. However, scaling the data affects the empirical FWER of the CAFB-test, as illustrated in Figure~\ref{fig_2_fwer}. Specifically, for the homoscedastic case, the CAFB-test tends to be too conservative when the data is scaled, while it performs quite accurately without scaling. Furthermore, the empirical power of the CAFB-test is not particularly high compared to its competitors, as seen in Figures~\ref{fig_5} and \ref{fig_6}. In addition, a similar effect to that observed for the L2b- and Fb-tests can be seen in Figure~\ref{fig_6} and in the supplement, where scaling leads to a notable loss of power for the CAFB-test, particularly under alternative~A1.

\paragraph{GPH- and mGPH-tests}
Finally, the performance of our proposed tests has to be considered. The GPH- and mGPH-tests are slightly conservative for all covariance scenarios, as indicated by the empirical FWERs being slightly less than 5\% (Figure~\ref{fig_1_fwer}). The mGPH-test tends to have a higher and, thus, more accurate empirical FWER than the GPH-test with Bonferroni correction.  
Due to the scale-invariance, scaling the data should not have an impact on the test result. This can also be checked in Figure~\ref{fig_2_fwer}: For the homoscedastic case, we can only notice small changes in the empirical FWERs for the GPH- and mGPH-test, respectively. Regarding the empirical power in Figure~\ref{fig_5} and \ref{fig_6}, it is observable that the power of both tests is quite high and does not change by scaling the data. The empirical power of the mGPH-test seems to be slightly higher than that of the GPH-test and has the highest empirical power of all methods in many scenarios, see Sections~S4.3 and S4.4 in the supplement for details. Thus, more power can be gained by using the proposed procedure from Section~\ref{sec:multiple} compared to simply applying the Bonferroni correction on our global test from Section~\ref{sec:testingProc}. Furthermore, our tests outperform the L2b-, Fb- and CAFB-tests 
for the scenarios with scaled data and at least do not perform notably worse without scaling considering the empirical power, as shown in Figures~\ref{fig_5} and \ref{fig_6}. 
In terms of empirical power, the Fmax- and the GPF-tests are the strongest competitors to our procedures in the case of homoscedasticity. However, for the heteroscedastic positive pairing scenario, the GPH and mGPH tests are more powerful. In the negative pairing the Fmax- and GPF-tests are too liberal, i.e., a high empirical power is neither meaningful nor convincing for the Fmax- and GPF-tests.

\paragraph{Recommentation} To sum up, the new GPH and mGPH tests are the only procedures, which control the type I error well and have sensible power in all scenarios. This covers bot, homoscedastic as well as heteroscedastic cases. They are also based on a scale-invariant test statistic, which results in not affecting their results by scaling the data. The mGPH test is recommended for practical use as it is at least slightly more powerful than the GPH test based on the Bonferroni correction. 


\section{Data Examples}
\label{sec:Data}
In this section, the proposed methods are illustrated by analyzing two real data examples. Additional simulations inspired by the data examples can be found in Section~S3 in the supplement and are summarized at the end of each of the following two subsections.

\subsection{Chlorine Concentration Data}
Our data analysis begins with the \textit{ChlorineConcentration} data set, obtained from \cite{UCRArchive}. The data set was produced by the simulation tool EPANET, which models the hydraulic and water quality behavior of water distribution piping systems, allowing for the tracking of water levels and pressures in tanks, water flow in pipes, and the concentration of chemical species throughout a given network over a simulated period of time. Here, simulated Chlorine concentration levels were measured at 166 pipe junctions over a period of 15 days, with one measurement taken every five minutes. This yields a total of 4310 measurements per observation. The final data set contains functional observations measured in 166 design time points (pipe junctions). Moreover, it is divided into three classes (groups). For illustrative purposes, we consider three samples of 25 functional observations each taken from the groups in the training data set. 


\begin{figure}[!b]
\centering
\includegraphics[width=
0.99\textwidth,height=0.25\textheight]{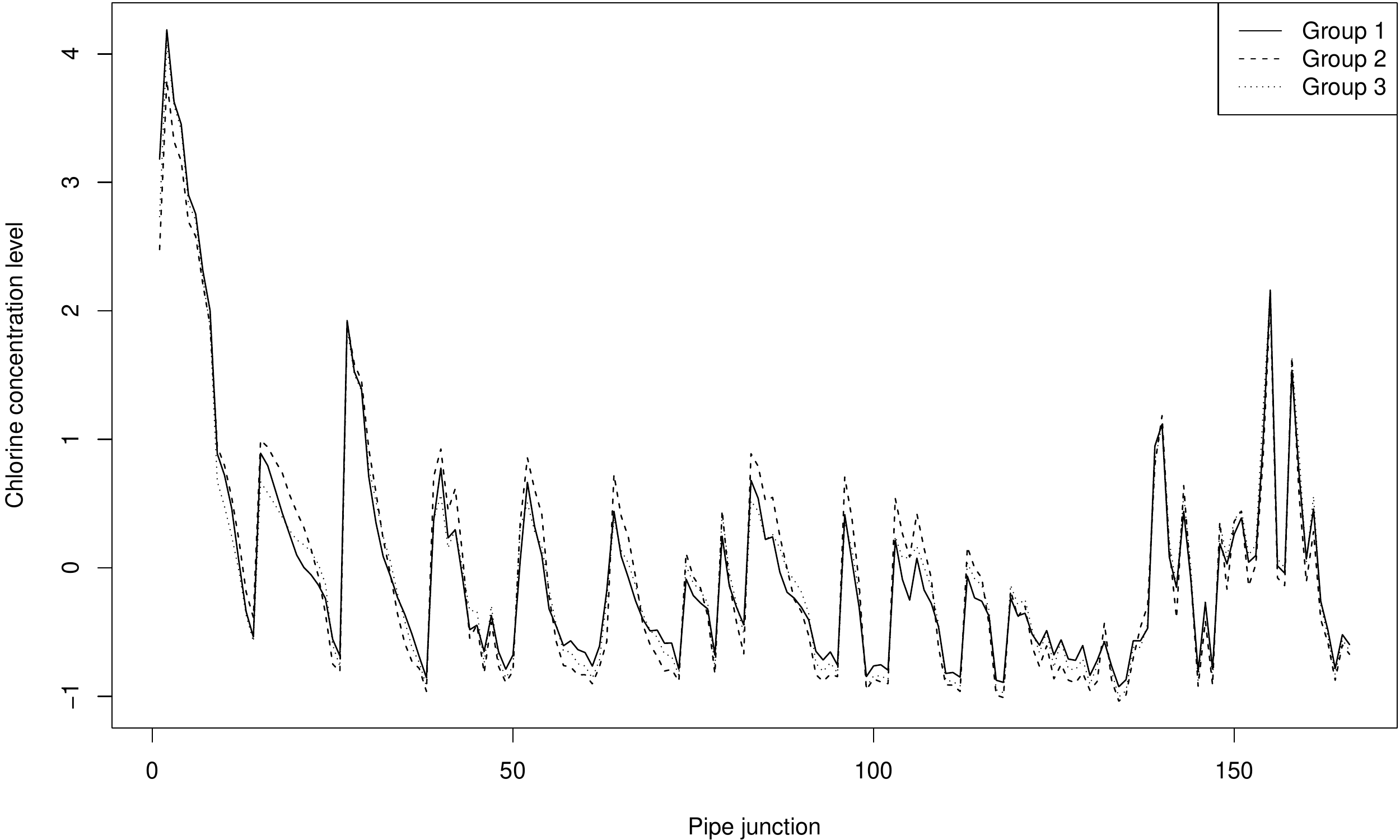}
\caption{Sample mean functions for the three groups in the Chlorine concentration data set}
\label{fig_cc}
\end{figure}

We are interested in verifying the possible differences in mean functions corresponding to groups. The sample mean functions for these three groups are presented in Figure~\ref{fig_cc}. In Figure~S1 in the supplement, 
the empirical covariance functions of the different groups are presented. By the naked eye, we can not detect any differences between the empirical covariance functions. Four standard tests on homoscedasticity \cite{Guoetal2018, GuoEtAl2019} can not reject the null hypothesis at level $5\%$, see Table~S1 in the supplement. Thus, it seems that we have a homoscedastic scenario in this example.

\begin{table}[!b]
    \centering
    \caption{Adjusted p-values in \% for the Chlorine Concentration data}
    \begin{tabular}{c|rrrrrrr}
        Hypothesis & Fmax & GPF & L2b & Fb & CAFB & GPH & mGPH  \\
        \hline
        $\mathcal H_{0,1}: \eta_1 \equiv \eta_2$ & {12.9} & {4.8} & {13.5}& {16.2}& {7.8}& {1.5}& {1.7}\\
        $\mathcal H_{0,2}: \eta_1 \equiv \eta_3$ & {100.0} & {90.0} & {100.0}& {100.0}& {100.0}& {83.1}& {53.4}\\
        $\mathcal H_{0,3}: \eta_2 \equiv \eta_3$ & {90.6} & {37.2} & {31.2}& {35.1}& 73.6& 38.1 & {28.7}  
    \end{tabular}
    \label{tab:tab1}
\end{table}

For comparing the different groups, we perform all-pair comparisons by using the Tukey-type contrast matrix with the level of significance $\alpha = 5\%$. The p-values of the global tests are adjusted by the Bonferroni correction. The resulting adjusted p-values given in Table~\ref{tab:tab1} 
show that only the GPF-, GPH- and mGPH-tests detect a significant difference regarding the mean functions between the first two groups. All other tests do not reject the global hypothesis that there is no difference between the mean functions.

To have a closer look at the results, we have conducted the simulation studies inspired by the Chlorine concentration data set, which results are presented in Section S3.3 in the supplement. The simulation results indicate that the power of the L2b-, Fb-, and CAFB-tests for the first hypothesis is less than 55\%. This is much smaller than the power of the other tests. Among the three tests that reject $\mathcal{H}_{0,1}$, the mGPH test is the most powerful. Interestingly, the Fmax-test has the largest empirical power with 70.2\% for the first hypothesis in the simulation but, however, can still not reject that the first two groups have the same mean function.

\subsection{Electrocardiogram Data}
The second data set, originally published in \cite{data}, is a 20-hour long electrocardiogram (ECG) obtained from Physionet. The data set underwent pre-processing in \cite{data2} consisting of two steps. Firstly, each heartbeat was extracted from the ECG and, secondly, the heartbeat lengths were adjusted using interpolation. Then, 5000 heartbeats were randomly chosen. The patient has severe congestive heart failure and automated annotation was used to obtain class values. The resulting groups are the following:
Normal, R-on-T Premature Ventricular Contraction, Premature Ventricular Contraction, Supraventricular premature or ectopic beat, and Unclassifiable. The focus of this study is on the first four groups as the last group size is very small. Finally, there are four samples of functional data measured in 141 design time points. The sample sizes are equal to 292, 177, 10, and 19 respectively, which results in an unbalanced design. 

\begin{figure}[!b]
\centering
\includegraphics[width=
0.99\textwidth,height=0.28\textheight]{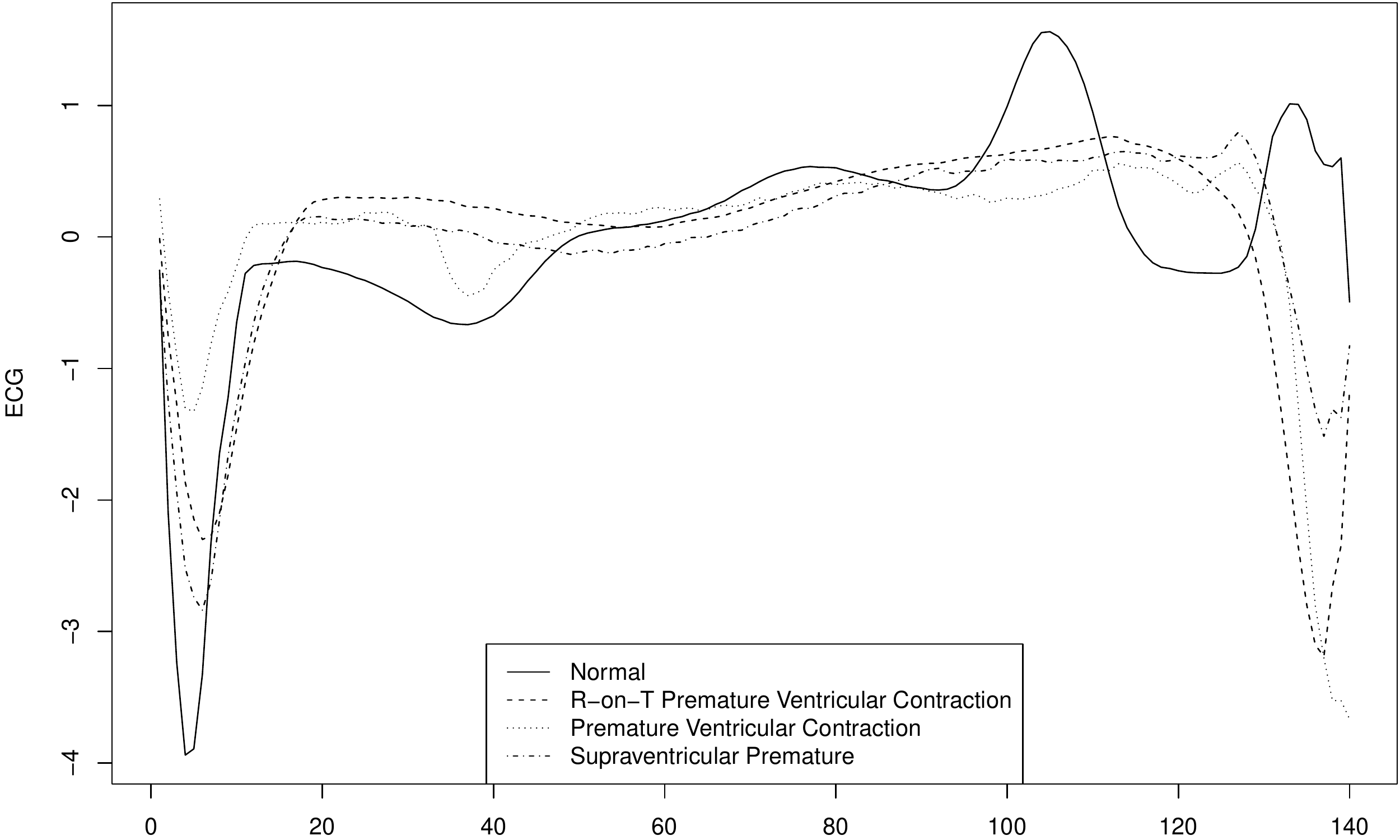}
\caption{Sample mean functions for the four groups in the Electrocardiogram data set}
\label{fig_ecg}
\end{figure}

The sample mean functions for the four samples are presented in Figure~\ref{fig_ecg}. The empirical covariance functions are illustrated in Figure~S2 
in the supplement. The covariance structures seem to differ among the different groups for the ECG data. 
Moreover, all four tests on homoscedasticity \cite{Guoetal2018, GuoEtAl2019} reject the null hypothesis that the covariance functions are equal at level $5\%$ (Table~S1 in the supplement). Thus, everything points in the direction of a heteroscedastic data setup. 
\begin{table}[!b]
    \centering
    \caption{Adjusted p-values in \% for the ECG data}
    \begin{tabular}{c|rrrrrrr}
        Hypothesis & Fmax & GPF & L2b & Fb & CAFB & GPH & mGPH  \\
        \hline
        $\mathcal H_{0,1}: \eta_1 \equiv \eta_2$ & {0.0} & {0.0} & {0.0}& {0.0}& {0.0}& {0.0}& {0.0}\\
        $\mathcal H_{0,2}: \eta_1 \equiv \eta_3$ & {0.0} & {0.0} & {0.0}& {0.0}& {0.0}& {0.0}& {0.0}\\
        $\mathcal H_{0,3}: \eta_1 \equiv \eta_4$ & {0.0} & {0.0} & {0.0}& {0.0}& {0.0}& {0.0}& {0.0}\\
        $\mathcal H_{0,4}: \eta_2 \equiv \eta_3$ & {0.0} & {0.0} & {1.8}& 39.0& {0.0}& 5.4& {3.8}\\
        $\mathcal H_{0,5}: \eta_2 \equiv \eta_4$ & {0.0} & {0.0} & {1.2}& {3.0}& {4.9}& 7.2& {4.9}\\
        $\mathcal H_{0,6}: \eta_3 \equiv \eta_4$ & {0.0} & 7.8 & {3.0}& {4.8}& {0.0}& 15.0& 9.6     
    \end{tabular}
    \label{tab:tab2}
\end{table}

Again, we compare all-pairs by using the Tukey-type contrast matrix at level $\alpha = 5\%$. In Table~\ref{tab:tab2}, 
the adjusted p-values can be found. Here, it is observable that the Fmax-, L2b- and CAFB-tests reject every single hypothesis simultaneously. The GPF- and mGPH-tests do not reject the hypothesis that the third and fourth groups have the same mean function. Furthermore, the Fb-test does not reject the equality of the mean functions of the second and third mean functions. The GPH-test can only reject the first three hypotheses where the mean function of the first group is compared to the other groups, respectively.

Let us look at these results more deeply. Once again we conducted the simulation study. Its results are given in Table S3 in the supplement. First of all, they indicate that the Fmax- and GPF-tests do not control the type-I error level well in this scenario. More precisely, they are usually too liberal, which coincides with the observations given in Section~\ref{sec:Simu} for the heteroscedastic case. The L2b test is also slightly too liberal. Thus, the empirical powers of these three tests are unfairly high. On the other hand, the Fb- and GPH-tests seem to have a conservative character with an empirical FWER less than 3.8\%, which results in smaller power. Thus, we see once again the advantage of the mGPH-test over the GPH-test. For this particular data set, the CAFB testing procedure seems to be the most powerful.

\section{Discussion and Outlook}
\label{sec:Discussion}
The analysis of functional data becomes more and more relevant in quantitative statistics and data science. In fact, as recently stated by \citeauthor{aneiros2019recent} \cite{aneiros2019recent} \textit{'functional data analysis is now one of the most active and relevant fields of investigation in data science.'} The toolbox especially covers  testing procedures for functional data. 
However, all existing methods seem to focus on global null hypothesis testing.
This does not reflect many applications, which are often interested in testing many 
individual hypotheses, e.g. in many-to-one or all-pairs comparisons. 
Hence, multiple tests for functional data are needed. In the present paper we closed this gap by proposing general multiple test procedures covering multiple contrast tests. The methods allow for coherent local and global test decisions and are valid for general heteroscedastic FANOVA designs. 

At their core, the methods are based upon integrated Hotelling-type 
statistics for each local hypotheses that are considerably designed for heteroscedastic settings. Carefully examining their dependency structure  and approximating their joint null distribution via parametric bootstrapping, led to novel multiple test procedures that can be flexibly applied in general FANOVA. The resuting methods 
are 
robust under certain transformations (orthogonal- and scale-invariance) and 
control the level of significance for both large and small to moderate sample sizes as demonstrated by asymptotic theory and finite sample simulations. 

So far, we only developed general multiple tests for univariate functional data. Multivariate functional data occur whenever a vector of multiple random functions is observed for each individual. An examples, where multivariate functional data occur in medicine, can, e.g., be found in \cite{multiEX}. Moreover, another example may be motivated from \cite{bathke2018testing} and leads to comparing functional EEG or SPECT data over time or just before and after treatment. 
Under the homoscedasticity assumption, \citeauthor{Gorecki} \cite{Gorecki}, \citeauthor{Multiv} \cite{Multiv} and \citeauthor{ZHU2022105095} \cite{ZHU2022105095} have already developed global tests for the functional MANOVA problem.  However, 
testing procedures for multivariate heteroscedastic functional data would be more desirable. Additionally, more general (linear) hypotheses and designs for multivariate functional data could be studied since only the one way design (equality of multivariate mean functions) seems to be considered for the multivariate setup in the literature up until now.

Furthermore, we only considered multiple tests for the mean functions of functional data in this paper. However, other effect measures could be of interest in practice. This includes a functional version of the sample median \cite{gervini_2008,baumeister2022quantile}, a functional version of the trimmed mean \cite{fraiman_muniz_2001}
or depth-based estimators \cite{cuevas_febrero_fraiman_2007, lopez-pintado_romo_2007}, to name just a few examples. 
Consequently, general multiple tests for further effect measures than the mean function could be investigated in future research.

\section*{Acknowledgments}
A part of calculations for the simulation study was made at the Pozna\'n Supercomputing and Networking Center (grant no. 382).

\bibliographystyle{myjmva}
\bibliography{sample}


\end{sloppypar}
\end{document}


\captionsetup{width=\textwidth}

\title{\Large \bf
Supplementary materials to\\
General multiple tests for functional data}
\author[1]{Merle Munko}
\author[2]{Marc Ditzhaus}
\author[3]{Markus Pauly}
\author[4]{\L ukasz Smaga}
\author[5]{Jin-Ting Zhang}

\address[1]{Faculty of Mathematics, Otto-von-Guericke University Magdeburg, Germany}
\address[2]{Faculty of Mathematics, Otto-von-Guericke University Magdeburg, Germany}
\address[3]{TU Dortmund University, Dortmund, Germany; Research Center Trustworthy Data Science and Security, UA Ruhr, Germany}
\address[4]{Faculty of Mathematics and Computer Science, Adam Mickiewicz University, Poland}
\address[5]{Faculty of Science, National University of Singapore, Singapore}

\date{Journal: ???}

\maketitle


This supplement contains all proofs and all results of simulation studies of Sections 6 and 7 of the main paper. We also present more details about the real data examples.

\tableofcontents

\section{Auxiliary Theorems}\label{App2}

In this section, some useful results are stated.
\\
In the following, a central limit theorem for triangular arrays of independent distributed random variables in a Hilbert space, which are not necessarily identical distributed, is presented.
We use this theorem in order to proof the consistency of the parametric bootstrap. 

\begin{theo}[\citeauthor{clt_paper} \cite{clt_paper}]
\label{clt}
Let $\mathbb{H}$ be a Hilbert space with inner product $\langle ., .\rangle$, induced norm $||.||$ and countable 
complete orthonormal system $\{{f}_m \mid m\in M\}$ for a countable 
index set~$M$. Furthermore, let $Z_{1,n},...,Z_{n,n}$ be independent random variables for all $n\in\mathbb{N}$ taking values in $\mathbb{H}$ and satisfying $\E\left[||Z_{j,n}||^2\right] < \infty$ for all $j\in\{1,...,n\}, n\in\mathbb{N}.$ Suppose
\begin{enumerate}
    \item $\E\left[\langle Z_{j,n}, f_m \rangle \right] = 0$ for all $j\in\{1,...,n\}, n\in\mathbb{N}, m\in M,$
    \item  $\sum\limits_{j=1}^n \E\left[\langle Z_{j,n}, f_m \rangle \langle Z_{j,n}, f_{\ell} \rangle \right] \to \sigma^2_{ml} $ as $n\to\infty$ for some $\sigma^2_{ml} \geq 0$ for all $m,{\ell}\in M$,
    \item $\sum\limits_{m \in M} \sum\limits_{j=1}^n \Var\left(\langle Z_{j,n}, f_m \rangle  \right) \to \sum\limits_{m \in M} \sigma^2_{mm} < \infty$ as $n\to\infty$,
    \item $\sum_{j=1}^n \E\left[  |\langle Z_{j,n}, f_m \rangle|^2 \mathbbm{1}\{ | \langle Z_{j,n}, f_m \rangle | > \varepsilon \}  \right] \to 0$ as $n\to\infty$ for all $\varepsilon > 0, m\in M.$
\end{enumerate}
Then, $\sum\limits_{j=1}^n Z_{j,n}$ converges in distribution to a centered Gaussian variable $G$ with covariance operator $C$ as $n\to\infty$, where the covariance operator $C$ is characterized by 
\begin{align} \label{eq:characterization}
     \langle Cf_{\ell}, f_m \rangle = \sigma^2_{ml} \quad\text{for all $m,l \in M$}.
\end{align}
\end{theo}

\begin{proof}[\textbf{Proof of Theorem~\ref{clt}}]
This theorem follows immediately from Theorem~1.1 in \cite[p.~266]{clt_paper}. 
Therefore, note that the covariance operator $C_n$ of $S_n := \sum\limits_{j=1}^n Z_{j,n}$ satisfies
\begin{align*}
    \langle C_n f_m, f_{\ell} \rangle &= \E\left[ \langle S_n, f_m  \rangle \langle S_n, f_{\ell} \rangle \right] 
    = \sum_{j=1}^n \E\left[ \langle Z_{j,n}, f_m  \rangle \langle Z_{j,n}, f_{\ell} \rangle \right]
\end{align*}
for all $n\in\mathbb{N}, m,{\ell} \in M$ by the independence and the centeredness of $Z_{1,n},...,Z_{n,n}$.
\end{proof}

\begin{rem} \label{Lyapunov}
The Lindeberg`s condition~(4) in Theorem~\ref{clt} can be replaced by the Lyapunov`s condition: There exists a $\delta > 0$ such that
$$\sum\limits_{j=1}^n \E\left[|\langle Z_{j,n}, f_m \rangle |^{2+\delta} \right] \to 0 \quad\text{as $n\to\infty$ \quad for all $m\in M$.}$$
In order to proof this, we firstly note that $| \langle Z_{j,n}, f_m \rangle | > \varepsilon$ implies ${| \langle Z_{j,n}, f_m \rangle |^{\delta}}/{\varepsilon^{\delta}} > 1$. Hence, it follows
\begin{align*}
    \sum_{j=1}^n \E\left[  |\langle Z_{j,n}, f_m \rangle|^2 \mathbbm{1}\{ | \langle Z_{j,n}, f_m \rangle | > \varepsilon \}  \right]
    &\leq \frac{1}{\varepsilon^{\delta}} \sum_{j=1}^n \E\left[  |\langle Z_{j,n}, f_m \rangle|^{2+\delta} \mathbbm{1}\{ | \langle Z_{j,n}, f_m \rangle | > \varepsilon \}  \right]
    \\&\leq \frac{1}{\varepsilon^{\delta}} \sum_{j=1}^n \E\left[  |\langle Z_{j,n}, f_m \rangle|^{2+\delta}  \right] \to 0
\end{align*}
as $n\to\infty$ for all $\varepsilon > 0, m\in M$.
\end{rem}

The following lemma is a rather technical result about the norm of a Moore-Penrose inverse.
\begin{lemma}\label{MyLemma}
    Let $r,k\in\mathbb{N}$ and $\mathbf{H}\in\R^{r\times k}$. Furthermore, let $\mathbf{A}\in\R^{k\times k}$ be a symmetric positive definite matrix. Then, we have $$|||(\mathbf{H}\mathbf{A}\mathbf{H}^{\top})^+||| \leq |||\mathbf{H}^+|||^2 |||\mathbf{A}^{-1}|||,$$ where here and throughout $|||.|||$ denotes the spectral norm. 
\end{lemma}
\begin{proof}[\textbf{Proof of Lemma~\ref{MyLemma}}]
    Obviously, $\mathbf{H}\mathbf{A}\mathbf{H}^{\top}$ and, therefore, its Moore-Penrose inverse is symmetric and positive semidefinite. Let $\lambda_{1}(\mathbf{B}) \geq \lambda_{2}(\mathbf{B}) \geq ... $ denote the ordered eigenvalues of a matrix $\mathbf{B}$.
    Then, it holds 
    \begin{align*}
        |||(\mathbf{H}\mathbf{A}\mathbf{H}^{\top})^+||| = \lambda_{1}\left( (\mathbf{H}\mathbf{A}\mathbf{H}^{\top})^+ \right).
    \end{align*}
    The nonzero eigenvalues of $(\mathbf{H}\mathbf{A}\mathbf{H}^{\top})^+$ are given as inverses of the nonzero eigenvalues of $(\mathbf{H}\mathbf{A}\mathbf{H}^{\top})$.
    Hence, $\lambda_{1}\left( (\mathbf{H}\mathbf{A}\mathbf{H}^{\top})^+ \right)$ is the inverse of the smallest nonzero eigenvalue of $(\mathbf{H}\mathbf{A}\mathbf{H}^{\top})$. Thus, for $h:=\text{rank}(\mathbf{H}) = \text{rank}(\mathbf{H}\mathbf{A}\mathbf{H}^{\top})$, it holds
    \begin{align*}
        |||(\mathbf{H}\mathbf{A}\mathbf{H}^{\top})^+||| = 1/\lambda_{h}\left( \mathbf{H}\mathbf{A}\mathbf{H}^{\top} \right).
    \end{align*}
    Furthermore, we have
    \begin{align*}
        \lambda_{h}\left( \mathbf{H}\mathbf{A}\mathbf{H}^{\top} \right) &= \min\limits_{\mathbf{x}\in \text{ker}(\mathbf{H}^{\top})^{\perp}\setminus\{\mathbf{0}_r\}} \frac{\mathbf{x}^{\top} \mathbf{H}\mathbf{A}\mathbf{H}^{\top}\mathbf{x}}{\mathbf{x}^{\top} \mathbf{x}}
        \\& \geq \min\limits_{\mathbf{y}\neq \mathbf{0}_k} \frac{\mathbf{y}^{\top} \mathbf{A}\mathbf{y}}{\mathbf{y}^{\top} \mathbf{y}} \min\limits_{\mathbf{x}\in \text{ker}(\mathbf{H}^{\top})^{\perp}\setminus\{\mathbf{0}_r\}} \frac{\mathbf{x}^{\top} \mathbf{H}\mathbf{H}^{\top}\mathbf{x}}{\mathbf{x}^{\top}\mathbf{x}}
        \\& = \lambda_k(\mathbf{A}) \lambda_h(\mathbf{H}\mathbf{H}^{\top}).
    \end{align*}
    Since $|||\mathbf{H}^+|||^2 = 1/\lambda_h(\mathbf{H}\mathbf{H}^{\top})$ and $|||\mathbf{A}^{-1}||| = 1/\lambda_k(\mathbf{A})$, the lemma is proved.
\end{proof}

\section{Proofs}
\label{App3}
In this section, the technical proofs of all stated theorems are presented.\\
Throughout this section,
we consider the Hilbert space $\mathcal{L}_2(\T)$ with inner product $$\langle ., .\rangle : \mathcal{L}_2(\T) \times \mathcal{L}_2(\T) \to \R, \quad \langle {g}, {h}\rangle:= \int_{\T} {g}(t){h}(t) \,\mathrm{ d }t, $$ induced norm $||.||_2$ and countable complete orthonormal system $\{{f}_m \mid m\in\mathbb{N}\}.$

First of all, we state an useful lemma, which gives us the convergence of the covariance function estimators uniformly over $\{(t,t)\in\T^2 \mid t\in\T\}$ in probability.

\begin{lemma}
\label{UniformConvLemma}
Under (A1), (A3) and (A6), we have 
$$  \sup_{t\in\T} \left|\left|\left|\widehat{\gamma}_i(t,t) - \gamma_i(t,t)\right|\right|\right| \xrightarrow{P} 0 \quad\text{as $n_i\to\infty$} $$
for all $i\in\{1,...,k\}.$
\end{lemma}
\begin{proof}[\textbf{Proof of Lemma~\ref{UniformConvLemma}}]
Let $i\in\{1,...,k\}$ be fixed but arbitrary. We aim to prove this statement by using Theorem~2.1 in \cite{newey1991}.
The pointwise convergence
$$ \hat{\gamma}_i(t,t) \xrightarrow{P} {\gamma}_i(t,t) \quad\text{as $n_i\to\infty$}$$
is well known for all $t\in\T$ under (A1), which can be proven by applying the weak law of large numbers. 
Since $\T$ is a compact interval and $\T\ni t\mapsto \gamma_i(t,t)$ is continuous under (A3), it is also uniformly continuous.  Note that uniform convergence and equicontinuity are equivalent for a singleton set.
Hence, the stochastic equicontinuity of the restriction of the covariance function estimator \mbox{$\T\ni t\mapsto \widehat{\gamma}_i(t,t)$} remains to show. For that purpose, let $\varepsilon, \eta > 0$ and $\delta > 0$ such that 
\begin{align}
\label{eq:Mschranke}
    \E_{\mathrm{out}}\left[\sup\limits_{\substack{t,s\in\T, \:|t-s|<\delta}} |||{v}_{i1}(t){v}_{ij}(t)-{v}_{i1}(s){v}_{ij}(s)|||\right] \leq \frac{\eta\varepsilon}{6}
\end{align}
    holds for $j\in\{1,2\}$. By (A6), such a $\delta>0$ exists.
     In the following, we just write $|t-s|<\delta$ in the index of the supremum for the sake of clarity.
    Inserting the definition of the covariance function estimator (3) 
    and applying the triangular inequality yields
    \begin{align*}
        \sup\limits_{|t-s|<\delta} \left|\left|\left| \hat{\gamma}_i(t,t) - \hat{\gamma}_i(s,s) \right|\right|\right| 
        &\leq    \frac{1}{n_i-1}\sum_{j=1}^{n_i} \sup\limits_{|t-s|<\delta} \left| \left| \left|{v}_{ij}^2(t)- {v}_{ij}^2(s)\right|\right|\right|
        \\&\quad + \frac{1}{(n_i-1)n_i}  \sum_{j_1,j_2=1}^{n_i}\sup\limits_{|t-s|<\delta} \left| \left| \left|{v}_{ij_1}(t){v}_{ij_2}(t)  - {v}_{ij_1}(s){v}_{ij_2}(s)\right|\right|\right|.
    \end{align*}
By Markov`s inequality, (A1) and (\ref{eq:Mschranke}), it follows
\begin{align*}
    \P_{\mathrm{out}}\left( \sup\limits_{|t-s|<\delta} \left|\left|\left| \hat{\gamma}_i(t,t) - \hat{\gamma}_i(s,s) \right|\right|\right|  > \varepsilon \right) &\leq \E_{\mathrm{out}}\left[  \sup\limits_{|t-s|<\delta} \left|\left|\left| \hat{\gamma}_i(t,t) - \hat{\gamma}_i(s,s) \right|\right|\right|   \right]/\varepsilon
   \\& \leq  \frac{n_i}{n_i-1}\frac{\eta}{6} +  \frac{n_i^2}{(n_i-1)n_i}\frac{\eta}{6} \to \frac{\eta}{3} < \eta
\end{align*}
 as $n_i \to \infty$.
The lemma is proven by applying Theorem~2.1 in \cite{newey1991}.
\end{proof}

The following lemma ensures that $\mathbf{H}\hat{\boldsymbol\eta}$ is asymptotically Gaussian.
\begin{lemma}
\label{asy_CLT}
Under (A1)-(A4), we have 
\begin{align*}
    \sqrt{n} (\mathbf{H}\hat{\boldsymbol\eta} - \mathbf{c})  \xrightarrow{d}
    \mathbf{z} \sim GP_{r}\left(\mathbf{0}_{r}, \mathbf{H}\boldsymbol\Sigma\mathbf{H}^{\top}\right) \quad\text{as $n\to\infty$}
\end{align*}
 in $\mathcal{L}^{r}_2(\T)$.
\end{lemma}
\begin{proof}[\textbf{Proof of Lemma~\ref{asy_CLT}}]
    Let $i\in\{1,...,k\}$ be arbitrary but fixed.
Then, the subject-effect functions ${v}_{i1},...,{v}_{in_i}$ take values in $\mathcal{L}_2(\T)$ under (A1) and (A2). Furthermore, it holds by Fubini`s theorem
\begin{align*}
    \E\left[ ||{v}_{i1}||_2^2 \right] 
    = \int_{\T}\E\left[  {v}_{i1}^{2}(t) \right] \,\mathrm{ d }t
    = \int_{\T}{\gamma}_i(t,t)\,\mathrm{ d }t.
\end{align*}
 Since $\T \ni t \mapsto {\gamma}_i(t,t)$ is continuous on the compact set $\T$ under (A3), it attains its maximum. In particular, we get 
 \begin{align}
 \label{eq:MEWertExistenz}
     \E\left[ ||{v}_{i1}||_2^2 \right] = \int_{\T}\boldsymbol{\gamma}_i(t,t) \,\mathrm{ d }t < \infty.
 \end{align}
Hence, Theorem~7.5.1 in \cite[p.~474]{lahaRohatgi} provides 
\begin{align*}
    \sqrt{n_i} (\hat{\eta}_i - \eta_i) = \frac{1}{\sqrt{n_i}} \sum\limits_{j=1}^{n_i} {v}_{ij} \xrightarrow{d} GP(0, \gamma_i) \quad \text{as $n\to\infty$}
\end{align*}
in $\mathcal{L}_2(\T)$ under (A1)-(A4). 
By Slutsky`s lemma as in \cite[p.~20]{vaartWellner1996}, it follows
\begin{align} \label{eq:fuerSpaeter}
    \sqrt{n} (\hat{\eta}_i - \eta_i) = \frac{\sqrt{n}}{\sqrt{n_i}}\sqrt{n_i} (\hat{\eta}_i - \eta_i) \xrightarrow{d} GP\left(0, \frac{1}{\tau_i}\gamma_i\right) \quad \text{as $n\to\infty$}
\end{align} in $\mathcal{L}_2(\T)$ under (A4).
Due to the independence of the samples of the different groups and the continuous mapping theorem as in \cite[p.~32]{vaartWellner1996}, the statement of the lemma holds.
\end{proof}

With the two previous lemmas, we are now able to proof Theorem~1.
\begin{proof}[\textbf{Proof of Theorem~1}]
Lemma~\ref{UniformConvLemma} provides that $\widehat{\boldsymbol{\Sigma}}$ converges uniformly over \mbox{$\{(t,t)\in\T^2 \mid t\in\T\}$} in probability to $\boldsymbol{\Sigma}$.
In the following, we also write $\widehat{\boldsymbol{\Sigma}}(t)$ and ${\boldsymbol{\Sigma}}(t)$ instead of $\widehat{\boldsymbol{\Sigma}}(t,t)$ and ${\boldsymbol{\Sigma}}(t,t)$, respectively, for all $t\in\T$.
Furthermore, Lemma~\ref{asy_CLT} ensures the convergence in distribution of $\sqrt{n} (\mathbf{H}\widehat{\boldsymbol\eta} - \mathbf{c})$ to $\mathbf{z}$.\\
Let 
\begin{align*}
    \mathbb{D} := \big\{ \widetilde{\boldsymbol\Sigma}:\T\to\R^{k\times k} \mid &\widetilde{\boldsymbol\Sigma} \text{ is continuous with 
$\widetilde{\boldsymbol\Sigma}(t)$ symmetric and positive semidefinite} \\&\text{for all $t\in\T$
}\big\}
\end{align*} be equipped with the $\mathcal{L}_{\infty}^{k \times k}(\T)$-norm.
Then, ${\boldsymbol\Sigma}\in\mathbb{D}$ holds due to (A3) and (A4). Moreover, by Remark~1~(5),
$\widehat{\boldsymbol{\Sigma}}$ takes values in $\mathbb{D}$ almost surely.
By the continuous mapping theorem, it suffices to show that the map
\begin{align*}
    h: \mathcal{L}_2^r(\T) \times \mathbb{D}
    \ni  (\widetilde{\mathbf{x}}, \widetilde{\boldsymbol\Sigma}) \mapsto \int_{\T} \widetilde{\mathbf{x}}^{\top}(t) \left(\mathbf{H} \widetilde{\boldsymbol\Sigma}(t) \mathbf{H}^{\top} \right)^+ \widetilde{\mathbf{x}}(t) \;\mathrm{d}t \in \R_0^+ \cup \{\infty\}
\end{align*}
is continuous on \mbox{$\mathcal{L}_2^r(\T) \times \{\boldsymbol\Sigma\}$.}
Consequently, let \mbox{$\widetilde{\mathbf{x}}_{\infty},\widetilde{\mathbf{x}}_1, \widetilde{\mathbf{x}}_2,... \in \mathcal{L}_2^r(\T) $} and $ \widetilde{\boldsymbol\Sigma}_1,\widetilde{\boldsymbol\Sigma}_2,... \in \mathbb{D}$ 
be functions with $\widetilde{\mathbf{x}}_n \to \widetilde{\mathbf{x}}_{\infty}$ in $\mathcal{L}_2^r(\T)$ and $\widetilde{\boldsymbol\Sigma}_n \to {\boldsymbol\Sigma}$ in $\mathcal{L}_{\infty}^{k \times k}(\T)$ as $n\to\infty.$
The triangular inequality implies
\begin{align*}
    \left|h(\widetilde{\mathbf{x}}_n,\widetilde{\boldsymbol\Sigma}_n) - h(\widetilde{\mathbf{x}}_{\infty},{\boldsymbol\Sigma})\right|
    & \leq 
    \left|\left|{\widetilde{\mathbf{x}}_n}^{\top} \left(\left(\mathbf{H} {\widetilde{\boldsymbol\Sigma}_n}\mathbf{H}^{\top} \right)^+-\left(\mathbf{H} {\boldsymbol\Sigma}\mathbf{H}^{\top} \right)^+\right) {\widetilde{\mathbf{x}}_n}\right|\right|_1 \\&\quad + \left|\left|{\widetilde{\mathbf{x}}_n}^{\top} \left(\mathbf{H} \boldsymbol\Sigma \mathbf{H}^{\top} \right)^+ (\widetilde{\mathbf{x}}_n-\widetilde{\mathbf{x}}_{\infty})\right|\right|_1
    + \left|\left|(\widetilde{\mathbf{x}}_n-\widetilde{\mathbf{x}}_{\infty})^{\top} \left(\mathbf{H} \boldsymbol\Sigma \mathbf{H}^{\top} \right)^+ {\widetilde{\mathbf{x}}}_{\infty}\right|\right|_1
\end{align*}
for all $n\in\mathbb{N}$.
We observe that \begin{align*}    || \mathbf{a}^{\top} \mathbf{A} \mathbf{b} ||_1 = \int_{\T} \left| \mathbf{a}^{\top}(t) \mathbf{A}(t) \mathbf{b}(t) \right| \;\mathrm{d}t \leq \int_{\T} \left|\left| \mathbf{a}(t) \right|\right|\, \left|\left| \mathbf{b}(t) \right|\right| \, ||| \mathbf{A}(t) ||| \;\mathrm{d}t \leq ||\mathbf{a}||_2 ||\mathbf{b}||_2 ||| \mathbf{A} |||_{\infty} \end{align*} holds for all measurable functions $\mathbf{A}:\T \to \R^{r\times r}$ and $\mathbf{a}, \mathbf{b}\in \mathcal{L}_2^r(\T)$  by the Cauchy-Schwarz inequality and the submultiplicativity of the spectral norm, where here and throughout $|||.|||_{\infty}$ denotes the $\mathcal{L}_{\infty}^{k \times k}(\T)$-norm.
The second triangular inequality yields
 $|| {\widetilde{\mathbf{x}}_n} ||_2 \to || \widetilde{\mathbf{x}}_{\infty} ||_2 < \infty$ as $n\to\infty$. 
Hence, it remains to show
\begin{align}\label{eq:toShow}
   \left|\left|\left| \left(\mathbf{H} {\widetilde{\boldsymbol\Sigma}_n} \mathbf{H}^{\top} \right)^+ - \left(\mathbf{H} {\boldsymbol\Sigma} \mathbf{H}^{\top} \right)^+\right|\right|\right|_{\infty} \to 0 \quad\text{as $n\to\infty$}
\end{align} to conclude the continuity of $h$.\\
Since $\boldsymbol\Sigma(t)$ is invertible for all $t\in\T$ and $\boldsymbol\Sigma$ is continuous, we can find an $\varepsilon > 0$ such that every $\mathbf{K} \in \R^{r \times r}$ with $|||\mathbf{K} - \boldsymbol\Sigma(t)||| \leq \varepsilon$ for some $t\in\T$ is invertible. 
We consider the set
$$\mathcal{K} := \{ \mathbf{H} \widetilde{\boldsymbol\Sigma} \mathbf{H}^{\top} \in \R^{r \times r} \mid \widetilde{\boldsymbol\Sigma}\in\R^{k \times k}, \inf\limits_{t\in\T}|||\widetilde{\boldsymbol\Sigma} - \boldsymbol\Sigma(t)||| \leq \varepsilon \}.$$
Since the function $j: \R^{k \times k} \ni \widetilde{\boldsymbol\Sigma} \mapsto \inf\limits_{t\in\T}|||\widetilde{\boldsymbol\Sigma} - \boldsymbol\Sigma(t)||| \in\R$ is continuous, it follows that the set $j^{-1}([0,\varepsilon])$ 
is a closed set. Moreover, it is bounded since $\sup\limits_{t\in\T}|||\boldsymbol\Sigma(t)||| < \infty$. 
Thus, $\mathcal{K}$ is compact as continuous image of a compact set. Hence, the continuous map $$ \mathcal{K} \ni \mathbf{K} \mapsto \mathbf{K}^{-1} = \mathbf{K}^+ \in \R^{k \times k} $$ is uniformly continuous on $\mathcal{K}$.
Let $\eta > 0$ be arbitrary and $\delta > 0$ such that \mbox{$ |||\mathbf{K}_1^+ - \mathbf{K}_2^+||| < \eta $} holds for all $\mathbf{K}_1,\mathbf{K}_2\in \mathcal{K}$ with $|||\mathbf{K}_1 - \mathbf{K}_2||| < \delta$. Furthermore, let $N\in\mathbb{N}$ such that
$\left|\left|\left| \widetilde{\boldsymbol\Sigma}_n - \boldsymbol\Sigma\right|\right|\right|_{\infty} < \min\{\varepsilon,\delta/|||\mathbf{H}|||^2\}$ for all $n \geq N$. Hence, it holds $$\left|\left|\left|\mathbf{H} \widetilde{\boldsymbol\Sigma}_n(t) \mathbf{H}^{\top} - \mathbf{H}\boldsymbol\Sigma(t) \mathbf{H}^{\top} \right|\right|\right|  = \left|\left|\left|\mathbf{H} (\widetilde{\boldsymbol\Sigma}_n(t) - \boldsymbol\Sigma(t)) \mathbf{H}^{\top}\right|\right|\right| \leq |||\mathbf{H}|||^2 \left|\left|\left|\widetilde{\boldsymbol\Sigma}_n(t) - \boldsymbol\Sigma(t)\right|\right|\right|  < \delta$$ by the submultiplicativity of the spectral norm and $ \mathbf{H}\widetilde{\boldsymbol\Sigma}_n(t)\mathbf{H}^{\top} \in \mathcal{K}$ for all $n \geq N$ and almost every $t\in\T$ and, therefore,
$ \left|\left|\left|(\mathbf{H} \widetilde{\boldsymbol\Sigma}_n \mathbf{H}^{\top})^+ - (\mathbf{H}\boldsymbol\Sigma \mathbf{H}^{\top})^+\right|\right|\right|_{\infty} < \eta $ for all $n \geq N$.
As a result, (\ref{eq:toShow}) follows and the theorem is proven.
\end{proof}

In the following, we generally condition on the functional data 
 \begin{align}\label{eq:funcData}
    (x_{i1},x_{i2},...)_{i\in\{1,...,k\}},
\end{align} if not otherwise specified.

An analogue of Lemma~\ref{asy_CLT} for the parametric bootstrap counterpart of the sample mean function is given by the following lemma.
\begin{lemma}
\label{Para_CLT}
Under (A1)-(A5), we have 
\begin{align*}
    \sqrt{n} \mathbf{H}\hat{\boldsymbol\eta}^{\mathcal{P}} \xrightarrow{d^*}
    \mathbf{z} \sim GP_{r}\left(\mathbf{0}_{r}, \mathbf{H}\boldsymbol\Sigma\mathbf{H}^{\top}\right) \quad\text{as $n\to\infty$}
\end{align*}
 in $\mathcal{L}^{r}_2(\T)$.
\end{lemma}
\begin{proof}[\textbf{Proof of Lemma~\ref{Para_CLT}}]
Let $i\in\{1,...,k\}$ be arbitrary but fixed. We aim to apply Theorem~\ref{clt} conditionally on the functional data (\ref{eq:funcData})
 with $Z_{j,n_i} := \frac{1}{\sqrt{n_i}}{x}_{ij}^{\mathcal{P}}$ for all $j\in\{1,...,n_i\}.$
Then, $Z_{1,n_i},...,Z_{n_i,n_i}$ are i.i.d.~conditionally on (\ref{eq:funcData}) by construction in Algorithm~1. 
Moreover, Fubini`s theorem provides
\begin{align*}   \E^*\left[ ||Z_{1,n_i}||_2^2 \right] &= \frac{1}{{n_i}} \int_{\T} \E^*\left[(x_{i1}^{\mathcal{P}}(t))^2 \right]  \;\mathrm{d}t   
= 
\frac{1}{{n_i}} \int_{\T} \widehat{\gamma}_i(t,t)   \;\mathrm{d}t   < \infty\end{align*}
 almost surely since the functional data take values in $\mathcal{L}_2(\T)$ under (A1), where here and throughout $\E^*$ denotes the conditional expectation given the functional data (\ref{eq:funcData}). It should be noted that all following statements about conditional
expectations hold just almost surely but we will not always add this throughout this supplement, for the sake of clarity.
Hence, $Z_{1,n_i},...,Z_{n_i,n_i}$ take values in $\mathcal{L}_2(\T)$ conditionally on the functional data (\ref{eq:funcData}).\\
Assumption~(1) of Theorem~\ref{clt} is satisfied conditionally on (\ref{eq:funcData}) since Fubini`s theorem provides
\begin{align*}
    \E^*\left[ \left\langle\frac{1}{\sqrt{n_i}}x_{i1}^{\mathcal{P}}, {f}_m \right\rangle \right] &= \int_{\T} \frac{1}{\sqrt{n_i}}\E^*\left[x_{i1}^{\mathcal{P}}(t)\right]{f}_m(t) \;\mathrm{d}t = 0
\end{align*} for all $m\in\mathbb{N}.$\\
Furthermore, due to the identical distribution of the parametric bootstrap observations and Fubini`s theorem, we have
\begin{align*} 
    \sum\limits_{j=1}^{n_i} \E^*\left[\left\langle \frac{1}{\sqrt{n_i}}x_{ij}^{\mathcal{P}}, {f}_m \right\rangle \left\langle \frac{1}{\sqrt{n_i}}x_{ij}^{\mathcal{P}}, {f}_{\ell} \right\rangle \right] 
    &= \E^*\left[\langle x_{i1}^{\mathcal{P}}, {f}_m \rangle \langle x_{i1}^{\mathcal{P}}, {f}_{\ell} \rangle \right]
    \\&= \E^*\left[\int_{\T^2} {f}_{m}(t)x_{i1}^{\mathcal{P}}(t)  x_{i1}^{\mathcal{P}}(s){f}_{l}(s) \;\mathrm{d}(t,s)\right]
    \\&= \int_{\T^2} {f}_{m}(t)\E^*\left[x_{i1}^{\mathcal{P}}(t)  x_{i1}^{\mathcal{P}}(s)\right]{f}_{l}(s) \;\mathrm{d}(t,s)
    \\&= \int_{\T^2} {f}_{m}(t)\widehat{\gamma}_i(t,s){f}_{l}(s) \;\mathrm{d}(t,s)
\end{align*} for all $m,{\ell}\in\mathbb{N}$.
 It holds \begin{align*}
     &\int_{\T^2}  {f}_{m}(t)\widehat{\gamma}_i (t,s) {f}_{l}(s)   \;\mathrm{d}(t,s) 
     \\&=\frac{1}{n_i-1}\sum\limits_{j=1}^{n_i} \int_{\T^2}  {f}_{m}(t) (x_{ij} - \hat{\eta}_i)(t)(x_{ij} - \hat{\eta}_i)(s) {f}_{l}(s)   \;\mathrm{d}(t,s)
    \\&=\frac{1}{n_i-1}\sum\limits_{j=1}^{n_i} \langle  {f}_{m}, x_{ij} - \hat{\eta}_i\rangle\langle  {f}_{l}, x_{ij} - \hat{\eta}_i\rangle
    \\&=\frac{1}{n_i-1}\sum\limits_{j=1}^{n_i} \left(  \langle{f}_{m}, x_{ij}\rangle  - \frac{1}{n_i}\sum\limits_{j_2=1}^{n_i}\langle {f}_{m}, x_{ij_2}\rangle\right) \left(  \langle {f}_{l}, x_{ij}\rangle  - \frac{1}{n_i}\sum\limits_{j_2=1}^{n_i}\langle {f}_{l}, x_{ij_2}\rangle\right) 
 \end{align*} for all $m,{\ell}\in\mathbb{N}$.
 Thus, the term $\int_{\T^2}  {f}_{m}(t)\widehat{\gamma}_i (t,s) {f}_{l}(s)   \;\mathrm{d}(t,s)  $ can be thought of as the sample covariance of $\langle {f}_{m}, x_{i1}\rangle$ and $\langle {f}_{l}, x_{i1}\rangle$. 
 Hence, assumption (2) of Theorem~\ref{clt} follows almost surely due to the strong consistency of the sample covariance, where $\sigma_{ml}^2$ denotes the covariance of $\langle {f}_{m}, x_{i1}\rangle$ and $\langle {f}_{l}, x_{i1}\rangle$ for all $m,{\ell}\in\mathbb{N}$. 
 By Fubini`s theorem, it holds
 \begin{align} \label{eq:sigma_ml}
     \sigma_{ml}^2 &= 
     \int_{\T^2}  {f}_{m}(t){\gamma}_i (t,s) {f}_{l}(s)   \;\mathrm{d}(t,s) 
 \end{align}
 for all $m,{\ell}\in\mathbb{N}$ with similar calculations as above.\\
 Analogously, we can calculate
 \begin{align} \begin{split}\label{eq:SumOfVariances}
    \sum\limits_{m\in\mathbb{N}}\sum\limits_{j=1}^{n_i} \Var^*\left(\left\langle \frac{1}{\sqrt{n_i}}x_{ij}^{\mathcal{P}}, {f}_m \right\rangle  \right) 
    &= 
   \sum\limits_{m\in\mathbb{N}} \int_{\T^2} {f}_{m}(t)\widehat{\gamma}_i(t,s){f}_{m}(s) \;\mathrm{d}(t,s)
  \\ &=
    \sum\limits_{m\in\mathbb{N}}\frac{1}{n_i-1}\sum\limits_{j=1}^{n_i} \langle  {f}_{m}, x_{ij} - \hat{\eta}_i\rangle^2,
 \end{split}\end{align}
 where here and throughout $\Var^*$ denotes the conditional variance given the functional data (\ref{eq:funcData}).
Hence, it follows
 \begin{align} \label{eq:ParaEwertSquared}\begin{split}
   \E\left[ \sum\limits_{m\in\mathbb{N}} \sum\limits_{j=1}^{n_i} \Var^*\left(\left\langle \frac{1}{\sqrt{n_i}}x_{ij}^{\mathcal{P}}, {f}_m \right\rangle  \right) \right]
    &= \sum\limits_{m\in\mathbb{N}} \int_{\T^2} {f}_{m}(t)\E\left[ \widehat{\gamma}_i(t,s) \right]{f}_{m}(s)\;\mathrm{d}(t,s)
    \\ &=  \sum\limits_{m\in\mathbb{N}}\int_{\T^2} {f}_{m}(t){\gamma}_i(t,s) {f}_{m}(s)\;\mathrm{d}(t,s)
    \\ &= \sum\limits_{m\in\mathbb{N}}\sigma_{mm}^2 \end{split}
 \end{align} by Fubini`s theorem. The sum is finite since Fubini`s theorem, the monotone convergence theorem and Parseval's identity \cite[p.~245]{hewitt} yield
  \begin{align} \label{eq:finite} \begin{split}
  \sum\limits_{m\in\mathbb{N}}\sigma_{mm}^2 &= \sum\limits_{m\in\mathbb{N}}\int_{\T^2} {f}_{m}(t){\gamma}_i(t,s) {f}_{m}(s)\;\mathrm{d}(t,s) 
  \\&=  \sum\limits_{m\in\mathbb{N}}\int_{\T^2} {f}_{m}(t)\E\left[ {v}_{i1}(t) {v}_{i1}(s) \right] {f}_{m}(s)\;\mathrm{d}(t,s) 
  \\&= \E\left[ \sum\limits_{m\in\mathbb{N}}\langle {f}_{m}, {v}_{i1} \rangle^2 \right]
  \\&= \E\left[ || {v}_{i1} ||_2^2 \right] < \infty \end{split}
 \end{align} due to (\ref{eq:MEWertExistenz}). In order to prove assumption (3) of Theorem~\ref{clt} in probability conditionally on (\ref{eq:funcData}), it remains to show that the variance of (\ref{eq:SumOfVariances}) vanishes asymptotically. 
 For the second moment of (\ref{eq:SumOfVariances}), we have
 \begin{align*}
     \E\left[ \left(\sum\limits_{m\in\mathbb{N}}\sum\limits_{j=1}^{n_i} \Var^*\left(\left\langle \frac{1}{\sqrt{n_i}}x_{ij}^{\mathcal{P}}, {f}_m \right\rangle  \right) \right)^2 \right]&=
     \E\left[ \left(
     \sum\limits_{m\in\mathbb{N}}\frac{1}{n_i-1}\sum\limits_{j=1}^{n_i} \langle  {f}_{m}, x_{ij} - \hat{\eta}_i\rangle^2
     \right)^2 \right].
 \end{align*}
 Rearranging the summands yields
 \begin{align*}
      \E\left[ \left(\sum\limits_{m\in\mathbb{N}}\sum\limits_{j=1}^{n_i} \Var^*\left(\left\langle \frac{1}{\sqrt{n_i}}x_{ij}^{\mathcal{P}}, {f}_m \right\rangle  \right) \right)^2 \right]
     &= \frac{1}{(n_i-1)^2}\sum\limits_{j_1,j_2=1}^{n_i}\E\left[
     \sum\limits_{m,{\ell}\in\mathbb{N}} \langle  {f}_{m}, x_{ij_1} - \hat{\eta}_i\rangle^2 \langle  {f}_{l}, x_{ij_2} - \hat{\eta}_i\rangle^2 \right].
 \end{align*}
 Here, the rearrangement is valid since all summands are non-negative.
By using the Parseval's identity \cite[p.~245]{hewitt} and (A1), the expectations on the right side can be expressed for $j_1 = j_2 \in\{1,...,n_i\}$ by
\begin{align}\label{eq:viertesMoment}
    \E\left[
    \left( \sum\limits_{m\in\mathbb{N}} \langle  {f}_{m}, x_{ij_1} - \hat{\eta}_i\rangle^2  \right)^2 \right]  = \E\left[
     || x_{ij_1} - \hat{\eta}_i ||_2^4   \right] = \E\left[
    || x_{i1} - \hat{\eta}_i ||_2^4   \right].
\end{align}
By applying the triangular inequality, we get
\begin{align*}
    || x_{i1} - \hat{\eta}_i ||_2 = \left|\left| \frac{n_i-1}{n_i} (x_{i1} - {\eta}_i) - \frac{1}{n_i} \sum\limits_{j=2}^{n_i}(x_{ij} - {\eta}_i) \right|\right|_2 \leq \sum\limits_{j=1}^{n_i} a_{j} || x_{ij} - {\eta}_i ||_2,
\end{align*} where $a_{1} := {(n_i-1)}/{n_i}$ and $a_{j} := {1}/{n_i}$ for all $j\in\{2,...,n_i\}.$
Under (A1) and (A5), Hölder`s inequality implies
\begin{align*}
    \E\left[ || x_{ij_1} - {\eta}_i ||_2 \, || x_{ij_2} - {\eta}_i ||_2 \, || x_{ij_3} - {\eta}_i ||_2\,  || x_{ij_4} - {\eta}_i ||_2 \right] \leq \E\left[ || x_{i1} - {\eta}_i ||_2^4 \right]  < \infty
\end{align*} for all $j_1,j_2,j_3,j_4\in\{1,...,n_i\}.$
Hence, the expectation in (\ref{eq:viertesMoment}) can be bounded uniformly over $n_i$ 
since
\begin{align*}
    \E\left[
    || x_{i1} - \hat{\eta}_i ||_2^4   \right] &\leq \sum\limits_{j_1,j_2,j_3,j_4 = 1}^{n_i} a_{j_1}a_{j_2}a_{j_3}a_{j_4} \E\left[ || x_{ij_1} - {\eta}_i ||_2 \, || x_{ij_2} - {\eta}_i ||_2\, || x_{ij_3} - {\eta}_i ||_2\, || x_{ij_4} - {\eta}_i ||_2 \right]
    \\& \leq \left(\sum\limits_{j = 1}^{n_i} a_{j}\right)^4 \E\left[ || x_{i1} - {\eta}_i ||_2^4 \right] \leq 2^4 \E\left[ || x_{i1} - {\eta}_i ||_2^4 \right] = 16 \E\left[ || x_{i1} - {\eta}_i ||_2^4 \right]
\end{align*} by definition of $a_{1},...,a_{n_i}$. Thus, it follows that the sum of the expectations with \mbox{$j_1=j_2\in\{1,...,n_i\}$,} that is 
\begin{align}\label{eq:part1}
    \frac{1}{(n_i-1)^2}\sum\limits_{j_1=1}^{n_i}\E\left[
     \sum\limits_{m,{\ell}\in\mathbb{N}} \langle  {f}_{m}, x_{ij_1} - \hat{\eta}_i\rangle^2 \langle  {f}_{l}, x_{ij_1} - \hat{\eta}_i\rangle^2 \right] = \frac{n_i}{(n_i-1)^2}\E\left[
    || x_{i1} - \hat{\eta}_i ||_2^4   \right],
\end{align}
vanishes asymptotically as $n\to\infty$.\\
 Furthermore, we have for the expectations with $j_1 \neq j_2, j_1, j_2 \in\{1,...,n_i\},$ 
 \begin{align*}
     \E\left[
     \sum\limits_{m,{\ell}\in\mathbb{N}} \langle  {f}_{m}, x_{ij_1} - \hat{\eta}_i\rangle^2 \langle  {f}_{l}, x_{ij_2} - \hat{\eta}_i\rangle^2 \right] 
     &=\E\left[
     \sum\limits_{m,{\ell}\in\mathbb{N}} \langle  {f}_{m}, x_{i1} - \hat{\eta}_i\rangle^2 \langle  {f}_{l}, x_{i2} - \hat{\eta}_i\rangle^2 \right] 
     \\&=  \sum\limits_{m,{\ell}\in\mathbb{N}} \E\left[
    \langle  {f}_{m}, {v}_{i1} - (\hat{\eta}_i - \eta_i )\rangle^2 \langle  {f}_{l}, {v}_{i2} - (\hat{\eta}_i - \eta_i )\rangle^2 \right] 
 \end{align*} due to (A1) and the monotone convergence theorem.
 By simplifying and applying the Parseval`s identity \cite[p.~245]{hewitt}, one can show
 \begin{align}\label{eq:part2}\begin{split}
     \E\left[
     \sum\limits_{m,{\ell}\in\mathbb{N}} \langle  {f}_{m}, x_{i1} - \hat{\eta}_i\rangle^2 \langle  {f}_{l}, x_{i2} - \hat{\eta}_i\rangle^2 \right] 
    &\to  \sum\limits_{m,{\ell}\in\mathbb{N}} \E\left[ \langle  {f}_{m}, {v}_{i1} \rangle^2\langle  {f}_{l}, {v}_{i2} \rangle^2 \right] 
     \\&= \left(\E\left[\sum\limits_{m\in\mathbb{N}} \langle  {f}_{m}, {v}_{i1} \rangle^2 \right]\right)^2 
     \\&= \left(  \sum\limits_{m\in\mathbb{N}} \sigma_{mm}^2  \right)^2 
 \end{split}\end{align} as $n\to\infty$ under (A5), where the last equality follows as in (\ref{eq:finite}). For the sake of clarity, we will not describe the calculation further.\\
Summing up the calculations above, we get
\begin{align*}
    \E\left[ \left(\sum\limits_{m\in\mathbb{N}}\sum\limits_{j=1}^{n_i} \Var^*\left(\left\langle \frac{1}{\sqrt{n_i}}x_{ij}^{\mathcal{P}}, {f}_m \right\rangle  \right) \right)^2 \right] &= o(1) + \frac{n_i}{n_i-1}\E\left[
     \sum\limits_{m,{\ell}\in\mathbb{N}} \langle  {f}_{m}, x_{i1} - \hat{\eta}_i\rangle^2 \langle  {f}_{l}, x_{i2} - \hat{\eta}_i\rangle^2 \right]
     \\&\to \left(  \sum\limits_{m\in\mathbb{N}} \sigma_{mm}^2  \right)^2 
\end{align*}
as $n\to\infty$ by (\ref{eq:part1}) and (\ref{eq:part2}), where and throughout $o(1)$ denotes convergence to $0$.
 Since
 \begin{align*}
     \left(  \sum\limits_{m\in\mathbb{N}} \sigma_{mm}^2  \right)^2  =  \E\left[ \sum\limits_{m\in\mathbb{N}} \sum\limits_{j=1}^{n_i} \Var^*\left(\left\langle \frac{1}{\sqrt{n_i}}x_{ij}^{\mathcal{P}}, {f}_m \right\rangle  \right) \right]^2
 \end{align*}
 holds by (\ref{eq:ParaEwertSquared}), we can conclude that the variance of (\ref{eq:SumOfVariances}) vanishes asymptotically. Hence, assumption (3) of Theorem~\ref{clt} holds in probability conditionally on (\ref{eq:funcData}).\\
 Finally, we aim to show assumption~(4) of Theorem~\ref{clt} by proving Lyapunov's condition in Remark~\ref{Lyapunov} with $\delta = 2.$ 
 By applying the Cauchy-Schwarz inequality several times and Fubini`s theorem, one can show
 \begin{align*}
     \sum\limits_{j=1}^{n_i} \E^*\left[\left\langle \frac{1}{\sqrt{n_i}}x_{ij}^{\mathcal{P}}, {f}_m \right\rangle^4 \right]
     &\leq 
      \frac{1}{n_i} \left( \int_{\T}\sqrt{\E^*\left[  (x_{i1}^{\mathcal{P}}(t))^4\right]}  \;\mathrm{d}t \right)^2
 \end{align*}
 Since $x_{i1}^{\mathcal{P}}(t)$ is normally distributed with mean $0$ and variance $\widehat{\gamma}_{i}(t,t)$ conditionally on (\ref{eq:funcData}), it follows
  \begin{align*}
     \sum\limits_{j=1}^{n_i} \E^*\left[\left\langle \frac{1}{\sqrt{n_i}}x_{ij}^{\mathcal{P}}, {f}_m \right\rangle^4 \right]
     \leq \frac{1}{n_i} \left( \sqrt{3}  \int_{\T} \widehat{\gamma}_{i}(t,t)   \;\mathrm{d}t\right)^2 
 \end{align*}
 for all $m\in\mathbb{N}$. Hence, Lemma~\ref{UniformConvLemma} and (A3) yield Lyapunov's condition in probability conditionally on (\ref{eq:funcData}) for all $\varepsilon > 0, m\in M$.  \\Assumptions (3) and (4) of Theorem~\ref{clt} hold in probability conditionally on (\ref{eq:funcData}). Thus, we can find for every subsequence of increasing sample sizes $n_i$ a further subsequence such that (3) and (4) hold almost surely along the latter subsequence by applying the subsequence criterion.
Consequently, Theorem~\ref{clt} provides that
$ \sqrt{n_i}\widehat{\eta}_i^{\mathcal{P}} = \sum\limits_{j=1}^{n_i} {Z}_{j,n_i} $ converges in distribution to a centered Gaussian process with covariance function ${\gamma}_{i}$ conditionally on (\ref{eq:funcData}) almost surely along the latter subsequence  as $n\to\infty$. 
Here, the covariance structure of the limiting process follows from the characterization (\ref{eq:characterization}) and equation (\ref{eq:sigma_ml}).\\
Since $ \sqrt{n_i}\widehat{\eta}_i^{\mathcal{P}}, i\in\{1,...,k\},$ are independent conditionally on $(x_{i1},...,x_{in_i})_{i\in\{1,...,k\}}$, it follows with Slutsky's lemma that
$$ \sqrt{n} \mathbf{H}\widehat{\boldsymbol\eta}^{\mathcal{P}}  =  \mathbf{H}\left( \frac{\sqrt{n}}{\sqrt{n_i}} \sqrt{n_i}\widehat{\eta}_i^{\mathcal{P}} \right)_{i\in\{1,...,k\}} $$ 
converges in distribution to $\mathbf{z} \sim GP_r(\mathbf{0}_r, \mathbf{H}\boldsymbol\Sigma \mathbf{H}^{\top})$ conditionally on (\ref{eq:funcData}) almost surely along the specified subsequence as $n\to\infty$. Hence, by applying the subsequence criterion again, it holds
$$ \sqrt{n} \mathbf{H}\widehat{\boldsymbol\eta}^{\mathcal{P}}  \xrightarrow{d^*} \mathbf{z}$$ as $n\to\infty$.
\end{proof}

Theorem~2 can now be proved by using Lemma~\ref{Para_CLT}.

\begin{proof}[\textbf{Proof of Theorem~2}]
By Lemma~\ref{Para_CLT} and the subsequence criterion, every subsequence has a further subsequence such that
$ \sqrt{n}\mathbf{H}\widehat{\boldsymbol\eta}^{\mathcal{P}}$ converges in distribution almost surely along the subsequence to $\mathbf{z}$ in $\mathcal{L}_2^r(\T)$. Applying the continuous mapping theorem 
yields 
that $$ \int_{\T}  n\left(\mathbf{H}\widehat{\boldsymbol\eta}^{\mathcal{P}}(t)\right)^{\top}\left(\mathbf{H}{\boldsymbol{\Sigma}}(t,t)\mathbf{H}^{\top}\right)^+\mathbf{H}\widehat{\boldsymbol\eta}^{\mathcal{P}}(t)\;\mathrm{d}t $$ converges in distribution almost surely along the latter subsequence to $\int_{\T} \mathbf{z}^{\top}(t)(\mathbf{H}\boldsymbol{\Sigma}(t,t)\mathbf{H}^{\top})^+\mathbf{z}(t) \,\mathrm{ d }t$. Hence, by applying the subsequence criterion again, it follows
\begin{align*}
    \int_{\T}  n\left(\mathbf{H}\widehat{\boldsymbol\eta}^{\mathcal{P}}(t)\right)^{\top}\left(\mathbf{H}{\boldsymbol{\Sigma}}(t,t)\mathbf{H}^{\top}\right)^+\mathbf{H}\widehat{\boldsymbol\eta}^{\mathcal{P}}(t)\;\mathrm{d}t \xrightarrow{d^*} \int_{\T} \mathbf{z}^{\top}(t)(\mathbf{H}\boldsymbol{\Sigma}(t,t)\mathbf{H}^{\top})^+\mathbf{z}(t) \,\mathrm{ d }t
\end{align*} as $n\to\infty$.\\
Consequently, it remains to show 
\begin{align}\label{eq:remains}
D_n^{\mathcal{P}}(\mathbf{H}) := \left|  T_n^{\mathcal{P}}(\mathbf{H}) - \int_{\T}  n\left(\mathbf{H}\widehat{\boldsymbol\eta}^{\mathcal{P}}(t)\right)^{\top}\left(\mathbf{H}{\boldsymbol{\Sigma}}(t,t)\mathbf{H}^{\top}\right)^+\mathbf{H}\widehat{\boldsymbol\eta}^{\mathcal{P}}(t)\;\mathrm{d}t\right|    \xrightarrow{P^*} 0
\end{align} as $n\to\infty,$ where $\xrightarrow{P^*}$ denotes convergence in outer probability conditionally on (\ref{eq:funcData}).
By the triangular inequality, Hölder's inequality and the submultiplicativity, one can show 
\begin{align*}
D_n^{\mathcal{P}}(\mathbf{H}) & 
\leq   \int_{\T}  n\left| \left(\mathbf{H}\widehat{\boldsymbol\eta}^{\mathcal{P}}(t)\right)^{\top}\left(\left(\mathbf{H}\widehat{\boldsymbol{\Sigma}}^{\mathcal P}(t,t)\mathbf{H}^{\top}\right)^+ - \left(\mathbf{H}{\boldsymbol{\Sigma}}(t,t)\mathbf{H}^{\top}\right)^+\right)\mathbf{H}\widehat{\boldsymbol\eta}^{\mathcal{P}}(t)\right| \;\mathrm{d}t
\\&\leq \left|\left|\left|\mathbf{H}\right|\right|\right|^2\left|\left|  \sqrt{n}\widehat{\boldsymbol\eta}^{\mathcal{P}}\right|\right|_4^2
\left(\int_{\T}\left|\left|\left|\left(\mathbf{H}\widehat{\boldsymbol{\Sigma}}^{\mathcal{P}}(t,t)\mathbf{H}^{\top}\right)^+ - \left(\mathbf{H}{\boldsymbol{\Sigma}}(t,t)\mathbf{H}^{\top}\right)^+\right|\right|\right|^2\;\mathrm{d}t\right)^{1/2}.
\end{align*}
Hence, by Markov's inequality and the Cauchy-Schwarz inequality, we have
\begin{align}\label{eq:DnP}
     \P^*_{\mathrm{out}}\left( D_n^{\mathcal{P}}(\mathbf{H}) > \varepsilon \right)
 \leq &
   \varepsilon^{-1}\left|\left|\left|\mathbf{H}\right|\right|\right|^2\sqrt{\E^*_{\mathrm{out}}\left[\left|\left|  \sqrt{n}\widehat{\boldsymbol\eta}^{\mathcal{P}}\right|\right|_4^4\right]}\cdot \\
   \label{eq:DnP2}&\sqrt{ \E^*_{\mathrm{out}}\left[\int_{\T}\left|\left|\left|\left(\mathbf{H}\widehat{\boldsymbol{\Sigma}}^{\mathcal{P}}(t,t)\mathbf{H}^{\top}\right)^+ - \left(\mathbf{H}{\boldsymbol{\Sigma}}(t,t)\mathbf{H}^{\top}\right)^+\right|\right|\right|^2\;\mathrm{d}t\right]}
\end{align}
for all $\varepsilon > 0,$ where here and throughout 
$\P^*_{\mathrm{out}}$ denotes the outer probability 
and $\E^*_{\mathrm{out}}$ denotes the outer expectation given  the functional data (\ref{eq:funcData}), respectively.
For all $t\in\T$, $i\in\{1,...,k\}$, it holds $\widehat\eta_i^{\mathcal{P}}(t) \sim \mathcal{N}\left(0,\widehat\gamma_i(t,t)/n_i\right)$ conditional on the data (\ref{eq:funcData}). Hence,
$$ \left|\left|  \widehat{\boldsymbol\eta}^{\mathcal{P}}(t)\right|\right|^2 = \sum\limits_{i=1}^k \widehat\eta_i^{\mathcal{P}}(t)^2 \overset{d}{=} \sum\limits_{i=1}^k \frac{\widehat\gamma_i(t,t)}{n_i} Z_i^2 $$
 conditional on the data (\ref{eq:funcData}), where $Z_1,...,Z_k $ are independent standard normal distributed and independent of the data.
Then, the expectation in (\ref{eq:DnP}) can be calculated as
\begin{align}\label{eq:togetherPart}\begin{split}
    \E^*_{\mathrm{out}}\left[\left|\left|  \sqrt{n}\widehat{\boldsymbol\eta}^{\mathcal{P}}\right|\right|_4^4\right] &=
    n^2\int_{\T}\E^*\left[ \left|\left|  \widehat{\boldsymbol\eta}^{\mathcal{P}}(t)\right|\right|^4\right]\;\mathrm{d}t
    \\&=
    n^2\int_{\T}\E^*\left[\left( \sum\limits_{i=1}^k \frac{\widehat\gamma_i(t,t)}{n_i} Z_i^2 \right)^2\right]\;\mathrm{d}t
    \\&=
    n^2\int_{\T}
    2\sum\limits_{i=1}^k \frac{\widehat\gamma_i(t,t)^2}{n_i^2} + \left( \sum\limits_{i=1}^k \frac{\widehat\gamma_i(t,t)}{n_i} \right)^2
    \;\mathrm{d}t
    \\ &\xrightarrow{P} \int_{\T}
    2\sum\limits_{i=1}^k \frac{\gamma_i(t,t)^2}{\tau_i^2} + \left( \sum\limits_{i=1}^k \frac{\gamma_i(t,t)}{\tau_i} \right)^2
    \;\mathrm{d}t < \infty
    \end{split}
\end{align}
as $n\to\infty$ by Fubini's theorem and Lemma~\ref{UniformConvLemma}.
Thus, it remains to show that the expectation in (\ref{eq:DnP2}) vanishes asymptotically in probability.
By Theorem~4.1 of \cite[p.~221]{mpinverse}, it holds
\begin{align*}
    &\left|\left|\left|\left(\mathbf{H}\widehat{\boldsymbol{\Sigma}}^{\mathcal{P}}(t,t)\mathbf{H}^{\top}\right)^+ - \left(\mathbf{H}{\boldsymbol{\Sigma}}(t,t)\mathbf{H}^{\top}\right)^+\right|\right|\right|\\& \leq \frac{1+\sqrt{5}}{2}\max\left\{  \left|\left|\left| \left(\mathbf{H}\widehat{\boldsymbol{\Sigma}}^{\mathcal{P}}(t,t)\mathbf{H}^{\top}\right)^+ \right|\right|\right|^2, \left|\left|\left| \left(\mathbf{H}{\boldsymbol{\Sigma}}(t,t)\mathbf{H}^{\top}\right)^+ \right|\right|\right|^2\right\} \left|\left|\left|\mathbf{H}\left(\widehat{\boldsymbol{\Sigma}}^{\mathcal{P}}(t,t)-{\boldsymbol{\Sigma}}(t,t)\right)\mathbf{H}^{\top}  \right|\right|\right|
\end{align*} for all $t\in\T.$\\
Since $\T\ni t\mapsto ||| (\mathbf{H}{\boldsymbol{\Sigma}}(t,t)\mathbf{H}^{\top})^+ |||^2$ is a continuous function, it attains its maximum over the compact set $\T$. Hence, there exists a $C>0$ such that $\left|\left|\left| \left(\mathbf{H}{\boldsymbol{\Sigma}}(t,t)\mathbf{H}^{\top}\right)^+ \right|\right|\right|^2 \leq C$ for all $t\in\T$. 
\\
Fubini's theorem, the submultiplicativity and the Cauchy-Schwarz inequality imply
\begin{align}\label{eq:fetteFormel}\begin{split}
    &\E^*_{\mathrm{out}}\left[\int_{\T}\left|\left|\left|\left(\mathbf{H}\widehat{\boldsymbol{\Sigma}}^{\mathcal{P}}(t,t)\mathbf{H}^{\top}\right)^+ - \left(\mathbf{H}{\boldsymbol{\Sigma}}(t,t)\mathbf{H}^{\top}\right)^+\right|\right|\right|^2\;\mathrm{d}t\right] 
    \\&\leq \left(\frac{1+\sqrt{5}}{2}\right)^2\left|\left|\left| \mathbf{H} \right|\right|\right|^4
    \sqrt{\int_{\T}  \E^*\left[  \left|\left|\left| \left(\mathbf{H}\widehat{\boldsymbol{\Sigma}}^{\mathcal{P}}(t,t)\mathbf{H}^{\top}\right)^+ \right|\right|\right|^8\right]+ C^4\;\mathrm{d}t} \cdot \\
    & \qquad \sqrt{\int_{\T}\E^*\left[\left|\left|\left|\widehat{\boldsymbol{\Sigma}}^{\mathcal{P}}(t,t)-{\boldsymbol{\Sigma}}(t,t)  \right|\right|\right|^4\right]\;\mathrm{d}t}.\end{split}
\end{align}
Let
$$ B_n := \left\{ \inf\limits_{t\in\T} \widehat\gamma_i(t,t) >  0 \quad \forall i\in\{1,...,k\} \right\}. $$
By Lemma~\ref{MyLemma}, it holds
\begin{align*}
    \E^*\left[  \left|\left|\left| \left(\mathbf{H}\widehat{\boldsymbol{\Sigma}}^{\mathcal{P}}(t,t)\mathbf{H}^{\top}\right)^+ \right|\right|\right|^8\right] &\leq ||| \mathbf{H}^+|||^{16} \E^*\left[  \left|\left|\left| \widehat{\boldsymbol{\Sigma}}^{\mathcal{P}}(t,t)^{-1} \right|\right|\right|^8\right]
    \\&\leq ||| \mathbf{H}^+|||^{16} \sum\limits_{i=1}^k \left(\frac{n_i}{n}\right)^{8}\E^*\left[  \left(\widehat{{\gamma}}_i^{\mathcal{P}}(t,t) \right)^{-8}\right]
\end{align*}  on $B_n$ for all $t\in\T$ since $\widehat{\boldsymbol{\Sigma}}^{\mathcal{P}}(t,t)$ is almost surely positive definite on $B_n$.
It is well known that $(n_i-1) {\widehat{{\gamma}}_i^{\mathcal{P}}(t,t)}/{\widehat{{\gamma}}_i(t,t) } \sim \chi^2_{n_i-1}$ for all $t\in\T, i\in\{1,...,k\}$ with $\widehat{{\gamma}}_i(t,t) > 0$. Thus, $\left((n_i-1) {\widehat{{\gamma}}_i^{\mathcal{P}}(t,t)}/{\widehat{{\gamma}}_i(t,t) }\right)^{-1}$ follows an inverse gamma distribution with parameters $(n_i-1)/2$ and $1/2$ for all $t\in\T$, $i\in\{1,...,k\}$ with $\widehat{{\gamma}}_i(t,t) > 0$ and its eight moment is given by 
$  \left({\left({n_i-3} \right)\left({n_i-5} \right)\cdot ...\cdot \left({n_i-17}  \right)}\right)^{-1}$.
Hence, it holds
\begin{align*}
    \E^*\left[  \left(\widehat{{\gamma}}_i^{\mathcal{P}}(t,t) \right)^{-8}\right] &= \left(\widehat{{\gamma}}_i(t,t) \right)^{-8} \frac{(n_i-1)^8}{\left({n_i-3} \right)\left({n_i-5} \right)\cdot ...\cdot \left({n_i-17}  \right)}
\end{align*}
 for all $t\in\T, i\in\{1,...,k\}$ with $\widehat{{\gamma}}_i(t,t) > 0$.
Therefore, we have
\begin{align*}\begin{split}
    &\int_{\T}  \E^*\left[  \left|\left|\left| \left(\mathbf{H}\widehat{\boldsymbol{\Sigma}}^{\mathcal{P}}(t,t)\mathbf{H}^{\top}\right)^+ \right|\right|\right|^8\right]\;\mathrm{d}t
    \\&\leq  ||| \mathbf{H}^+|||^{16} \sum\limits_{i=1}^k \left(\frac{n_i}{n}\right)^{8}\frac{(n_i-1)^8}{\left({n_i-3} \right)\left({n_i-5} \right)\cdot ...\cdot \left({n_i-17}  \right)}\int_{\T}  \left(\widehat{{\gamma}}_i(t,t) \right)^{-8} \;\mathrm{d}t
    \\&\xrightarrow{P} ||| \mathbf{H}^+|||^{16} \sum\limits_{i=1}^k \tau_i^{8}\int_{\T}  \left({{\gamma}}_i(t,t) \right)^{-8} \;\mathrm{d}t =: G < \infty
\end{split}
\end{align*} on $B_n.$
Due to Lemma~\ref{UniformConvLemma}, it holds $ \P_{\mathrm{out}}\left( B_n \right) \to 1 $ 
and, thus, 
\begin{align}\label{eq:part1a}
    \P_{\mathrm{out}}\left( \int_{\T}  \E^*\left[  \left|\left|\left| \left(\mathbf{H}\widehat{\boldsymbol{\Sigma}}^{\mathcal{P}}(t,t)\mathbf{H}^{\top}\right)^+ \right|\right|\right|^8\right]\;\mathrm{d}t > G + \delta \right) \to 0
\end{align}
as $n\to\infty$ for all $\delta>0$.
\\Furthermore, it can be shown
\begin{align}\label{eq:Formel16}
    \E^*\left[\left|\left|\left|\widehat{\boldsymbol{\Sigma}}^{\mathcal{P}}(t,t)-{\boldsymbol{\Sigma}}(t,t)  \right|\right|\right|^4\right]
  &\leq 16\E^*\left[\left|\left|\left|\widehat{\boldsymbol{\Sigma}}^{\mathcal{P}}(t,t)-\widehat{\boldsymbol{\Sigma}}(t,t)  \right|\right|\right|^4  \right]
+ 16\left|\left|\left|\widehat{\boldsymbol{\Sigma}}(t,t)-{\boldsymbol{\Sigma}}(t,t)  \right|\right|\right|^4
\end{align}
for all $t\in\T$, where the second summand vanishes uniformly over $\T$ in probability as $n\to\infty$ due to Lemma~\ref{UniformConvLemma}.
Moreover, we have
\begin{align*}
    \E^*\left[\left|\left|\left|\widehat{\boldsymbol{\Sigma}}^{\mathcal{P}}(t,t)-\widehat{\boldsymbol{\Sigma}}(t,t)  \right|\right|\right|^4  \right] &= \E^*\left[\max\limits_{i\in\{1,...,k\}}\left\{\frac{n}{n_i}\left| \widehat{{\gamma}}_i^{\mathcal{P}}(t,t)-\widehat{{\gamma}}_i(t,t)  \right|\right\}^4  \right]
    \\ &\leq \sum\limits_{i=1}^k\left(\frac{n}{n_i}\right)^4\E^*\left[\left| \widehat{{\gamma}}_i^{\mathcal{P}}(t,t)-\widehat{{\gamma}}_i(t,t)  \right|^4  \right]
    \\ &\leq \sum\limits_{\substack{i=1\\ \widehat{\gamma}_i(t,t) > 0}}^k\left(\frac{n}{n_i}\right)^4\E^*\left[\left| \frac{\widehat{{\gamma}}_i^{\mathcal{P}}(t,t)}{\widehat{{\gamma}}_i(t,t) }-1 \right|^4  \right]\widehat{{\gamma}}_i(t,t)^4
\end{align*}
for all $t\in\T$  since $\widehat{\gamma}_i^{\mathcal{P}}(t,t)$ is almost surely zero whenever $\widehat{\gamma}_i(t,t) = 0$. 
Recall that $(n_i-1) {\widehat{{\gamma}}_i^{\mathcal{P}}(t,t)}/{\widehat{{\gamma}}_i(t,t) } \sim \chi^2_{n_i-1}$ for all $t\in\T, i\in\{1,...,k\}$ with $\widehat\gamma_i(t,t) > 0$. Since the fourth central moment of a $\chi^2_{\nu}$-distributed random variable is $12\nu(\nu+4)$ for $\nu\in\mathbb{N}$, it follows
$$ \E^*\left[\left| \frac{\widehat{{\gamma}}_i^{\mathcal{P}}(t,t)}{\widehat{{\gamma}}_i(t,t) }-1 \right|^4  \right] = 12\frac{n_i+3}{(n_i-1)^3} $$ for all $t\in\T, i\in\{1,...,k\}$ with $\widehat\gamma_i(t,t) > 0$. Hence, by the uniform convergence of $\widehat\gamma_i$ in probability, see Lemma~\ref{UniformConvLemma},
it follows
\begin{align*}
    \int_{\T}\E^*\left[\left|\left|\left|\widehat{\boldsymbol{\Sigma}}^{\mathcal{P}}(t,t)-\widehat{\boldsymbol{\Sigma}}(t,t)  \right|\right|\right|^4\right]\;\mathrm{d}t 
    &\leq
    12\sum\limits_{i=1}^k\left(\frac{n}{n_i}\right)^4\frac{n_i+3}{(n_i-1)^3}\int_{\T}\widehat{{\gamma}}_i(t,t)^4\;\mathrm{d}t 
    \xrightarrow{P} 0
\end{align*}
as $n\to\infty$ and, thus,
\begin{align}\label{eq:part2a}\begin{split}
    \int_{\T} \E^*\left[\left|\left|\left|\widehat{\boldsymbol{\Sigma}}^{\mathcal{P}}(t,t)-{\boldsymbol{\Sigma}}(t,t)  \right|\right|\right|^4\right]
  &\leq 16 \int_{\T}\E^*\left[\left|\left|\left|\widehat{\boldsymbol{\Sigma}}^{\mathcal{P}}(t,t)-\widehat{\boldsymbol{\Sigma}}(t,t)  \right|\right|\right|^4  \right]\;\mathrm{d}t 
+ 16 \int_{\T}\left|\left|\left|\widehat{\boldsymbol{\Sigma}}(t,t)-{\boldsymbol{\Sigma}}(t,t)  \right|\right|\right|^4\;\mathrm{d}t 
\\& \xrightarrow{P} 0
\end{split}
\end{align}
as $n\to\infty$ by (\ref{eq:Formel16}).
\\
Combining (\ref{eq:part1a}) and (\ref{eq:part2a}) in (\ref{eq:fetteFormel}), we get
\begin{align*}
    \E^*_{\mathrm{out}}\left[\int_{\T}\left|\left|\left|\left(\mathbf{H}\widehat{\boldsymbol{\Sigma}}^{\mathcal{P}}(t,t)\mathbf{H}^{\top}\right)^+ - \left(\mathbf{H}{\boldsymbol{\Sigma}}(t,t)\mathbf{H}^{\top}\right)^+\right|\right|\right|^2\;\mathrm{d}t\right] \xrightarrow{P} 0
\end{align*} as $n\to\infty$.
Together with (\ref{eq:DnP})--(\ref{eq:togetherPart}), it follows
\begin{align*}
   \P^*_{\mathrm{out}}\left( D_n^{\mathcal{P}}(\mathbf{H}) > \varepsilon \right) \xrightarrow{P} 0
\end{align*}
as $n\to\infty$  for all $\varepsilon > 0$ and, thus, \eqref{eq:remains}.
 \end{proof}

With minor adjustments, the proofs of Theorem~3 and 4 can be obtained analogously as the proofs of Theorem~1 and 2, respectively.

\begin{proof}[\textbf{Proof of Theorem~3}]
Lemma~\ref{UniformConvLemma} provides that $\widehat{\boldsymbol{\Sigma}}$ converges uniformly over \mbox{$\{(t,t)\in\T^2 \mid t\in\T\}$} in probability to $\boldsymbol{\Sigma}$.
Furthermore, Lemma~\ref{asy_CLT} ensures the convergence in distribution of $\sqrt{n} (\mathbf{H}\hat{\boldsymbol\eta} - \mathbf{c})$ to $\mathbf{z}$.
    By the continuous mapping theorem, it suffices to show that the map 
    $$ \mathbf{g}: \mathcal{L}_2^r(\T) \times \mathbb{D} \to (\R_0^+ \cup \{\infty\})^R, \quad \mathbf{g}(\widetilde{\mathbf{x}}, \widetilde{\boldsymbol\Sigma}) := \left(\int_{\T} \widetilde{\mathbf{x}}_{\ell}^{\top}(t) \left(\mathbf{H}_{\ell} \widetilde{\boldsymbol\Sigma}(t) \mathbf{H}_{\ell}^{\top} \right)^+ \widetilde{\mathbf{x}}_{\ell}(t) \;\mathrm{d}t\right)_{\ell\in\{1,...,R\}} $$ is continuous on $ \mathcal{L}_2^r(\T) \times \{\boldsymbol\Sigma \}$ with $\mathbb{D}$ as in the proof of Theorem~1.
    This holds whenever its component functions are continuous. The continuity of the component functions follows analogously as in the proof of Theorem~1.
\end{proof}

\begin{proof}[\textbf{Proof of Theorem~4}]
By Lemma~\ref{Para_CLT} and the subsequence criterion, every subsequence has a further subsequence such that
$ \sqrt{n}\mathbf{H}\widehat{\boldsymbol\eta}^{\mathcal{P}}$ converges in distribution almost surely along the subsequence to $\mathbf{z}$ in $\mathcal{L}_2^r(\T)$. Applying the continuous mapping theorem 
yields 
that $$ \left(\int_{\T}  n\left(\mathbf{H}_{\ell}{\boldsymbol\eta}^{\mathcal{P}}(t)\right)^{\top}\left(\mathbf{H}_{\ell}{\boldsymbol{\Sigma}}(t,t)\mathbf{H}_{\ell}^{\top}\right)^+\mathbf{H}_{\ell}\widehat{\boldsymbol\eta}^{\mathcal{P}}(t)\;\mathrm{d}t \right)_{\ell\in\{1,...,R\}}$$ converges in distribution almost surely along the latter subsequence to $ \left(\int_{\T} \mathbf{z}_{\ell}^{\top}(t)(\mathbf{H}_{\ell}\boldsymbol{\Sigma}(t,t)\mathbf{H}_{\ell}^{\top})^+\mathbf{z}_{\ell}(t) \,\mathrm{ d }t\right)_{\ell\in\{1,...,R\}}$. Hence, by applying the subsequence criterion again, it follows
\begin{align*}
    &\left(\int_{\T}  n\left(\mathbf{H}_{\ell}\widehat{\boldsymbol\eta}^{\mathcal{P}}(t)\right)^{\top}\left(\mathbf{H}_{\ell}{\boldsymbol{\Sigma}}(t,t)\mathbf{H}_{\ell}^{\top}\right)^+\mathbf{H}_{\ell}\widehat{\boldsymbol\eta}^{\mathcal{P}}(t)\;\mathrm{d}t\right)_{\ell\in\{1,...,R\}} \\&\xrightarrow{d^*} \left(\int_{\T} \mathbf{z}_{\ell}^{\top}(t)(\mathbf{H}_{\ell}\boldsymbol{\Sigma}(t,t)\mathbf{H}_{\ell}^{\top})^+\mathbf{z}_{\ell}(t) \,\mathrm{ d }t\right)_{\ell\in\{1,...,R\}}
\end{align*} as $n\to\infty$.\\
Consequently, it remains to show 
\begin{align*}
D_{n}^{\mathcal{P}}(\mathbf{H}_{\ell}) = \left|  T_n^{\mathcal{P}}(\mathbf{H}_{\ell}) - \int_{\T}  n\left(\mathbf{H}_{\ell}\widehat{\boldsymbol\eta}^{\mathcal{P}}(t)\right)^{\top}\left(\mathbf{H}_{\ell}{\boldsymbol{\Sigma}}(t,t)\mathbf{H}_{\ell}^{\top}\right)^+\mathbf{H}_{\ell}\widehat{\boldsymbol\eta}^{\mathcal{P}}(t)\;\mathrm{d}t\right|   \xrightarrow{P^*} 0
\end{align*} as $n\to\infty$ for all $\ell\in\{1,...,R\}.$ This can be shown analogously as in the proof of Theorem~2.
\end{proof}

\FloatBarrier
\section{More Details on the Data Examples}\label{more-details-on-the-data-examples}

In this section, we first present the plots of the empirical covariance functions for Example~1 based on the chlorine concentration data (Section~\ref{plots-of-the-empirical-covariances-example-1}) and for Example~2 based on the electrocardiogram data (Section~\ref{plots-of-the-empirical-covariances-example-2}). For the chlorine concentration data, the sample covariance functions seem to be very similar, which can not be said for the electrocardiogram data. To verify these issues, we used the tests for equality of covariance functions by \cite{Guoetal2018, Guoetal2019}. The results are presented in Table~\ref{table_cov_fun} and confirm our observations.

\begin{table}[!h]
\caption[]{P-values of tests for equality of covariance functions by Guo et al. (2018, $T_{\max,rp}$) and by Guo et al. (2019, GPF$_{nv}$, GPF$_{rp}$, $F_{\max,rp}$) for chlorine concentration and electrocardiogram data}\label{table_cov_fun}
\centering
\begin{tabular}{l|rrrr}\hline
Data set&$T_{\max,rp}$&GPF$_{nv}$&GPF$_{rp}$&$F_{\max,rp}$\\\hline
Chlorine concentration&0.515&0.310&0.401&0.427\\
Electrocardiogram&0.000&0.000&0.000&0.000\\
\hline
\end{tabular}
\end{table}

Now we describe and present the results of the simulation studies based on both real data examples. To mimic the data given in the sets, we generated the simulation data from the multivariate normal distribution with the following specifications:
\begin{itemize}
\item the samples sizes from the data example, i.e., $n_i=25$, $i=1,2,3$ for the chlorine concentration data, and $n_1=292$, $n_2=177$, $n_3=10$, and $n_4=19$ for the electrocardiogram data,
\item for checking the type-I error control, in each group, the mean and covariance matrix were set to the sample mean function and sample covariance function of the pooled data,
\item for power investigation, the mean and covariance matrix in the $i$-th group was equal to the sample mean function and sample covariance function for the $i$-th sample from the data set.
\end{itemize}
We have calculated the empirical FWER, sizes and powers for multiple contrast tests. We considered the Tukey-type contrast matrix. The results of these simulation studies are presented in Sections~\ref{simulation-study-inspired-by-example-1}-\ref{simulation-study-inspired-by-example-2}.

\newpage
\FloatBarrier
\subsection{Plots of the Empirical Covariances -- Example
1}\label{plots-of-the-empirical-covariances-example-1}

\begin{figure}[!h]
\centering
\includegraphics[width= 0.95\textwidth,height=0.9\textheight]{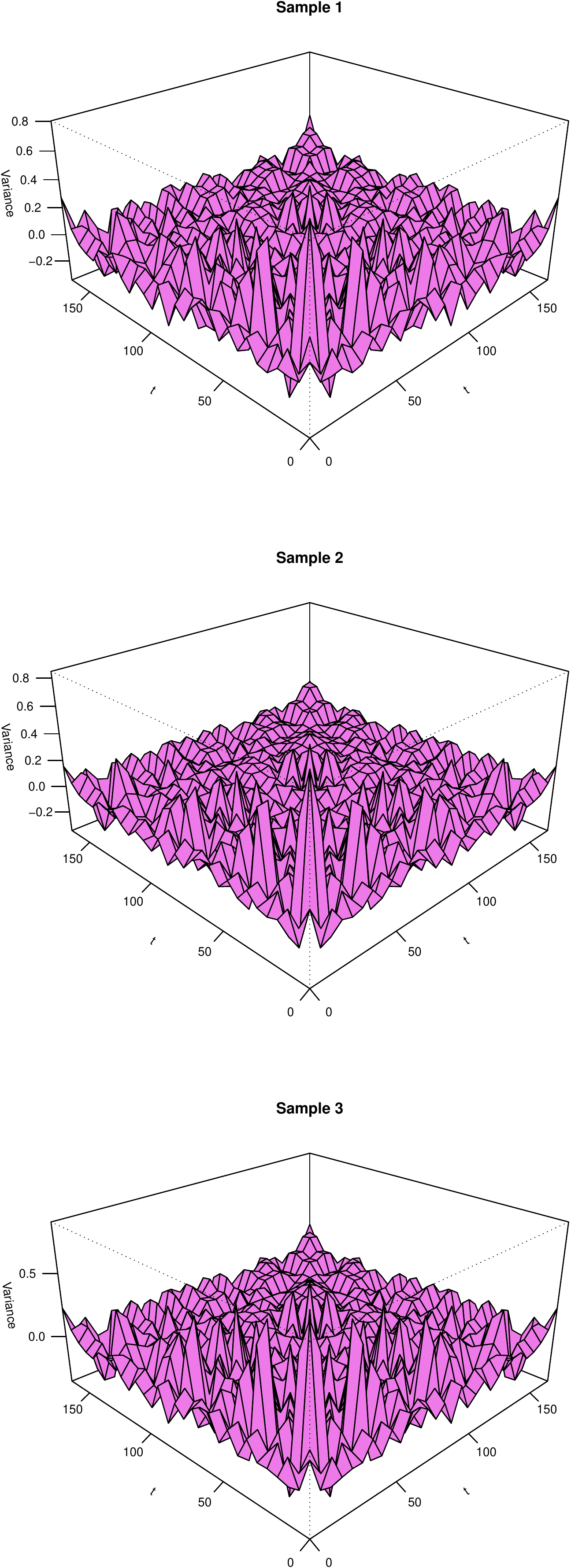}
\caption{Empirical covariance functions for the chlorine concentration data}
\end{figure}

\newpage
\FloatBarrier
\subsection{Plots of the Empirical Covariances -- Example
2}\label{plots-of-the-empirical-covariances-example-2}

\begin{figure}[!h]
\centering
\includegraphics[width= 0.95\textwidth,height=0.9\textheight]{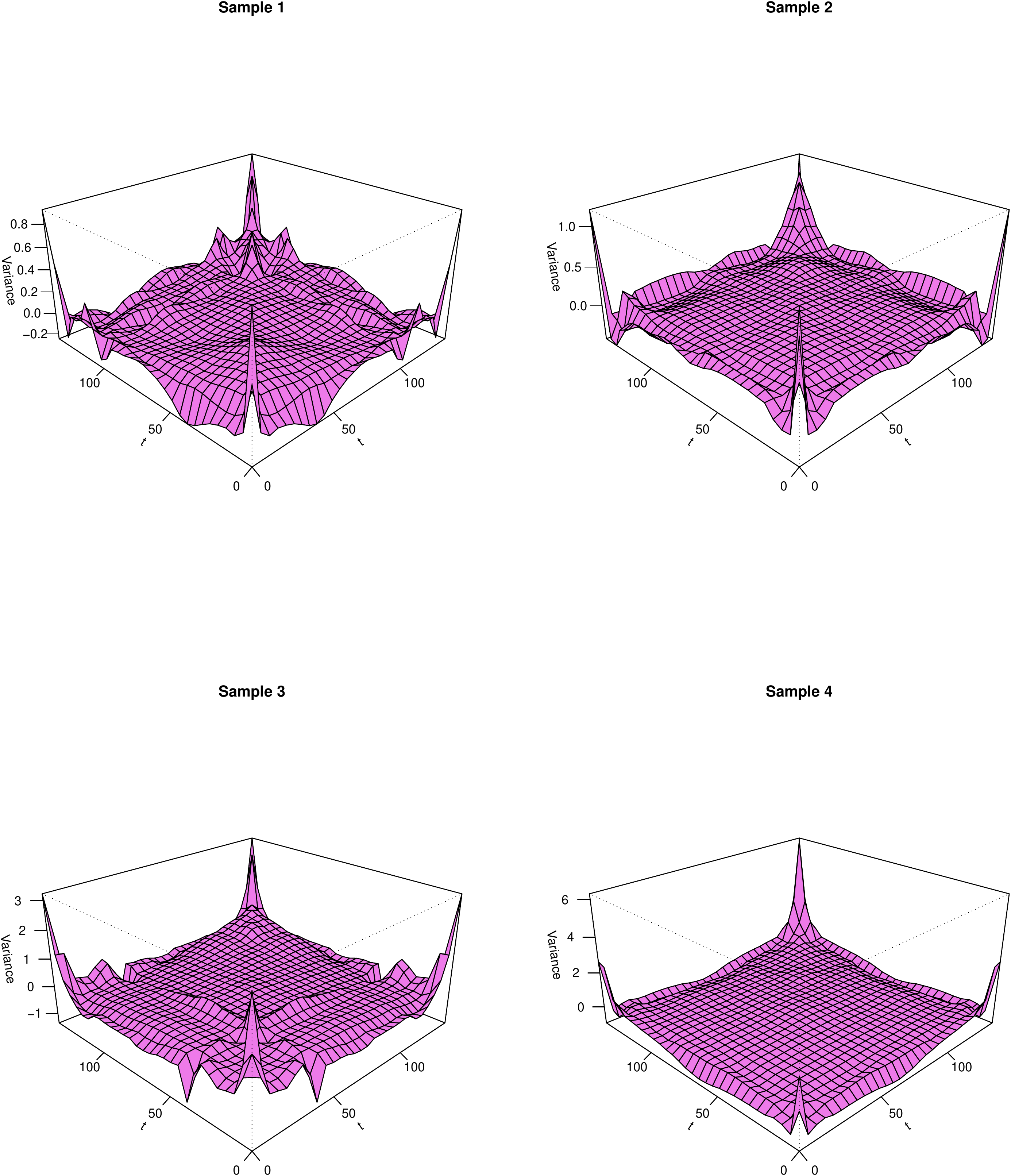}
\caption{Empirical covariance functions for the electrocardiogram data}
\end{figure}

\newpage
\FloatBarrier
\subsection{Simulation Study inspired by Example
1}\label{simulation-study-inspired-by-example-1}

\begin{longtable}[t]{rr|rrrrrrr}
\caption[Empirical FWER, sizes and powers of all tests obtained for the Tukey constrasts in the simulation based on the chlorine concentration data]{\label{tab:unnamed-chunk-4}Empirical FWER, sizes and powers (as percentages) of all tests obtained for the Tukey constrasts in the simulation based on the chlorine concentration data ($\mathcal H_{0,1}: \eta_1 \equiv \eta_2$, $\mathcal H_{0,2}: \eta_1 \equiv \eta_3$, $\mathcal H_{0,3}: \eta_2 \equiv \eta_3$)}\\
\hline
&&Fmax&GPF&L2b&Fb&CAFB&GPH&mGPH\\
\hline
\endfirsthead
\caption[]{Empirical FWER, sizes and powers (as percentages) of all tests obtained for the Tukey constrasts in the simulation based on the chlorine concentration data ($\mathcal H_{0,1}: \eta_1 \equiv \eta_2$, $\mathcal H_{0,2}: \eta_1 \equiv \eta_3$, $\mathcal H_{0,3}: \eta_2 \equiv \eta_3$) \textit{(continued)}}\\
\hline
&&Fmax&GPF&L2b&Fb&CAFB&GPH&mGPH\\
\hline
\endhead
FWER&&4.95&4.95&6.35&5.15&4.20&4.80&5.35\\
Empirical sizes&$\mathcal{H}_{0,1}$&2.10&1.65&2.20&1.80&1.55&1.50&1.70\\
&$\mathcal{H}_{0,2}$&2.20&2.10&2.80&2.00&1.75&2.10&2.35\\
&$\mathcal{H}_{0,3}$&1.10&1.55&2.05&1.80&1.30&1.45&1.60\\
Empirical powers&$\mathcal{H}_{0,1}$&70.20&66.10&40.70&35.00&54.30&65.95&67.10\\
&$\mathcal{H}_{0,2}$&26.90&16.65&10.25&7.60&17.50&15.90&17.35\\
&$\mathcal{H}_{0,3}$&30.10&26.95&28.55&25.10&21.00&27.65&28.75\\
\hline
\end{longtable}

\subsection{Simulation Study inspired by Example
2}\label{simulation-study-inspired-by-example-2}

\begin{longtable}[t]{rr|rrrrrrr}
\caption[Empirical FWER, sizes and powers of all tests obtained for the Tukey constrasts in the simulation based on the electrocardiogram data]{\label{tab:unnamed-chunk-5}Empirical FWER, sizes and powers (as percentages) of all tests obtained for the Tukey constrasts in the simulation based on the electrocardiogram data ($\mathcal H_{0,1}: \eta_1 \equiv \eta_2$, $\mathcal H_{0,2}: \eta_1 \equiv \eta_3$, $\mathcal H_{0,3}: \eta_1 \equiv \eta_4$, $\mathcal H_{0,4}: \eta_2 \equiv \eta_3$, $\mathcal H_{0,5}: \eta_2 \equiv \eta_4$, $\mathcal H_{0,6}: \eta_3 \equiv \eta_4$)}\\
\hline
&&Fmax&GPF&L2b&Fb&CAFB&GPH&mGPH\\
\hline
\endfirsthead
\caption[]{Empirical FWER, sizes and powers (as percentages) of all tests obtained for the Tukey constrasts in the simulation based on the electrocardiogram data ($\mathcal H_{0,1}: \eta_1 \equiv \eta_2$, $\mathcal H_{0,2}: \eta_1 \equiv \eta_3$, $\mathcal H_{0,3}: \eta_1 \equiv \eta_4$, $\mathcal H_{0,4}: \eta_2 \equiv \eta_3$, $\mathcal H_{0,5}: \eta_2 \equiv \eta_4$, $\mathcal H_{0,6}: \eta_3 \equiv \eta_4$) \textit{(continued)}}\\
\hline
&&Fmax&GPF&L2b&Fb&CAFB&GPH&mGPH\\
\hline
\endhead
FWER&&57.50&35.80&6.90&3.40&5.20&3.75&5.45\\
Empirical sizes&$\mathcal{H}_{0,1}$&1.05&0.70&0.90&0.85&0.60&0.85&1.20\\
&$\mathcal{H}_{0,2}$&26.15&17.90&2.45&0.45&1.45&1.00&1.45\\
&$\mathcal{H}_{0,3}$&26.95&16.30&2.45&1.55&1.00&1.35&1.90\\
&$\mathcal{H}_{0,4}$&38.65&21.45&2.20&0.35&1.65&1.00&1.30\\
&$\mathcal{H}_{0,5}$&14.30&10.55&2.45&1.25&0.90&1.00&1.90\\
&$\mathcal{H}_{0,6}$&2.35&1.85&1.45&0.60&1.05&0.95&1.45\\
Empirical powers&$\mathcal{H}_{0,1}$&100.00&100.00&100.00&100.00&100.00&100.00&100.00\\
&$\mathcal{H}_{0,2}$&100.00&100.00&100.00&99.65&100.00&99.30&99.90\\
&$\mathcal{H}_{0,3}$&100.00&100.00&100.00&99.60&100.00&100.00&100.00\\
&$\mathcal{H}_{0,4}$&100.00&100.00&85.00&45.05&97.25&55.30&65.75\\
&$\mathcal{H}_{0,5}$&93.90&89.60&71.15&60.00&69.20&55.60&62.15\\
&$\mathcal{H}_{0,6}$&99.85&57.50&76.25&53.55&96.50&40.85&49.45\\
\hline
\end{longtable}

\section{Detailed Results from the Simulation
Study}\label{detailed-results-from-the-simulation-study}

In this section, we present all alternative hypotheses considered and the results of the simulation studies. 

The scenarios in the six alternative hypotheses are as follows:
\begin{itemize}
    \item[A1.] $\mu_1(t) = \mu_2(t) = \mu_3(t) = 0$ and $\mu_4(t)=2\frac{t-1}{J-1}$,
    \item[A2.] $\mu_1(t) = \mu_2(t) = \mu_3(t) = 0$ and $\mu_4(t)=1.5$,
    \item[A3.] $\mu_1(t) = 0, \mu_2(t) = 0.75, \mu_3(t) = 1.5$ and $\mu_4(t)=2.25$,
    \item[A4.] $\mu_1(t) = 0, \mu_2(t) = \frac{t-1}{J-1}, \mu_3(t) = 2\frac{t-1}{J-1}$ and $\mu_4(t)=3\frac{t-1}{J-1}$,
    \item[A5.] $\mu_1(t) = \mu_2(t) = \mu_3(t) = 0$ and $\mu_4(t)=2\frac{J-t}{J}$,
    \item[A6.] $\mu_1(t) = 0, \mu_2(t) = \frac{J-t}{J}, \mu_3(t) = 2\frac{J-t}{J}$ and $\mu_4(t)=3\frac{J-t}{J}$  
\end{itemize}
for all $t\in[0,1]$.

The results of the simulation studies are presented in tables and figures. The lists of them are given on the next page (Sections~\ref{list_tables}-\ref{list_figures}). The results for the usual simulation (i.e., without scaling function) are depicted in Section~\ref{results-without-scaling-function}, while those with scaling function are given in Section~\ref{results-with-scaling-function}. In column $\mathcal{H}$ in the tables, $\mathcal{H}_i$ represents the $i$-th hypothesis for given kind of contrasts, i.e.,
\begin{itemize}
\item for Dunnett contrasts: $\mathcal H_{0,1}: \eta_1 \equiv \eta_2$, $\mathcal H_{0,2}: \eta_1 \equiv \eta_3$, $\mathcal H_{0,3}: \eta_2 \equiv \eta_3$,
\item for Tukey contrasts: $\mathcal H_{0,1}: \eta_1 \equiv \eta_2$, $\mathcal H_{0,2}: \eta_1 \equiv \eta_3$, $\mathcal H_{0,3}: \eta_1 \equiv \eta_4$, $\mathcal H_{0,4}: \eta_2 \equiv \eta_3$, $\mathcal H_{0,5}: \eta_2 \equiv \eta_4$, $\mathcal H_{0,6}: \eta_3 \equiv \eta_4$.
\end{itemize}
Moreover, the column D denotes the distribution of the generated data, i.e., N - the normal distribution, t - the $t_5$-distribution and $\chi^2$ - the $\chi_t^2$-distribution.

\subsection{List of Tables}
\label{list_tables}

\listoftables

\subsection{List of Figures}
\label{list_figures}

\listoffigures

\subsection{Results without Scaling Function}\label{results-without-scaling-function}

\begin{longtable}[t]{rrrr|rrrrrrr}
\caption[Empirical FWER and sizes of all tests obtained for the Dunnett constrasts]{\label{tab:unnamed-chunk-10}Empirical FWER and sizes (as percentages) of all tests obtained for the Dunnett constrasts (D - distribution, $(\lambda_1,\lambda_2,\lambda_3,\lambda_4)$: (1,1,1,1) - homoscedastic case, (1,1.25,1.5,1.75) - heteroscedastic case (positive pairing), (1.75,1.5,1.25,1) - heteroscedastic case (negative pairing))}\\
\hline
D&$(n_1,n_2,n_3,n_4)$&$(\lambda_1,\lambda_2,\lambda_3,\lambda_4)$&$\mathcal{H}$&Fmax&GPF&L2b&Fb&CAFB&GPH&mGPH\\
\hline
\endfirsthead
\caption[]{Empirical FWER and sizes (as percentages) of all tests obtained for the Dunnett constrasts (D - distribution, $(\lambda_1,\lambda_2,\lambda_3,\lambda_4)$: (1,1,1,1) - homoscedastic case, (1,1.25,1.5,1.75) - heteroscedastic case (positive pairing), (1.75,1.5,1.25,1) - heteroscedastic case (negative pairing)) \textit{(continued)}}\\
\hline
D&$(n_1,n_2,n_3,n_4)$&$(\lambda_1,\lambda_2,\lambda_3,\lambda_4)$&$\mathcal{H}$&Fmax&GPF&L2b&Fb&CAFB&GPH&mGPH\\
\hline
\endhead
N&(15,20,25,30)&(1,1,1,1)&FWER&4.20&3.15&4.85&3.00&6.00&2.85&3.70\\
&&&$\mathcal{H}_{0,1}$&1.40&1.15&1.80&0.95&2.05&1.20&1.55\\
&&&$\mathcal{H}_{0,2}$&1.70&1.55&2.20&1.30&2.05&1.40&1.70\\
&&&$\mathcal{H}_{0,3}$&1.45&1.20&2.00&1.25&2.60&1.00&1.35\\
N&(15,20,25,30)&(1,1.25,1.5,1.75)&FWER&3.15&1.85&5.60&3.95&5.85&4.30&4.80\\
&&&$\mathcal{H}_{0,1}$&2.20&1.20&2.35&1.45&2.00&1.55&1.90\\
&&&$\mathcal{H}_{0,2}$&0.90&0.60&1.70&1.15&2.65&1.40&1.50\\
&&&$\mathcal{H}_{0,3}$&0.35&0.10&2.15&1.65&1.45&1.70&1.85\\
N&(15,20,25,30)&(1.75,1.5,1.25,1)&FWER&12.90&8.80&4.40&2.25&5.65&2.40&3.70\\
&&&$\mathcal{H}_{0,1}$&1.75&2.15&2.35&1.25&1.90&1.25&1.90\\
&&&$\mathcal{H}_{0,2}$&4.00&2.85&1.45&0.85&2.05&1.00&1.45\\
&&&$\mathcal{H}_{0,3}$&9.75&6.65&1.75&0.70&2.20&1.00&1.25\\
N&(30,40,50,60)&(1,1,1,1)&FWER&4.80&3.55&3.90&3.15&5.05&3.45&4.05\\
&&&$\mathcal{H}_{0,1}$&1.95&1.60&1.85&1.50&1.90&1.40&1.70\\
&&&$\mathcal{H}_{0,2}$&1.75&1.40&1.55&1.25&2.10&1.30&1.55\\
&&&$\mathcal{H}_{0,3}$&1.75&1.15&1.35&1.10&1.50&1.30&1.50\\
N&(30,40,50,60)&(1,1.25,1.5,1.75)&FWER&1.40&1.45&4.75&3.90&5.40&3.90&4.30\\
&&&$\mathcal{H}_{0,1}$&0.80&0.75&1.85&1.40&1.75&1.40&1.45\\
&&&$\mathcal{H}_{0,2}$&0.35&0.55&1.90&1.65&2.25&1.45&1.65\\
&&&$\mathcal{H}_{0,3}$&0.25&0.25&1.65&1.45&1.70&1.50&1.70\\
N&(30,40,50,60)&(1.75,1.5,1.25,1)&FWER&11.35&10.25&3.45&2.60&4.45&2.85&4.15\\
&&&$\mathcal{H}_{0,1}$&2.10&1.55&1.50&1.20&1.90&1.20&1.90\\
&&&$\mathcal{H}_{0,2}$&3.55&3.40&1.75&1.25&1.60&1.25&1.75\\
&&&$\mathcal{H}_{0,3}$&8.80&8.55&1.55&1.10&1.75&1.35&1.95\\
N&(60,80,100,120)&(1,1,1,1)&FWER&4.20&4.35&4.45&3.80&4.70&4.35&5.30\\
&&&$\mathcal{H}_{0,1}$&1.85&1.65&1.85&1.60&2.10&1.75&2.15\\
&&&$\mathcal{H}_{0,2}$&1.65&1.50&1.40&1.25&1.60&1.45&1.80\\
&&&$\mathcal{H}_{0,3}$&1.45&1.80&2.00&1.60&1.35&1.85&2.30\\
N&(60,80,100,120)&(1,1.25,1.5,1.75)&FWER&1.80&1.20&4.00&3.65&3.90&3.80&3.95\\
&&&$\mathcal{H}_{0,1}$&1.25&0.75&1.45&1.20&1.05&1.20&1.25\\
&&&$\mathcal{H}_{0,2}$&0.45&0.45&1.10&1.05&1.50&1.25&1.35\\
&&&$\mathcal{H}_{0,3}$&0.10&0.00&1.75&1.70&1.45&1.50&1.55\\
N&(60,80,100,120)&(1.75,1.5,1.25,1)&FWER&12.30&10.90&3.75&3.40&4.50&3.10&4.90\\
&&&$\mathcal{H}_{0,1}$&1.90&1.65&1.30&1.10&1.80&1.15&1.75\\
&&&$\mathcal{H}_{0,2}$&3.70&4.05&2.10&1.80&1.70&1.45&2.45\\
&&&$\mathcal{H}_{0,3}$&9.95&8.35&1.55&1.30&1.40&1.50&2.30\\
t&(15,20,25,30)&(1,1,1,1)&FWER&4.05&3.25&4.20&2.45&4.55&2.90&3.40\\
&&&$\mathcal{H}_{0,1}$&1.20&1.05&1.80&0.75&1.35&0.90&1.15\\
&&&$\mathcal{H}_{0,2}$&1.50&1.45&1.60&0.90&1.95&0.90&1.30\\
&&&$\mathcal{H}_{0,3}$&1.65&1.40&1.60&1.20&1.65&1.40&1.55\\
t&(15,20,25,30)&(1,1.25,1.5,1.75)&FWER&2.65&1.55&4.65&2.95&4.25&3.25&3.60\\
&&&$\mathcal{H}_{0,1}$&1.35&1.00&1.85&1.00&1.50&1.30&1.35\\
&&&$\mathcal{H}_{0,2}$&0.80&0.45&1.45&0.90&1.55&1.20&1.30\\
&&&$\mathcal{H}_{0,3}$&0.55&0.20&1.80&1.35&1.45&1.05&1.30\\
t&(15,20,25,30)&(1.75,1.5,1.25,1)&FWER&11.80&9.30&3.25&1.75&5.55&2.05&3.30\\
&&&$\mathcal{H}_{0,1}$&1.90&1.45&1.30&0.60&1.85&0.80&1.40\\
&&&$\mathcal{H}_{0,2}$&4.85&3.50&1.50&0.70&1.90&0.75&1.40\\
&&&$\mathcal{H}_{0,3}$&8.70&6.95&1.35&0.80&2.25&0.90&1.30\\
t&(30,40,50,60)&(1,1,1,1)&FWER&4.70&3.50&4.10&3.35&4.65&3.10&4.10\\
&&&$\mathcal{H}_{0,1}$&1.30&1.30&1.55&1.20&1.50&1.25&1.55\\
&&&$\mathcal{H}_{0,2}$&2.00&1.40&1.70&1.30&1.85&1.20&1.60\\
&&&$\mathcal{H}_{0,3}$&1.80&1.25&1.20&1.05&1.55&1.10&1.45\\
t&(30,40,50,60)&(1,1.25,1.5,1.75)&FWER&1.85&1.40&4.35&3.60&5.15&4.00&4.25\\
&&&$\mathcal{H}_{0,1}$&1.10&0.95&1.30&1.15&1.75&1.20&1.30\\
&&&$\mathcal{H}_{0,2}$&0.60&0.35&1.55&1.20&1.75&1.45&1.60\\
&&&$\mathcal{H}_{0,3}$&0.20&0.15&1.75&1.45&1.80&1.55&1.60\\
t&(30,40,50,60)&(1.75,1.5,1.25,1)&FWER&12.90&10.10&3.25&2.40&4.50&2.50&3.75\\
&&&$\mathcal{H}_{0,1}$&2.05&1.40&1.10&0.90&1.35&1.05&1.45\\
&&&$\mathcal{H}_{0,2}$&4.45&3.80&1.60&1.15&1.80&1.20&1.70\\
&&&$\mathcal{H}_{0,3}$&9.30&8.10&1.45&1.00&1.65&0.95&1.50\\
t&(60,80,100,120)&(1,1,1,1)&FWER&4.25&4.15&4.40&3.95&4.20&3.90&4.50\\
&&&$\mathcal{H}_{0,1}$&1.65&1.85&2.05&1.90&1.90&1.85&2.20\\
&&&$\mathcal{H}_{0,2}$&1.80&1.50&1.70&1.50&1.20&1.50&1.65\\
&&&$\mathcal{H}_{0,3}$&1.35&1.45&1.45&1.30&1.40&1.35&1.60\\
t&(60,80,100,120)&(1,1.25,1.5,1.75)&FWER&2.25&1.30&4.25&3.85&3.95&3.95&4.45\\
&&&$\mathcal{H}_{0,1}$&1.15&0.90&1.50&1.40&0.95&1.35&1.50\\
&&&$\mathcal{H}_{0,2}$&0.85&0.35&1.40&1.30&1.70&1.30&1.40\\
&&&$\mathcal{H}_{0,3}$&0.25&0.05&1.60&1.30&1.40&1.50&1.75\\
t&(60,80,100,120)&(1.75,1.5,1.25,1)&FWER&11.55&10.40&4.15&3.60&4.00&3.75&4.75\\
&&&$\mathcal{H}_{0,1}$&1.65&1.95&1.70&1.35&1.40&1.40&1.80\\
&&&$\mathcal{H}_{0,2}$&3.80&3.80&1.90&1.80&1.60&1.65&2.25\\
&&&$\mathcal{H}_{0,3}$&8.55&7.80&1.40&1.20&1.25&1.55&2.05\\
$\chi^2$&(15,20,25,30)&(1,1,1,1)&FWER&4.10&3.25&4.90&3.05&6.15&2.95&3.90\\
&&&$\mathcal{H}_{0,1}$&1.80&1.35&1.90&1.25&1.75&1.15&1.60\\
&&&$\mathcal{H}_{0,2}$&1.20&1.35&1.95&1.05&2.45&1.10&1.30\\
&&&$\mathcal{H}_{0,3}$&1.45&1.05&1.70&0.95&2.65&1.05&1.45\\
$\chi^2$&(15,20,25,30)&(1,1.25,1.5,1.75)&FWER&2.05&1.60&5.20&3.35&5.40&3.55&4.15\\
&&&$\mathcal{H}_{0,1}$&1.10&1.15&2.25&1.55&2.00&1.55&1.75\\
&&&$\mathcal{H}_{0,2}$&0.80&0.35&1.50&0.85&1.90&1.00&1.10\\
&&&$\mathcal{H}_{0,3}$&0.20&0.15&1.90&1.00&1.75&1.15&1.45\\
$\chi^2$&(15,20,25,30)&(1.75,1.5,1.25,1)&FWER&12.50&8.85&3.75&2.20&5.65&2.15&3.30\\
&&&$\mathcal{H}_{0,1}$&2.20&1.35&1.70&1.00&1.45&0.75&1.60\\
&&&$\mathcal{H}_{0,2}$&4.30&2.80&1.80&1.10&2.30&1.15&1.65\\
&&&$\mathcal{H}_{0,3}$&9.60&7.40&1.70&0.90&2.90&1.10&1.35\\
$\chi^2$&(30,40,50,60)&(1,1,1,1)&FWER&4.20&3.80&4.60&3.85&5.15&4.10&4.65\\
&&&$\mathcal{H}_{0,1}$&1.40&1.25&1.70&1.35&1.80&1.55&1.80\\
&&&$\mathcal{H}_{0,2}$&1.60&1.70&2.10&1.80&1.90&1.65&1.95\\
&&&$\mathcal{H}_{0,3}$&1.75&1.60&1.90&1.45&1.80&1.50&1.70\\
$\chi^2$&(30,40,50,60)&(1,1.25,1.5,1.75)&FWER&1.80&1.30&4.40&3.75&4.15&3.90&4.20\\
&&&$\mathcal{H}_{0,1}$&0.85&0.80&1.40&1.15&1.50&1.20&1.35\\
&&&$\mathcal{H}_{0,2}$&0.75&0.25&1.70&1.25&1.70&1.35&1.45\\
&&&$\mathcal{H}_{0,3}$&0.25&0.25&1.55&1.50&1.20&1.45&1.50\\
$\chi^2$&(30,40,50,60)&(1.75,1.5,1.25,1)&FWER&11.85&9.85&3.35&2.80&5.30&2.95&4.20\\
&&&$\mathcal{H}_{0,1}$&1.95&1.65&1.35&1.00&1.95&1.15&1.65\\
&&&$\mathcal{H}_{0,2}$&4.75&3.90&1.70&1.30&1.80&1.45&2.35\\
&&&$\mathcal{H}_{0,3}$&8.95&7.75&1.40&1.25&1.95&1.30&1.90\\
$\chi^2$&(60,80,100,120)&(1,1,1,1)&FWER&4.45&4.00&4.15&3.60&4.05&4.20&5.15\\
&&&$\mathcal{H}_{0,1}$&1.65&1.70&2.05&1.80&1.40&1.85&2.25\\
&&&$\mathcal{H}_{0,2}$&1.75&1.35&1.30&1.15&1.20&1.20&1.65\\
&&&$\mathcal{H}_{0,3}$&1.60&1.45&1.45&1.20&1.70&1.60&1.80\\
$\chi^2$&(60,80,100,120)&(1,1.25,1.5,1.75)&FWER&1.55&1.55&4.70&4.05&4.00&3.55&4.05\\
&&&$\mathcal{H}_{0,1}$&1.00&0.80&1.35&1.15&1.45&0.95&1.10\\
&&&$\mathcal{H}_{0,2}$&0.30&0.50&1.80&1.55&1.45&1.40&1.45\\
&&&$\mathcal{H}_{0,3}$&0.25&0.25&1.85&1.55&1.25&1.45&1.75\\
$\chi^2$&(60,80,100,120)&(1.75,1.5,1.25,1)&FWER&12.40&10.55&3.45&3.15&4.10&3.15&4.55\\
&&&$\mathcal{H}_{0,1}$&2.40&1.30&1.50&1.45&1.15&1.15&1.80\\
&&&$\mathcal{H}_{0,2}$&4.55&3.95&1.50&1.30&1.95&1.50&2.45\\
&&&$\mathcal{H}_{0,3}$&9.15&8.20&1.55&1.45&1.70&1.55&1.90\\
\hline
\end{longtable}

\begin{figure}
\centering
\includegraphics[width= 0.95\textwidth,height=0.9\textheight]{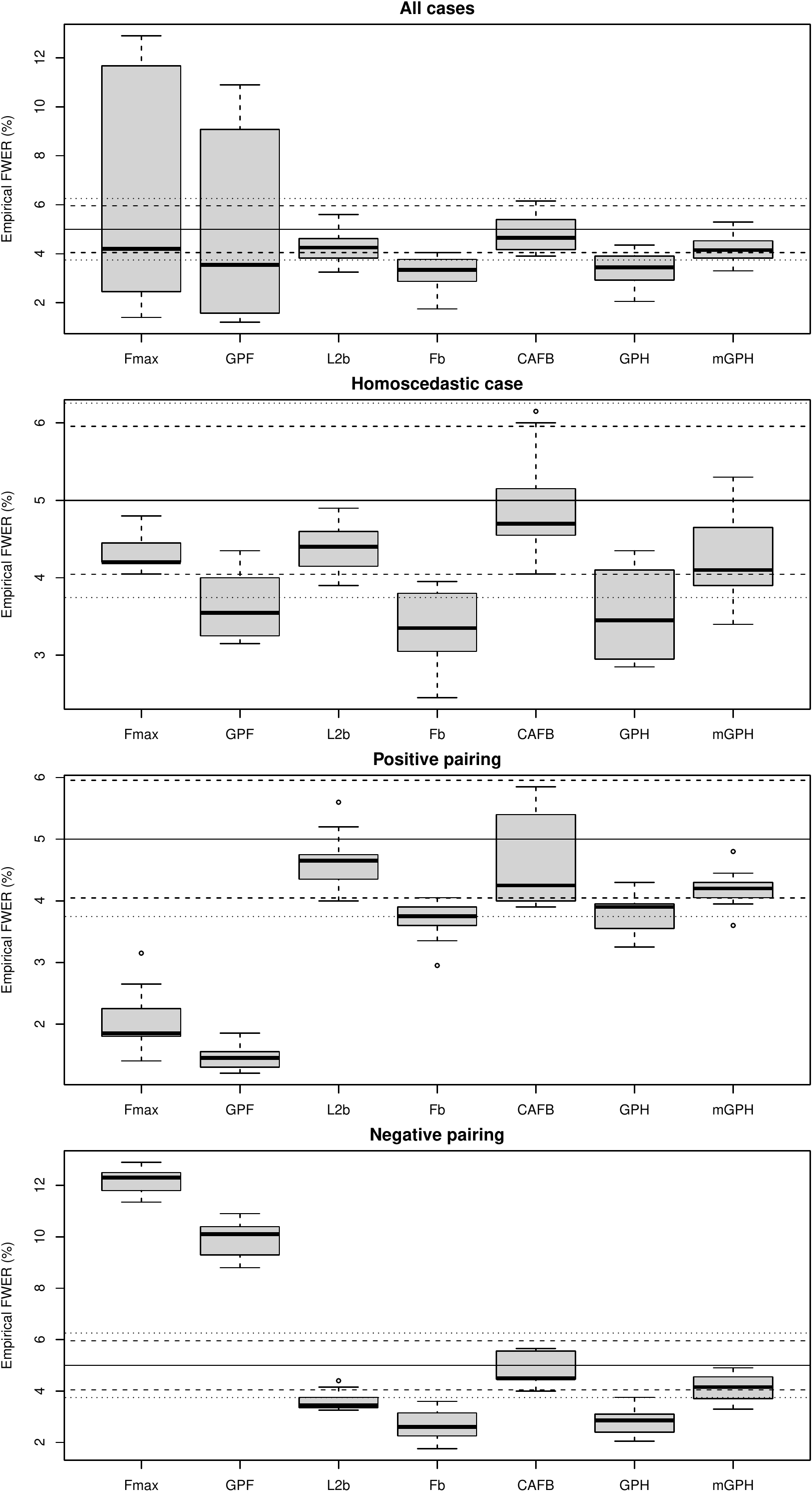}
\caption[Box-and-whisker plots for the empirical FWER of all tests obtained for the Dunnett constrasts for homoscedastic and
heteroscedastic cases]{Box-and-whisker plots for the empirical FWER (as percentages) of all tests obtained for the Dunnett constrasts for homoscedastic and heteroscedastic cases}
\end{figure}

\begin{figure}
\centering
\includegraphics[width= 0.95\textwidth,height=0.9\textheight]{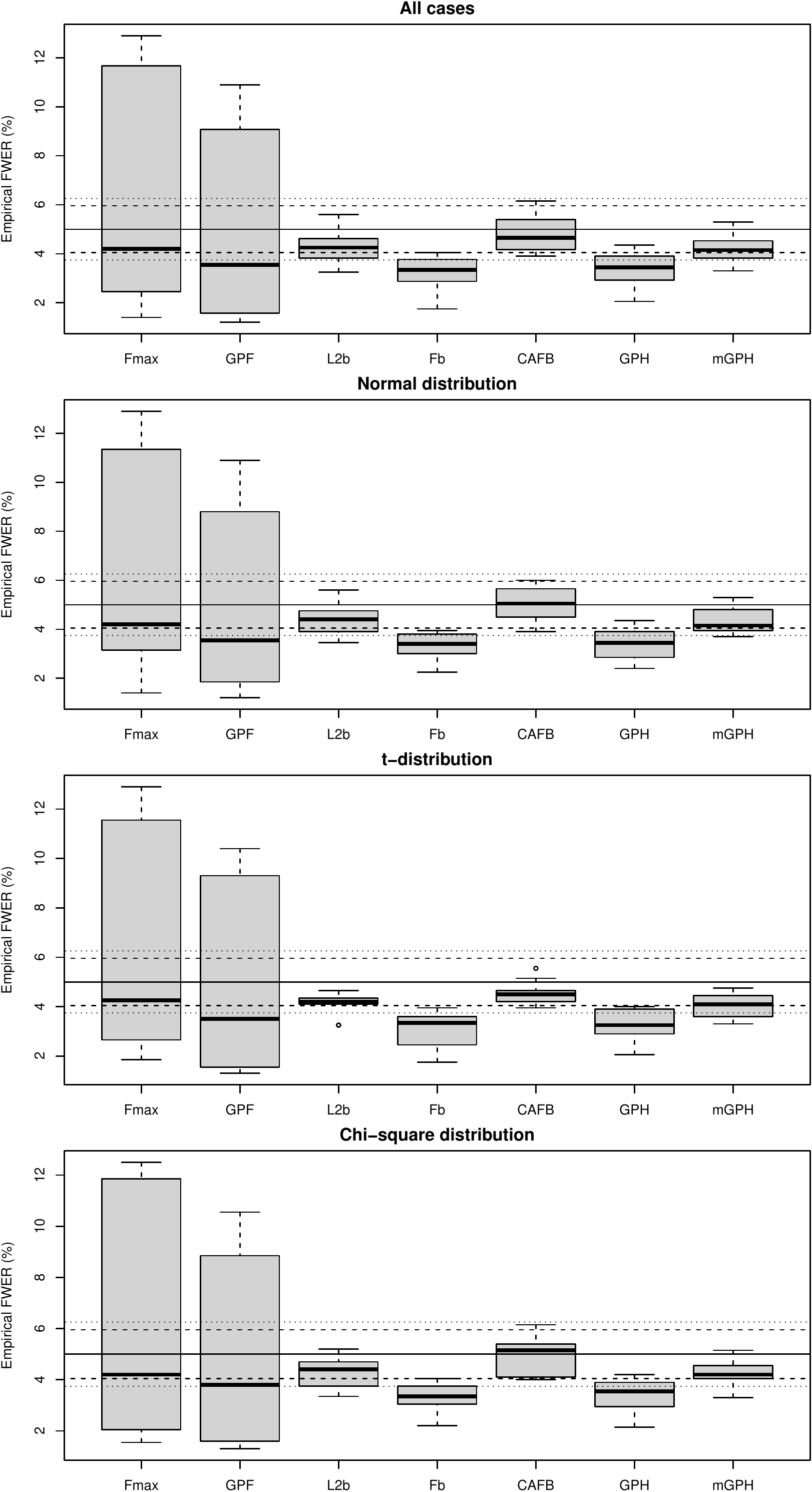}
\caption[Box-and-whisker plots for the empirical FWER of all tests obtained for the Dunnett constrasts for different
distributions]{Box-and-whisker plots for the empirical FWER (as percentages) of all tests obtained for the Dunnett constrasts for different distributions}
\end{figure}

\newpage

\begin{longtable}[t]{rrrr|rrrrrrr}
\caption[Empirical sizes and powers of all tests obtained under alternative A1 for the Dunnett constrasts]{\label{tab:unnamed-chunk-20}Empirical sizes ($\mathcal{H}_{0,1}$, $\mathcal{H}_{0,2}$) and powers ($\mathcal{H}_{0,3}$) (as percentages) of all tests obtained under alternative A1 for the Dunnett constrasts (D - distribution, $(\lambda_1,\lambda_2,\lambda_3,\lambda_4)$: (1,1,1,1) - homoscedastic case, (1,1.25,1.5,1.75) - heteroscedastic case (positive pairing), (1.75,1.5,1.25,1) - heteroscedastic case (negative pairing))}\\
\hline
D&$(n_1,n_2,n_3,n_4)$&$(\lambda_1,\lambda_2,\lambda_3,\lambda_4)$&$\mathcal{H}$&Fmax&GPF&L2b&Fb&CAFB&GPH&mGPH\\
\hline
\endfirsthead
\caption[]{Empirical sizes ($\mathcal{H}_{0,1}$, $\mathcal{H}_{0,2}$) and powers ($\mathcal{H}_{0,3}$) (as percentages) of all tests obtained under alternative A1 for the Dunnett constrasts (D - distribution, $(\lambda_1,\lambda_2,\lambda_3,\lambda_4)$: (1,1,1,1) - homoscedastic case, (1,1.25,1.5,1.75) - heteroscedastic case (positive pairing), (1.75,1.5,1.25,1) - heteroscedastic case (negative pairing)) \textit{(continued)}}\\
\hline
D&$(n_1,n_2,n_3,n_4)$&$(\lambda_1,\lambda_2,\lambda_3,\lambda_4)$&$\mathcal{H}$&Fmax&GPF&L2b&Fb&CAFB&GPH&mGPH\\
\hline
\endhead
N&(15,20,25,30)&(1,1,1,1)&$\mathcal{H}_{0,1}$&1.35&1.35&2.05&0.95&2.35&1.45&1.70\\
&&&$\mathcal{H}_{0,2}$&1.70&1.05&1.70&0.75&2.05&1.00&1.25\\
&&&$\mathcal{H}_{0,3}$&82.65&89.55&85.50&78.95&74.70&86.70&89.25\\
N&(15,20,25,30)&(1,1.25,1.5,1.75)&$\mathcal{H}_{0,1}$&1.15&0.70&1.75&1.10&1.55&1.25&1.45\\
&&&$\mathcal{H}_{0,2}$&0.55&0.50&2.05&1.60&2.20&1.60&1.85\\
&&&$\mathcal{H}_{0,3}$&24.40&28.75&56.45&49.35&45.40&58.35&60.85\\
N&(15,20,25,30)&(1.75,1.5,1.25,1)&$\mathcal{H}_{0,1}$&2.30&1.85&2.15&1.25&2.05&1.25&1.95\\
&&&$\mathcal{H}_{0,2}$&3.45&3.20&1.60&0.75&2.00&0.80&1.50\\
&&&$\mathcal{H}_{0,3}$&66.55&68.20&34.90&24.55&29.30&32.25&39.00\\
t&(15,20,25,30)&(1,1,1,1)&$\mathcal{H}_{0,1}$&1.65&0.95&2.10&1.10&2.05&1.20&1.40\\
&&&$\mathcal{H}_{0,2}$&1.80&1.05&1.40&0.90&1.75&1.00&1.35\\
&&&$\mathcal{H}_{0,3}$&53.05&60.25&56.05&46.95&47.10&56.90&61.35\\
t&(15,20,25,30)&(1,1.25,1.5,1.75)&$\mathcal{H}_{0,1}$&1.40&0.75&1.60&1.10&1.95&0.90&0.95\\
&&&$\mathcal{H}_{0,2}$&0.60&0.45&1.85&1.25&1.60&1.15&1.25\\
&&&$\mathcal{H}_{0,3}$&9.60&9.50&29.45&24.75&23.55&30.10&31.75\\
t&(15,20,25,30)&(1.75,1.5,1.25,1)&$\mathcal{H}_{0,1}$&2.65&1.95&2.20&1.30&2.45&1.40&1.90\\
&&&$\mathcal{H}_{0,2}$&3.65&2.90&1.75&0.80&2.10&1.00&1.30\\
&&&$\mathcal{H}_{0,3}$&43.70&44.00&16.85&10.70&14.45&14.55&18.50\\
$\chi^2$&(15,20,25,30)&(1,1,1,1)&$\mathcal{H}_{0,1}$&1.55&1.00&1.50&0.75&1.70&1.15&1.35\\
&&&$\mathcal{H}_{0,2}$&1.95&0.85&1.40&0.80&1.80&0.75&1.10\\
&&&$\mathcal{H}_{0,3}$&82.15&88.60&83.30&78.00&74.50&85.20&87.85\\
$\chi^2$&(15,20,25,30)&(1,1.25,1.5,1.75)&$\mathcal{H}_{0,1}$&1.15&0.85&1.50&0.95&1.65&1.00&1.20\\
&&&$\mathcal{H}_{0,2}$&0.85&0.20&1.85&1.15&1.60&1.05&1.20\\
&&&$\mathcal{H}_{0,3}$&23.10&25.25&54.50&48.75&47.65&57.65&60.05\\
$\chi^2$&(15,20,25,30)&(1.75,1.5,1.25,1)&$\mathcal{H}_{0,1}$&1.80&1.80&1.50&0.85&2.10&1.30&1.65\\
&&&$\mathcal{H}_{0,2}$&4.05&3.25&1.60&0.90&2.20&1.20&1.70\\
&&&$\mathcal{H}_{0,3}$&66.40&67.85&35.70&26.45&31.10&33.50&39.25\\
\hline
\end{longtable}

\begin{figure}
\centering
\includegraphics[width= 0.95\textwidth,height=0.9\textheight]{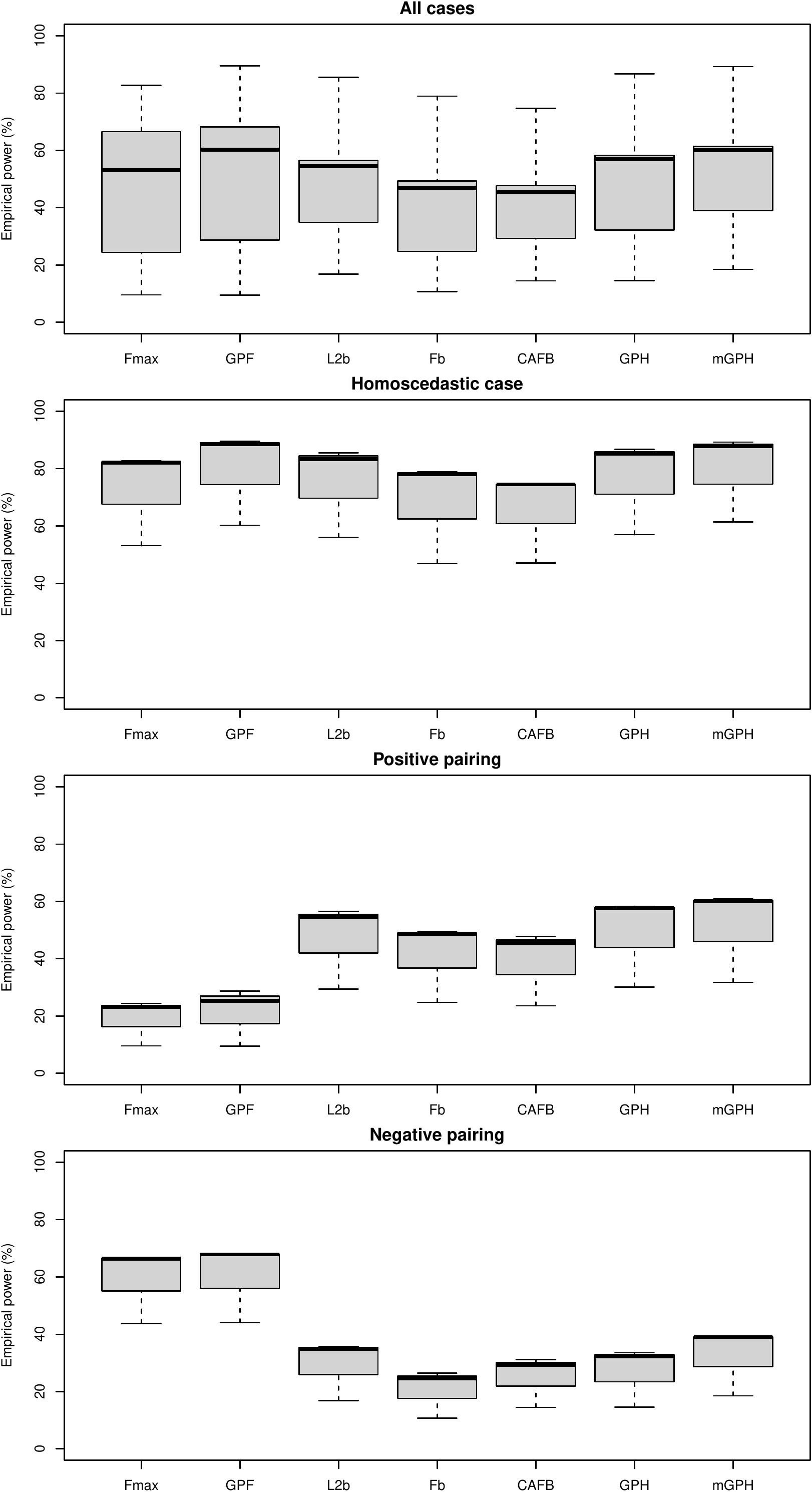}
\caption[Box-and-whisker plots for the empirical powers of all tests obtained under alternative A1 for the Dunnett constrasts and homoscedastic and heteroscedastic cases]{Box-and-whisker plots for the empirical powers (as percentages) of all tests obtained under alternative A1 for the Dunnett constrasts and homoscedastic and heteroscedastic cases}
\end{figure}

\begin{figure}
\centering
\includegraphics[width= 0.95\textwidth,height=0.9\textheight]{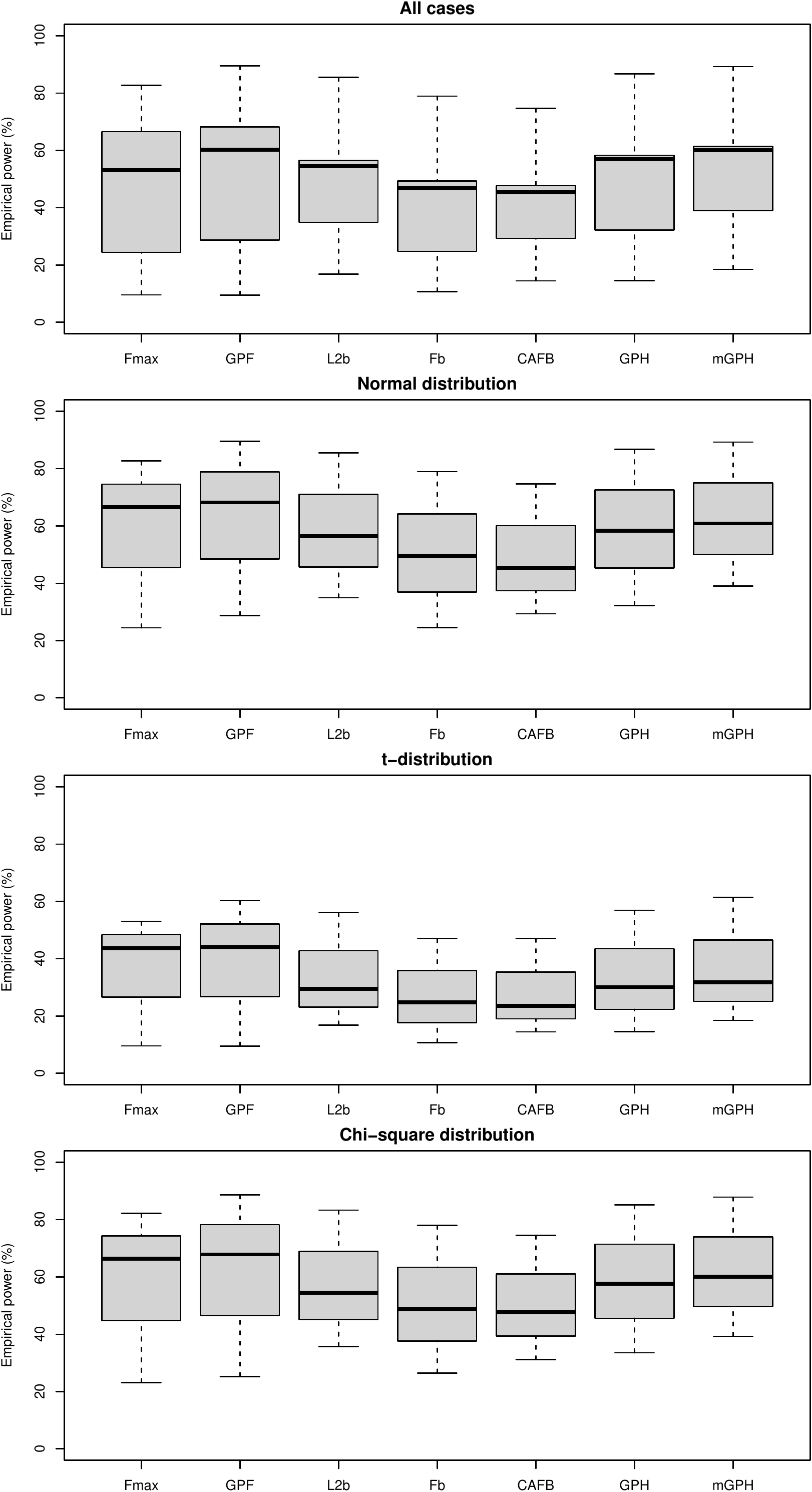}
\caption[Box-and-whisker plots for the empirical powers of all tests obtained under alternative A1 for the Dunnett constrasts and different distributions]{Box-and-whisker plots for the empirical powers (as percentages) of all tests obtained under alternative A1 for the Dunnett constrasts and different distributions}
\end{figure}

\newpage

\begin{longtable}[t]{rrrr|rrrrrrr}
\caption[Empirical sizes and powers of all tests obtained under alternative A2 for the Dunnett constrasts]{\label{tab:unnamed-chunk-15}Empirical sizes ($\mathcal{H}_{0,1}$, $\mathcal{H}_{0,2}$) and powers ($\mathcal{H}_{0,3}$) (as percentages) of all tests obtained under alternative A2 for the Dunnett constrasts (D - distribution, $(\lambda_1,\lambda_2,\lambda_3,\lambda_4)$: (1,1,1,1) - homoscedastic case, (1,1.25,1.5,1.75) - heteroscedastic case (positive pairing), (1.75,1.5,1.25,1) - heteroscedastic case (negative pairing))}\\
\hline
D&$(n_1,n_2,n_3,n_4)$&$(\lambda_1,\lambda_2,\lambda_3,\lambda_4)$&$\mathcal{H}$&Fmax&GPF&L2b&Fb&CAFB&GPH&mGPH\\
\hline
\endfirsthead
\caption[]{Empirical sizes ($\mathcal{H}_{0,1}$, $\mathcal{H}_{0,2}$) and powers ($\mathcal{H}_{0,3}$) (as percentages) of all tests obtained under alternative A2 for the Dunnett constrasts (D - distribution, $(\lambda_1,\lambda_2,\lambda_3,\lambda_4)$: (1,1,1,1) - homoscedastic case, (1,1.25,1.5,1.75) - heteroscedastic case (positive pairing), (1.75,1.5,1.25,1) - heteroscedastic case (negative pairing)) \textit{(continued)}}\\
D&$(n_1,n_2,n_3,n_4)$&$(\lambda_1,\lambda_2,\lambda_3,\lambda_4)$&$\mathcal{H}$&Fmax&GPF&L2b&Fb&CAFB&GPH&mGPH\\
\endhead
N&(15,20,25,30)&(1,1,1,1)&$\mathcal{H}_{0,1}$&1.80&1.45&2.20&1.35&2.30&1.65&2.00\\
&&&$\mathcal{H}_{0,2}$&1.70&1.00&1.80&0.90&1.70&1.10&1.20\\
&&&$\mathcal{H}_{0,3}$&80.20&98.50&97.85&96.75&91.95&98.10&98.50\\
N&(15,20,25,30)&(1,1.25,1.5,1.75)&$\mathcal{H}_{0,1}$&1.25&1.05&1.55&1.10&1.90&1.25&1.35\\
&&&$\mathcal{H}_{0,2}$&0.85&0.40&1.75&1.30&1.90&1.40&1.55\\
&&&$\mathcal{H}_{0,3}$&21.50&58.55&84.15&79.50&66.00&84.30&85.20\\
N&(15,20,25,30)&(1.75,1.5,1.25,1)&$\mathcal{H}_{0,1}$&1.90&1.55&1.80&1.10&1.75&1.25&1.65\\
&&&$\mathcal{H}_{0,2}$&4.05&3.40&1.80&0.80&2.35&1.05&1.50\\
&&&$\mathcal{H}_{0,3}$&68.05&85.10&61.20&49.35&43.90&56.40&62.85\\
t&(15,20,25,30)&(1,1,1,1)&$\mathcal{H}_{0,1}$&2.00&1.15&1.55&1.10&2.10&1.30&1.35\\
&&&$\mathcal{H}_{0,2}$&2.00&1.45&2.35&1.25&1.90&1.35&1.60\\
&&&$\mathcal{H}_{0,3}$&52.20&86.35&83.55&77.90&68.05&83.20&87.10\\
t&(15,20,25,30)&(1,1.25,1.5,1.75)&$\mathcal{H}_{0,1}$&0.80&0.70&1.15&0.70&2.05&0.80&0.90\\
&&&$\mathcal{H}_{0,2}$&0.75&0.50&1.50&1.10&1.90&1.05&1.20\\
&&&$\mathcal{H}_{0,3}$&11.00&27.75&56.05&50.40&41.60&55.75&57.80\\
t&(15,20,25,30)&(1.75,1.5,1.25,1)&$\mathcal{H}_{0,1}$&1.95&1.90&2.05&1.05&2.15&1.40&2.20\\
&&&$\mathcal{H}_{0,2}$&4.10&2.90&1.50&0.95&2.10&0.90&1.50\\
&&&$\mathcal{H}_{0,3}$&49.15&64.10&34.75&26.00&25.30&30.85&37.00\\
$\chi^2$&(15,20,25,30)&(1,1,1,1)&$\mathcal{H}_{0,1}$&1.95&1.05&1.90&1.10&1.25&1.10&1.30\\
&&&$\mathcal{H}_{0,2}$&1.55&1.60&1.85&1.10&2.00&1.45&1.85\\
&&&$\mathcal{H}_{0,3}$&81.25&98.30&96.25&94.40&90.65&96.35&97.20\\
$\chi^2$&(15,20,25,30)&(1,1.25,1.5,1.75)&$\mathcal{H}_{0,1}$&1.40&0.75&1.75&1.05&1.75&0.85&0.90\\
&&&$\mathcal{H}_{0,2}$&0.90&0.40&1.75&1.35&1.90&1.20&1.30\\
&&&$\mathcal{H}_{0,3}$&20.30&60.80&84.75&80.45&67.70&86.05&86.85\\
$\chi^2$&(15,20,25,30)&(1.75,1.5,1.25,1)&$\mathcal{H}_{0,1}$&2.10&1.75&1.30&0.75&1.70&1.15&1.80\\
&&&$\mathcal{H}_{0,2}$&4.15&2.60&1.35&0.65&2.30&0.75&0.95\\
&&&$\mathcal{H}_{0,3}$&71.10&86.50&61.65&53.35&51.90&59.65&65.40\\
\hline
\end{longtable}

\begin{figure}
\centering
\includegraphics[width= 0.95\textwidth,height=0.9\textheight]{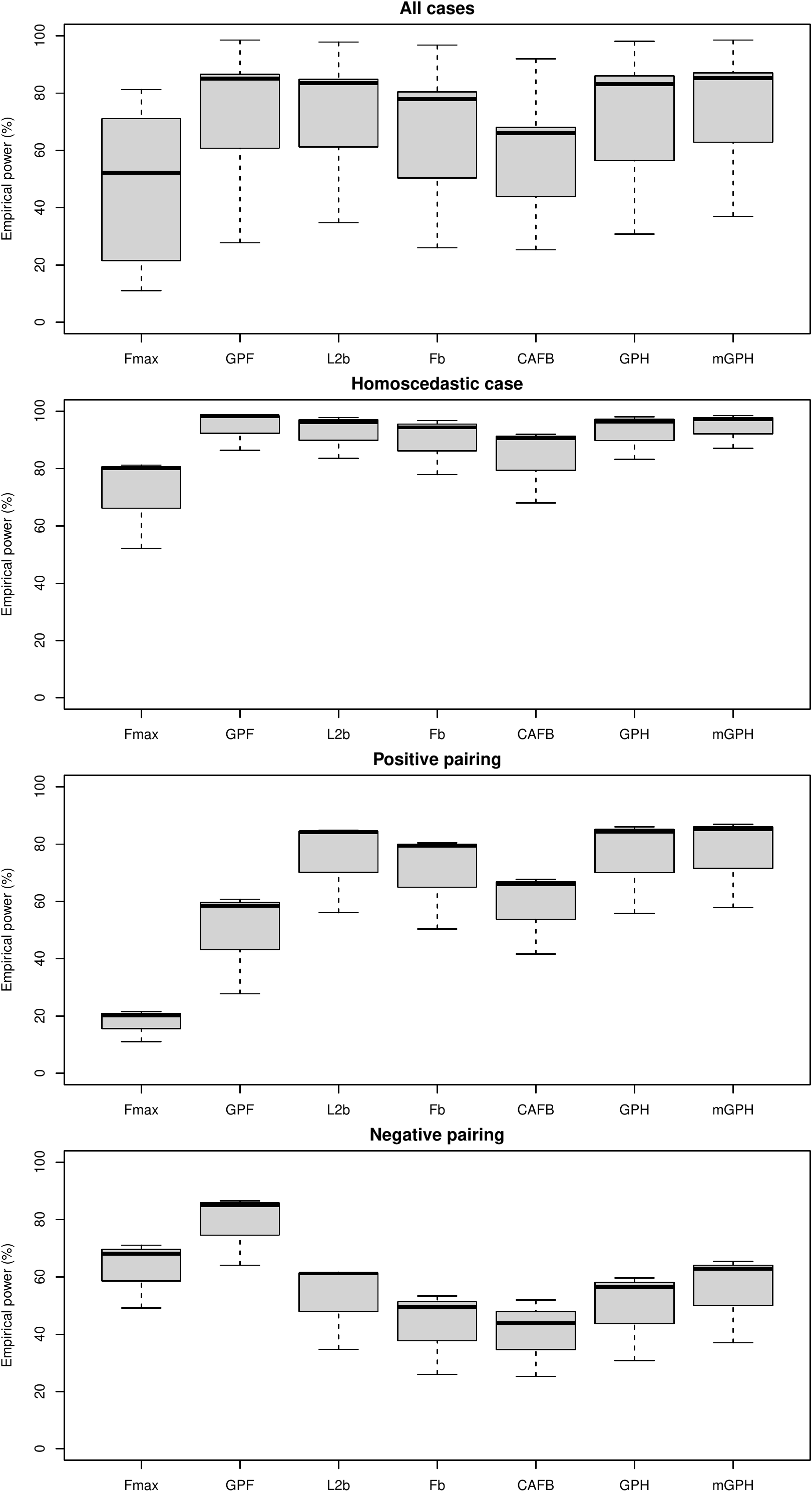}
\caption[Box-and-whisker plots for the empirical powers of all tests obtained under alternative A2 for the Dunnett constrasts and homoscedastic and heteroscedastic cases]{Box-and-whisker plots for the empirical powers (as percentages) of all tests obtained under alternative A2 for the Dunnett constrasts and homoscedastic and heteroscedastic cases}
\end{figure}

\begin{figure}
\centering
\includegraphics[width= 0.95\textwidth,height=0.9\textheight]{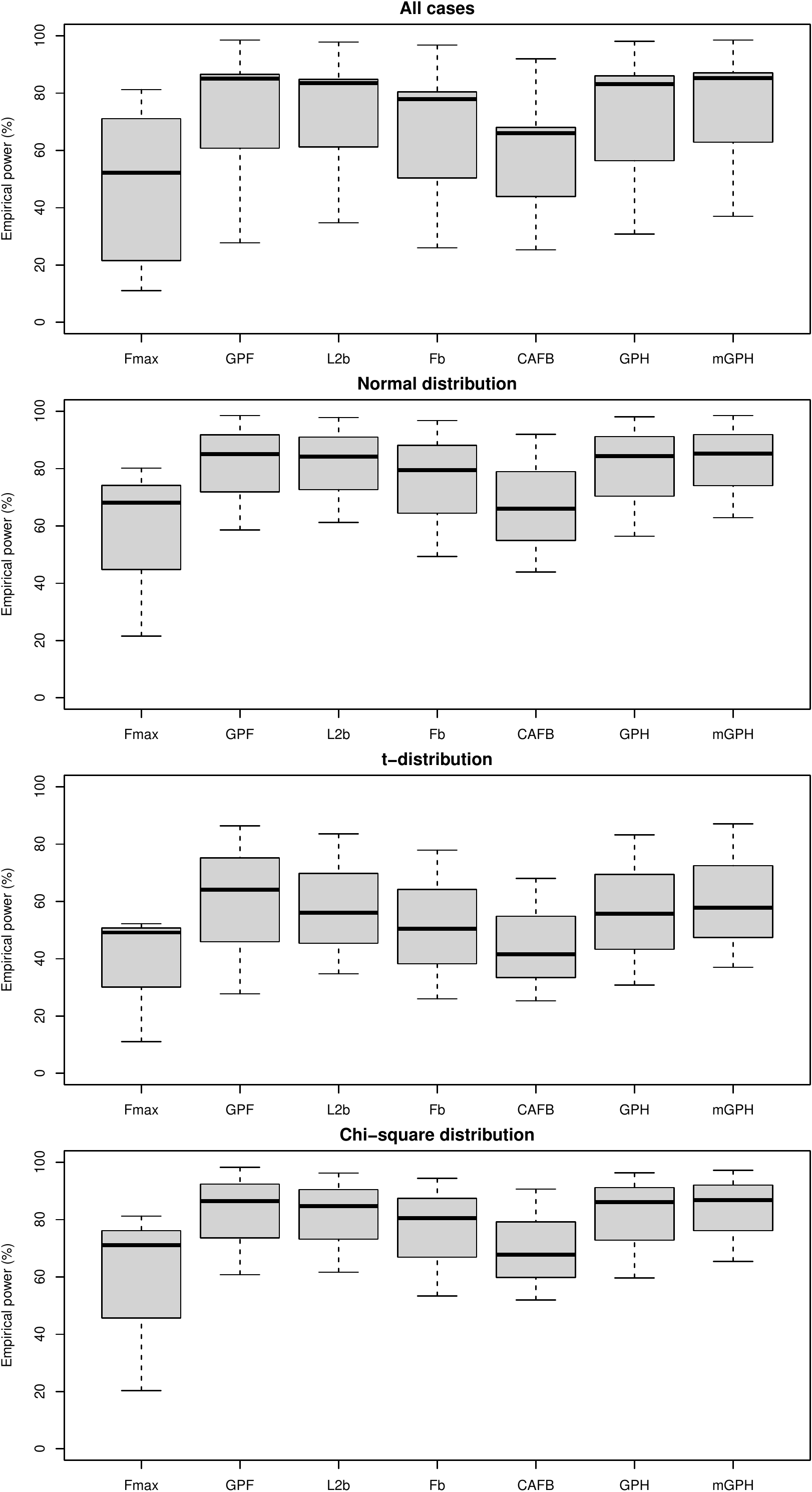}
\caption[Box-and-whisker plots for the empirical powers of all tests obtained under alternative A2 for the Dunnett constrasts and different distributions]{Box-and-whisker plots for the empirical powers (as percentages) of all tests obtained under alternative A2 for the Dunnett constrasts and different distributions}
\end{figure}

\newpage

\begin{longtable}[t]{rrrr|rrrrrrr}
\caption[Empirical powers of all tests obtained under alternative A3 for the Dunnett constrasts]{\label{tab:unnamed-chunk-25}Empirical powers (as percentages) of all tests obtained under alternative A3 for the Dunnett constrasts (D - distribution, $(\lambda_1,\lambda_2,\lambda_3,\lambda_4)$: (1,1,1,1) - homoscedastic case, (1,1.25,1.5,1.75) - heteroscedastic case (positive pairing), (1.75,1.5,1.25,1) - heteroscedastic case (negative pairing))}\\
\hline
D&$(n_1,n_2,n_3,n_4)$&$(\lambda_1,\lambda_2,\lambda_3,\lambda_4)$&$\mathcal{H}$&Fmax&GPF&L2b&Fb&CAFB&GPH&mGPH\\
\hline
\endfirsthead
\caption[]{Empirical powers (as percentages) of all tests obtained under alternative A3 for the Dunnett constrasts (D - distribution, $(\lambda_1,\lambda_2,\lambda_3,\lambda_4)$: (1,1,1,1) - homoscedastic case, (1,1.25,1.5,1.75) - heteroscedastic case (positive pairing), (1.75,1.5,1.25,1) - heteroscedastic case (negative pairing)) \textit{(continued)}}\\
D&$(n_1,n_2,n_3,n_4)$&$(\lambda_1,\lambda_2,\lambda_3,\lambda_4)$&$\mathcal{H}$&Fmax&GPF&L2b&Fb&CAFB&GPH&mGPH\\
\endhead
N&(15,20,25,30)&(1,1,1,1)&$\mathcal{H}_{0,1}$&14.90&26.90&29.70&23.05&19.85&25.95&29.05\\
&&&$\mathcal{H}_{0,2}$&76.10&97.45&96.85&94.80&88.65&96.80&97.55\\
&&&$\mathcal{H}_{0,3}$&99.75&100.00&100.00&100.00&100.00&100.00&100.00\\
N&(15,20,25,30)&(1,1.25,1.5,1.75)&$\mathcal{H}_{0,1}$&9.50&17.25&23.15&17.80&14.80&20.65&22.15\\
&&&$\mathcal{H}_{0,2}$&37.00&76.95&86.85&82.65&72.40&86.90&88.30\\
&&&$\mathcal{H}_{0,3}$&70.35&98.70&99.70&99.65&98.05&99.75&99.80\\
N&(15,20,25,30)&(1.75,1.5,1.25,1)&$\mathcal{H}_{0,1}$&7.15&8.65&8.85&5.95&7.20&6.70&9.40\\
&&&$\mathcal{H}_{0,2}$&45.05&68.90&53.65&44.10&37.85&51.35&58.40\\
&&&$\mathcal{H}_{0,3}$&95.90&99.80&96.90&93.65&84.80&96.35&97.85\\
t&(15,20,25,30)&(1,1,1,1)&$\mathcal{H}_{0,1}$&8.75&14.95&16.90&12.65&11.40&13.80&16.50\\
&&&$\mathcal{H}_{0,2}$&48.30&80.90&78.10&73.10&63.40&78.70&82.20\\
&&&$\mathcal{H}_{0,3}$&94.00&99.70&99.25&98.85&97.30&99.60&99.70\\
t&(15,20,25,30)&(1,1.25,1.5,1.75)&$\mathcal{H}_{0,1}$&6.65&9.05&12.50&8.80&9.25&10.25&11.25\\
&&&$\mathcal{H}_{0,2}$&19.10&42.75&59.85&53.20&43.45&59.90&61.75\\
&&&$\mathcal{H}_{0,3}$&37.40&80.75&93.95&92.75&85.25&94.75&95.60\\
t&(15,20,25,30)&(1.75,1.5,1.25,1)&$\mathcal{H}_{0,1}$&4.75&5.15&5.15&3.40&5.15&3.90&5.45\\
&&&$\mathcal{H}_{0,2}$&27.35&44.35&30.80&22.95&22.95&27.65&33.70\\
&&&$\mathcal{H}_{0,3}$&82.40&94.65&77.55&68.90&63.25&76.15&82.15\\
$\chi^2$&(15,20,25,30)&(1,1,1,1)&$\mathcal{H}_{0,1}$&15.95&29.30&32.00&25.25&21.85&29.25&32.30\\
&&&$\mathcal{H}_{0,2}$&76.85&97.05&94.95&92.60&87.50&95.55&96.30\\
&&&$\mathcal{H}_{0,3}$&99.75&100.00&99.95&99.80&99.80&100.00&100.00\\
$\chi^2$&(15,20,25,30)&(1,1.25,1.5,1.75)&$\mathcal{H}_{0,1}$&9.45&14.75&21.70&16.30&15.45&18.60&20.15\\
&&&$\mathcal{H}_{0,2}$&37.55&77.15&85.65&81.60&71.70&85.85&86.90\\
&&&$\mathcal{H}_{0,3}$&70.00&99.10&99.85&99.60&97.60&99.90&100.00\\
$\chi^2$&(15,20,25,30)&(1.75,1.5,1.25,1)&$\mathcal{H}_{0,1}$&6.80&9.25&9.70&6.60&8.80&8.10&10.25\\
&&&$\mathcal{H}_{0,2}$&47.95&70.30&55.30&46.65&42.75&53.60&59.35\\
&&&$\mathcal{H}_{0,3}$&96.25&99.25&93.95&89.90&84.80&94.40&95.85\\
\hline
\end{longtable}

\begin{figure}
\centering
\includegraphics[width= 0.95\textwidth,height=0.9\textheight]{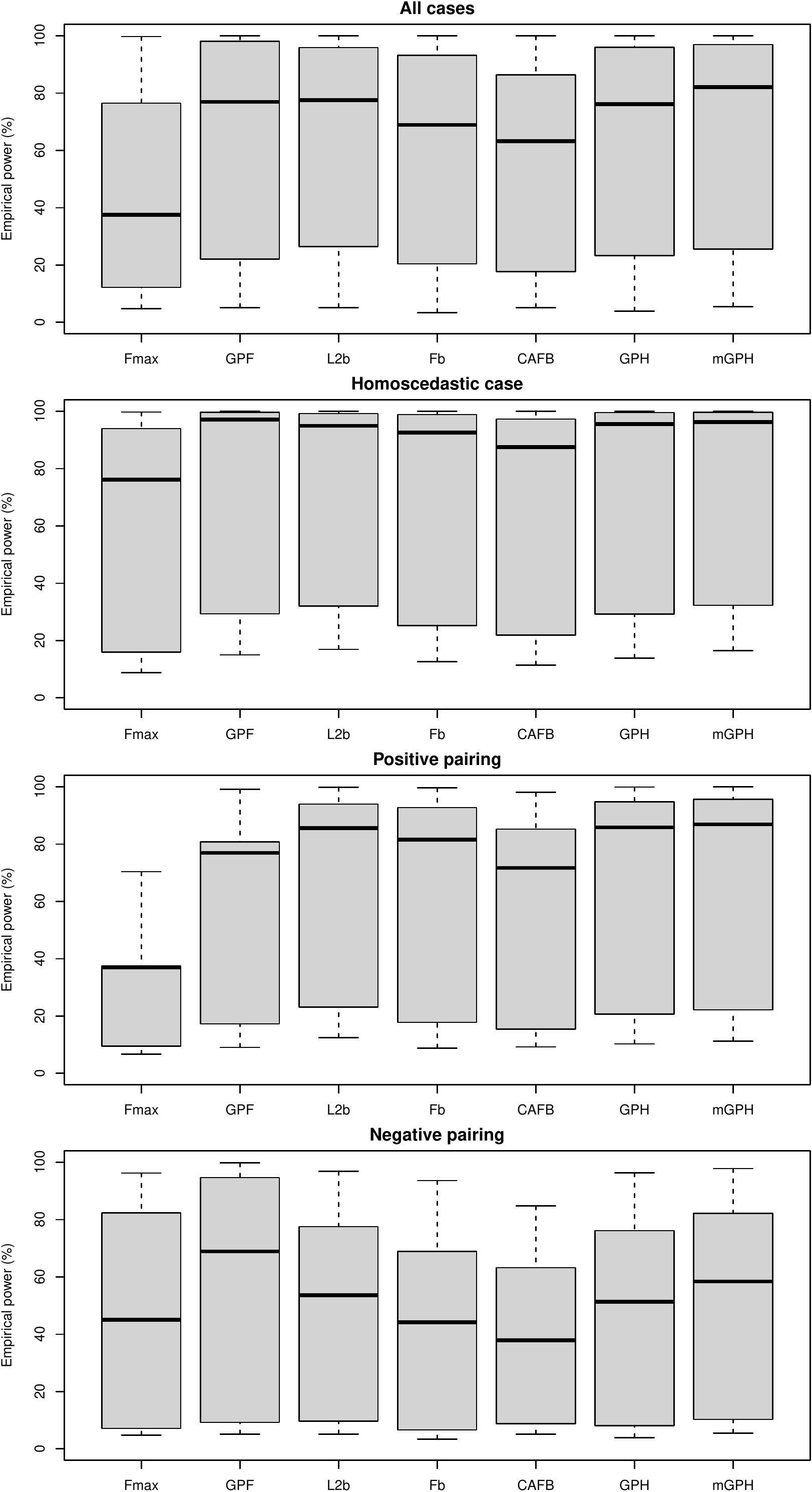}
\caption[Box-and-whisker plots for the empirical powers of all tests obtained under alternative A3 for the Dunnett constrasts and homoscedastic and heteroscedastic cases]{Box-and-whisker plots for the empirical powers (as percentages) of all tests obtained under alternative A3 for the Dunnett constrasts and homoscedastic and heteroscedastic cases}
\end{figure}

\begin{figure}
\centering
\includegraphics[width= 0.95\textwidth,height=0.9\textheight]{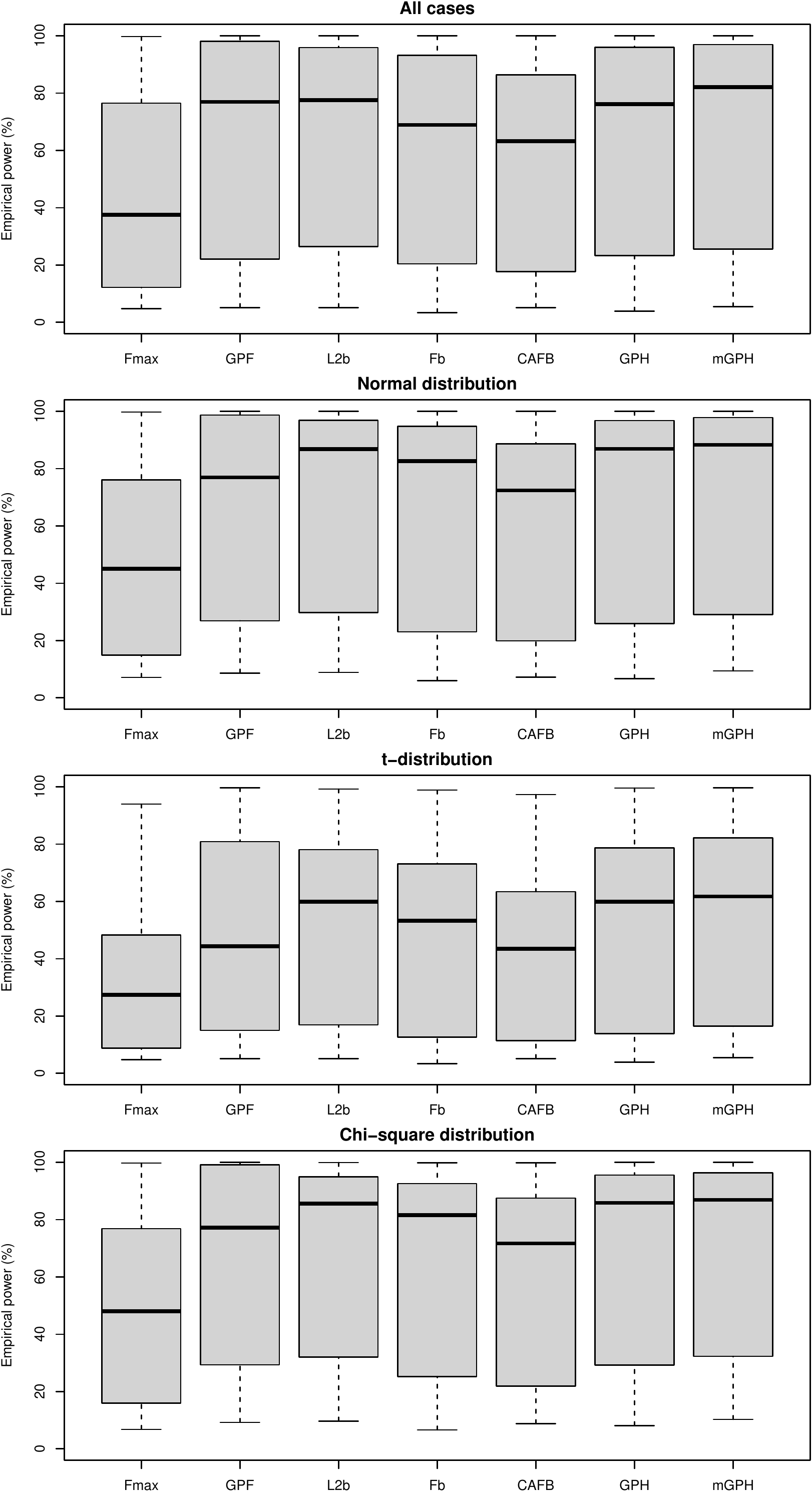}
\caption[Box-and-whisker plots for the empirical powers of all tests obtained under alternative A3 for the Dunnett constrasts and different distributions]{Box-and-whisker plots for the empirical powers (as percentages) of all tests obtained under alternative A3 for the Dunnett constrasts and different distributions}
\end{figure}

\newpage

\begin{longtable}[t]{rrrr|rrrrrrr}
\caption[Empirical powers of all tests obtained under alternative A4 for the Dunnett constrasts]{\label{tab:unnamed-chunk-28}Empirical powers (as percentages) of all tests obtained under alternative A4 for the Dunnett constrasts (D - distribution, $(\lambda_1,\lambda_2,\lambda_3,\lambda_4)$: (1,1,1,1) - homoscedastic case, (1,1.25,1.5,1.75) - heteroscedastic case (positive pairing), (1.75,1.5,1.25,1) - heteroscedastic case (negative pairing))}\\
\hline
D&$(n_1,n_2,n_3,n_4)$&$(\lambda_1,\lambda_2,\lambda_3,\lambda_4)$&$\mathcal{H}$&Fmax&GPF&L2b&Fb&CAFB&GPH&mGPH\\
\hline
\endfirsthead
\caption[]{Empirical powers (as percentages) of all tests obtained under alternative A4 for the Dunnett constrasts (D - distribution, $(\lambda_1,\lambda_2,\lambda_3,\lambda_4)$: (1,1,1,1) - homoscedastic case, (1,1.25,1.5,1.75) - heteroscedastic case (positive pairing), (1.75,1.5,1.25,1) - heteroscedastic case (negative pairing)) \textit{(continued)}}\\
\hline
D&$(n_1,n_2,n_3,n_4)$&$(\lambda_1,\lambda_2,\lambda_3,\lambda_4)$&$\mathcal{H}$&Fmax&GPF&L2b&Fb&CAFB&GPH&mGPH\\
\hline
\endhead
N&(15,20,25,30)&(1,1,1,1)&$\mathcal{H}_{0,1}$&12.95&13.75&16.20&11.10&12.65&13.55&15.65\\
&&&$\mathcal{H}_{0,2}$&77.15&85.55&82.25&75.00&70.50&83.75&86.85\\
&&&$\mathcal{H}_{0,3}$&99.80&100.00&99.85&99.80&98.90&100.00&100.00\\
N&(15,20,25,30)&(1,1.25,1.5,1.75)&$\mathcal{H}_{0,1}$&8.70&8.50&12.40&8.95&9.30&10.75&11.70\\
&&&$\mathcal{H}_{0,2}$&37.20&44.05&60.80&52.90&49.90&62.20&64.65\\
&&&$\mathcal{H}_{0,3}$&77.40&85.25&96.25&94.15&89.05&97.65&98.00\\
N&(15,20,25,30)&(1.75,1.5,1.25,1)&$\mathcal{H}_{0,1}$&6.10&5.10&5.25&3.40&5.00&4.35&5.60\\
&&&$\mathcal{H}_{0,2}$&43.05&46.15&31.00&21.60&24.30&28.60&35.15\\
&&&$\mathcal{H}_{0,3}$&95.95&97.70&81.65&71.10&66.10&81.50&86.85\\
t&(15,20,25,30)&(1,1,1,1)&$\mathcal{H}_{0,1}$&7.70&7.00&8.30&6.25&8.00&7.05&8.40\\
&&&$\mathcal{H}_{0,2}$&47.05&53.45&49.30&42.65&40.70&49.70&54.10\\
&&&$\mathcal{H}_{0,3}$&94.10&96.80&93.95&90.85&87.90&95.25&96.65\\
t&(15,20,25,30)&(1,1.25,1.5,1.75)&$\mathcal{H}_{0,1}$&4.80&4.75&7.15&4.75&6.05&6.20&6.95\\
&&&$\mathcal{H}_{0,2}$&17.85&20.05&32.40&26.25&26.00&32.50&34.05\\
&&&$\mathcal{H}_{0,3}$&40.60&47.65&73.60&69.30&62.50&76.10&77.70\\
t&(15,20,25,30)&(1.75,1.5,1.25,1)&$\mathcal{H}_{0,1}$&2.60&3.40&3.50&2.00&3.25&2.45&3.60\\
&&&$\mathcal{H}_{0,2}$&25.15&26.15&16.00&10.70&13.10&13.85&17.85\\
&&&$\mathcal{H}_{0,3}$&79.95&81.75&49.95&38.70&41.95&48.00&55.60\\
$\chi^2$&(15,20,25,30)&(1,1,1,1)&$\mathcal{H}_{0,1}$&13.45&14.35&15.90&11.45&13.05&13.90&16.40\\
&&&$\mathcal{H}_{0,2}$&78.30&84.60&79.30&73.40&71.40&82.15&85.00\\
&&&$\mathcal{H}_{0,3}$&99.85&100.00&99.55&99.25&98.65&99.75&99.80\\
$\chi^2$&(15,20,25,30)&(1,1.25,1.5,1.75)&$\mathcal{H}_{0,1}$&8.05&8.45&12.60&9.30&9.40&11.15&12.10\\
&&&$\mathcal{H}_{0,2}$&38.85&44.40&61.00&53.75&49.15&61.25&63.40\\
&&&$\mathcal{H}_{0,3}$&76.45&86.35&95.25&93.55&88.55&97.15&97.35\\
$\chi^2$&(15,20,25,30)&(1.75,1.5,1.25,1)&$\mathcal{H}_{0,1}$&5.70&5.40&6.00&3.80&4.75&4.85&6.20\\
&&&$\mathcal{H}_{0,2}$&44.75&49.40&32.75&25.10&28.40&31.75&37.55\\
&&&$\mathcal{H}_{0,3}$&95.95&96.50&77.40&68.90&68.70&79.15&84.05\\
\hline
\end{longtable}

\begin{figure}
\centering
\includegraphics[width= 0.95\textwidth,height=0.9\textheight]{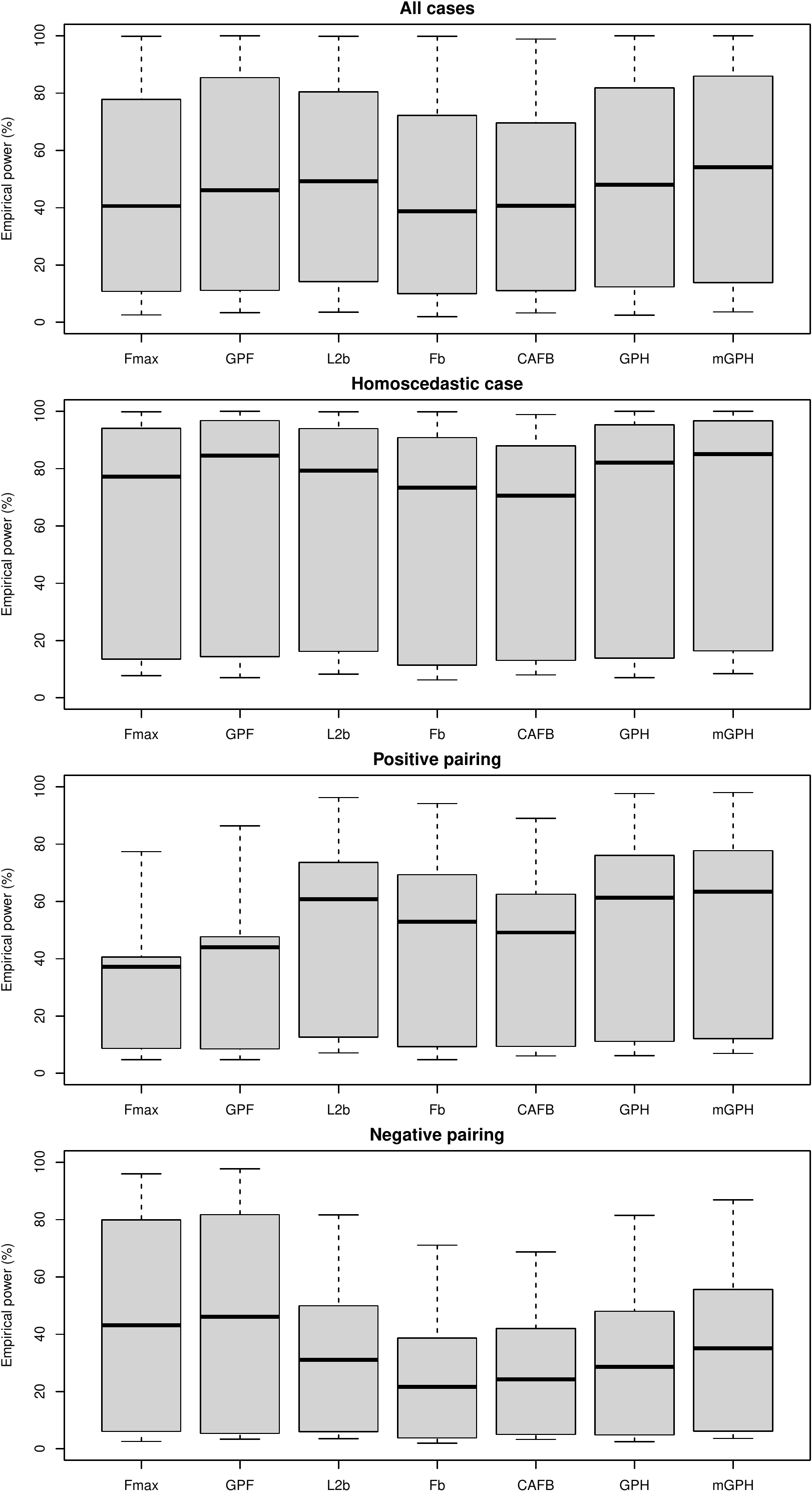}
\caption[Box-and-whisker plots for the empirical powers of all tests obtained under alternative A4 for the Dunnett constrasts and homoscedastic and heteroscedastic cases]{Box-and-whisker plots for the empirical powers (as percentages) of all tests obtained under alternative A4 for the Dunnett constrasts and homoscedastic and heteroscedastic cases}
\end{figure}

\begin{figure}
\centering
\includegraphics[width= 0.95\textwidth,height=0.9\textheight]{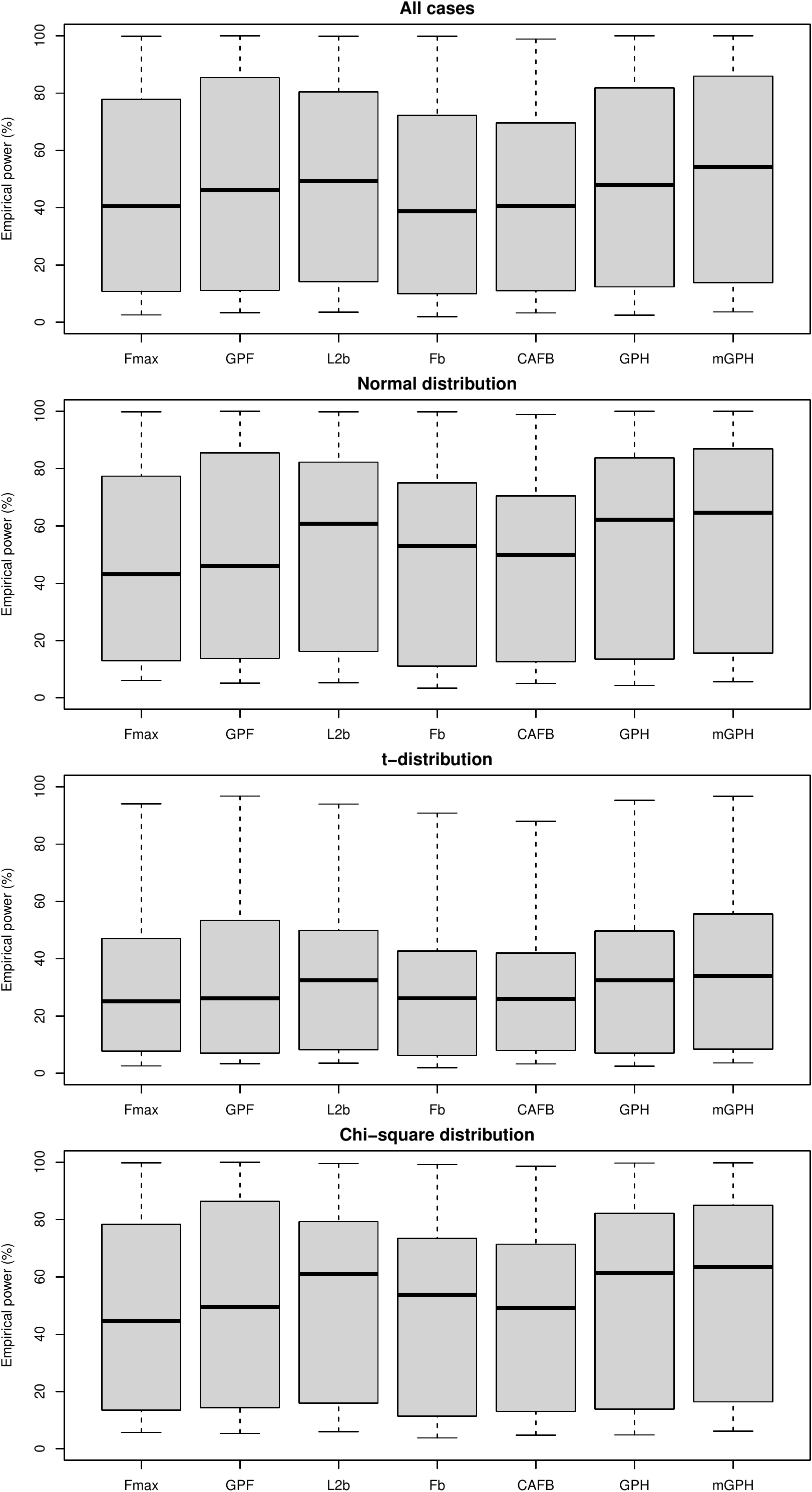}
\caption[Box-and-whisker plots for the empirical powers of all tests obtained under alternative A4 for the Dunnett constrasts and different distributions]{Box-and-whisker plots for the empirical powers (as percentages) of all tests obtained under alternative A4 for the Dunnett constrasts and different distributions}
\end{figure}

\newpage

\begin{longtable}[t]{rrrr|rrrrrrr}
\caption[Empirical sizes and powers of all tests obtained under alternative A5 for the Dunnett constrasts]{\label{tab:unnamed-chunk-31}Empirical sizes ($\mathcal{H}_{0,1}$, $\mathcal{H}_{0,2}$) and powers ($\mathcal{H}_{0,3}$) (as percentages) of all tests obtained under alternative A5 for the Dunnett constrasts (D - distribution, $(\lambda_1,\lambda_2,\lambda_3,\lambda_4)$: (1,1,1,1) - homoscedastic case, (1,1.25,1.5,1.75) - heteroscedastic case (positive pairing), (1.75,1.5,1.25,1) - heteroscedastic case (negative pairing))}\\
\hline
D&$(n_1,n_2,n_3,n_4)$&$(\lambda_1,\lambda_2,\lambda_3,\lambda_4)$&$\mathcal{H}$&Fmax&GPF&L2b&Fb&CAFB&GPH&mGPH\\
\hline
\endfirsthead
\caption[]{Empirical sizes ($\mathcal{H}_{0,1}$, $\mathcal{H}_{0,2}$) and powers ($\mathcal{H}_{0,3}$) (as percentages) of all tests obtained under alternative A5 for the Dunnett constrasts (D - distribution, $(\lambda_1,\lambda_2,\lambda_3,\lambda_4)$: (1,1,1,1) - homoscedastic case, (1,1.25,1.5,1.75) - heteroscedastic case (positive pairing), (1.75,1.5,1.25,1) - heteroscedastic case (negative pairing)) \textit{(continued)}}\\
\hline
D&$(n_1,n_2,n_3,n_4)$&$(\lambda_1,\lambda_2,\lambda_3,\lambda_4)$&$\mathcal{H}$&Fmax&GPF&L2b&Fb&CAFB&GPH&mGPH\\
\hline
\endhead
N&(15,20,25,30)&(1,1,1,1)&$\mathcal{H}_{0,1}$&1.30&1.35&2.25&1.40&2.55&1.50&1.80\\
&&&$\mathcal{H}_{0,2}$&2.10&1.25&2.45&1.40&1.95&1.40&1.85\\
&&&$\mathcal{H}_{0,3}$&79.00&86.10&82.70&76.60&71.00&83.25&86.25\\
N&(15,20,25,30)&(1,1.25,1.5,1.75)&$\mathcal{H}_{0,1}$&1.45&0.70&1.50&0.90&2.35&0.85&0.85\\
&&&$\mathcal{H}_{0,2}$&1.20&0.75&2.65&1.65&2.45&1.85&2.00\\
&&&$\mathcal{H}_{0,3}$&21.70&25.40&53.35&47.10&43.45&55.60&57.75\\
N&(15,20,25,30)&(1.75,1.5,1.25,1)&$\mathcal{H}_{0,1}$&1.95&1.85&2.55&1.20&2.05&1.10&2.10\\
&&&$\mathcal{H}_{0,2}$&4.65&3.50&1.70&1.05&2.20&1.20&1.80\\
&&&$\mathcal{H}_{0,3}$&66.10&68.15&33.55&23.20&27.70&30.65&37.20\\
t&(15,20,25,30)&(1,1,1,1)&$\mathcal{H}_{0,1}$&1.65&1.40&2.00&1.05&1.85&1.25&1.60\\
&&&$\mathcal{H}_{0,2}$&1.65&1.00&1.45&0.70&2.05&0.85&1.30\\
&&&$\mathcal{H}_{0,3}$&51.20&58.05&56.00&47.65&45.95&55.30&59.60\\
t&(15,20,25,30)&(1,1.25,1.5,1.75)&$\mathcal{H}_{0,1}$&1.50&1.05&1.80&1.00&1.95&1.20&1.35\\
&&&$\mathcal{H}_{0,2}$&0.60&0.30&1.45&1.05&1.90&1.10&1.25\\
&&&$\mathcal{H}_{0,3}$&9.75&9.15&25.85&21.50&22.00&26.05&27.50\\
t&(15,20,25,30)&(1.75,1.5,1.25,1)&$\mathcal{H}_{0,1}$&2.30&1.60&1.40&0.75&1.95&1.10&1.80\\
&&&$\mathcal{H}_{0,2}$&3.70&3.25&1.45&0.90&2.10&1.05&1.55\\
&&&$\mathcal{H}_{0,3}$&44.50&44.15&15.55&10.70&15.05&14.20&18.05\\
$\chi^2$&(15,20,25,30)&(1,1,1,1)&$\mathcal{H}_{0,1}$&2.05&1.35&1.45&1.00&2.10&1.35&1.55\\
&&&$\mathcal{H}_{0,2}$&1.25&0.90&1.50&0.80&2.00&0.95&1.15\\
&&&$\mathcal{H}_{0,3}$&82.75&87.75&81.20&75.05&73.20&83.65&85.95\\
$\chi^2$&(15,20,25,30)&(1,1.25,1.5,1.75)&$\mathcal{H}_{0,1}$&1.55&0.85&1.70&1.05&2.10&1.15&1.25\\
&&&$\mathcal{H}_{0,2}$&0.85&0.45&2.10&1.25&1.85&1.65&1.85\\
&&&$\mathcal{H}_{0,3}$&19.85&21.50&52.75&47.35&42.75&55.75&57.95\\
$\chi^2$&(15,20,25,30)&(1.75,1.5,1.25,1)&$\mathcal{H}_{0,1}$&2.00&1.70&1.95&1.25&1.80&1.10&1.85\\
&&&$\mathcal{H}_{0,2}$&4.30&3.70&2.00&1.25&2.65&1.50&2.00\\
&&&$\mathcal{H}_{0,3}$&65.55&67.05&34.90&25.95&33.55&33.90&39.85\\
\hline
\end{longtable}

\begin{figure}
\centering
\includegraphics[width= 0.95\textwidth,height=0.9\textheight]{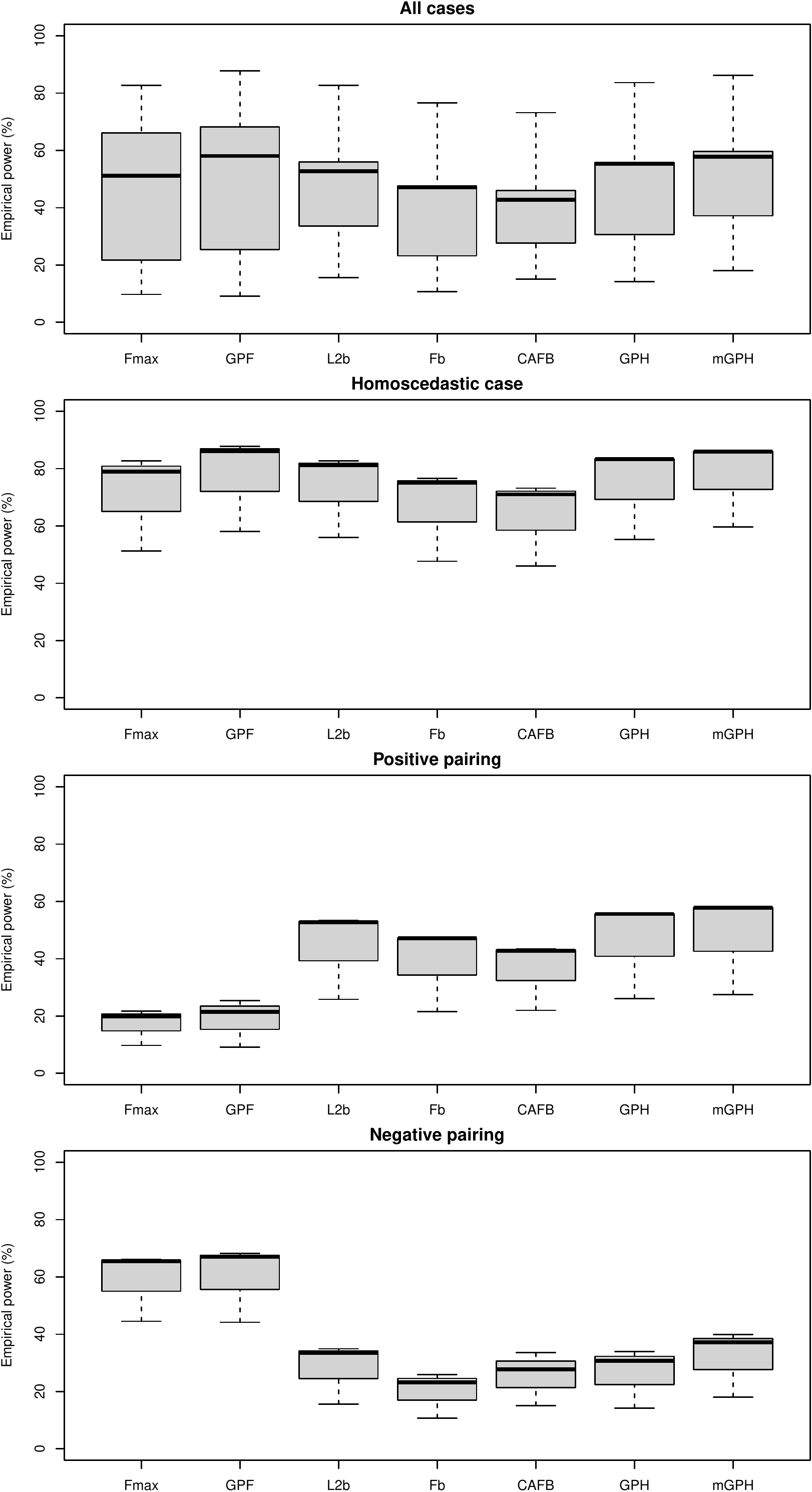}
\caption[Box-and-whisker plots for the empirical powers of all tests obtained under alternative A5 for the Dunnett constrasts and homoscedastic and heteroscedastic cases]{Box-and-whisker plots for the empirical powers (as percentages) of all tests obtained under alternative A5 for the Dunnett constrasts and homoscedastic and heteroscedastic cases}
\end{figure}

\begin{figure}
\centering
\includegraphics[width= 0.95\textwidth,height=0.9\textheight]{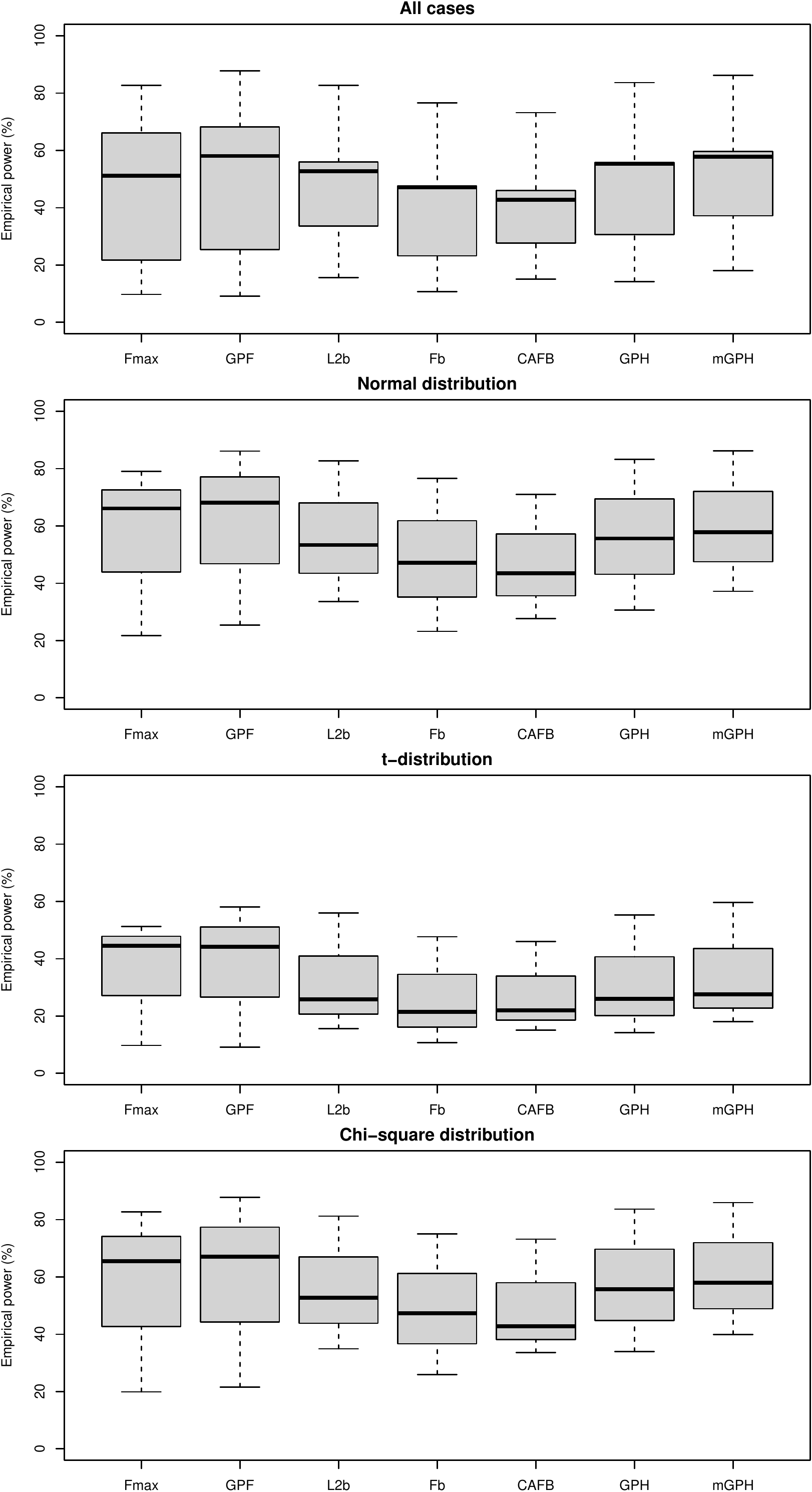}
\caption[Box-and-whisker plots for the empirical powers of all tests obtained under alternative A5 for the Dunnett constrasts and different distributions]{Box-and-whisker plots for the empirical powers (as percentages) of all tests obtained under alternative A5 for the Dunnett constrasts and different distributions}
\end{figure}

\newpage

\begin{longtable}[t]{rrrr|rrrrrrr}
\caption[Empirical powers of all tests obtained under alternative A6 for the Dunnett constrasts]{\label{tab:unnamed-chunk-36}Empirical powers (as percentages) of all tests obtained under alternative A6 for the Dunnett constrasts (D - distribution, $(\lambda_1,\lambda_2,\lambda_3,\lambda_4)$: (1,1,1,1) - homoscedastic case, (1,1.25,1.5,1.75) - heteroscedastic case (positive pairing), (1.75,1.5,1.25,1) - heteroscedastic case (negative pairing))}\\
\hline
D&$(n_1,n_2,n_3,n_4)$&$(\lambda_1,\lambda_2,\lambda_3,\lambda_4)$&$\mathcal{H}$&Fmax&GPF&L2b&Fb&CAFB&GPH&mGPH\\
\hline
\endfirsthead
\caption[]{Empirical powers (as percentages) of all tests obtained under alternative A6 for the Dunnett constrasts (D - distribution, $(\lambda_1,\lambda_2,\lambda_3,\lambda_4)$: (1,1,1,1) - homoscedastic case, (1,1.25,1.5,1.75) - heteroscedastic case (positive pairing), (1.75,1.5,1.25,1) - heteroscedastic case (negative pairing)) \textit{(continued)}}\\
\hline
D&$(n_1,n_2,n_3,n_4)$&$(\lambda_1,\lambda_2,\lambda_3,\lambda_4)$&$\mathcal{H}$&Fmax&GPF&L2b&Fb&CAFB&GPH&mGPH\\
\hline
\endhead
N&(15,20,25,30)&(1,1,1,1)&$\mathcal{H}_{0,1}$&10.90&12.45&14.60&9.80&11.75&12.20&14.70\\
&&&$\mathcal{H}_{0,2}$&73.70&81.15&78.95&70.90&66.90&80.30&83.00\\
&&&$\mathcal{H}_{0,3}$&100.00&99.95&99.90&99.75&98.90&99.90&99.95\\
N&(15,20,25,30)&(1,1.25,1.5,1.75)&$\mathcal{H}_{0,1}$&6.10&6.65&10.00&7.05&8.45&8.20&9.10\\
&&&$\mathcal{H}_{0,2}$&36.45&41.55&57.20&50.50&46.10&57.55&59.80\\
&&&$\mathcal{H}_{0,3}$&72.75&82.15&95.55&93.60&87.90&96.60&97.10\\
N&(15,20,25,30)&(1.75,1.5,1.25,1)&$\mathcal{H}_{0,1}$&5.20&4.65&4.85&3.30&4.40&3.75&5.15\\
&&&$\mathcal{H}_{0,2}$&40.30&45.65&29.70&20.65&23.90&25.45&32.25\\
&&&$\mathcal{H}_{0,3}$&94.95&96.35&80.20&69.85&64.90&80.50&84.65\\
t&(15,20,25,30)&(1,1,1,1)&$\mathcal{H}_{0,1}$&7.25&7.45&8.85&5.35&6.95&7.20&8.70\\
&&&$\mathcal{H}_{0,2}$&46.85&54.25&51.00&43.70&41.25&51.95&55.60\\
&&&$\mathcal{H}_{0,3}$&92.90&96.25&93.00&90.20&86.55&94.40&95.80\\
t&(15,20,25,30)&(1,1.25,1.5,1.75)&$\mathcal{H}_{0,1}$&5.00&4.70&6.65&4.85&5.45&5.80&6.40\\
&&&$\mathcal{H}_{0,2}$&17.75&18.35&32.65&26.45&24.35&32.30&34.00\\
&&&$\mathcal{H}_{0,3}$&39.75&45.00&71.15&67.10&60.70&74.40&75.70\\
t&(15,20,25,30)&(1.75,1.5,1.25,1)&$\mathcal{H}_{0,1}$&5.30&4.10&3.80&2.30&3.45&2.90&4.45\\
&&&$\mathcal{H}_{0,2}$&25.00&26.45&16.65&11.60&13.10&14.35&17.95\\
&&&$\mathcal{H}_{0,3}$&78.15&80.95&47.30&36.50&38.95&45.25&51.65\\
$\chi^2$&(15,20,25,30)&(1,1,1,1)&$\mathcal{H}_{0,1}$&11.95&10.85&14.05&9.50&12.85&11.55&13.75\\
&&&$\mathcal{H}_{0,2}$&75.45&82.35&77.05&69.55&67.75&79.85&82.70\\
&&&$\mathcal{H}_{0,3}$&99.80&99.95&99.30&98.55&98.20&99.55&99.75\\
$\chi^2$&(15,20,25,30)&(1,1.25,1.5,1.75)&$\mathcal{H}_{0,1}$&8.15&7.55&11.55&8.65&9.35&10.05&11.25\\
&&&$\mathcal{H}_{0,2}$&35.10&40.90&57.00&50.65&47.20&58.80&60.55\\
&&&$\mathcal{H}_{0,3}$&75.45&84.15&95.00&93.05&87.65&96.35&97.00\\
$\chi^2$&(15,20,25,30)&(1.75,1.5,1.25,1)&$\mathcal{H}_{0,1}$&6.60&5.40&5.55&3.70&5.40&4.40&6.25\\
&&&$\mathcal{H}_{0,2}$&44.00&47.35&31.05&24.15&31.05&30.30&36.40\\
&&&$\mathcal{H}_{0,3}$&94.70&95.70&76.30&66.80&69.30&77.75&82.90\\
\hline
\end{longtable}

\begin{figure}
\centering
\includegraphics[width= 0.95\textwidth,height=0.9\textheight]{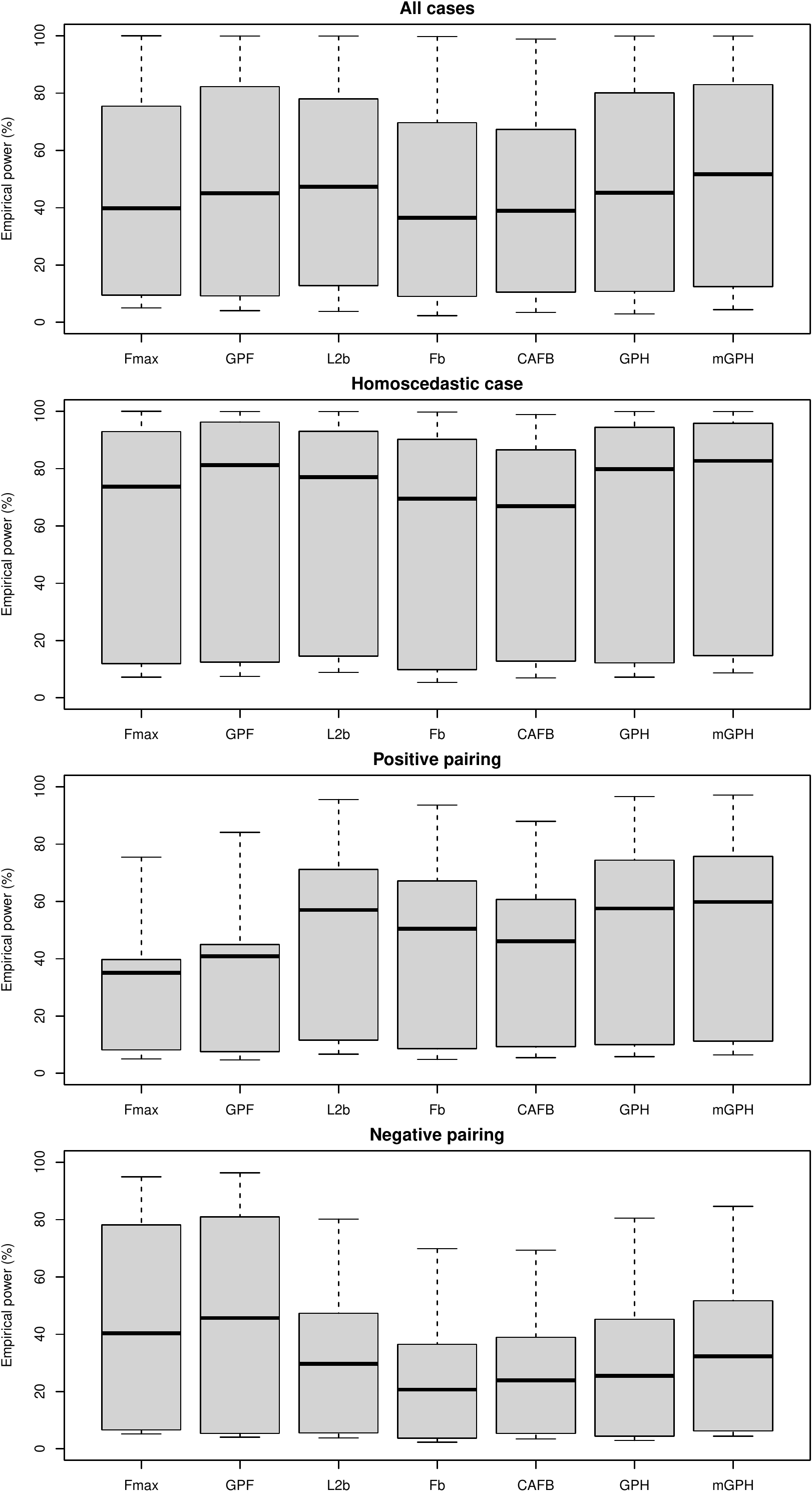}
\caption[Box-and-whisker plots for the empirical powers of all tests obtained under alternative A6 for the Dunnett constrasts and homoscedastic and heteroscedastic cases]{Box-and-whisker plots for the empirical powers (as percentages) of all tests obtained under alternative A6 for the Dunnett constrasts and homoscedastic and heteroscedastic cases}
\end{figure}

\begin{figure}
\centering
\includegraphics[width= 0.95\textwidth,height=0.9\textheight]{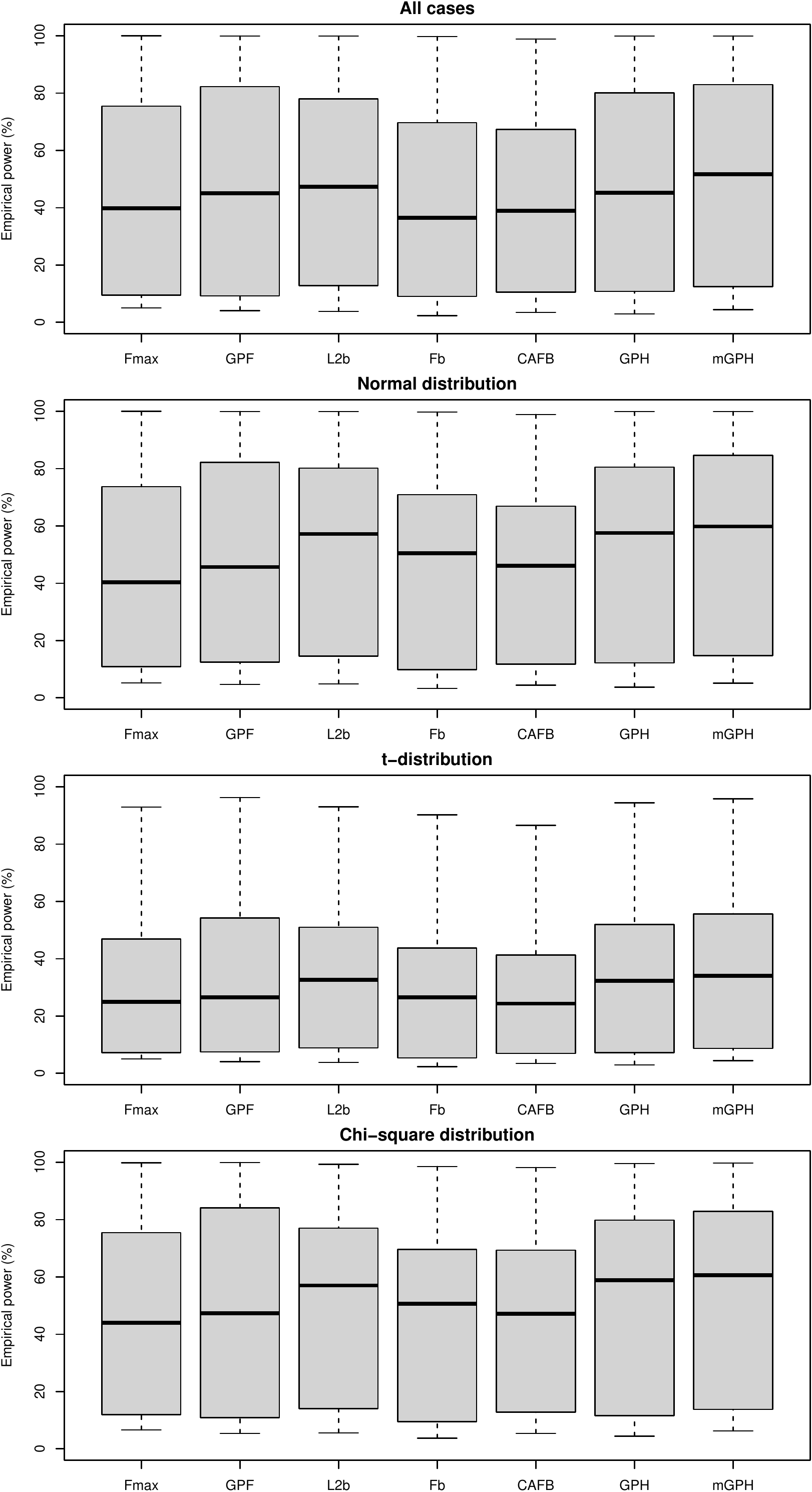}
\caption[Box-and-whisker plots for the empirical powers of all tests obtained under alternative A6 for the Dunnett constrasts and different distributions]{Box-and-whisker plots for the empirical powers (as percentages) of all tests obtained under alternative A6 for the Dunnett constrasts and different distributions}
\end{figure}

\newpage

\begin{longtable}[t]{rrrr|rrrrrrr}
\caption[Empirical FWER and sizes of all tests obtained for the Tukey constrasts]{\label{tab:unnamed-chunk-39}Empirical FWER and sizes (as percentages) of all tests obtained for the Tukey constrasts (D - distribution, $(\lambda_1,\lambda_2,\lambda_3,\lambda_4)$: (1,1,1,1) - homoscedastic case, (1,1.25,1.5,1.75) - heteroscedastic case (positive pairing), (1.75,1.5,1.25,1) - heteroscedastic case (negative pairing))}\\
\hline
D&$(n_1,n_2,n_3,n_4)$&$(\lambda_1,\lambda_2,\lambda_3,\lambda_4)$&$\mathcal{H}$&Fmax&GPF&L2b&Fb&CAFB&GPH&mGPH\\
\hline
\endfirsthead
\caption[]{Empirical FWER and sizes (as percentages) of all tests obtained for the Tukey constrasts (D - distribution, $(\lambda_1,\lambda_2,\lambda_3,\lambda_4)$: (1,1,1,1) - homoscedastic case, (1,1.25,1.5,1.75) - heteroscedastic case (positive pairing), (1.75,1.5,1.25,1) - heteroscedastic case (negative pairing)) \textit{(continued)}}\\
\hline
D&$(n_1,n_2,n_3,n_4)$&$(\lambda_1,\lambda_2,\lambda_3,\lambda_4)$&$\mathcal{H}$&Fmax&GPF&L2b&Fb&CAFB&GPH&mGPH\\
\hline
\endhead
N&(15,20,25,30)&(1,1,1,1)&FWER&4.50&3.35&5.25&2.90&5.95&3.65&4.25\\
&&&$\mathcal{H}_{0,1}$&0.70&0.55&1.00&0.55&1.00&0.80&0.90\\
&&&$\mathcal{H}_{0,2}$&0.95&0.65&1.30&0.60&1.10&0.65&0.95\\
&&&$\mathcal{H}_{0,3}$&0.75&0.55&1.00&0.60&1.70&0.70&0.75\\
&&&$\mathcal{H}_{0,4}$&0.80&0.65&0.80&0.50&1.05&0.65&0.80\\
&&&$\mathcal{H}_{0,5}$&1.05&0.90&1.00&0.50&0.75&0.75&0.80\\
&&&$\mathcal{H}_{0,6}$&0.80&0.75&1.05&0.80&0.90&0.80&0.95\\
N&(15,20,25,30)&(1,1.25,1.5,1.75)&FWER&3.65&2.10&5.15&3.60&5.70&3.65&4.65\\
&&&$\mathcal{H}_{0,1}$&1.15&0.55&1.30&1.00&1.00&0.75&1.00\\
&&&$\mathcal{H}_{0,2}$&0.45&0.20&0.90&0.65&1.20&0.95&0.95\\
&&&$\mathcal{H}_{0,3}$&0.15&0.10&1.45&1.00&0.80&0.60&0.85\\
&&&$\mathcal{H}_{0,4}$&0.90&0.60&1.10&0.55&1.05&0.75&0.85\\
&&&$\mathcal{H}_{0,5}$&0.60&0.15&0.70&0.35&0.90&0.50&0.75\\
&&&$\mathcal{H}_{0,6}$&0.80&0.55&0.60&0.50&0.95&0.60&0.85\\
N&(15,20,25,30)&(1.75,1.5,1.25,1)&FWER&11.65&8.60&4.85&2.90&6.00&3.05&3.75\\
&&&$\mathcal{H}_{0,1}$&0.60&0.80&1.20&0.60&1.15&0.60&0.85\\
&&&$\mathcal{H}_{0,2}$&2.40&1.85&0.90&0.55&0.95&0.55&0.65\\
&&&$\mathcal{H}_{0,3}$&6.15&4.40&0.80&0.30&1.20&0.45&0.55\\
&&&$\mathcal{H}_{0,4}$&1.00&1.05&1.15&0.80&0.85&0.80&1.05\\
&&&$\mathcal{H}_{0,5}$&2.25&1.95&0.90&0.55&1.60&0.60&0.80\\
&&&$\mathcal{H}_{0,6}$&1.05&1.00&0.90&0.60&0.80&0.65&0.75\\
N&(30,40,50,60)&(1,1,1,1)&FWER&5.25&4.30&4.70&3.80&5.40&4.40&4.80\\
&&&$\mathcal{H}_{0,1}$&0.95&1.00&1.05&0.95&1.20&0.90&1.00\\
&&&$\mathcal{H}_{0,2}$&1.05&0.95&1.00&0.80&1.15&0.80&0.85\\
&&&$\mathcal{H}_{0,3}$&0.95&0.60&0.75&0.60&0.95&0.70&0.70\\
&&&$\mathcal{H}_{0,4}$&0.90&0.70&0.95&0.80&1.10&1.15&1.15\\
&&&$\mathcal{H}_{0,5}$&1.20&0.90&0.80&0.60&1.35&0.85&0.90\\
&&&$\mathcal{H}_{0,6}$&0.85&1.10&1.10&0.85&0.45&0.90&1.15\\
N&(30,40,50,60)&(1,1.25,1.5,1.75)&FWER&2.40&2.20&5.10&4.15&4.85&3.80&4.30\\
&&&$\mathcal{H}_{0,1}$&0.30&0.45&1.05&0.80&0.90&0.60&0.80\\
&&&$\mathcal{H}_{0,2}$&0.20&0.30&1.15&1.00&1.00&0.75&0.85\\
&&&$\mathcal{H}_{0,3}$&0.15&0.10&1.05&0.90&0.95&0.80&0.90\\
&&&$\mathcal{H}_{0,4}$&0.90&1.00&1.45&1.25&1.00&1.20&1.25\\
&&&$\mathcal{H}_{0,5}$&0.45&0.20&0.55&0.40&0.75&0.45&0.55\\
&&&$\mathcal{H}_{0,6}$&0.60&0.45&0.85&0.60&0.70&0.55&0.65\\
N&(30,40,50,60)&(1.75,1.5,1.25,1)&FWER&10.90&10.15&4.70&3.70&5.15&3.55&4.70\\
&&&$\mathcal{H}_{0,1}$&1.15&1.10&1.05&0.75&0.85&0.90&1.15\\
&&&$\mathcal{H}_{0,2}$&2.00&1.70&1.00&0.70&0.50&0.75&0.90\\
&&&$\mathcal{H}_{0,3}$&5.80&5.40&1.05&0.65&1.10&0.65&0.80\\
&&&$\mathcal{H}_{0,4}$&1.30&1.20&1.00&0.75&1.20&0.85&1.05\\
&&&$\mathcal{H}_{0,5}$&2.30&2.60&1.15&1.00&1.20&0.85&1.30\\
&&&$\mathcal{H}_{0,6}$&1.05&0.95&1.00&0.85&0.70&0.75&0.90\\
N&(60,80,100,120)&(1,1,1,1)&FWER&4.65&4.40&4.40&4.00&4.40&4.60&5.25\\
&&&$\mathcal{H}_{0,1}$&0.95&1.00&0.95&0.80&1.05&0.95&1.10\\
&&&$\mathcal{H}_{0,2}$&0.90&0.65&0.55&0.50&0.75&0.70&0.80\\
&&&$\mathcal{H}_{0,3}$&0.70&1.05&1.05&0.85&0.70&1.05&1.20\\
&&&$\mathcal{H}_{0,4}$&0.90&0.75&0.85&0.75&0.75&0.90&1.05\\
&&&$\mathcal{H}_{0,5}$&1.25&0.85&0.90&0.85&0.90&0.80&0.95\\
&&&$\mathcal{H}_{0,6}$&0.85&0.70&0.75&0.75&0.60&0.75&0.80\\
N&(60,80,100,120)&(1,1.25,1.5,1.75)&FWER&2.90&2.15&4.35&4.00&4.70&3.90&4.60\\
&&&$\mathcal{H}_{0,1}$&0.60&0.50&0.55&0.55&0.50&0.60&0.65\\
&&&$\mathcal{H}_{0,2}$&0.20&0.15&0.75&0.70&0.95&0.65&0.80\\
&&&$\mathcal{H}_{0,3}$&0.05&0.00&0.90&0.75&0.70&0.90&1.05\\
&&&$\mathcal{H}_{0,4}$&0.85&0.60&0.85&0.80&0.80&0.70&0.95\\
&&&$\mathcal{H}_{0,5}$&0.35&0.40&1.05&0.85&1.05&0.80&0.95\\
&&&$\mathcal{H}_{0,6}$&1.00&0.55&0.70&0.75&0.95&0.75&0.90\\
N&(60,80,100,120)&(1.75,1.5,1.25,1)&FWER&10.45&9.35&4.40&3.80&3.85&3.20&4.30\\
&&&$\mathcal{H}_{0,1}$&1.15&1.15&0.95&0.90&0.90&0.70&0.80\\
&&&$\mathcal{H}_{0,2}$&2.25&2.20&1.05&0.75&0.90&0.60&0.80\\
&&&$\mathcal{H}_{0,3}$&6.65&4.95&0.70&0.55&0.65&0.70&1.15\\
&&&$\mathcal{H}_{0,4}$&1.15&1.00&0.85&0.70&0.80&0.60&0.75\\
&&&$\mathcal{H}_{0,5}$&1.35&2.10&0.85&0.85&0.45&0.65&0.85\\
&&&$\mathcal{H}_{0,6}$&0.95&0.95&0.80&0.80&0.55&0.70&1.10\\
t&(15,20,25,30)&(1,1,1,1)&FWER&4.30&3.45&4.00&2.85&4.90&3.00&3.60\\
&&&$\mathcal{H}_{0,1}$&0.55&0.50&0.70&0.45&0.60&0.40&0.45\\
&&&$\mathcal{H}_{0,2}$&0.55&0.75&0.90&0.55&0.75&0.45&0.60\\
&&&$\mathcal{H}_{0,3}$&0.90&0.85&1.15&0.90&1.00&0.75&0.75\\
&&&$\mathcal{H}_{0,4}$&0.80&0.70&0.65&0.50&1.15&0.65&0.65\\
&&&$\mathcal{H}_{0,5}$&0.95&0.60&0.80&0.55&0.90&0.60&0.80\\
&&&$\mathcal{H}_{0,6}$&0.85&0.40&0.35&0.25&1.10&0.45&0.65\\
t&(15,20,25,30)&(1,1.25,1.5,1.75)&FWER&2.90&1.90&4.10&2.80&4.00&3.20&3.60\\
&&&$\mathcal{H}_{0,1}$&0.70&0.40&0.95&0.55&0.75&0.60&0.70\\
&&&$\mathcal{H}_{0,2}$&0.40&0.25&0.65&0.60&0.50&0.65&0.75\\
&&&$\mathcal{H}_{0,3}$&0.25&0.15&0.90&0.55&0.80&0.85&0.95\\
&&&$\mathcal{H}_{0,4}$&0.65&0.45&0.70&0.45&0.75&0.55&0.60\\
&&&$\mathcal{H}_{0,5}$&0.45&0.25&0.90&0.55&0.85&0.40&0.50\\
&&&$\mathcal{H}_{0,6}$&0.70&0.70&0.80&0.60&0.60&0.75&0.80\\
t&(15,20,25,30)&(1.75,1.5,1.25,1)&FWER&12.10&8.65&4.10&2.30&6.20&2.85&3.55\\
&&&$\mathcal{H}_{0,1}$&1.05&0.65&0.55&0.30&1.00&0.50&0.55\\
&&&$\mathcal{H}_{0,2}$&3.00&1.30&0.70&0.35&1.05&0.45&0.55\\
&&&$\mathcal{H}_{0,3}$&5.65&4.45&0.75&0.30&1.40&0.60&0.70\\
&&&$\mathcal{H}_{0,4}$&1.60&1.10&1.05&0.70&0.90&0.60&0.80\\
&&&$\mathcal{H}_{0,5}$&2.70&2.00&0.70&0.45&1.15&0.40&0.55\\
&&&$\mathcal{H}_{0,6}$&1.10&1.10&0.95&0.70&1.10&0.70&0.95\\
t&(30,40,50,60)&(1,1,1,1)&FWER&3.95&3.30&4.30&3.65&5.10&3.15&3.65\\
&&&$\mathcal{H}_{0,1}$&0.55&0.65&0.80&0.70&0.80&0.75&0.80\\
&&&$\mathcal{H}_{0,2}$&0.75&0.65&0.95&0.80&1.05&0.60&0.80\\
&&&$\mathcal{H}_{0,3}$&0.75&0.50&0.80&0.60&0.75&0.45&0.50\\
&&&$\mathcal{H}_{0,4}$&0.80&0.65&0.70&0.50&1.05&0.55&0.65\\
&&&$\mathcal{H}_{0,5}$&0.60&0.50&0.60&0.40&1.00&0.40&0.50\\
&&&$\mathcal{H}_{0,6}$&0.85&0.85&0.95&1.00&0.80&0.75&0.95\\
t&(30,40,50,60)&(1,1.25,1.5,1.75)&FWER&2.95&2.35&4.45&3.70&4.55&3.90&4.80\\
&&&$\mathcal{H}_{0,1}$&0.45&0.45&0.80&0.60&0.95&0.65&0.85\\
&&&$\mathcal{H}_{0,2}$&0.25&0.20&0.80&0.55&0.80&0.65&0.70\\
&&&$\mathcal{H}_{0,3}$&0.15&0.10&1.10&0.85&0.65&0.70&1.00\\
&&&$\mathcal{H}_{0,4}$&0.80&0.60&0.75&0.70&0.50&0.80&0.85\\
&&&$\mathcal{H}_{0,5}$&0.70&0.35&0.80&0.70&1.05&0.80&1.10\\
&&&$\mathcal{H}_{0,6}$&0.90&0.85&0.80&0.75&0.95&0.75&1.00\\
t&(30,40,50,60)&(1.75,1.5,1.25,1)&FWER&11.55&10.15&3.70&3.10&5.00&2.90&4.25\\
&&&$\mathcal{H}_{0,1}$&1.10&0.80&0.60&0.60&0.70&0.35&0.70\\
&&&$\mathcal{H}_{0,2}$&2.20&2.15&0.65&0.45&1.20&0.65&0.75\\
&&&$\mathcal{H}_{0,3}$&6.05&5.05&0.75&0.45&0.70&0.35&0.65\\
&&&$\mathcal{H}_{0,4}$&1.20&0.70&0.45&0.35&0.65&0.60&0.65\\
&&&$\mathcal{H}_{0,5}$&2.00&2.30&0.75&0.70&1.05&0.60&0.95\\
&&&$\mathcal{H}_{0,6}$&1.75&1.50&1.30&1.15&1.05&0.95&1.40\\
t&(60,80,100,120)&(1,1,1,1)&FWER&4.70&4.20&4.35&4.10&4.25&4.05&4.45\\
&&&$\mathcal{H}_{0,1}$&0.90&1.15&1.05&1.20&1.10&1.15&1.20\\
&&&$\mathcal{H}_{0,2}$&0.80&0.55&0.85&0.80&0.65&0.85&1.00\\
&&&$\mathcal{H}_{0,3}$&0.60&0.70&0.90&0.60&0.70&0.50&0.65\\
&&&$\mathcal{H}_{0,4}$&1.05&0.80&0.85&0.85&0.60&0.75&0.75\\
&&&$\mathcal{H}_{0,5}$&0.50&0.70&0.75&0.80&0.75&0.70&0.85\\
&&&$\mathcal{H}_{0,6}$&1.35&1.15&1.00&0.90&0.90&1.05&1.20\\
t&(60,80,100,120)&(1,1.25,1.5,1.75)&FWER&2.70&2.25&4.30&4.05&4.10&3.85&4.60\\
&&&$\mathcal{H}_{0,1}$&0.45&0.55&0.75&0.75&0.60&0.45&0.60\\
&&&$\mathcal{H}_{0,2}$&0.25&0.10&0.70&0.60&0.85&0.70&0.80\\
&&&$\mathcal{H}_{0,3}$&0.05&0.00&0.70&0.60&0.65&0.55&0.80\\
&&&$\mathcal{H}_{0,4}$&1.00&0.85&1.10&1.10&0.70&1.15&1.30\\
&&&$\mathcal{H}_{0,5}$&0.35&0.30&0.80&0.70&1.00&0.75&0.85\\
&&&$\mathcal{H}_{0,6}$&0.75&0.70&1.00&0.90&0.60&0.75&0.85\\
t&(60,80,100,120)&(1.75,1.5,1.25,1)&FWER&11.10&9.80&4.45&3.90&4.05&4.10&5.35\\
&&&$\mathcal{H}_{0,1}$&1.10&1.05&0.70&0.65&0.40&0.70&0.90\\
&&&$\mathcal{H}_{0,2}$&2.15&2.35&0.90&0.80&0.85&0.85&1.20\\
&&&$\mathcal{H}_{0,3}$&5.95&4.55&0.85&0.65&0.50&0.80&1.15\\
&&&$\mathcal{H}_{0,4}$&0.95&1.00&0.65&0.50&0.75&0.80&0.95\\
&&&$\mathcal{H}_{0,5}$&1.95&2.05&1.10&0.95&0.80&0.75&0.90\\
&&&$\mathcal{H}_{0,6}$&1.30&1.40&1.25&1.15&1.10&1.05&1.35\\
$\chi^2$&(15,20,25,30)&(1,1,1,1)&FWER&4.90&3.25&4.25&2.60&5.45&3.05&3.55\\
&&&$\mathcal{H}_{0,1}$&1.00&0.75&1.20&0.75&0.75&0.60&0.70\\
&&&$\mathcal{H}_{0,2}$&0.60&0.70&0.95&0.35&1.35&0.50&0.75\\
&&&$\mathcal{H}_{0,3}$&0.90&0.45&0.75&0.40&1.10&0.50&0.60\\
&&&$\mathcal{H}_{0,4}$&1.25&0.60&0.80&0.45&0.75&0.60&0.80\\
&&&$\mathcal{H}_{0,5}$&0.70&0.45&0.55&0.30&1.15&0.45&0.45\\
&&&$\mathcal{H}_{0,6}$&0.80&0.75&0.60&0.60&1.00&0.55&0.70\\
$\chi^2$&(15,20,25,30)&(1,1.25,1.5,1.75)&FWER&2.95&2.00&5.40&3.40&4.70&3.35&4.15\\
&&&$\mathcal{H}_{0,1}$&0.60&0.45&1.45&0.80&0.70&0.75&0.80\\
&&&$\mathcal{H}_{0,2}$&0.35&0.20&0.85&0.35&0.70&0.35&0.65\\
&&&$\mathcal{H}_{0,3}$&0.05&0.05&0.75&0.50&1.10&0.55&0.70\\
&&&$\mathcal{H}_{0,4}$&0.90&0.70&1.05&0.75&1.20&0.85&1.05\\
&&&$\mathcal{H}_{0,5}$&0.50&0.20&0.85&0.60&0.70&0.55&0.60\\
&&&$\mathcal{H}_{0,6}$&0.65&0.60&0.90&0.70&0.75&0.70&0.85\\
$\chi^2$&(15,20,25,30)&(1.75,1.5,1.25,1)&FWER&12.35&8.40&4.30&2.55&6.60&2.95&3.95\\
&&&$\mathcal{H}_{0,1}$&1.25&0.70&1.00&0.50&0.85&0.55&0.60\\
&&&$\mathcal{H}_{0,2}$&2.10&1.60&1.00&0.40&1.55&0.55&0.85\\
&&&$\mathcal{H}_{0,3}$&6.50&4.40&1.10&0.30&1.60&0.60&0.85\\
&&&$\mathcal{H}_{0,4}$&1.60&1.05&1.05&0.75&1.20&0.80&1.10\\
&&&$\mathcal{H}_{0,5}$&2.00&1.40&0.55&0.35&1.10&0.45&0.65\\
&&&$\mathcal{H}_{0,6}$&1.40&1.30&0.90&0.70&1.05&0.85&1.10\\
$\chi^2$&(30,40,50,60)&(1,1,1,1)&FWER&4.65&3.85&4.35&3.45&5.45&3.40&4.05\\
&&&$\mathcal{H}_{0,1}$&0.70&0.85&0.90&0.70&0.75&0.70&0.85\\
&&&$\mathcal{H}_{0,2}$&0.95&1.05&1.00&0.70&1.25&0.70&0.80\\
&&&$\mathcal{H}_{0,3}$&0.80&0.85&1.00&0.70&1.05&0.80&1.00\\
&&&$\mathcal{H}_{0,4}$&0.70&0.35&0.55&0.50&0.55&0.35&0.45\\
&&&$\mathcal{H}_{0,5}$&1.05&0.80&0.70&0.65&1.10&0.60&0.75\\
&&&$\mathcal{H}_{0,6}$&0.90&0.75&0.85&0.70&0.90&0.75&0.85\\
$\chi^2$&(30,40,50,60)&(1,1.25,1.5,1.75)&FWER&2.50&2.00&4.20&3.45&4.45&3.20&4.15\\
&&&$\mathcal{H}_{0,1}$&0.45&0.40&0.65&0.40&0.70&0.45&0.85\\
&&&$\mathcal{H}_{0,2}$&0.45&0.20&0.80&0.70&0.95&0.55&0.70\\
&&&$\mathcal{H}_{0,3}$&0.05&0.15&0.90&0.70&0.80&0.70&0.95\\
&&&$\mathcal{H}_{0,4}$&0.65&0.45&0.85&0.70&0.65&0.45&0.50\\
&&&$\mathcal{H}_{0,5}$&0.55&0.25&0.60&0.55&0.85&0.65&0.80\\
&&&$\mathcal{H}_{0,6}$&0.45&0.65&0.85&0.75&0.85&0.55&0.75\\
$\chi^2$&(30,40,50,60)&(1.75,1.5,1.25,1)&FWER&11.85&8.75&3.80&3.00&4.85&3.35&3.65\\
&&&$\mathcal{H}_{0,1}$&1.15&0.90&0.85&0.60&1.05&0.75&0.90\\
&&&$\mathcal{H}_{0,2}$&2.60&2.30&0.70&0.55&0.90&0.60&0.70\\
&&&$\mathcal{H}_{0,3}$&5.75&4.85&0.65&0.50&0.70&0.90&0.95\\
&&&$\mathcal{H}_{0,4}$&1.15&0.60&0.70&0.40&0.80&0.45&0.45\\
&&&$\mathcal{H}_{0,5}$&2.45&1.80&0.60&0.45&0.65&0.55&0.65\\
&&&$\mathcal{H}_{0,6}$&1.25&1.20&1.00&0.90&1.05&0.80&0.85\\
$\chi^2$&(60,80,100,120)&(1,1,1,1)&FWER&4.25&4.75&4.35&4.15&4.05&4.50&4.90\\
&&&$\mathcal{H}_{0,1}$&0.70&1.00&1.15&1.05&0.75&0.90&1.10\\
&&&$\mathcal{H}_{0,2}$&0.80&0.85&0.75&0.70&0.50&0.70&0.85\\
&&&$\mathcal{H}_{0,3}$&0.80&1.10&0.90&0.85&1.05&0.90&1.00\\
&&&$\mathcal{H}_{0,4}$&0.90&0.75&0.85&0.65&0.60&0.95&1.00\\
&&&$\mathcal{H}_{0,5}$&0.70&0.75&0.50&0.50&1.05&0.90&1.05\\
&&&$\mathcal{H}_{0,6}$&0.75&1.15&1.10&1.05&0.50&1.10&1.25\\
$\chi^2$&(60,80,100,120)&(1,1.25,1.5,1.75)&FWER&2.55&2.10&4.55&4.10&4.05&4.10&4.80\\
&&&$\mathcal{H}_{0,1}$&0.35&0.35&0.65&0.60&0.65&0.55&0.65\\
&&&$\mathcal{H}_{0,2}$&0.15&0.10&0.90&0.70&0.85&0.90&1.00\\
&&&$\mathcal{H}_{0,3}$&0.05&0.20&0.65&0.60&0.60&0.65&0.85\\
&&&$\mathcal{H}_{0,4}$&1.00&0.80&1.15&1.00&0.45&0.70&0.95\\
&&&$\mathcal{H}_{0,5}$&0.35&0.25&1.00&0.80&0.75&0.95&1.05\\
&&&$\mathcal{H}_{0,6}$&0.70&0.70&0.80&0.90&0.80&0.95&1.05\\
$\chi^2$&(60,80,100,120)&(1.75,1.5,1.25,1)&FWER&10.65&9.45&4.00&3.75&3.75&3.60&4.60\\
&&&$\mathcal{H}_{0,1}$&1.25&0.90&0.95&0.90&0.50&0.65&0.80\\
&&&$\mathcal{H}_{0,2}$&2.55&2.20&0.50&0.55&0.90&0.85&1.00\\
&&&$\mathcal{H}_{0,3}$&5.80&4.85&1.10&1.00&0.80&0.80&1.00\\
&&&$\mathcal{H}_{0,4}$&1.00&0.95&0.80&0.65&0.60&0.70&0.90\\
&&&$\mathcal{H}_{0,5}$&1.60&1.90&0.85&0.85&0.40&0.80&0.95\\
&&&$\mathcal{H}_{0,6}$&0.85&0.85&0.60&0.60&1.05&0.70&0.90\\
\hline
\end{longtable}

\begin{figure}
\centering
\includegraphics[width= 0.95\textwidth,height=0.9\textheight]{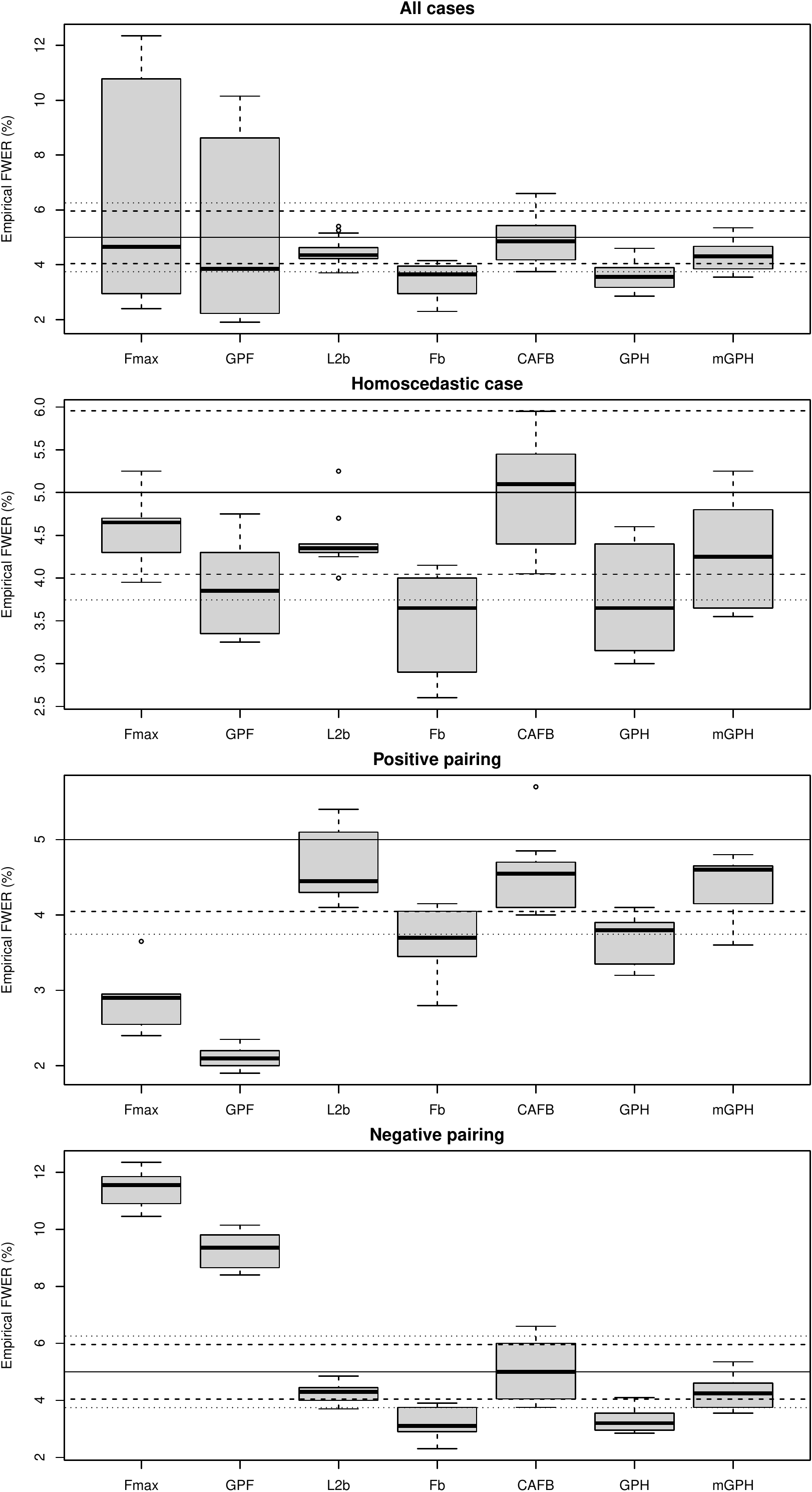}
\caption[Box-and-whisker plots for the empirical FWER of all tests obtained for the Tukey constrasts and homoscedastic and heteroscedastic cases]{Box-and-whisker plots for the empirical FWER (as percentages) of all tests obtained for the Tukey constrasts and homoscedastic and heteroscedastic cases}
\end{figure}

\begin{figure}
\centering
\includegraphics[width= 0.95\textwidth,height=0.9\textheight]{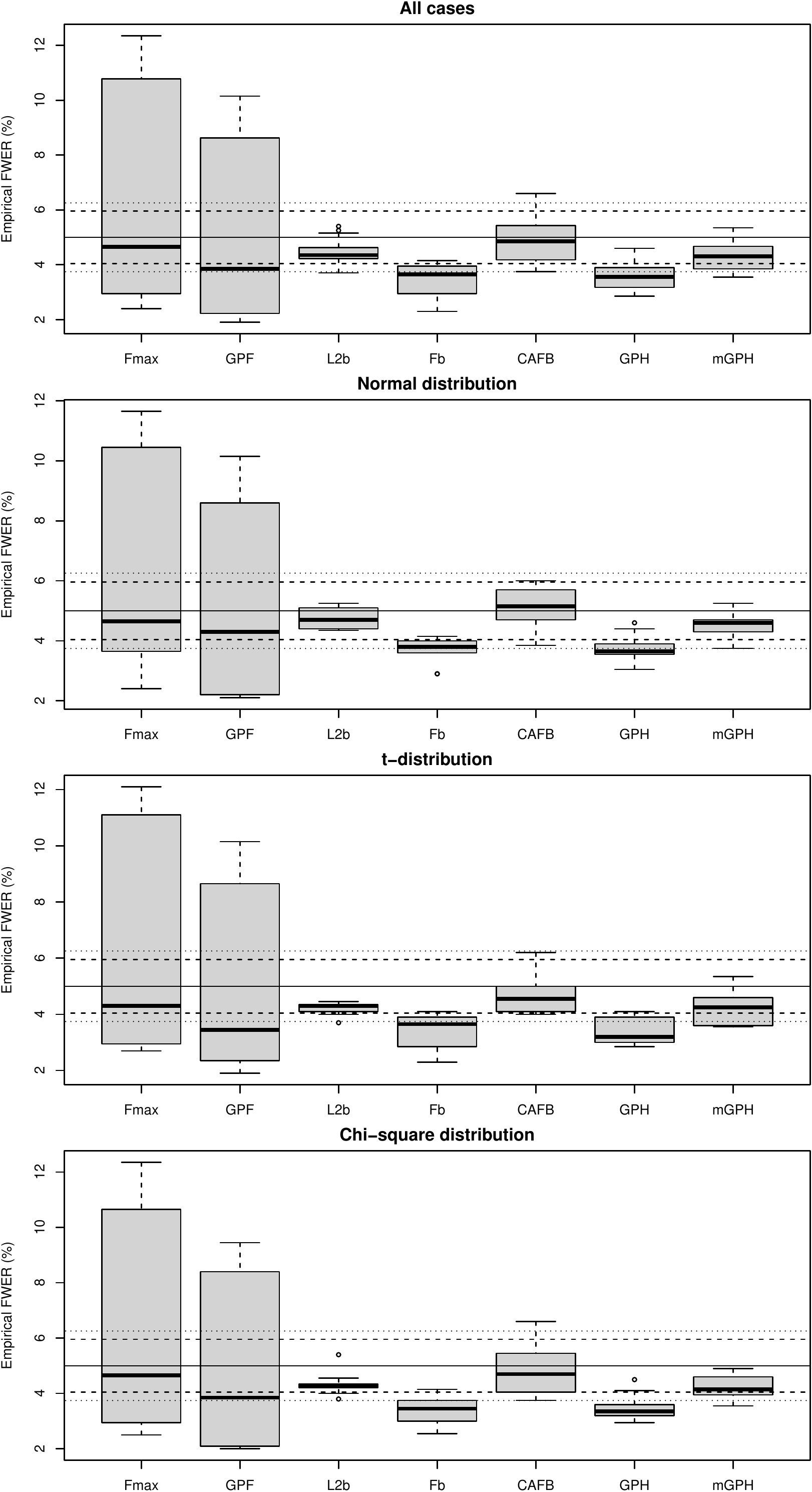}
\caption[Box-and-whisker plots for the empirical FWER of all tests obtained for the Tukey constrasts and different
distributions]{Box-and-whisker plots for the empirical FWER (as percentages) of all tests obtained for the Tukey constrasts and different distributions}
\end{figure}

\newpage

\begin{longtable}[t]{rrrr|rrrrrrr}
\caption[Empirical sizes and powers of all tests obtained under alternative A1 for the Tukey constrasts]{\label{tab:unnamed-chunk-49}Empirical sizes ($\mathcal{H}_{0,1}$, $\mathcal{H}_{0,2}$, $\mathcal{H}_{0,4}$) and powers ($\mathcal{H}_{0,3}$, $\mathcal{H}_{0,5}$, $\mathcal{H}_{0,6}$) (as percentages) of all tests obtained under alternative A1 for the Tukey constrasts (D - distribution, $(\lambda_1,\lambda_2,\lambda_3,\lambda_4)$: (1,1,1,1) - homoscedastic case, (1,1.25,1.5,1.75) - heteroscedastic case (positive pairing), (1.75,1.5,1.25,1) - heteroscedastic case (negative pairing))}\\
\hline
D&$(n_1,n_2,n_3,n_4)$&$(\lambda_1,\lambda_2,\lambda_3,\lambda_4)$&$\mathcal{H}$&Fmax&GPF&L2b&Fb&CAFB&GPH&mGPH\\
\hline
\endfirsthead
\caption[]{Empirical sizes ($\mathcal{H}_{0,1}$, $\mathcal{H}_{0,2}$, $\mathcal{H}_{0,4}$) and powers ($\mathcal{H}_{0,3}$, $\mathcal{H}_{0,5}$, $\mathcal{H}_{0,6}$) (as percentages) of all tests obtained under alternative A1 for the Tukey constrasts (D - distribution, $(\lambda_1,\lambda_2,\lambda_3,\lambda_4)$: (1,1,1,1) - homoscedastic case, (1,1.25,1.5,1.75) - heteroscedastic case (positive pairing), (1.75,1.5,1.25,1) - heteroscedastic case (negative pairing)) \textit{(continued)}}\\
\hline
D&$(n_1,n_2,n_3,n_4)$&$(\lambda_1,\lambda_2,\lambda_3,\lambda_4)$&$\mathcal{H}$&Fmax&GPF&L2b&Fb&CAFB&GPH&mGPH\\
\hline
\endhead
N&(15,20,25,30)&(1,1,1,1)&$\mathcal{H}_{0,1}$&0.55&0.60&0.95&0.65&1.20&0.65&0.75\\
&&&$\mathcal{H}_{0,2}$&0.90&0.60&0.70&0.35&1.05&0.30&0.40\\
&&&$\mathcal{H}_{0,3}$&73.20&82.35&77.90&69.00&64.85&79.25&81.20\\
&&&$\mathcal{H}_{0,4}$&0.80&1.15&1.20&0.85&1.30&1.00&1.15\\
&&&$\mathcal{H}_{0,5}$&84.30&92.35&88.50&84.55&76.85&91.50&92.40\\
&&&$\mathcal{H}_{0,6}$&90.45&96.50&94.35&91.45&83.25&96.65&97.05\\
N&(15,20,25,30)&(1,1.25,1.5,1.75)&$\mathcal{H}_{0,1}$&0.60&0.25&0.90&0.45&0.55&0.50&0.60\\
&&&$\mathcal{H}_{0,2}$&0.05&0.20&1.40&0.75&1.10&0.90&1.10\\
&&&$\mathcal{H}_{0,3}$&15.75&18.55&44.75&37.70&34.75&47.10&49.90\\
&&&$\mathcal{H}_{0,4}$&0.85&0.55&0.75&0.45&0.85&0.65&0.85\\
&&&$\mathcal{H}_{0,5}$&28.85&34.15&43.60&37.60&33.10&46.10&49.20\\
&&&$\mathcal{H}_{0,6}$&34.60&40.70&39.75&35.25&31.80&42.70&45.50\\
N&(15,20,25,30)&(1.75,1.5,1.25,1)&$\mathcal{H}_{0,1}$&1.10&1.05&1.30&0.60&1.10&0.50&0.80\\
&&&$\mathcal{H}_{0,2}$&1.90&1.65&0.80&0.40&1.00&0.55&0.55\\
&&&$\mathcal{H}_{0,3}$&56.05&59.00&24.80&15.25&21.25&23.75&26.55\\
&&&$\mathcal{H}_{0,4}$&1.75&1.35&1.50&0.80&0.95&1.00&1.10\\
&&&$\mathcal{H}_{0,5}$&64.95&72.00&51.80&43.85&39.85&53.85&57.85\\
&&&$\mathcal{H}_{0,6}$&80.75&87.05&79.95&75.40&67.80&84.00&86.65\\
t&(15,20,25,30)&(1,1,1,1)&$\mathcal{H}_{0,1}$&0.80&0.45&0.85&0.30&1.10&0.50&0.50\\
&&&$\mathcal{H}_{0,2}$&0.85&0.50&0.80&0.55&0.80&0.60&0.75\\
&&&$\mathcal{H}_{0,3}$&41.90&49.40&44.95&36.90&35.20&44.85&48.50\\
&&&$\mathcal{H}_{0,4}$&0.95&1.05&1.00&0.60&0.85&1.05&1.15\\
&&&$\mathcal{H}_{0,5}$&55.50&63.80&56.65&50.40&45.60&61.30&64.50\\
&&&$\mathcal{H}_{0,6}$&65.50&72.80&65.35&61.60&54.50&71.15&73.80\\
t&(15,20,25,30)&(1,1.25,1.5,1.75)&$\mathcal{H}_{0,1}$&0.85&0.30&0.80&0.45&1.00&0.40&0.50\\
&&&$\mathcal{H}_{0,2}$&0.30&0.15&0.75&0.50&0.75&0.80&0.85\\
&&&$\mathcal{H}_{0,3}$&5.65&5.50&21.05&16.65&17.00&20.80&22.95\\
&&&$\mathcal{H}_{0,4}$&0.95&0.60&1.10&0.75&0.85&0.75&0.85\\
&&&$\mathcal{H}_{0,5}$&11.30&11.85&18.10&14.30&14.60&18.60&20.45\\
&&&$\mathcal{H}_{0,6}$&16.00&18.15&17.10&14.50&14.25&18.95&21.25\\
t&(15,20,25,30)&(1.75,1.5,1.25,1)&$\mathcal{H}_{0,1}$&1.35&1.00&1.10&0.45&1.30&0.65&0.90\\
&&&$\mathcal{H}_{0,2}$&2.55&1.40&0.75&0.35&1.35&0.40&0.50\\
&&&$\mathcal{H}_{0,3}$&34.30&33.85&10.15&6.70&8.85&8.75&10.75\\
&&&$\mathcal{H}_{0,4}$&1.05&0.70&0.75&0.50&1.05&0.60&0.85\\
&&&$\mathcal{H}_{0,5}$&37.80&42.10&24.40&19.35&20.40&25.30&28.55\\
&&&$\mathcal{H}_{0,6}$&47.60&56.60&46.55&41.35&37.55&50.10&54.60\\
$\chi^2$&(15,20,25,30)&(1,1,1,1)&$\mathcal{H}_{0,1}$&1.00&0.50&0.65&0.25&1.05&0.45&0.55\\
&&&$\mathcal{H}_{0,2}$&1.15&0.45&0.55&0.20&0.95&0.45&0.45\\
&&&$\mathcal{H}_{0,3}$&73.80&83.10&75.75&69.45&65.55&79.00&81.20\\
&&&$\mathcal{H}_{0,4}$&0.75&0.50&0.60&0.30&0.65&0.45&0.50\\
&&&$\mathcal{H}_{0,5}$&85.55&92.30&86.90&83.60&76.40&90.30&91.95\\
&&&$\mathcal{H}_{0,6}$&91.35&95.30&91.05&89.50&84.25&94.60&95.55\\
$\chi^2$&(15,20,25,30)&(1,1.25,1.5,1.75)&$\mathcal{H}_{0,1}$&0.40&0.45&0.85&0.60&0.85&0.55&0.60\\
&&&$\mathcal{H}_{0,2}$&0.35&0.05&0.85&0.55&0.80&0.35&0.45\\
&&&$\mathcal{H}_{0,3}$&14.55&16.75&44.10&38.90&36.45&46.90&49.75\\
&&&$\mathcal{H}_{0,4}$&0.70&0.70&1.10&0.70&1.40&0.80&0.95\\
&&&$\mathcal{H}_{0,5}$&27.25&32.05&41.45&37.25&33.45&43.35&46.10\\
&&&$\mathcal{H}_{0,6}$&32.50&39.20&37.70&33.90&31.10&40.95&43.40\\
$\chi^2$&(15,20,25,30)&(1.75,1.5,1.25,1)&$\mathcal{H}_{0,1}$&0.90&0.90&0.60&0.45&1.00&0.55&0.70\\
&&&$\mathcal{H}_{0,2}$&2.15&1.90&0.95&0.40&1.30&0.55&0.85\\
&&&$\mathcal{H}_{0,3}$&56.90&57.65&27.10&19.30&22.25&24.80&27.55\\
&&&$\mathcal{H}_{0,4}$&0.90&0.70&0.70&0.35&0.95&0.35&0.50\\
&&&$\mathcal{H}_{0,5}$&64.35&71.20&52.70&44.95&42.00&53.45&57.85\\
&&&$\mathcal{H}_{0,6}$&80.90&88.75&80.65&77.20&71.20&84.45&87.15\\
\hline
\end{longtable}

\begin{figure}
\centering
\includegraphics[width= 0.95\textwidth,height=0.9\textheight]{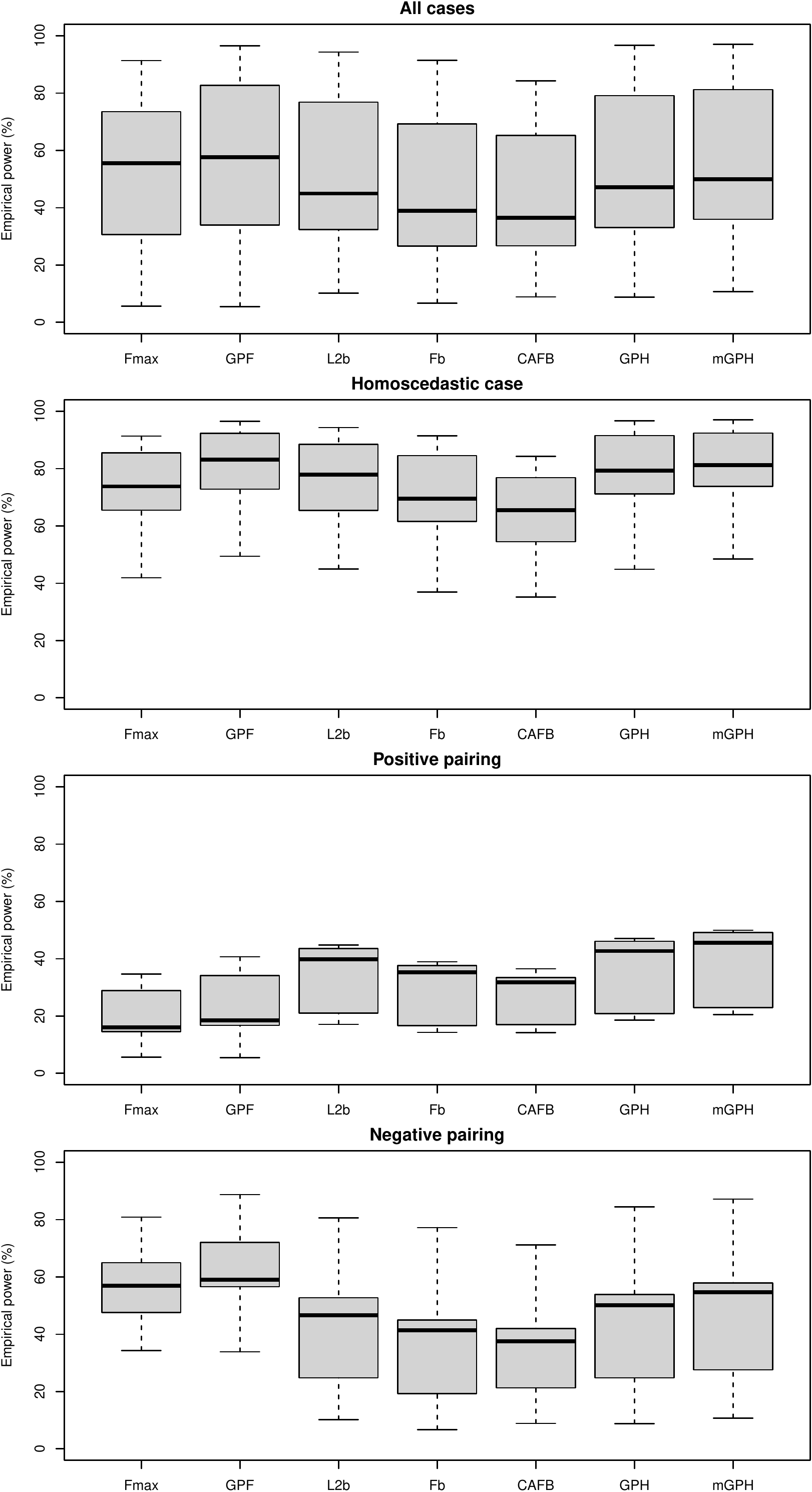}
\caption[Box-and-whisker plots for the empirical powers of all tests obtained under alternative A1 for the Tukey constrasts and homoscedastic and heteroscedastic cases]{Box-and-whisker plots for the empirical powers (as percentages) of all tests obtained under alternative A1 for the Tukey constrasts and homoscedastic and heteroscedastic cases}
\end{figure}

\begin{figure}
\centering
\includegraphics[width= 0.95\textwidth,height=0.9\textheight]{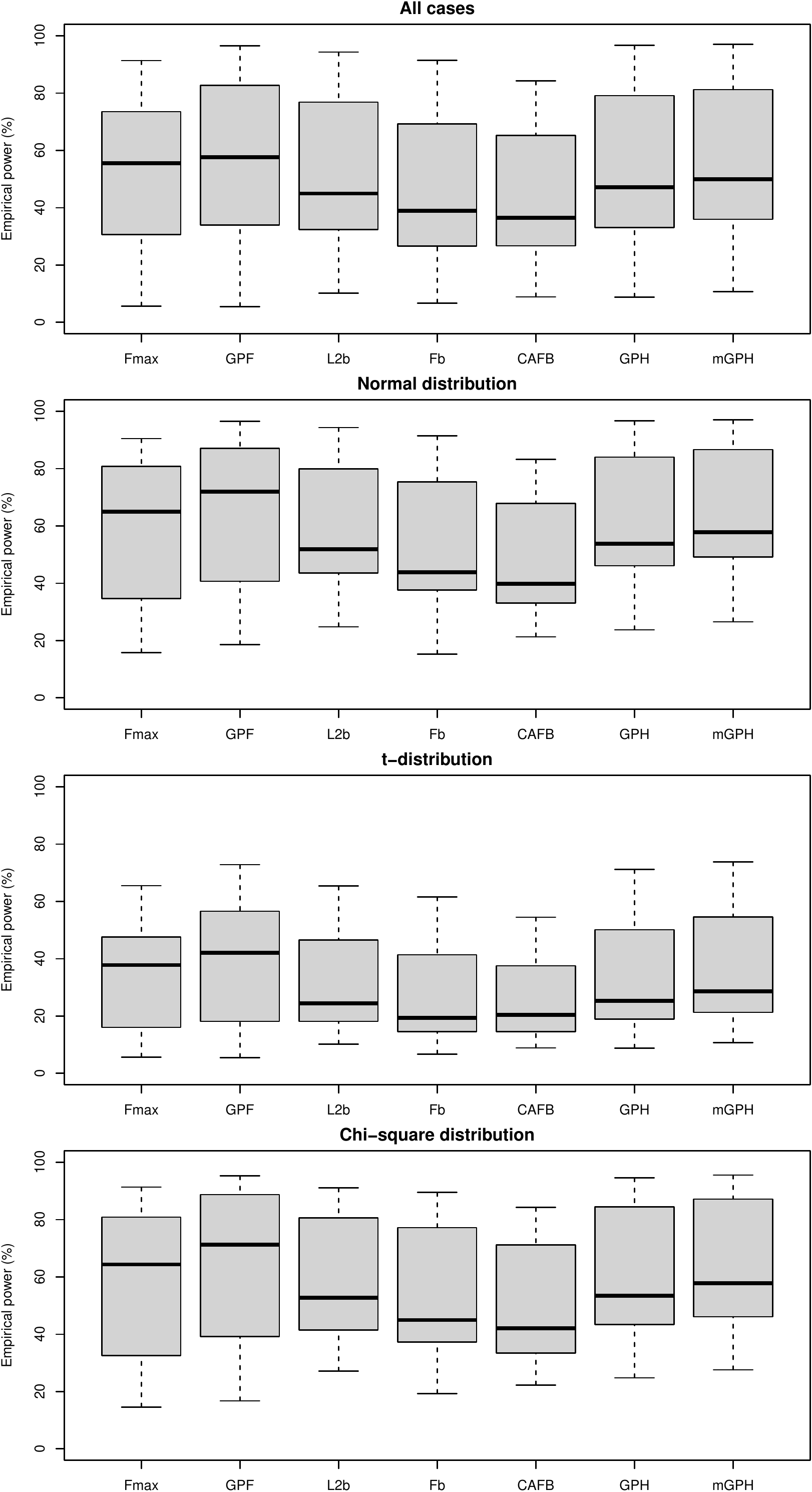}
\caption[Box-and-whisker plots for the empirical powers of all tests obtained under alternative A1 for the Tukey constrasts and different distributions]{Box-and-whisker plots for the empirical powers (as percentages) of all tests obtained under alternative A1 for the Tukey constrasts and different distributions}
\end{figure}

\newpage

\begin{longtable}[t]{rrrr|rrrrrrr}
\caption[Empirical sizes and powers of all tests obtained under alternative A2 for the Tukey constrasts]{\label{tab:unnamed-chunk-44}Empirical sizes ($\mathcal{H}_{0,1}$, $\mathcal{H}_{0,2}$, $\mathcal{H}_{0,4}$) and powers ($\mathcal{H}_{0,3}$, $\mathcal{H}_{0,5}$, $\mathcal{H}_{0,6}$) (as percentages) of all tests obtained under alternative A2 for the Tukey constrasts (D - distribution, $(\lambda_1,\lambda_2,\lambda_3,\lambda_4)$: (1,1,1,1) - homoscedastic case, (1,1.25,1.5,1.75) - heteroscedastic case (positive pairing), (1.75,1.5,1.25,1) - heteroscedastic case (negative pairing))}\\
\hline
D&$(n_1,n_2,n_3,n_4)$&$(\lambda_1,\lambda_2,\lambda_3,\lambda_4)$&$\mathcal{H}$&Fmax&GPF&L2b&Fb&CAFB&GPH&mGPH\\
\hline
\endfirsthead
\caption[]{Empirical sizes ($\mathcal{H}_{0,1}$, $\mathcal{H}_{0,2}$, $\mathcal{H}_{0,4}$) and powers ($\mathcal{H}_{0,3}$, $\mathcal{H}_{0,5}$, $\mathcal{H}_{0,6}$) (as percentages) of all tests obtained under alternative A2 for the Tukey constrasts (D - distribution, $(\lambda_1,\lambda_2,\lambda_3,\lambda_4)$: (1,1,1,1) - homoscedastic case, (1,1.25,1.5,1.75) - heteroscedastic case (positive pairing), (1.75,1.5,1.25,1) - heteroscedastic case (negative pairing)) \textit{(continued)}}\\
\hline
D&$(n_1,n_2,n_3,n_4)$&$(\lambda_1,\lambda_2,\lambda_3,\lambda_4)$&$\mathcal{H}$&Fmax&GPF&L2b&Fb&CAFB&GPH&mGPH\\
\hline
\endhead
N&(15,20,25,30)&(1,1,1,1)&$\mathcal{H}_{0,1}$&0.65&0.80&1.15&0.70&1.15&0.70&0.95\\
&&&$\mathcal{H}_{0,2}$&0.60&0.70&0.80&0.45&1.00&0.60&0.65\\
&&&$\mathcal{H}_{0,3}$&69.90&97.25&96.50&93.80&86.65&96.25&96.60\\
&&&$\mathcal{H}_{0,4}$&1.45&0.85&1.20&0.80&1.40&0.95&1.15\\
&&&$\mathcal{H}_{0,5}$&81.70&99.20&99.10&98.35&93.80&99.25&99.45\\
&&&$\mathcal{H}_{0,6}$&88.60&99.60&99.60&99.35&96.90&99.60&99.70\\
N&(15,20,25,30)&(1,1.25,1.5,1.75)&$\mathcal{H}_{0,1}$&0.50&0.45&0.95&0.70&0.95&0.80&1.10\\
&&&$\mathcal{H}_{0,2}$&0.50&0.40&1.15&0.55&1.15&0.75&0.90\\
&&&$\mathcal{H}_{0,3}$&13.90&47.15&76.90&70.95&55.50&76.50&78.55\\
&&&$\mathcal{H}_{0,4}$&0.60&0.45&0.90&0.55&0.80&0.55&0.65\\
&&&$\mathcal{H}_{0,5}$&23.80&61.95&71.95&66.25&53.25&73.00&75.50\\
&&&$\mathcal{H}_{0,6}$&30.40&66.95&67.35&62.90&47.70&68.25&71.25\\
N&(15,20,25,30)&(1.75,1.5,1.25,1)&$\mathcal{H}_{0,1}$&1.10&1.00&1.10&0.50&0.95&0.70&0.85\\
&&&$\mathcal{H}_{0,2}$&2.45&2.05&0.90&0.40&1.25&0.60&0.75\\
&&&$\mathcal{H}_{0,3}$&57.05&79.95&49.75&37.50&34.40&45.35&49.70\\
&&&$\mathcal{H}_{0,4}$&1.10&1.00&1.20&0.65&0.95&0.55&0.65\\
&&&$\mathcal{H}_{0,5}$&64.00&91.80&82.15&76.55&62.35&82.00&84.75\\
&&&$\mathcal{H}_{0,6}$&76.95&98.25&97.05&95.90&88.45&97.75&98.20\\
t&(15,20,25,30)&(1,1,1,1)&$\mathcal{H}_{0,1}$&1.10&0.70&1.05&0.55&1.00&0.70&0.80\\
&&&$\mathcal{H}_{0,2}$&0.80&0.80&1.15&0.75&1.35&0.80&0.90\\
&&&$\mathcal{H}_{0,3}$&39.05&78.70&75.70&68.70&57.05&74.95&77.55\\
&&&$\mathcal{H}_{0,4}$&0.50&0.85&0.85&0.65&1.20&0.80&0.95\\
&&&$\mathcal{H}_{0,5}$&50.05&87.75&84.40&81.85&70.20&86.45&88.20\\
&&&$\mathcal{H}_{0,6}$&58.20&92.85&90.65&88.90&77.90&92.40&93.25\\
t&(15,20,25,30)&(1,1.25,1.5,1.75)&$\mathcal{H}_{0,1}$&0.55&0.20&0.55&0.20&0.75&0.35&0.55\\
&&&$\mathcal{H}_{0,2}$&0.30&0.25&0.90&0.65&1.10&0.55&0.80\\
&&&$\mathcal{H}_{0,3}$&6.60&19.30&45.85&40.25&32.20&45.35&48.35\\
&&&$\mathcal{H}_{0,4}$&0.65&0.30&0.60&0.40&0.55&0.45&0.75\\
&&&$\mathcal{H}_{0,5}$&12.95&32.20&40.80&36.90&27.65&41.00&44.50\\
&&&$\mathcal{H}_{0,6}$&15.80&37.30&36.55&33.30&26.60&38.55&41.40\\
t&(15,20,25,30)&(1.75,1.5,1.25,1)&$\mathcal{H}_{0,1}$&1.10&1.05&1.00&0.50&0.75&0.55&0.80\\
&&&$\mathcal{H}_{0,2}$&2.15&1.65&0.80&0.40&1.35&0.45&0.70\\
&&&$\mathcal{H}_{0,3}$&38.00&56.35&26.00&17.70&17.65&22.20&25.40\\
&&&$\mathcal{H}_{0,4}$&1.35&0.65&0.65&0.35&0.90&0.60&0.60\\
&&&$\mathcal{H}_{0,5}$&40.90&67.95&50.90&44.30&37.35&52.05&55.20\\
&&&$\mathcal{H}_{0,6}$&47.60&82.60&76.95&73.75&61.80&79.05&81.55\\
$\chi^2$&(15,20,25,30)&(1,1,1,1)&$\mathcal{H}_{0,1}$&1.00&0.55&1.00&0.45&0.65&0.45&0.60\\
&&&$\mathcal{H}_{0,2}$&0.80&0.75&0.90&0.60&0.90&0.60&0.95\\
&&&$\mathcal{H}_{0,3}$&69.85&96.75&93.35&91.45&85.45&93.90&94.50\\
&&&$\mathcal{H}_{0,4}$&0.70&0.65&0.95&0.55&0.70&0.70&0.80\\
&&&$\mathcal{H}_{0,5}$&81.60&99.30&98.10&97.00&93.10&98.85&99.00\\
&&&$\mathcal{H}_{0,6}$&89.80&99.80&99.45&99.20&96.50&99.65&99.75\\
$\chi^2$&(15,20,25,30)&(1,1.25,1.5,1.75)&$\mathcal{H}_{0,1}$&0.80&0.20&0.95&0.65&0.75&0.50&0.60\\
&&&$\mathcal{H}_{0,2}$&0.45&0.30&1.25&0.60&1.25&0.75&0.90\\
&&&$\mathcal{H}_{0,3}$&13.35&48.35&76.75&72.20&57.75&78.25&80.80\\
&&&$\mathcal{H}_{0,4}$&0.65&0.50&0.80&0.55&0.75&0.45&0.55\\
&&&$\mathcal{H}_{0,5}$&24.05&62.35&70.65&65.90&52.90&73.20&75.60\\
&&&$\mathcal{H}_{0,6}$&29.00&67.05&66.15&62.00&48.95&68.95&71.30\\
$\chi^2$&(15,20,25,30)&(1.75,1.5,1.25,1)&$\mathcal{H}_{0,1}$&1.15&0.80&0.60&0.35&0.95&0.50&0.60\\
&&&$\mathcal{H}_{0,2}$&2.20&1.30&0.70&0.20&0.95&0.25&0.40\\
&&&$\mathcal{H}_{0,3}$&61.25&80.75&53.50&43.55&42.45&50.15&54.40\\
&&&$\mathcal{H}_{0,4}$&1.10&1.05&0.95&0.60&1.10&0.70&0.80\\
&&&$\mathcal{H}_{0,5}$&67.50&90.90&79.45&74.50&66.30&80.90&83.45\\
&&&$\mathcal{H}_{0,6}$&80.05&98.30&95.75&94.80&88.85&96.90&97.70\\
\hline
\end{longtable}

\begin{figure}
\centering
\includegraphics[width= 0.95\textwidth,height=0.9\textheight]{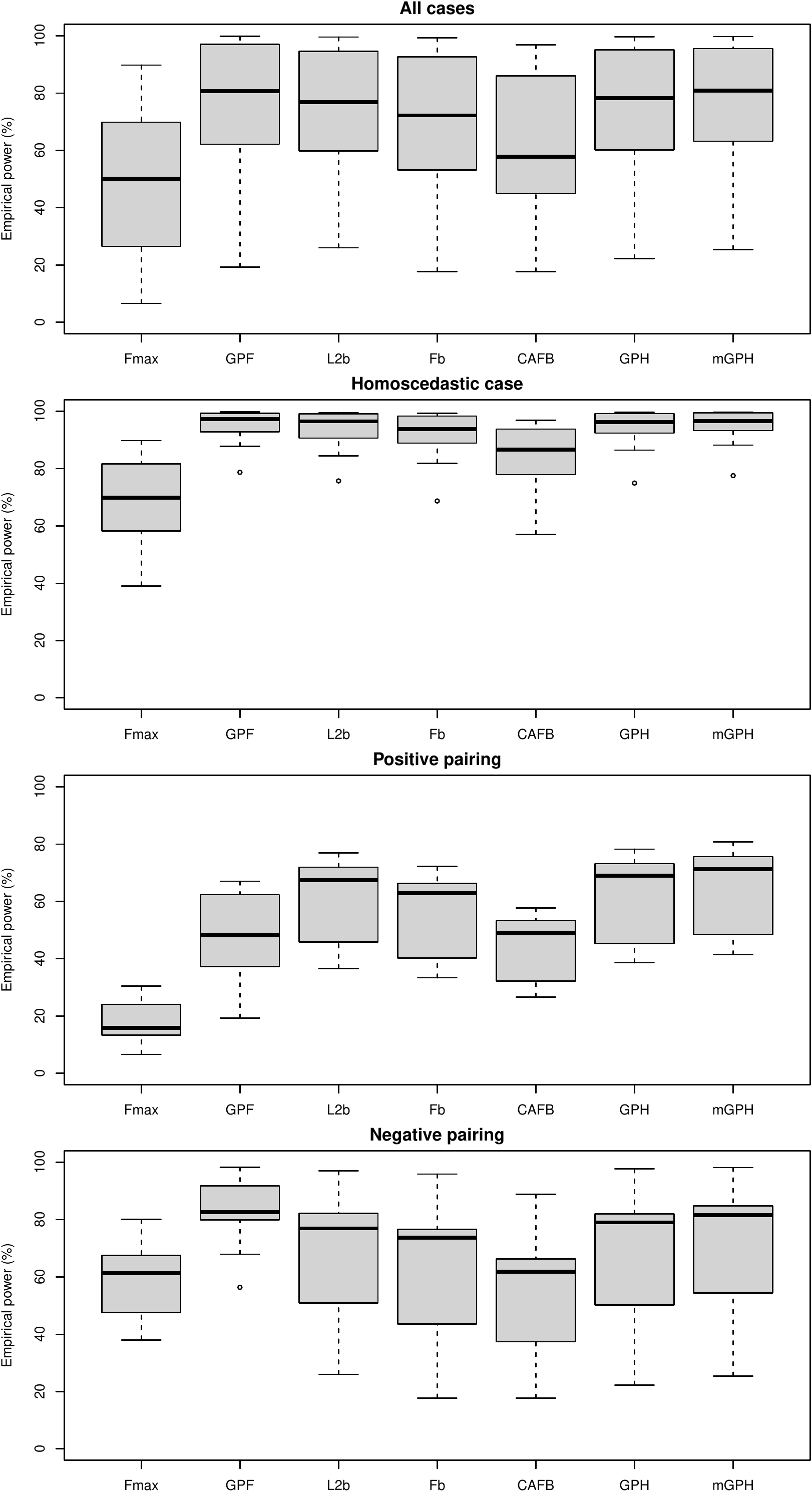}
\caption[Box-and-whisker plots for the empirical powers of all tests obtained under alternative A2 for the Tukey constrasts and homoscedastic and heteroscedastic cases]{Box-and-whisker plots for the empirical powers (as percentages) of all tests obtained under alternative A2 for the Tukey constrasts and homoscedastic and heteroscedastic cases}
\end{figure}

\begin{figure}
\centering
\includegraphics[width= 0.95\textwidth,height=0.9\textheight]{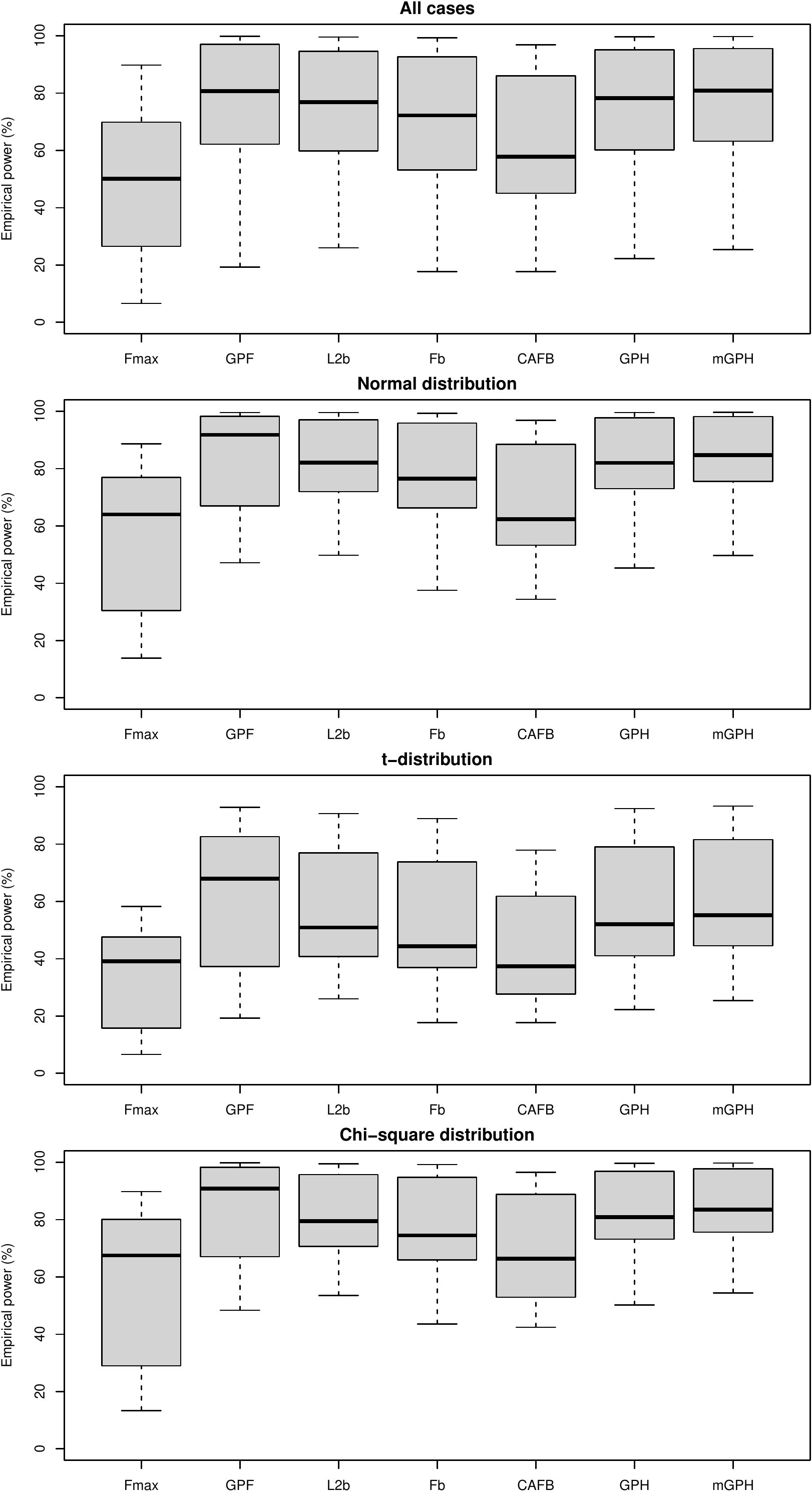}
\caption[Box-and-whisker plots for the empirical powers of all tests obtained under alternative A2 for the Tukey constrasts and different distributions]{Box-and-whisker plots for the empirical powers (as percentages) of all tests obtained under alternative A2 for the Tukey constrasts and different distributions}
\end{figure}

\newpage

\begin{longtable}[t]{rrrr|rrrrrrr}
\caption[Empirical powers of all tests obtained under alternative A3 for the Tukey constrasts]{\label{tab:unnamed-chunk-54}Empirical powers (as percentages) of all tests obtained under alternative A3 for the Tukey constrasts (D - distribution, $(\lambda_1,\lambda_2,\lambda_3,\lambda_4)$: (1,1,1,1) - homoscedastic case, (1,1.25,1.5,1.75) - heteroscedastic case (positive pairing), (1.75,1.5,1.25,1) - heteroscedastic case (negative pairing))}\\
\hline
D&$(n_1,n_2,n_3,n_4)$&$(\lambda_1,\lambda_2,\lambda_3,\lambda_4)$&$\mathcal{H}$&Fmax&GPF&L2b&Fb&CAFB&GPH&mGPH\\
\hline
\endfirsthead
\caption[]{Empirical powers (as percentages) of all tests obtained under alternative A3 for the Tukey constrasts (D - distribution, $(\lambda_1,\lambda_2,\lambda_3,\lambda_4)$: (1,1,1,1) - homoscedastic case, (1,1.25,1.5,1.75) - heteroscedastic case (positive pairing), (1.75,1.5,1.25,1) - heteroscedastic case (negative pairing)) \textit{(continued)}}\\
\hline
D&$(n_1,n_2,n_3,n_4)$&$(\lambda_1,\lambda_2,\lambda_3,\lambda_4)$&$\mathcal{H}$&Fmax&GPF&L2b&Fb&CAFB&GPH&mGPH\\
\hline
\endhead
N&(15,20,25,30)&(1,1,1,1)&$\mathcal{H}_{0,1}$&8.95&19.20&22.05&16.20&13.40&19.25&20.90\\
&&&$\mathcal{H}_{0,2}$&64.40&95.45&94.45&91.40&82.90&94.60&95.50\\
&&&$\mathcal{H}_{0,3}$&99.10&100.00&100.00&100.00&99.80&100.00&100.00\\
&&&$\mathcal{H}_{0,4}$&13.75&31.80&32.35&26.60&21.35&31.10&33.25\\
&&&$\mathcal{H}_{0,5}$&81.60&99.55&98.85&98.50&93.75&99.45&99.55\\
&&&$\mathcal{H}_{0,6}$&19.80&43.80&43.25&38.40&29.00&42.90&46.10\\
N&(15,20,25,30)&(1,1.25,1.5,1.75)&$\mathcal{H}_{0,1}$&5.75&10.75&17.10&11.45&9.25&14.05&15.40\\
&&&$\mathcal{H}_{0,2}$&25.50&67.65&80.65&75.05&62.25&80.50&81.85\\
&&&$\mathcal{H}_{0,3}$&56.80&97.10&99.50&99.10&96.40&99.50&99.60\\
&&&$\mathcal{H}_{0,4}$&5.55&10.30&13.25&10.35&8.00&11.55&13.05\\
&&&$\mathcal{H}_{0,5}$&25.80&63.65&72.30&67.40&54.35&73.85&76.50\\
&&&$\mathcal{H}_{0,6}$&5.15&10.25&11.50&9.70&8.20&11.05&12.05\\
N&(15,20,25,30)&(1.75,1.5,1.25,1)&$\mathcal{H}_{0,1}$&4.25&5.90&5.50&3.50&4.15&4.35&5.30\\
&&&$\mathcal{H}_{0,2}$&33.75&59.60&43.55&33.15&28.50&39.80&44.40\\
&&&$\mathcal{H}_{0,3}$&92.30&99.50&93.95&88.35&78.30&93.25&94.70\\
&&&$\mathcal{H}_{0,4}$&6.50&13.20&12.50&9.75&7.65&11.25&13.45\\
&&&$\mathcal{H}_{0,5}$&64.70&91.45&82.35&77.40&65.50&82.50&85.25\\
&&&$\mathcal{H}_{0,6}$&15.30&33.50&30.65&25.70&19.35&30.10&33.00\\
t&(15,20,25,30)&(1,1,1,1)&$\mathcal{H}_{0,1}$&5.00&10.20&11.50&8.10&6.85&9.65&10.65\\
&&&$\mathcal{H}_{0,2}$&35.10&73.85&70.10&63.90&53.10&70.95&73.60\\
&&&$\mathcal{H}_{0,3}$&87.60&99.40&98.40&97.80&94.90&99.05&99.40\\
&&&$\mathcal{H}_{0,4}$&7.50&14.25&14.10&11.35&9.35&13.70&14.90\\
&&&$\mathcal{H}_{0,5}$&50.40&87.90&84.85&81.90&69.95&86.20&88.00\\
&&&$\mathcal{H}_{0,6}$&10.70&22.15&21.45&18.65&14.15&21.65&23.95\\
t&(15,20,25,30)&(1,1.25,1.5,1.75)&$\mathcal{H}_{0,1}$&3.85&5.60&8.20&5.95&5.45&7.05&7.75\\
&&&$\mathcal{H}_{0,2}$&12.95&32.25&48.70&41.70&34.05&48.35&51.55\\
&&&$\mathcal{H}_{0,3}$&25.85&72.55&90.50&87.80&77.30&90.65&92.40\\
&&&$\mathcal{H}_{0,4}$&3.20&4.40&5.90&4.50&4.65&5.00&5.60\\
&&&$\mathcal{H}_{0,5}$&12.25&31.45&39.00&34.75&27.25&41.30&44.45\\
&&&$\mathcal{H}_{0,6}$&3.40&5.35&5.70&4.90&4.10&5.20&6.45\\
t&(15,20,25,30)&(1.75,1.5,1.25,1)&$\mathcal{H}_{0,1}$&2.65&3.30&3.35&2.20&3.00&2.35&3.10\\
&&&$\mathcal{H}_{0,2}$&18.90&34.95&21.85&15.00&15.80&18.70&22.20\\
&&&$\mathcal{H}_{0,3}$&74.00&90.80&68.50&58.55&52.35&66.95&71.00\\
&&&$\mathcal{H}_{0,4}$&4.65&6.90&6.05&4.30&4.70&4.95&6.05\\
&&&$\mathcal{H}_{0,5}$&41.65&69.55&53.50&47.10&37.00&52.95&57.05\\
&&&$\mathcal{H}_{0,6}$&9.10&17.00&14.10&11.45&8.65&13.35&15.95\\
$\chi^2$&(15,20,25,30)&(1,1,1,1)&$\mathcal{H}_{0,1}$&9.30&20.95&23.35&18.20&14.90&20.85&22.90\\
&&&$\mathcal{H}_{0,2}$&64.55&95.20&91.80&88.60&80.85&93.25&94.15\\
&&&$\mathcal{H}_{0,3}$&99.15&100.00&99.85&99.70&99.45&100.00&100.00\\
&&&$\mathcal{H}_{0,4}$&13.90&32.75&32.80&27.80&21.35&31.40&34.15\\
&&&$\mathcal{H}_{0,5}$&82.55&99.40&98.35&97.25&92.85&98.65&99.00\\
&&&$\mathcal{H}_{0,6}$&18.80&42.85&41.55&37.75&29.05&42.25&45.40\\
$\chi^2$&(15,20,25,30)&(1,1.25,1.5,1.75)&$\mathcal{H}_{0,1}$&5.90&9.80&14.45&10.30&9.80&12.90&13.95\\
&&&$\mathcal{H}_{0,2}$&26.70&66.85&78.15&73.70&62.05&79.95&81.90\\
&&&$\mathcal{H}_{0,3}$&56.15&97.55&99.45&99.15&96.25&99.75&99.90\\
&&&$\mathcal{H}_{0,4}$&5.90&11.85&15.05&12.20&9.70&13.65&14.90\\
&&&$\mathcal{H}_{0,5}$&26.70&66.45&74.10&69.45&54.40&75.90&78.30\\
&&&$\mathcal{H}_{0,6}$&4.95&8.40&11.75&9.25&7.65&9.25&10.80\\
$\chi^2$&(15,20,25,30)&(1.75,1.5,1.25,1)&$\mathcal{H}_{0,1}$&4.10&5.80&6.20&4.25&5.25&5.00&5.80\\
&&&$\mathcal{H}_{0,2}$&35.65&61.55&45.15&36.25&33.85&43.50&47.55\\
&&&$\mathcal{H}_{0,3}$&92.20&98.90&89.40&83.65&78.85&90.50&92.15\\
&&&$\mathcal{H}_{0,4}$&8.85&13.85&13.20&10.20&9.40&12.05&13.85\\
&&&$\mathcal{H}_{0,5}$&67.30&91.00&79.15&74.90&64.90&81.80&84.55\\
&&&$\mathcal{H}_{0,6}$&16.85&35.10&29.40&26.35&20.95&30.85&33.65\\
\hline
\end{longtable}

\begin{figure}
\centering
\includegraphics[width= 0.95\textwidth,height=0.9\textheight]{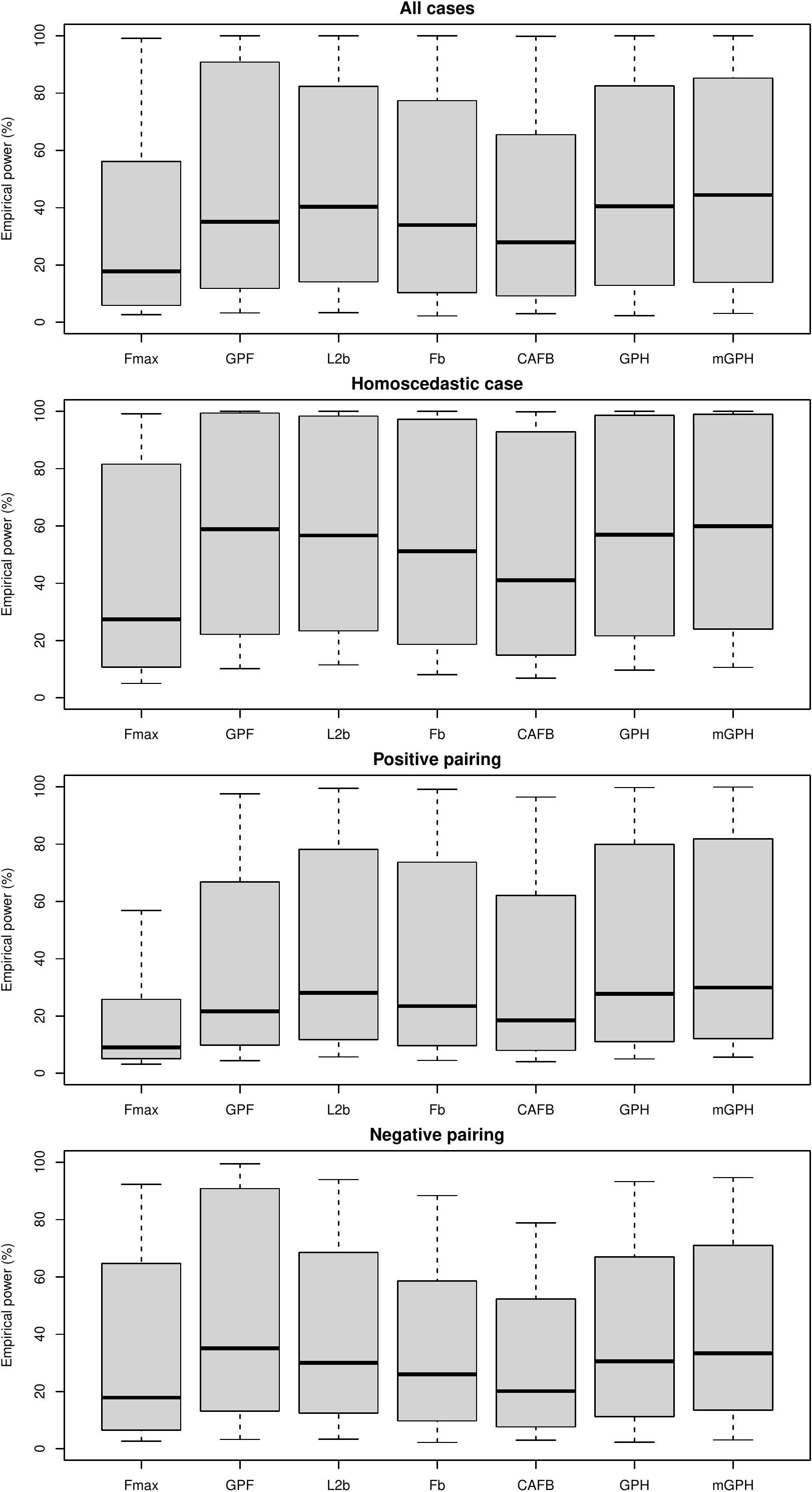}
\caption[Box-and-whisker plots for the empirical powers of all tests obtained under alternative A3 for the Tukey constrasts and homoscedastic and heteroscedastic cases]{Box-and-whisker plots for the empirical powers (as percentages) of all tests obtained under alternative A3 for the Tukey constrasts and homoscedastic and heteroscedastic cases}
\end{figure}

\begin{figure}
\centering
\includegraphics[width= 0.95\textwidth,height=0.9\textheight]{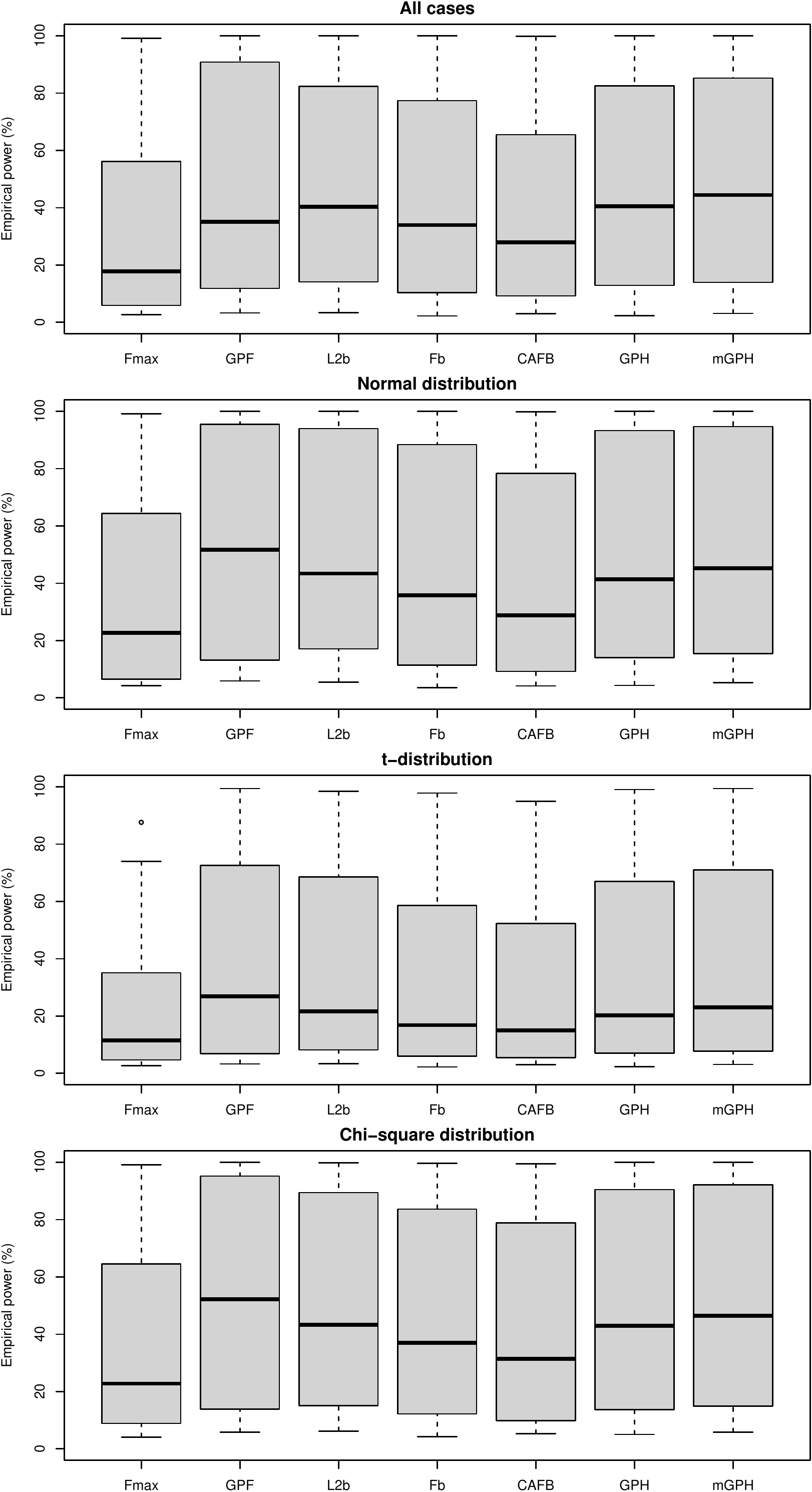}
\caption[Box-and-whisker plots for the empirical powers of all tests obtained under alternative A3 for the Tukey constrasts and different distributions]{Box-and-whisker plots for the empirical powers (as percentages) of all tests obtained under alternative A3 for the Tukey constrasts and different distributions}
\end{figure}

\newpage

\begin{longtable}[t]{rrrr|rrrrrrr}
\caption[Empirical powers of all tests obtained under alternative A4 for the Tukey constrasts]{\label{tab:unnamed-chunk-57}Empirical powers (as percentages) of all tests obtained under alternative A4 for the Tukey constrasts (D - distribution, $(\lambda_1,\lambda_2,\lambda_3,\lambda_4)$: (1,1,1,1) - homoscedastic case, (1,1.25,1.5,1.75) - heteroscedastic case (positive pairing), (1.75,1.5,1.25,1) - heteroscedastic case (negative pairing))}\\
\hline
D&$(n_1,n_2,n_3,n_4)$&$(\lambda_1,\lambda_2,\lambda_3,\lambda_4)$&$\mathcal{H}$&Fmax&GPF&L2b&Fb&CAFB&GPH&mGPH\\
\hline
\endfirsthead
\caption[]{Empirical powers (as percentages) of all tests obtained under alternative A4 for the Tukey constrasts (D - distribution, $(\lambda_1,\lambda_2,\lambda_3,\lambda_4)$: (1,1,1,1) - homoscedastic case, (1,1.25,1.5,1.75) - heteroscedastic case (positive pairing), (1.75,1.5,1.25,1) - heteroscedastic case (negative pairing)) \textit{(continued)}}\\
\hline
D&$(n_1,n_2,n_3,n_4)$&$(\lambda_1,\lambda_2,\lambda_3,\lambda_4)$&$\mathcal{H}$&Fmax&GPF&L2b&Fb&CAFB&GPH&mGPH\\
\hline
\endhead
N&(15,20,25,30)&(1,1,1,1)&$\mathcal{H}_{0,1}$&8.30&9.15&10.35&6.80&7.75&8.75&9.80\\
&&&$\mathcal{H}_{0,2}$&67.15&77.50&73.75&65.85&59.30&75.00&78.30\\
&&&$\mathcal{H}_{0,3}$&99.60&99.95&99.70&99.25&97.50&99.90&99.95\\
&&&$\mathcal{H}_{0,4}$&12.85&13.90&13.70&10.20&10.20&13.85&15.40\\
&&&$\mathcal{H}_{0,5}$&84.60&92.05&88.35&84.45&77.55&90.20&92.15\\
&&&$\mathcal{H}_{0,6}$&16.80&19.75&18.55&15.00&14.55&20.00&22.05\\
N&(15,20,25,30)&(1,1.25,1.5,1.75)&$\mathcal{H}_{0,1}$&5.20&5.50&7.90&5.40&5.30&7.25&7.80\\
&&&$\mathcal{H}_{0,2}$&27.20&34.55&50.40&41.90&37.95&50.85&53.70\\
&&&$\mathcal{H}_{0,3}$&66.85&75.60&93.25&89.60&82.35&94.95&95.80\\
&&&$\mathcal{H}_{0,4}$&4.95&4.25&6.30&4.60&5.50&5.15&6.10\\
&&&$\mathcal{H}_{0,5}$&24.65&30.35&39.95&33.70&31.50&42.30&44.70\\
&&&$\mathcal{H}_{0,6}$&4.45&4.35&4.95&4.00&4.40&5.00&5.65\\
N&(15,20,25,30)&(1.75,1.5,1.25,1)&$\mathcal{H}_{0,1}$&3.35&3.20&3.50&2.10&2.85&2.60&3.35\\
&&&$\mathcal{H}_{0,2}$&33.25&36.50&21.80&14.60&16.80&20.75&23.85\\
&&&$\mathcal{H}_{0,3}$&92.80&95.95&71.55&59.05&55.55&71.20&75.30\\
&&&$\mathcal{H}_{0,4}$&6.70&6.20&5.05&3.00&4.20&4.35&6.05\\
&&&$\mathcal{H}_{0,5}$&64.85&71.10&52.30&43.75&40.00&52.90&57.60\\
&&&$\mathcal{H}_{0,6}$&14.05&14.45&13.00&10.50&9.35&12.70&14.70\\
t&(15,20,25,30)&(1,1,1,1)&$\mathcal{H}_{0,1}$&4.75&4.55&5.35&3.45&4.60&4.05&4.80\\
&&&$\mathcal{H}_{0,2}$&36.60&41.90&39.60&31.70&32.10&39.90&42.40\\
&&&$\mathcal{H}_{0,3}$&89.55&94.05&89.40&85.60&81.05&91.40&92.95\\
&&&$\mathcal{H}_{0,4}$&6.65&6.95&6.70&5.05&5.55&5.95&6.80\\
&&&$\mathcal{H}_{0,5}$&52.85&61.00&54.95&50.00&47.15&59.05&61.80\\
&&&$\mathcal{H}_{0,6}$&8.25&9.80&9.00&7.55&8.00&9.55&10.45\\
t&(15,20,25,30)&(1,1.25,1.5,1.75)&$\mathcal{H}_{0,1}$&3.30&2.90&3.80&2.55&3.75&3.10&3.85\\
&&&$\mathcal{H}_{0,2}$&12.00&13.00&23.45&18.45&18.35&23.35&26.05\\
&&&$\mathcal{H}_{0,3}$&30.40&36.00&64.30&57.55&52.80&67.75&70.30\\
&&&$\mathcal{H}_{0,4}$&2.40&2.40&2.80&2.25&3.05&2.75&3.15\\
&&&$\mathcal{H}_{0,5}$&12.05&12.85&18.50&15.75&15.40&18.90&21.10\\
&&&$\mathcal{H}_{0,6}$&3.20&2.70&3.15&2.10&2.40&2.75&3.30\\
t&(15,20,25,30)&(1.75,1.5,1.25,1)&$\mathcal{H}_{0,1}$&1.40&1.60&1.90&1.00&1.50&1.25&1.45\\
&&&$\mathcal{H}_{0,2}$&17.50&19.05&9.80&6.55&8.75&8.50&10.30\\
&&&$\mathcal{H}_{0,3}$&71.60&73.85&38.55&29.00&32.00&37.75&41.75\\
&&&$\mathcal{H}_{0,4}$&3.65&3.20&2.90&2.05&2.35&2.10&2.85\\
&&&$\mathcal{H}_{0,5}$&38.10&42.80&25.55&20.95&21.00&25.30&29.55\\
&&&$\mathcal{H}_{0,6}$&7.85&8.50&6.70&5.45&5.00&6.30&7.35\\
$\chi^2$&(15,20,25,30)&(1,1,1,1)&$\mathcal{H}_{0,1}$&9.15&9.05&10.50&7.40&8.30&9.00&10.75\\
&&&$\mathcal{H}_{0,2}$&67.85&77.20&70.90&64.20&61.60&73.85&76.70\\
&&&$\mathcal{H}_{0,3}$&99.75&99.85&99.05&98.20&97.00&99.65&99.65\\
&&&$\mathcal{H}_{0,4}$&13.00&12.95&13.70&10.85&11.55&13.15&15.00\\
&&&$\mathcal{H}_{0,5}$&86.45&92.25&87.10&83.95&78.10&90.35&91.40\\
&&&$\mathcal{H}_{0,6}$&19.00&22.75&20.60&17.85&15.85&22.25&23.75\\
$\chi^2$&(15,20,25,30)&(1,1.25,1.5,1.75)&$\mathcal{H}_{0,1}$&4.55&4.60&7.65&4.85&6.15&6.85&7.90\\
&&&$\mathcal{H}_{0,2}$&28.50&33.15&48.80&41.45&39.20&51.85&54.50\\
&&&$\mathcal{H}_{0,3}$&66.25&76.75&92.05&89.40&82.50&94.55&95.55\\
&&&$\mathcal{H}_{0,4}$&5.50&4.00&5.35&3.70&4.75&5.60&6.45\\
&&&$\mathcal{H}_{0,5}$&28.05&31.95&39.45&35.10&32.35&43.30&45.80\\
&&&$\mathcal{H}_{0,6}$&3.65&4.35&5.00&4.10&5.10&5.10&5.80\\
$\chi^2$&(15,20,25,30)&(1.75,1.5,1.25,1)&$\mathcal{H}_{0,1}$&2.85&2.95&3.45&2.10&2.30&2.50&3.50\\
&&&$\mathcal{H}_{0,2}$&34.80&38.70&24.15&16.80&19.85&22.60&26.35\\
&&&$\mathcal{H}_{0,3}$&92.80&93.90&68.90&58.30&58.05&70.10&74.40\\
&&&$\mathcal{H}_{0,4}$&5.35&6.45&5.95&4.55&4.80&5.00&5.90\\
&&&$\mathcal{H}_{0,5}$&66.65&71.45&51.95&44.50&43.70&54.30&58.40\\
&&&$\mathcal{H}_{0,6}$&15.85&16.80&14.45&12.25&11.35&14.70&16.80\\
\hline
\end{longtable}

\begin{figure}
\centering
\includegraphics[width= 0.95\textwidth,height=0.9\textheight]{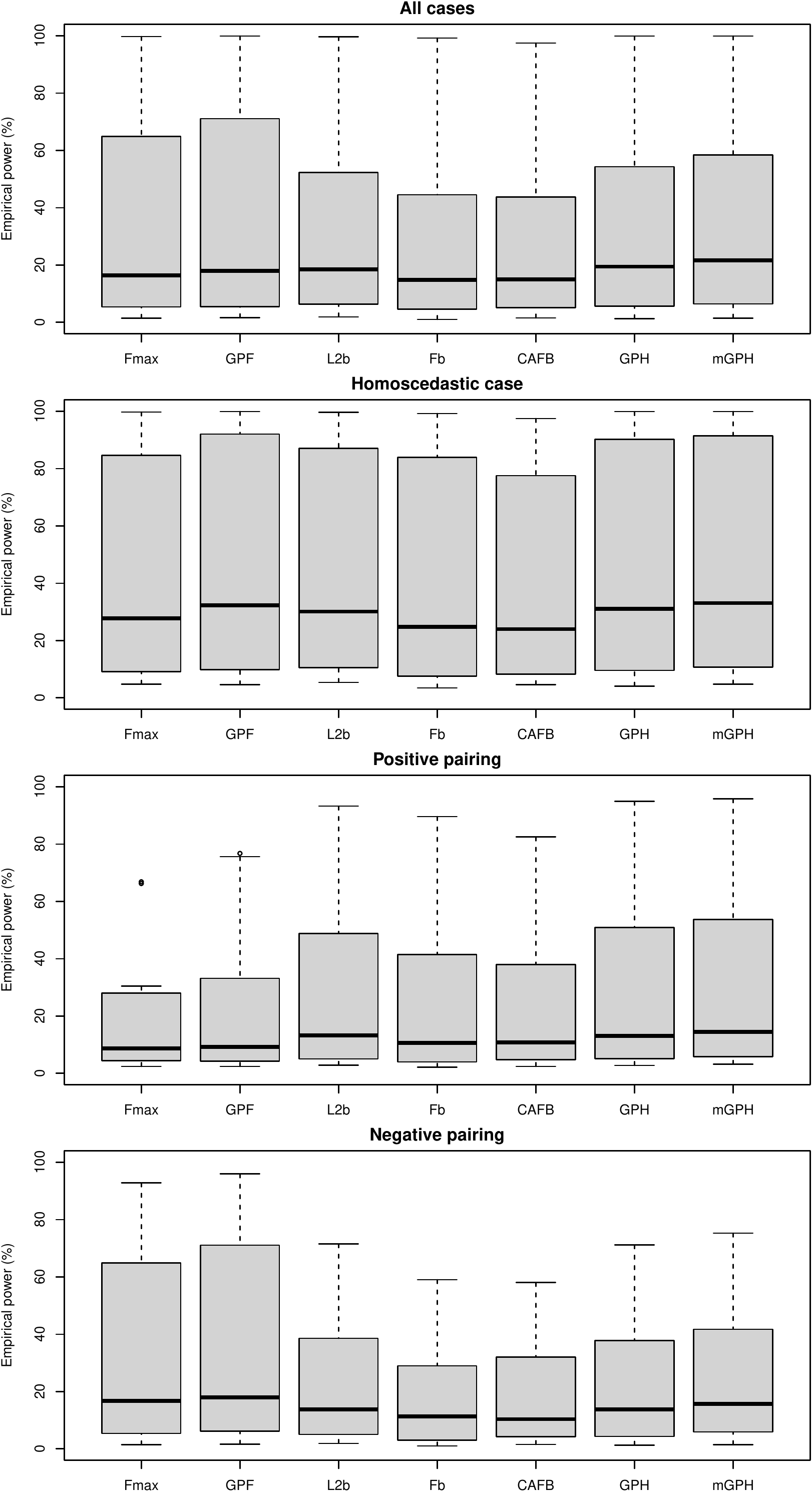}
\caption[Box-and-whisker plots for the empirical powers of all tests obtained under alternative A4 for the Tukey constrasts and homoscedastic and heteroscedastic cases]{Box-and-whisker plots for the empirical powers (as percentages) of all tests obtained under alternative A4 for the Tukey constrasts and homoscedastic and heteroscedastic cases}
\end{figure}

\begin{figure}
\centering
\includegraphics[width= 0.95\textwidth,height=0.9\textheight]{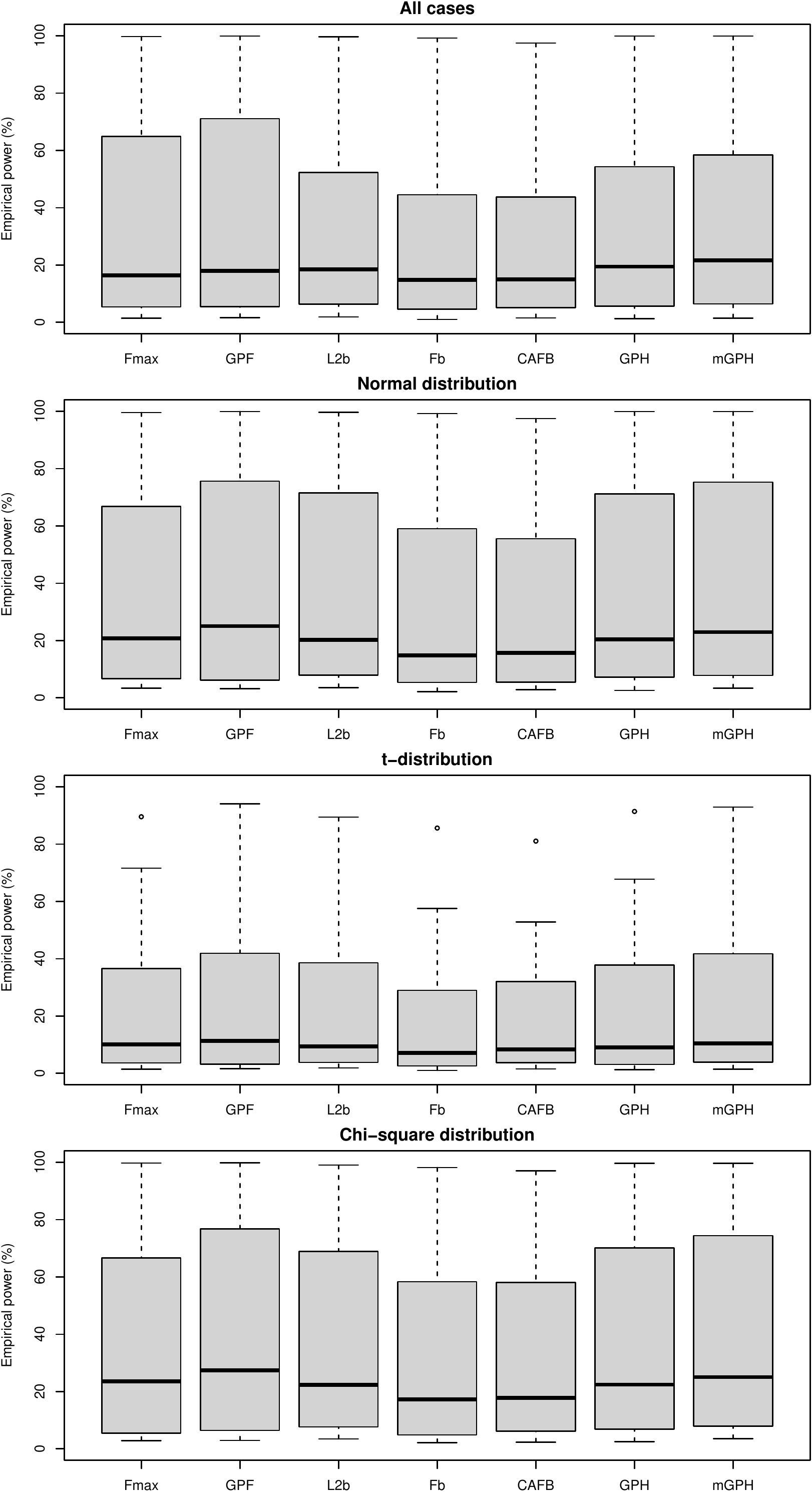}
\caption[Box-and-whisker plots for the empirical powers of all tests obtained under alternative A4 for the Tukey constrasts and different distributions]{Box-and-whisker plots for the empirical powers (as percentages) of all tests obtained under alternative A4 for the Tukey constrasts and different distributions}
\end{figure}

\newpage

\begin{longtable}[t]{rrrr|rrrrrrr}
\caption[Empirical sizes and powers of all tests obtained under alternative A5 for the Tukey constrasts]{\label{tab:unnamed-chunk-60}Empirical sizes ($\mathcal{H}_{0,1}$, $\mathcal{H}_{0,2}$, $\mathcal{H}_{0,4}$) and powers ($\mathcal{H}_{0,3}$, $\mathcal{H}_{0,5}$, $\mathcal{H}_{0,6}$) (as percentages) of all tests obtained under alternative A5 for the Tukey constrasts (D - distribution, $(\lambda_1,\lambda_2,\lambda_3,\lambda_4)$: (1,1,1,1) - homoscedastic case, (1,1.25,1.5,1.75) - heteroscedastic case (positive pairing), (1.75,1.5,1.25,1) - heteroscedastic case (negative pairing))}\\
\hline
D&$(n_1,n_2,n_3,n_4)$&$(\lambda_1,\lambda_2,\lambda_3,\lambda_4)$&$\mathcal{H}$&Fmax&GPF&L2b&Fb&CAFB&GPH&mGPH\\
\hline
\endfirsthead
\caption[]{Empirical sizes ($\mathcal{H}_{0,1}$, $\mathcal{H}_{0,2}$, $\mathcal{H}_{0,4}$) and powers ($\mathcal{H}_{0,3}$, $\mathcal{H}_{0,5}$, $\mathcal{H}_{0,6}$) (as percentages) of all tests obtained under alternative A5 for the Tukey constrasts (D - distribution, $(\lambda_1,\lambda_2,\lambda_3,\lambda_4)$: (1,1,1,1) - homoscedastic case, (1,1.25,1.5,1.75) - heteroscedastic case (positive pairing), (1.75,1.5,1.25,1) - heteroscedastic case (negative pairing)) \textit{(continued)}}\\
\hline
D&$(n_1,n_2,n_3,n_4)$&$(\lambda_1,\lambda_2,\lambda_3,\lambda_4)$&$\mathcal{H}$&Fmax&GPF&L2b&Fb&CAFB&GPH&mGPH\\
\hline
\endhead
N&(15,20,25,30)&(1,1,1,1)&$\mathcal{H}_{0,1}$&0.65&0.55&1.25&0.60&1.35&0.65&0.80\\
&&&$\mathcal{H}_{0,2}$&0.90&0.65&1.10&0.60&1.00&0.65&0.80\\
&&&$\mathcal{H}_{0,3}$&69.85&79.15&74.85&66.70&60.35&76.75&78.45\\
&&&$\mathcal{H}_{0,4}$&0.90&0.65&1.25&0.85&1.00&0.75&0.95\\
&&&$\mathcal{H}_{0,5}$&82.35&90.35&87.60&83.05&76.05&90.10&91.30\\
&&&$\mathcal{H}_{0,6}$&89.30&95.10&92.60&90.55&83.70&94.75&95.70\\
N&(15,20,25,30)&(1,1.25,1.5,1.75)&$\mathcal{H}_{0,1}$&0.60&0.30&0.80&0.45&1.20&0.55&0.60\\
&&&$\mathcal{H}_{0,2}$&0.55&0.60&1.35&1.00&1.10&1.20&1.25\\
&&&$\mathcal{H}_{0,3}$&14.05&17.20&43.10&36.05&33.20&45.20&48.20\\
&&&$\mathcal{H}_{0,4}$&0.90&0.60&0.95&0.60&0.80&0.75&0.90\\
&&&$\mathcal{H}_{0,5}$&25.70&29.90&39.15&32.20&29.45&41.10&44.20\\
&&&$\mathcal{H}_{0,6}$&29.05&37.20&36.25&31.20&28.40&39.25&42.60\\
N&(15,20,25,30)&(1.75,1.5,1.25,1)&$\mathcal{H}_{0,1}$&1.15&0.80&1.10&0.35&0.95&0.45&0.70\\
&&&$\mathcal{H}_{0,2}$&2.30&1.95&1.15&0.55&1.05&0.70&0.85\\
&&&$\mathcal{H}_{0,3}$&56.35&58.40&24.00&15.90&19.35&21.70&25.15\\
&&&$\mathcal{H}_{0,4}$&1.30&0.75&0.75&0.45&0.85&0.40&0.80\\
&&&$\mathcal{H}_{0,5}$&62.50&69.10&50.75&41.90&40.00&51.75&56.20\\
&&&$\mathcal{H}_{0,6}$&79.00&86.00&79.05&74.20&68.15&82.85&85.60\\
t&(15,20,25,30)&(1,1,1,1)&$\mathcal{H}_{0,1}$&1.05&0.70&1.15&0.60&0.85&0.80&0.90\\
&&&$\mathcal{H}_{0,2}$&0.70&0.50&0.45&0.25&0.80&0.45&0.60\\
&&&$\mathcal{H}_{0,3}$&40.55&47.65&45.55&35.80&37.15&45.05&48.60\\
&&&$\mathcal{H}_{0,4}$&0.90&0.85&1.00&0.60&0.75&0.80&0.90\\
&&&$\mathcal{H}_{0,5}$&50.40&60.40&56.35&50.10&45.60&59.85&62.90\\
&&&$\mathcal{H}_{0,6}$&59.10&69.10&62.30&58.05&51.25&67.95&70.75\\
t&(15,20,25,30)&(1,1.25,1.5,1.75)&$\mathcal{H}_{0,1}$&0.80&0.50&0.95&0.65&0.75&0.45&0.80\\
&&&$\mathcal{H}_{0,2}$&0.30&0.25&0.95&0.75&0.80&0.60&0.65\\
&&&$\mathcal{H}_{0,3}$&6.20&5.35&18.30&14.70&15.30&18.55&20.40\\
&&&$\mathcal{H}_{0,4}$&0.70&0.45&0.90&0.60&0.90&0.65&0.75\\
&&&$\mathcal{H}_{0,5}$&10.90&12.10&16.70&13.60&15.00&17.50&19.25\\
&&&$\mathcal{H}_{0,6}$&13.95&14.25&14.85&12.20&13.50&15.20&17.00\\
t&(15,20,25,30)&(1.75,1.5,1.25,1)&$\mathcal{H}_{0,1}$&1.15&0.70&0.85&0.30&1.15&0.40&0.85\\
&&&$\mathcal{H}_{0,2}$&2.30&1.65&0.90&0.25&1.15&0.50&0.70\\
&&&$\mathcal{H}_{0,3}$&35.05&35.35&10.90&6.85&10.65&9.45&10.85\\
&&&$\mathcal{H}_{0,4}$&0.80&0.50&0.80&0.35&0.85&0.25&0.50\\
&&&$\mathcal{H}_{0,5}$&39.70&41.70&25.15&18.95&19.80&24.90&28.45\\
&&&$\mathcal{H}_{0,6}$&47.75&55.45&44.05&39.60&37.30&49.30&53.60\\
$\chi^2$&(15,20,25,30)&(1,1,1,1)&$\mathcal{H}_{0,1}$&1.15&0.50&1.00&0.55&1.10&0.75&0.85\\
&&&$\mathcal{H}_{0,2}$&0.90&0.25&0.60&0.35&1.05&0.50&0.60\\
&&&$\mathcal{H}_{0,3}$&72.60&79.95&72.50&65.65&63.80&76.40&78.50\\
&&&$\mathcal{H}_{0,4}$&0.55&0.45&0.75&0.50&0.60&0.50&0.50\\
&&&$\mathcal{H}_{0,5}$&84.00&89.20&83.70&79.80&77.25&87.20&88.75\\
&&&$\mathcal{H}_{0,6}$&89.10&94.70&90.25&88.70&83.55&94.45&95.25\\
$\chi^2$&(15,20,25,30)&(1,1.25,1.5,1.75)&$\mathcal{H}_{0,1}$&0.80&0.45&0.90&0.55&1.10&0.60&0.65\\
&&&$\mathcal{H}_{0,2}$&0.30&0.30&0.90&0.55&0.80&0.65&0.85\\
&&&$\mathcal{H}_{0,3}$&12.15&13.35&42.60&36.20&32.30&44.30&47.55\\
&&&$\mathcal{H}_{0,4}$&1.00&0.55&0.85&0.55&0.80&0.50&0.70\\
&&&$\mathcal{H}_{0,5}$&23.40&29.15&40.35&34.85&30.90&42.20&44.60\\
&&&$\mathcal{H}_{0,6}$&29.35&36.30&34.90&30.45&28.95&38.25&40.55\\
$\chi^2$&(15,20,25,30)&(1.75,1.5,1.25,1)&$\mathcal{H}_{0,1}$&1.00&0.80&0.90&0.25&0.90&0.40&0.65\\
&&&$\mathcal{H}_{0,2}$&2.40&2.45&1.05&0.60&1.65&0.70&0.85\\
&&&$\mathcal{H}_{0,3}$&56.50&59.35&25.40&17.75&25.40&24.05&27.85\\
&&&$\mathcal{H}_{0,4}$&1.25&0.80&0.95&0.50&1.00&0.50&0.70\\
&&&$\mathcal{H}_{0,5}$&64.25&69.65&50.15&44.10&42.10&52.70&57.05\\
&&&$\mathcal{H}_{0,6}$&79.90&85.80&77.20&72.85&68.25&81.20&84.10\\
\hline
\end{longtable}

\begin{figure}
\centering
\includegraphics[width= 0.95\textwidth,height=0.9\textheight]{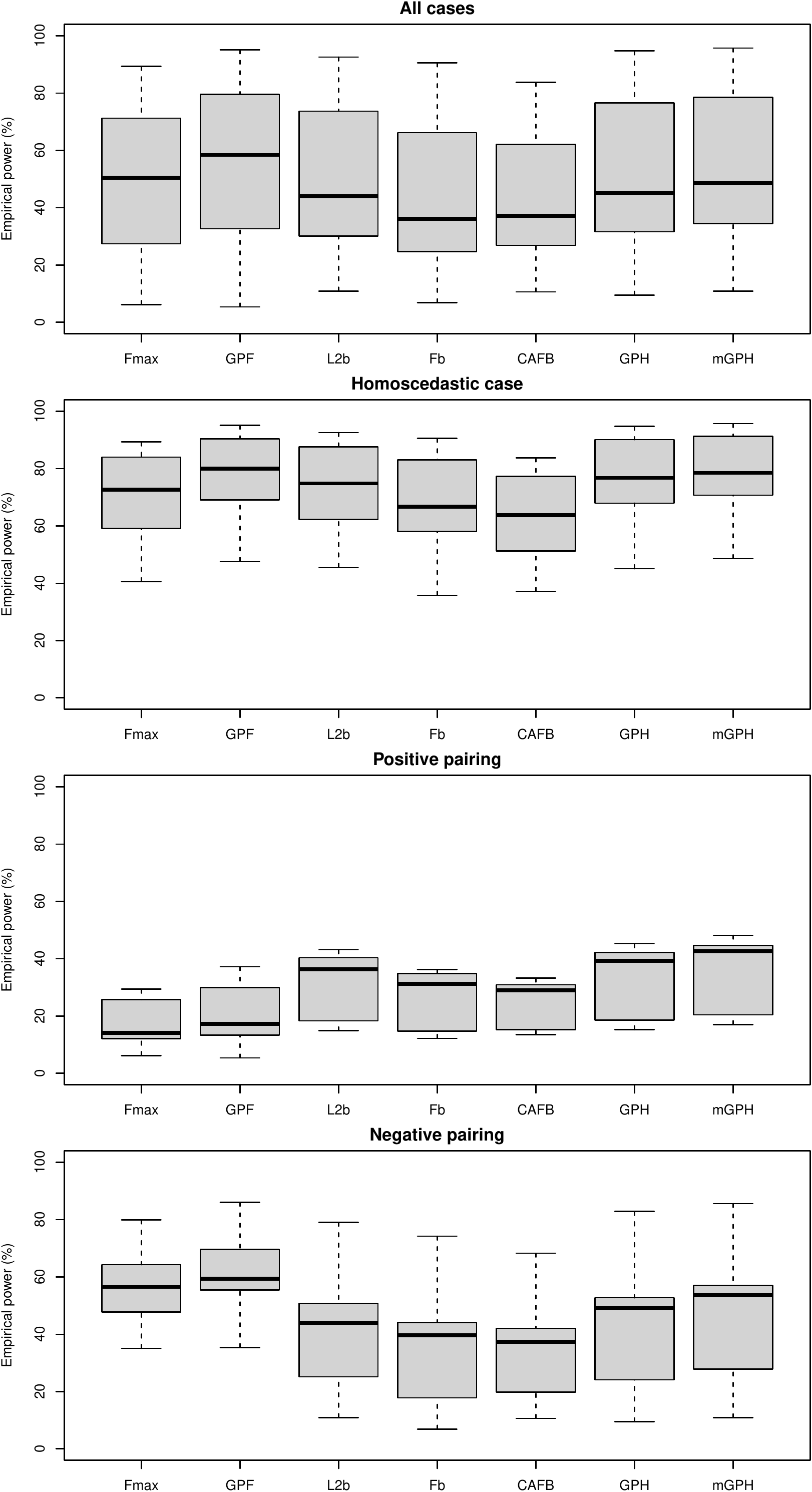}
\caption[Box-and-whisker plots for the empirical powers of all tests obtained under alternative A5 for the Tukey constrasts and homoscedastic and heteroscedastic cases]{Box-and-whisker plots for the empirical powers (as percentages) of all tests obtained under alternative A5 for the Tukey constrasts and homoscedastic and heteroscedastic cases}
\end{figure}

\begin{figure}
\centering
\includegraphics[width= 0.95\textwidth,height=0.9\textheight]{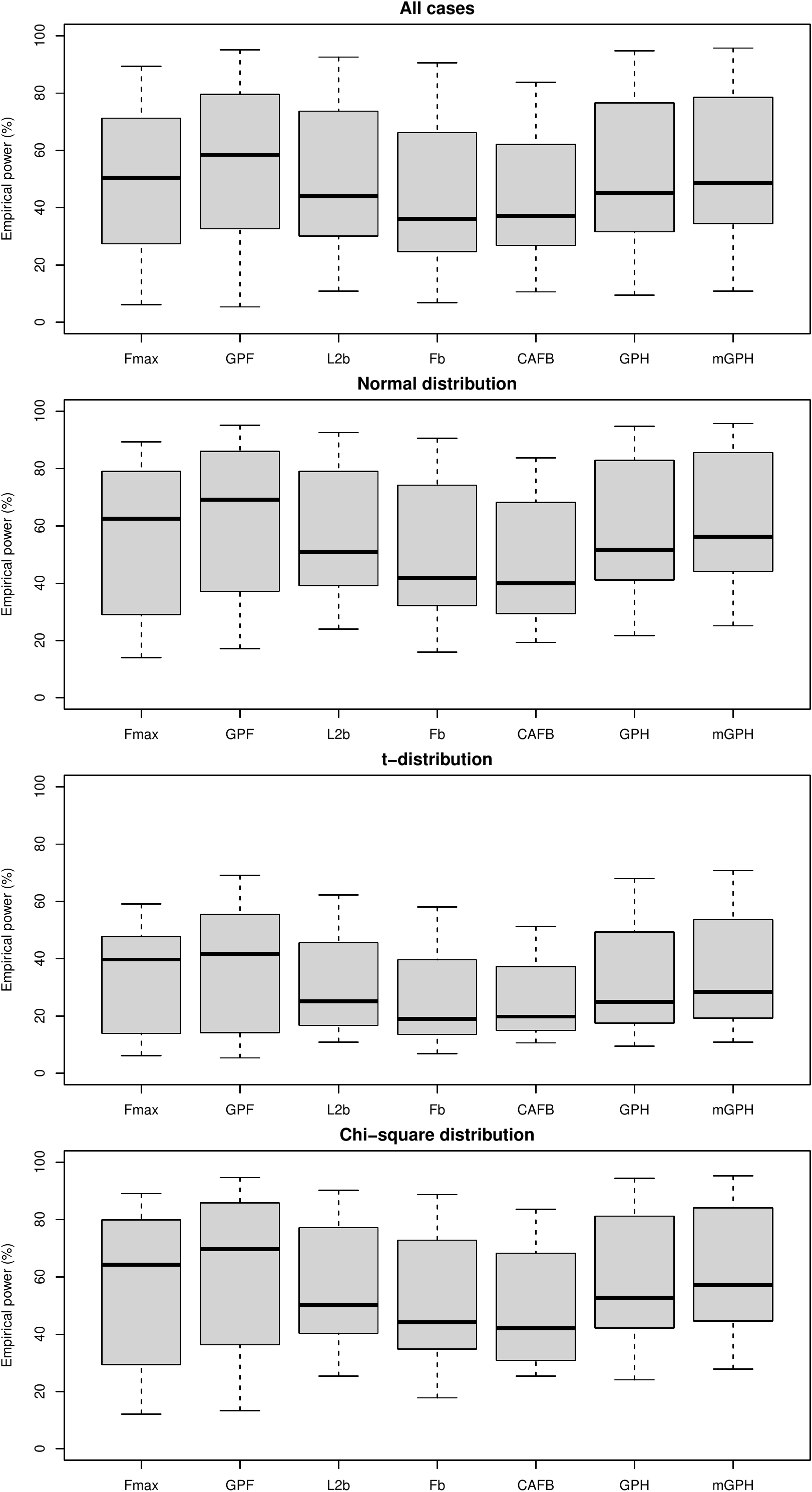}
\caption[Box-and-whisker plots for the empirical powers of all tests obtained under alternative A5 for the Tukey constrasts and different distributions]{Box-and-whisker plots for the empirical powers (as percentages) of all tests obtained under alternative A5 for the Tukey constrasts and different distributions}
\end{figure}

\newpage

\begin{longtable}[t]{rrrr|rrrrrrr}
\caption[Empirical powers of all tests obtained under alternative A6 for the Tukey constrasts]{\label{tab:unnamed-chunk-65}Empirical powers (as percentages) of all tests obtained under alternative A6 for the Tukey constrasts (D - distribution, $(\lambda_1,\lambda_2,\lambda_3,\lambda_4)$: (1,1,1,1) - homoscedastic case, (1,1.25,1.5,1.75) - heteroscedastic case (positive pairing), (1.75,1.5,1.25,1) - heteroscedastic case (negative pairing))}\\
\hline
D&$(n_1,n_2,n_3,n_4)$&$(\lambda_1,\lambda_2,\lambda_3,\lambda_4)$&$\mathcal{H}$&Fmax&GPF&L2b&Fb&CAFB&GPH&mGPH\\
\hline
\endfirsthead
\caption[]{Empirical powers (as percentages) of all tests obtained under alternative A6 for the Tukey constrasts (D - distribution, $(\lambda_1,\lambda_2,\lambda_3,\lambda_4)$: (1,1,1,1) - homoscedastic case, (1,1.25,1.5,1.75) - heteroscedastic case (positive pairing), (1.75,1.5,1.25,1) - heteroscedastic case (negative pairing)) \textit{(continued)}}\\
\hline
D&$(n_1,n_2,n_3,n_4)$&$(\lambda_1,\lambda_2,\lambda_3,\lambda_4)$&$\mathcal{H}$&Fmax&GPF&L2b&Fb&CAFB&GPH&mGPH\\
\hline
\endhead
N&(15,20,25,30)&(1,1,1,1)&$\mathcal{H}_{0,1}$&7.45&7.65&9.25&5.30&6.95&7.40&8.55\\
&&&$\mathcal{H}_{0,2}$&63.70&73.20&69.05&59.75&56.65&70.45&73.95\\
&&&$\mathcal{H}_{0,3}$&99.80&99.90&99.65&99.15&97.80&99.70&99.80\\
&&&$\mathcal{H}_{0,4}$&11.10&11.70&12.70&9.55&11.25&11.75&13.65\\
&&&$\mathcal{H}_{0,5}$&82.25&90.15&86.95&81.65&74.35&88.90&90.45\\
&&&$\mathcal{H}_{0,6}$&17.00&18.65&18.30&14.35&13.40&18.80&21.25\\
N&(15,20,25,30)&(1,1.25,1.5,1.75)&$\mathcal{H}_{0,1}$&3.75&3.95&6.30&4.05&4.85&5.10&5.95\\
&&&$\mathcal{H}_{0,2}$&27.10&31.35&47.20&40.35&35.15&46.55&50.30\\
&&&$\mathcal{H}_{0,3}$&62.00&71.75&92.05&88.70&81.75&93.85&94.70\\
&&&$\mathcal{H}_{0,4}$&4.60&4.70&5.85&4.10&5.45&5.70&6.25\\
&&&$\mathcal{H}_{0,5}$&24.70&29.95&39.10&32.10&30.05&42.05&44.30\\
&&&$\mathcal{H}_{0,6}$&3.50&3.00&4.30&3.15&3.90&3.65&4.25\\
N&(15,20,25,30)&(1.75,1.5,1.25,1)&$\mathcal{H}_{0,1}$&2.80&3.10&3.15&1.70&2.55&2.25&2.70\\
&&&$\mathcal{H}_{0,2}$&29.60&34.40&20.60&12.55&16.30&17.30&20.60\\
&&&$\mathcal{H}_{0,3}$&90.20&94.20&70.60&58.00&53.50&70.90&75.00\\
&&&$\mathcal{H}_{0,4}$&6.40&6.60&5.85&3.60&4.50&4.65&6.00\\
&&&$\mathcal{H}_{0,5}$&63.35&69.00&49.25&40.85&38.05&51.55&56.10\\
&&&$\mathcal{H}_{0,6}$&13.70&14.60&12.65&9.50&9.65&12.05&14.30\\
t&(15,20,25,30)&(1,1,1,1)&$\mathcal{H}_{0,1}$&4.50&4.40&4.75&3.05&4.10&4.15&5.35\\
&&&$\mathcal{H}_{0,2}$&35.85&43.35&40.30&31.85&31.85&40.25&43.45\\
&&&$\mathcal{H}_{0,3}$&88.05&93.90&87.90&83.90&80.10&90.75&92.15\\
&&&$\mathcal{H}_{0,4}$&6.55&6.80&7.25&5.35&5.85&6.60&7.60\\
&&&$\mathcal{H}_{0,5}$&50.30&58.10&53.10&48.40&45.20&57.40&60.65\\
&&&$\mathcal{H}_{0,6}$&8.45&8.85&8.85&7.45&7.20&8.45&9.60\\
t&(15,20,25,30)&(1,1.25,1.5,1.75)&$\mathcal{H}_{0,1}$&3.20&2.10&4.20&2.45&3.25&2.90&3.35\\
&&&$\mathcal{H}_{0,2}$&11.25&12.35&23.20&17.85&16.35&22.45&25.55\\
&&&$\mathcal{H}_{0,3}$&28.65&33.85&62.20&56.25&50.35&64.85&67.75\\
&&&$\mathcal{H}_{0,4}$&2.70&2.15&3.10&2.00&2.95&2.50&3.05\\
&&&$\mathcal{H}_{0,5}$&11.85&13.40&19.30&15.70&14.50&19.95&22.20\\
&&&$\mathcal{H}_{0,6}$&3.25&2.45&2.70&2.15&2.85&2.70&3.00\\
t&(15,20,25,30)&(1.75,1.5,1.25,1)&$\mathcal{H}_{0,1}$&2.50&2.10&2.00&0.95&1.95&1.45&1.95\\
&&&$\mathcal{H}_{0,2}$&16.70&19.45&10.65&6.55&8.95&8.80&10.85\\
&&&$\mathcal{H}_{0,3}$&70.30&72.65&36.05&25.75&29.95&34.60&38.65\\
&&&$\mathcal{H}_{0,4}$&3.70&3.85&3.45&2.15&2.95&2.50&3.55\\
&&&$\mathcal{H}_{0,5}$&36.65&39.60&23.45&18.45&20.15&23.85&27.40\\
&&&$\mathcal{H}_{0,6}$&6.75&7.15&5.40&4.25&4.15&5.65&6.70\\
$\chi^2$&(15,20,25,30)&(1,1,1,1)&$\mathcal{H}_{0,1}$&7.30&6.60&9.10&5.05&7.65&6.90&8.15\\
&&&$\mathcal{H}_{0,2}$&64.85&74.10&66.90&59.95&57.60&70.20&73.55\\
&&&$\mathcal{H}_{0,3}$&99.35&99.70&98.40&97.35&96.35&99.20&99.35\\
&&&$\mathcal{H}_{0,4}$&12.70&12.75&13.05&10.00&10.85&13.00&14.55\\
&&&$\mathcal{H}_{0,5}$&83.40&90.75&83.80&79.85&75.40&87.60&90.00\\
&&&$\mathcal{H}_{0,6}$&17.70&18.90&19.80&16.35&15.45&19.75&22.15\\
$\chi^2$&(15,20,25,30)&(1,1.25,1.5,1.75)&$\mathcal{H}_{0,1}$&5.30&4.70&7.95&4.85&6.05&6.75&7.65\\
&&&$\mathcal{H}_{0,2}$&25.65&29.80&46.45&39.85&35.40&47.60&51.25\\
&&&$\mathcal{H}_{0,3}$&62.85&75.15&90.45&87.20&80.85&93.20&94.10\\
&&&$\mathcal{H}_{0,4}$&3.60&3.60&5.85&4.05&4.20&5.10&5.50\\
&&&$\mathcal{H}_{0,5}$&22.65&28.30&37.35&32.15&29.60&40.80&44.15\\
&&&$\mathcal{H}_{0,6}$&3.90&3.25&4.00&3.30&3.60&3.95&4.55\\
$\chi^2$&(15,20,25,30)&(1.75,1.5,1.25,1)&$\mathcal{H}_{0,1}$&4.45&2.80&3.20&2.10&3.10&2.40&2.95\\
&&&$\mathcal{H}_{0,2}$&34.95&37.85&23.55&16.80&22.75&22.60&24.45\\
&&&$\mathcal{H}_{0,3}$&90.85&93.05&67.00&56.40&59.90&67.95&72.80\\
&&&$\mathcal{H}_{0,4}$&7.85&6.60&5.95&4.30&5.30&6.00&6.95\\
&&&$\mathcal{H}_{0,5}$&63.85&69.15&48.75&42.15&42.45&50.95&55.45\\
&&&$\mathcal{H}_{0,6}$&15.80&16.50&13.85&11.55&12.40&14.70&17.35\\
\hline
\end{longtable}

\begin{figure}
\centering
\includegraphics[width= 0.95\textwidth,height=0.9\textheight]{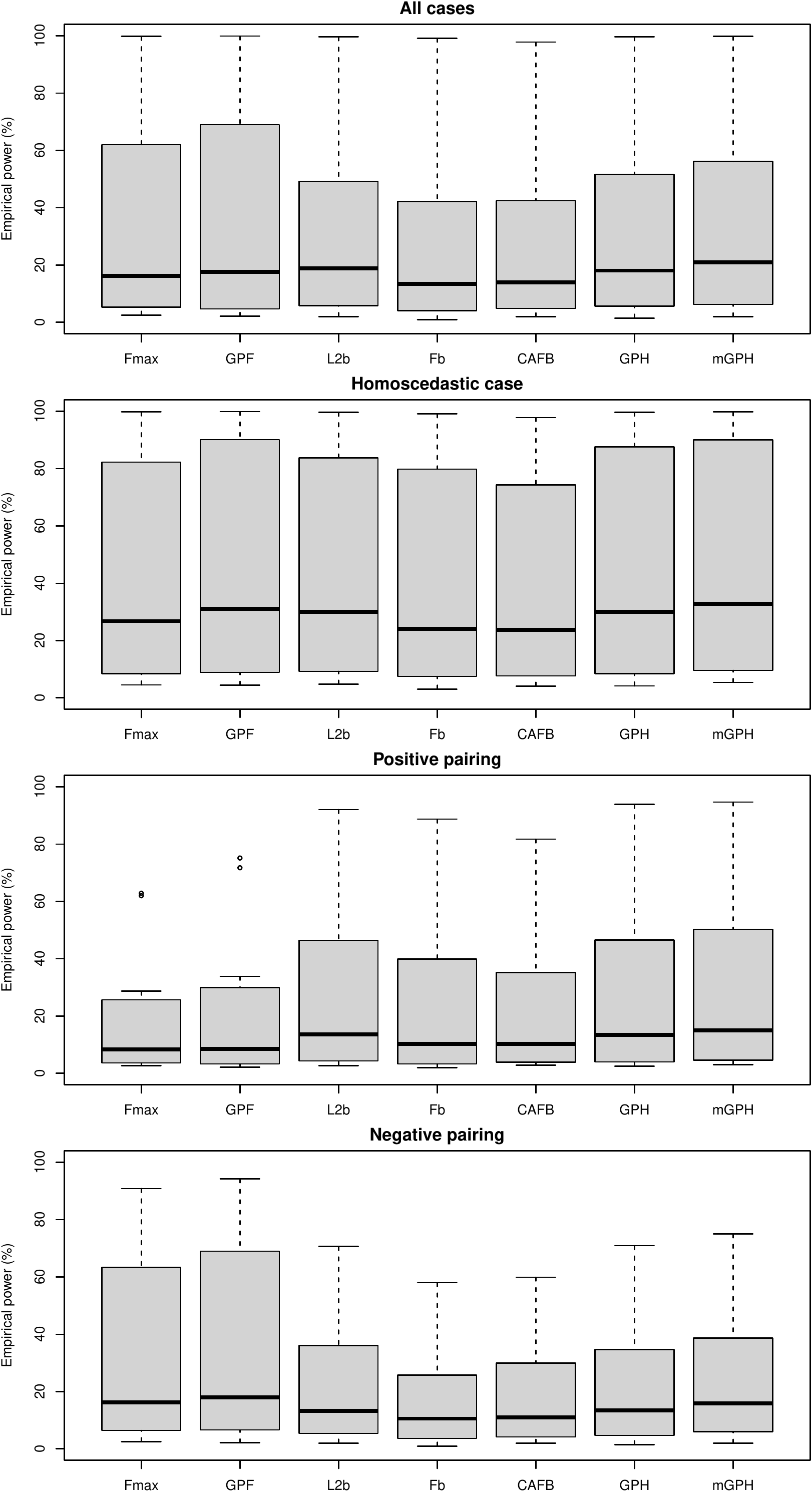}
\caption[Box-and-whisker plots for the empirical powers of all tests obtained under alternative A6 for the Tukey constrasts and homoscedastic and heteroscedastic cases]{Box-and-whisker plots for the empirical powers (as percentages) of all tests obtained under alternative A6 for the Tukey constrasts and homoscedastic and heteroscedastic cases}
\end{figure}

\begin{figure}
\centering
\includegraphics[width= 0.95\textwidth,height=0.9\textheight]{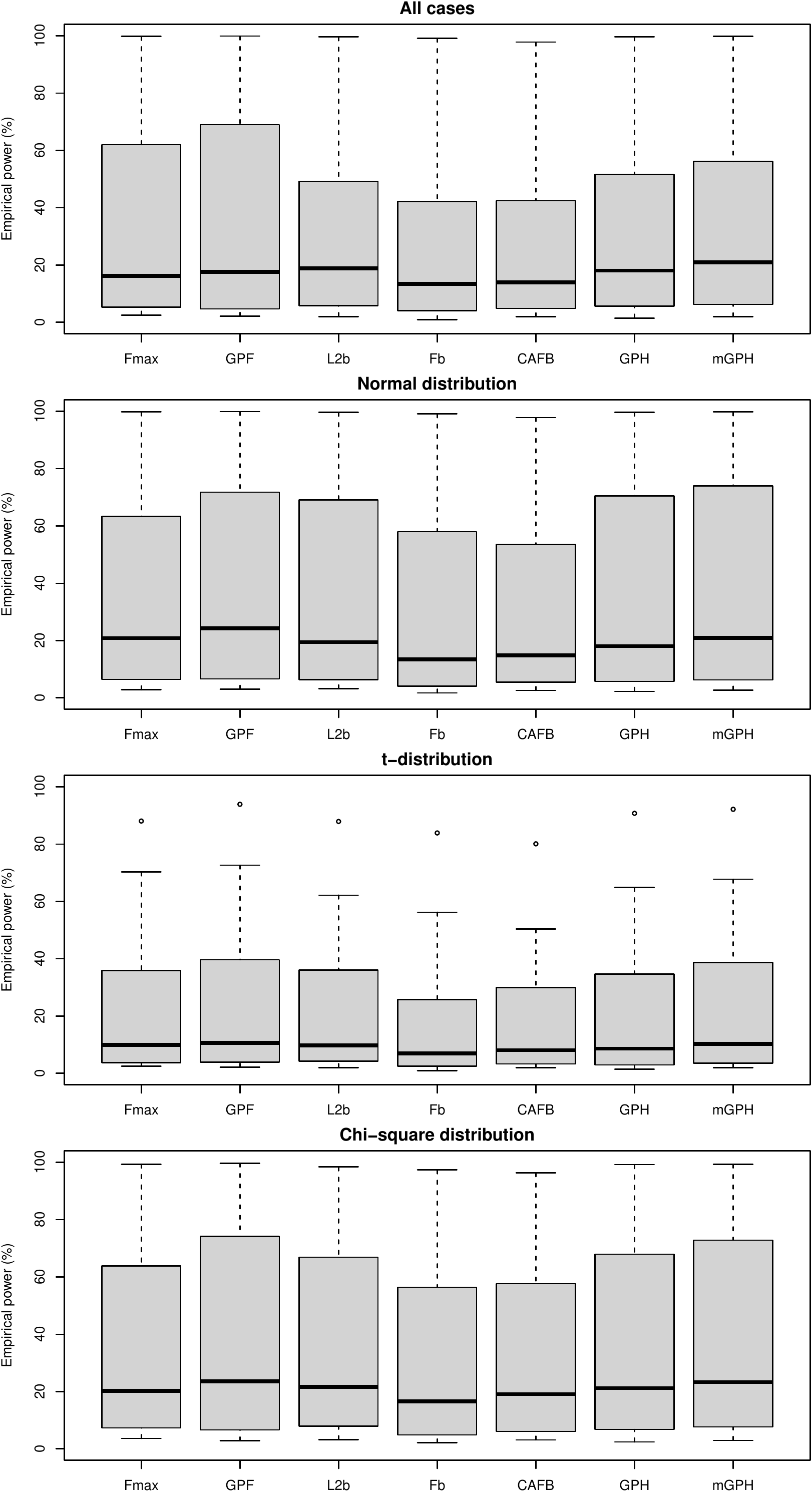}
\caption[Box-and-whisker plots for the empirical powers of all tests obtained under alternative A6 for the Tukey constrasts and different distributions]{Box-and-whisker plots for the empirical powers (as percentages) of all tests obtained under alternative A6 for the Tukey constrasts and different distributions}
\end{figure}

\newpage

\subsection{Results with Scaling Function}\label{results-with-scaling-function}

\begin{longtable}[t]{rrrr|rrrrrrr}
\caption[Empirical FWER and sizes of all tests obtained for the Dunnett constrasts with scaling function]{\label{tab:unnamed-chunk-69}Empirical FWER and sizes (as percentages) of all tests obtained for the Dunnett constrasts with scaling function (D - distribution, $(\lambda_1,\lambda_2,\lambda_3,\lambda_4)$: (1,1,1,1) - homoscedastic case, (1,1.25,1.5,1.75) - heteroscedastic case (positive pairing), (1.75,1.5,1.25,1) - heteroscedastic case (negative pairing))}\\
\hline
D&$(n_1,n_2,n_3,n_4)$&$(\lambda_1,\lambda_2,\lambda_3,\lambda_4)$&$\mathcal{H}$&Fmax&GPF&L2b&Fb&CAFB&GPH&mGPH\\
\hline
\endfirsthead
\caption[]{Empirical FWER and sizes (as percentages) of all tests obtained for the Dunnett constrasts with scaling function (D - distribution, $(\lambda_1,\lambda_2,\lambda_3,\lambda_4)$: (1,1,1,1) - homoscedastic case, (1,1.25,1.5,1.75) - heteroscedastic case (positive pairing), (1.75,1.5,1.25,1) - heteroscedastic case (negative pairing)) \textit{(continued)}}\\
\hline
D&$(n_1,n_2,n_3,n_4)$&$(\lambda_1,\lambda_2,\lambda_3,\lambda_4)$&$\mathcal{H}$&Fmax&GPF&L2b&Fb&CAFB&GPH&mGPH\\
\hline
\endhead
N&(15,20,25,30)&(1,1,1,1)&FWER&3.90&3.30&6.35&4.55&3.85&3.25&3.95\\
&&&$\mathcal{H}_{0,1}$&1.55&1.35&2.85&2.25&1.70&1.30&1.50\\
&&&$\mathcal{H}_{0,2}$&1.35&1.35&2.40&1.40&1.60&1.40&1.50\\
&&&$\mathcal{H}_{0,3}$&1.25&1.30&2.75&1.95&1.45&1.15&1.70\\
N&(15,20,25,30)&(1,1.25,1.5,1.75)&FWER&2.10&1.40&5.50&3.60&3.40&3.55&4.25\\
&&&$\mathcal{H}_{0,1}$&1.15&1.00&2.00&1.30&1.05&1.20&1.45\\
&&&$\mathcal{H}_{0,2}$&0.55&0.30&1.75&1.20&1.25&1.25&1.35\\
&&&$\mathcal{H}_{0,3}$&0.40&0.15&2.40&1.55&1.25&1.30&1.65\\
N&(15,20,25,30)&(1.75,1.5,1.25,1)&FWER&13.10&9.60&6.20&4.00&4.20&2.90&3.75\\
&&&$\mathcal{H}_{0,1}$&2.00&1.50&2.80&1.95&1.80&1.35&1.60\\
&&&$\mathcal{H}_{0,2}$&4.30&3.40&3.35&2.10&1.65&0.90&1.45\\
&&&$\mathcal{H}_{0,3}$&10.45&8.05&2.85&1.45&1.65&1.35&1.95\\
N&(30,40,50,60)&(1,1,1,1)&FWER&5.95&4.75&5.75&4.70&3.75&4.85&6.05\\
&&&$\mathcal{H}_{0,1}$&2.00&1.45&2.40&1.85&1.40&1.55&2.00\\
&&&$\mathcal{H}_{0,2}$&2.15&1.70&2.15&1.85&1.60&1.80&2.25\\
&&&$\mathcal{H}_{0,3}$&2.35&2.35&2.45&2.00&1.30&2.20&2.65\\
N&(30,40,50,60)&(1,1.25,1.5,1.75)&FWER&1.80&1.30&5.80&4.55&4.00&4.25&4.50\\
&&&$\mathcal{H}_{0,1}$&1.25&0.90&2.05&1.40&1.45&1.20&1.25\\
&&&$\mathcal{H}_{0,2}$&0.50&0.35&2.25&1.70&1.60&1.75&1.90\\
&&&$\mathcal{H}_{0,3}$&0.10&0.10&2.00&1.70&1.15&1.40&1.50\\
N&(30,40,50,60)&(1.75,1.5,1.25,1)&FWER&12.85&10.00&5.35&4.05&3.10&2.80&3.95\\
&&&$\mathcal{H}_{0,1}$&2.00&2.00&2.20&1.60&1.30&1.65&2.20\\
&&&$\mathcal{H}_{0,2}$&4.20&3.30&2.85&2.05&1.15&1.15&1.70\\
&&&$\mathcal{H}_{0,3}$&9.85&7.75&2.80&2.15&1.25&1.05&1.80\\
N&(60,80,100,120)&(1,1,1,1)&FWER&3.10&3.90&5.15&4.80&3.00&3.80&4.50\\
&&&$\mathcal{H}_{0,1}$&1.15&1.40&2.10&1.90&0.95&1.35&1.50\\
&&&$\mathcal{H}_{0,2}$&0.95&1.45&1.75&1.55&0.90&1.45&1.85\\
&&&$\mathcal{H}_{0,3}$&1.30&1.60&2.05&1.85&1.30&1.60&1.95\\
N&(60,80,100,120)&(1,1.25,1.5,1.75)&FWER&1.80&1.65&5.35&4.85&3.10&4.10&4.70\\
&&&$\mathcal{H}_{0,1}$&1.00&0.90&1.95&1.65&1.00&1.35&1.55\\
&&&$\mathcal{H}_{0,2}$&0.70&0.55&2.00&1.80&1.25&1.60&1.85\\
&&&$\mathcal{H}_{0,3}$&0.10&0.25&1.75&1.70&1.10&1.60&1.90\\
N&(60,80,100,120)&(1.75,1.5,1.25,1)&FWER&12.55&11.20&5.00&4.30&2.95&3.15&4.25\\
&&&$\mathcal{H}_{0,1}$&2.55&2.00&2.65&2.20&1.60&1.55&2.20\\
&&&$\mathcal{H}_{0,2}$&3.85&3.20&1.80&1.70&0.95&1.30&1.70\\
&&&$\mathcal{H}_{0,3}$&9.60&9.05&2.30&1.90&1.05&1.30&1.60\\
t&(15,20,25,30)&(1,1,1,1)&FWER&4.05&3.50&5.30&3.20&3.40&3.00&3.65\\
&&&$\mathcal{H}_{0,1}$&1.65&1.10&1.70&0.70&0.90&1.15&1.50\\
&&&$\mathcal{H}_{0,2}$&1.25&1.30&2.05&1.40&1.50&1.15&1.30\\
&&&$\mathcal{H}_{0,3}$&1.65&1.55&2.65&1.60&1.30&1.40&1.60\\
t&(15,20,25,30)&(1,1.25,1.5,1.75)&FWER&2.50&1.55&6.25&3.55&3.75&4.00&4.55\\
&&&$\mathcal{H}_{0,1}$&1.20&1.00&2.50&1.30&1.10&1.40&1.65\\
&&&$\mathcal{H}_{0,2}$&1.05&0.50&1.75&1.05&1.55&1.55&1.65\\
&&&$\mathcal{H}_{0,3}$&0.30&0.10&2.85&1.75&1.20&1.25&1.45\\
t&(15,20,25,30)&(1.75,1.5,1.25,1)&FWER&12.80&10.25&6.20&3.90&3.60&2.00&3.50\\
&&&$\mathcal{H}_{0,1}$&2.10&1.50&2.25&1.25&1.05&0.80&1.50\\
&&&$\mathcal{H}_{0,2}$&4.20&3.30&3.05&1.85&1.75&1.00&1.55\\
&&&$\mathcal{H}_{0,3}$&9.95&8.95&3.20&1.95&1.70&0.80&1.50\\
t&(30,40,50,60)&(1,1,1,1)&FWER&4.55&4.15&4.35&3.45&3.20&3.50&3.90\\
&&&$\mathcal{H}_{0,1}$&1.65&1.50&1.70&1.45&1.10&1.35&1.40\\
&&&$\mathcal{H}_{0,2}$&1.45&1.45&1.55&1.20&1.15&1.25&1.45\\
&&&$\mathcal{H}_{0,3}$&2.00&1.65&1.80&1.45&1.35&1.40&1.60\\
t&(30,40,50,60)&(1,1.25,1.5,1.75)&FWER&2.20&1.70&5.55&4.95&3.80&3.60&3.95\\
&&&$\mathcal{H}_{0,1}$&1.50&1.10&2.35&2.00&1.65&1.20&1.25\\
&&&$\mathcal{H}_{0,2}$&0.55&0.35&1.35&1.35&1.15&1.20&1.30\\
&&&$\mathcal{H}_{0,3}$&0.20&0.30&2.35&2.05&1.20&1.50&1.70\\
t&(30,40,50,60)&(1.75,1.5,1.25,1)&FWER&12.20&11.05&4.60&3.55&3.10&3.30&5.20\\
&&&$\mathcal{H}_{0,1}$&2.60&1.95&1.65&1.45&1.15&1.15&2.10\\
&&&$\mathcal{H}_{0,2}$&3.80&4.25&1.80&1.40&1.40&1.45&2.30\\
&&&$\mathcal{H}_{0,3}$&9.40&8.60&2.55&1.60&1.15&1.65&2.15\\
t&(60,80,100,120)&(1,1,1,1)&FWER&4.50&4.10&4.75&4.25&3.05&3.95&4.65\\
&&&$\mathcal{H}_{0,1}$&1.45&1.55&1.60&1.55&1.10&1.30&1.60\\
&&&$\mathcal{H}_{0,2}$&2.20&1.90&1.70&1.30&1.35&2.05&2.40\\
&&&$\mathcal{H}_{0,3}$&1.40&1.55&1.90&1.90&0.85&1.55&1.70\\
t&(60,80,100,120)&(1,1.25,1.5,1.75)&FWER&1.60&1.80&5.10&4.95&3.55&4.10&4.60\\
&&&$\mathcal{H}_{0,1}$&0.90&1.10&2.00&1.85&1.20&1.60&1.80\\
&&&$\mathcal{H}_{0,2}$&0.50&0.60&1.90&1.85&1.10&1.35&1.50\\
&&&$\mathcal{H}_{0,3}$&0.25&0.20&2.10&2.10&1.40&1.40&1.60\\
t&(60,80,100,120)&(1.75,1.5,1.25,1)&FWER&12.50&11.35&4.60&4.10&2.40&4.00&5.15\\
&&&$\mathcal{H}_{0,1}$&3.00&2.45&1.90&1.65&0.90&2.10&2.65\\
&&&$\mathcal{H}_{0,2}$&4.80&4.85&2.30&2.00&0.95&1.75&2.40\\
&&&$\mathcal{H}_{0,3}$&9.20&8.40&2.40&2.25&1.10&1.50&1.95\\
$\chi^2$&(15,20,25,30)&(1,1,1,1)&FWER&3.85&3.15&7.05&4.35&3.55&2.90&3.85\\
&&&$\mathcal{H}_{0,1}$&1.05&1.35&2.80&1.75&1.10&1.05&1.40\\
&&&$\mathcal{H}_{0,2}$&1.15&1.10&2.95&1.70&1.60&1.20&1.45\\
&&&$\mathcal{H}_{0,3}$&1.90&1.20&2.50&1.50&1.05&1.00&1.60\\
$\chi^2$&(15,20,25,30)&(1,1.25,1.5,1.75)&FWER&2.35&1.55&6.35&4.35&3.85&3.60&4.05\\
&&&$\mathcal{H}_{0,1}$&1.20&1.10&2.85&1.90&1.40&1.25&1.45\\
&&&$\mathcal{H}_{0,2}$&1.00&0.30&2.10&1.35&1.20&1.10&1.25\\
&&&$\mathcal{H}_{0,3}$&0.20&0.20&1.75&1.15&1.30&1.35&1.50\\
$\chi^2$&(15,20,25,30)&(1.75,1.5,1.25,1)&FWER&12.40&9.55&5.55&3.20&4.10&2.15&3.50\\
&&&$\mathcal{H}_{0,1}$&1.95&1.60&2.30&1.35&1.20&1.00&1.35\\
&&&$\mathcal{H}_{0,2}$&3.90&3.65&2.25&1.55&1.80&0.95&1.70\\
&&&$\mathcal{H}_{0,3}$&9.40&7.60&2.85&1.70&1.95&1.00&1.75\\
$\chi^2$&(30,40,50,60)&(1,1,1,1)&FWER&4.30&3.30&4.85&4.25&2.85&3.10&3.85\\
&&&$\mathcal{H}_{0,1}$&1.60&1.40&2.05&1.65&0.85&1.20&1.65\\
&&&$\mathcal{H}_{0,2}$&1.60&1.45&1.90&1.75&1.05&1.40&1.65\\
&&&$\mathcal{H}_{0,3}$&1.45&1.00&1.70&1.45&1.35&1.00&1.20\\
$\chi^2$&(30,40,50,60)&(1,1.25,1.5,1.75)&FWER&1.30&1.45&5.55&4.60&2.85&3.70&3.80\\
&&&$\mathcal{H}_{0,1}$&0.90&0.95&1.90&1.60&1.10&1.15&1.15\\
&&&$\mathcal{H}_{0,2}$&0.25&0.35&2.05&1.70&0.75&1.20&1.25\\
&&&$\mathcal{H}_{0,3}$&0.15&0.15&1.85&1.50&1.05&1.50&1.60\\
$\chi^2$&(30,40,50,60)&(1.75,1.5,1.25,1)&FWER&12.45&11.05&5.25&4.25&3.45&3.30&4.90\\
&&&$\mathcal{H}_{0,1}$&2.30&1.95&2.30&1.85&1.50&1.45&2.00\\
&&&$\mathcal{H}_{0,2}$&3.90&3.35&2.05&1.60&1.35&1.30&1.85\\
&&&$\mathcal{H}_{0,3}$&10.15&8.85&2.70&2.35&1.70&1.30&2.55\\
$\chi^2$&(60,80,100,120)&(1,1,1,1)&FWER&4.35&4.45&4.40&3.95&2.80&4.00&4.85\\
&&&$\mathcal{H}_{0,1}$&1.40&1.85&1.85&1.40&1.20&1.75&2.05\\
&&&$\mathcal{H}_{0,2}$&1.80&1.50&1.40&1.35&1.00&1.15&1.50\\
&&&$\mathcal{H}_{0,3}$&1.60&1.65&1.75&1.65&0.80&1.70&1.95\\
$\chi^2$&(60,80,100,120)&(1,1.25,1.5,1.75)&FWER&2.20&1.40&5.10&4.80&3.90&4.15&4.55\\
&&&$\mathcal{H}_{0,1}$&1.45&0.85&2.35&2.10&1.50&1.45&1.65\\
&&&$\mathcal{H}_{0,2}$&0.55&0.45&1.65&1.65&1.30&2.00&2.10\\
&&&$\mathcal{H}_{0,3}$&0.30&0.10&1.80&1.80&1.60&1.20&1.30\\
$\chi^2$&(60,80,100,120)&(1.75,1.5,1.25,1)&FWER&11.45&10.35&4.15&3.85&2.85&4.10&5.00\\
&&&$\mathcal{H}_{0,1}$&1.85&1.50&2.10&1.95&1.25&1.40&1.65\\
&&&$\mathcal{H}_{0,2}$&4.55&3.65&1.80&1.70&1.00&1.95&2.45\\
&&&$\mathcal{H}_{0,3}$&8.95&8.95&1.75&1.65&1.25&2.10&2.75\\
\hline
\end{longtable}

\begin{figure}
\centering
\includegraphics[width= 0.95\textwidth,height=0.9\textheight]{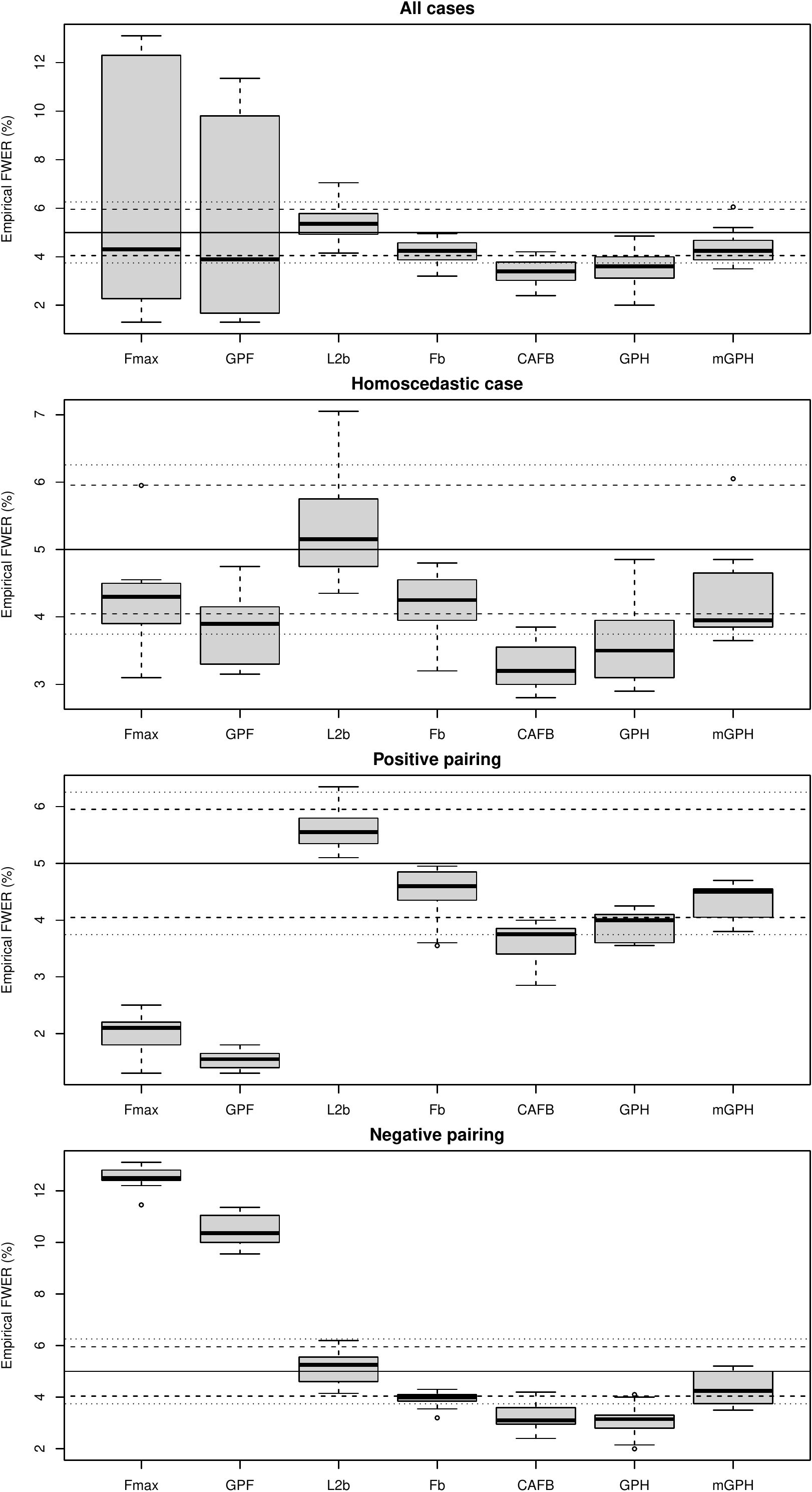}
\caption[Box-and-whisker plots for the empirical FWER of all tests obtained for the Dunnett constrasts with scaling function
for homoscedastic and heteroscedastic cases]{Box-and-whisker plots for the empirical FWER (as percentages) of all tests obtained for the Dunnett constrasts with scaling function for homoscedastic and heteroscedastic cases}
\end{figure}

\begin{figure}
\centering
\includegraphics[width= 0.95\textwidth,height=0.9\textheight]{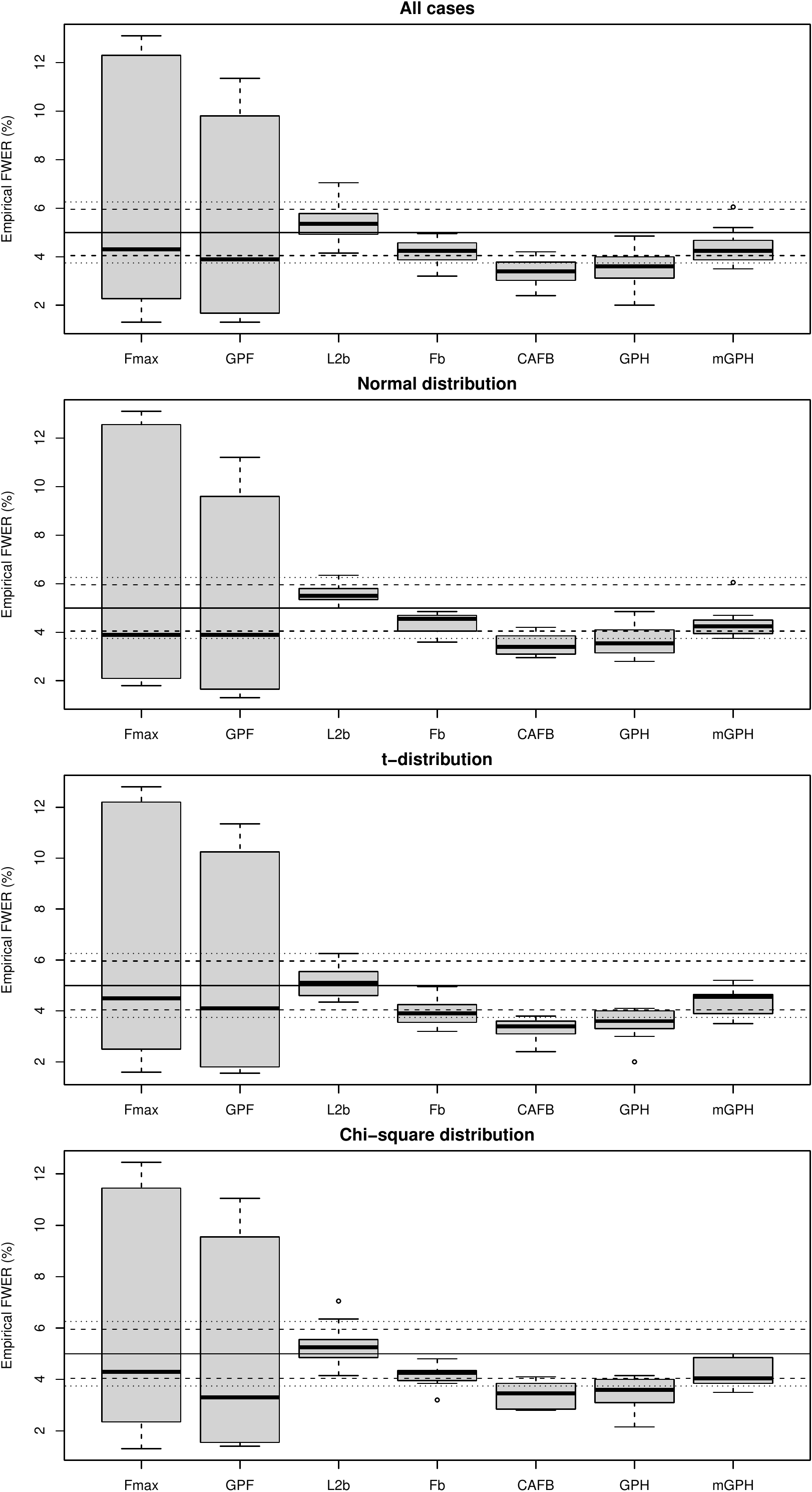}
\caption[Box-and-whisker plots for the empirical FWER of all tests obtained for the Dunnett constrasts with scaling function
for different distributions]{Box-and-whisker plots for the empirical FWER (as percentages) of all tests obtained for the Dunnett constrasts with scaling function for different distributions}
\end{figure}

\newpage

\begin{longtable}[t]{rrrr|rrrrrrr}
\caption[Empirical sizes and powers of all tests obtained under alternative A1 for the Dunnett constrasts with scaling function]{\label{tab:unnamed-chunk-79}Empirical sizes ($\mathcal{H}_{0,1}$, $\mathcal{H}_{0,2}$) and powers ($\mathcal{H}_{0,3}$) (as percentages) of all tests obtained under alternative A1 for the Dunnett constrasts with scaling function (D - distribution, $(\lambda_1,\lambda_2,\lambda_3,\lambda_4)$: (1,1,1,1) - homoscedastic case, (1,1.25,1.5,1.75) - heteroscedastic case (positive pairing), (1.75,1.5,1.25,1) - heteroscedastic case (negative pairing))}\\
\hline
D&$(n_1,n_2,n_3,n_4)$&$(\lambda_1,\lambda_2,\lambda_3,\lambda_4)$&$\mathcal{H}$&Fmax&GPF&L2b&Fb&CAFB&GPH&mGPH\\
\hline
\endfirsthead
\caption[]{Empirical sizes ($\mathcal{H}_{0,1}$, $\mathcal{H}_{0,2}$) and powers ($\mathcal{H}_{0,3}$) (as percentages) of all tests obtained under alternative A1 for the Dunnett constrasts with scaling function (D - distribution, $(\lambda_1,\lambda_2,\lambda_3,\lambda_4)$: (1,1,1,1) - homoscedastic case, (1,1.25,1.5,1.75) - heteroscedastic case (positive pairing), (1.75,1.5,1.25,1) - heteroscedastic case (negative pairing)) \textit{(continued)}}\\
\hline
D&$(n_1,n_2,n_3,n_4)$&$(\lambda_1,\lambda_2,\lambda_3,\lambda_4)$&$\mathcal{H}$&Fmax&GPF&L2b&Fb&CAFB&GPH&mGPH\\
\hline
\endhead
N&(15,20,25,30)&(1,1,1,1)&$\mathcal{H}_{0,1}$&1.85&1.25&3.00&1.65&1.50&1.15&1.70\\
&&&$\mathcal{H}_{0,2}$&1.80&1.45&2.95&2.10&1.30&1.40&1.90\\
&&&$\mathcal{H}_{0,3}$&82.30&88.55&2.70&1.95&5.50&86.60&88.70\\
N&(15,20,25,30)&(1,1.25,1.5,1.75)&$\mathcal{H}_{0,1}$&1.30&0.80&2.70&1.75&1.35&1.20&1.60\\
&&&$\mathcal{H}_{0,2}$&0.70&0.40&3.25&2.00&1.45&1.35&1.35\\
&&&$\mathcal{H}_{0,3}$&21.60&25.15&2.15&1.60&3.65&56.65&58.60\\
N&(15,20,25,30)&(1.75,1.5,1.25,1)&$\mathcal{H}_{0,1}$&1.70&1.40&3.15&1.75&1.75&1.10&1.50\\
&&&$\mathcal{H}_{0,2}$&4.05&2.60&2.75&1.65&1.40&1.05&1.45\\
&&&$\mathcal{H}_{0,3}$&65.85&68.50&3.55&2.15&3.80&31.45&38.10\\
t&(15,20,25,30)&(1,1,1,1)&$\mathcal{H}_{0,1}$&1.70&1.15&3.05&1.90&1.25&1.20&1.35\\
&&&$\mathcal{H}_{0,2}$&2.05&1.45&2.55&1.65&1.30&1.20&1.55\\
&&&$\mathcal{H}_{0,3}$&53.60&61.90&3.05&2.00&4.75&59.10&62.55\\
t&(15,20,25,30)&(1,1.25,1.5,1.75)&$\mathcal{H}_{0,1}$&1.65&1.10&2.55&1.40&1.40&1.35&1.50\\
&&&$\mathcal{H}_{0,2}$&0.75&0.45&2.45&1.80&1.25&1.25&1.40\\
&&&$\mathcal{H}_{0,3}$&10.55&9.65&2.00&1.50&2.80&30.25&32.20\\
t&(15,20,25,30)&(1.75,1.5,1.25,1)&$\mathcal{H}_{0,1}$&2.20&1.50&2.65&1.40&1.50&0.75&1.20\\
&&&$\mathcal{H}_{0,2}$&3.85&3.35&2.85&1.90&1.75&0.70&1.50\\
&&&$\mathcal{H}_{0,3}$&44.55&45.35&3.70&2.20&2.65&14.80&18.95\\
$\chi^2$&(15,20,25,30)&(1,1,1,1)&$\mathcal{H}_{0,1}$&1.45&1.10&2.40&1.70&1.10&0.80&1.15\\
&&&$\mathcal{H}_{0,2}$&1.20&1.40&2.25&1.45&1.50&1.15&1.25\\
&&&$\mathcal{H}_{0,3}$&82.90&88.70&2.90&1.65&6.90&84.15&87.05\\
$\chi^2$&(15,20,25,30)&(1,1.25,1.5,1.75)&$\mathcal{H}_{0,1}$&1.45&0.55&2.55&1.65&1.55&0.65&0.80\\
&&&$\mathcal{H}_{0,2}$&0.80&0.40&2.25&1.85&1.00&0.80&1.00\\
&&&$\mathcal{H}_{0,3}$&22.55&23.60&2.35&1.75&3.20&56.50&57.95\\
$\chi^2$&(15,20,25,30)&(1.75,1.5,1.25,1)&$\mathcal{H}_{0,1}$&1.55&1.35&2.65&1.50&1.85&0.95&1.45\\
&&&$\mathcal{H}_{0,2}$&4.40&2.60&2.80&1.60&1.60&0.75&1.20\\
&&&$\mathcal{H}_{0,3}$&66.95&68.10&3.75&2.00&3.80&33.05&39.60\\
\hline
\end{longtable}

\begin{figure}
\centering
\includegraphics[width= 0.95\textwidth,height=0.9\textheight]{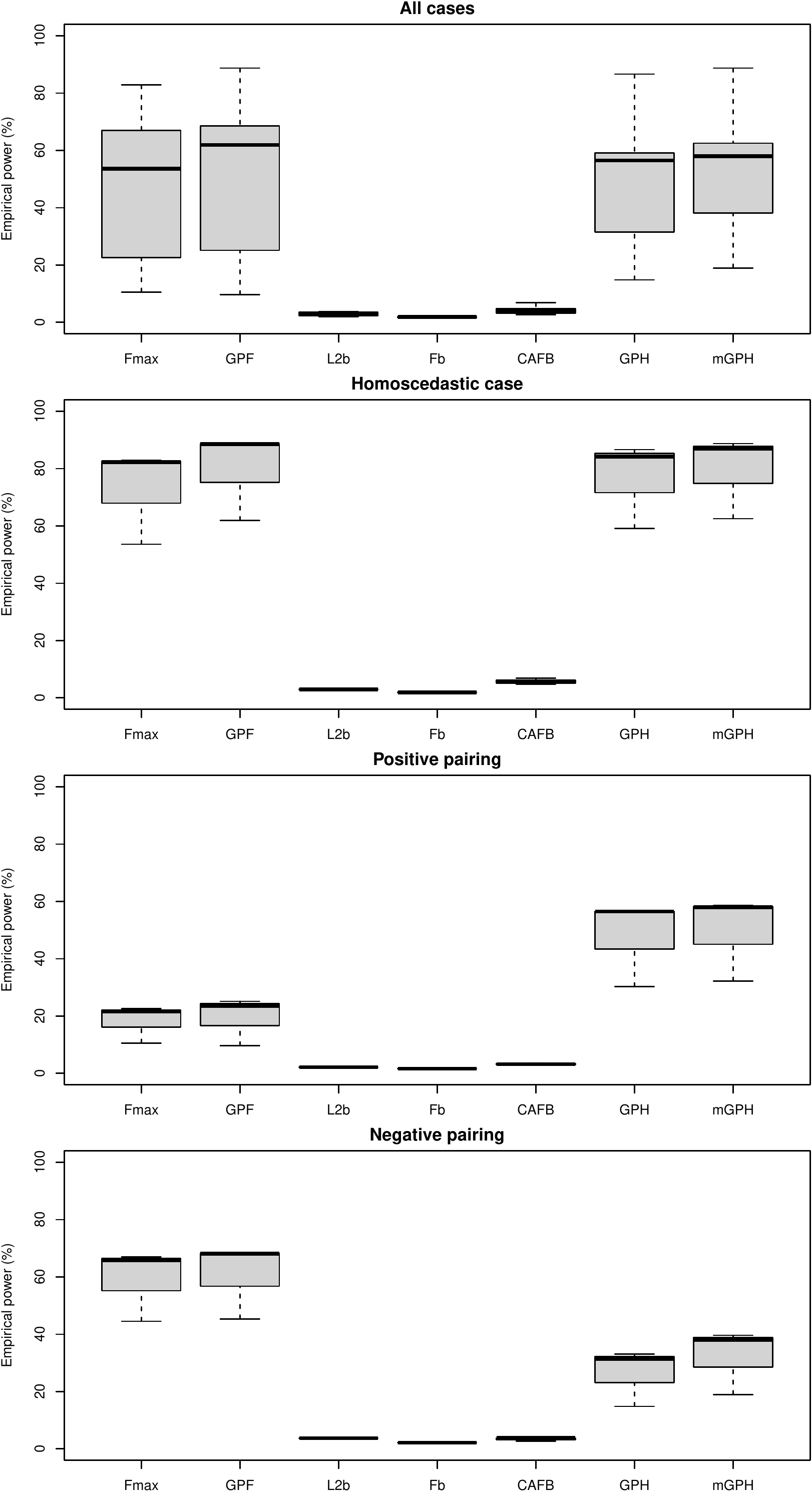}
\caption[Box-and-whisker plots for the empirical powers of all tests obtained under alternative A1 for the Dunnett constrasts with scaling function and homoscedastic and heteroscedastic cases]{Box-and-whisker plots for the empirical powers (as percentages) of all tests obtained under alternative A1 for the Dunnett constrasts with scaling function and homoscedastic and heteroscedastic cases}
\end{figure}

\begin{figure}
\centering
\includegraphics[width= 0.95\textwidth,height=0.9\textheight]{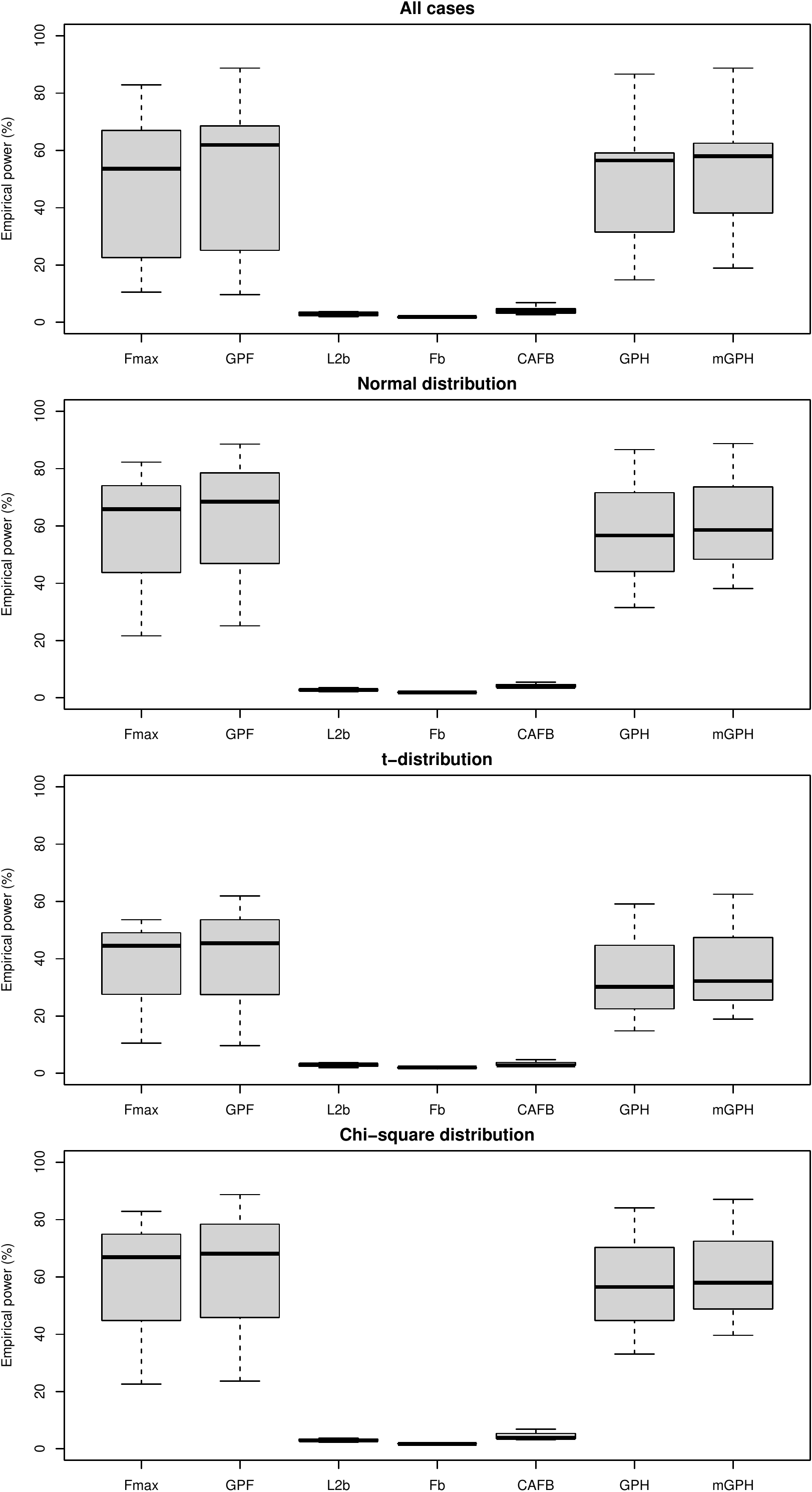}
\caption[Box-and-whisker plots for the empirical powers of all tests obtained under alternative A1 for the Dunnett constrasts with scaling function and different distributions]{Box-and-whisker plots for the empirical powers (as percentages) of all tests obtained under alternative A1 for the Dunnett constrasts with scaling function and different distributions}
\end{figure}

\newpage

\begin{longtable}[t]{rrrr|rrrrrrr}
\caption[Empirical sizes and powers of all tests obtained under alternative A2 for the Dunnett constrasts with scaling function]{\label{tab:unnamed-chunk-74}Empirical sizes ($\mathcal{H}_{0,1}$, $\mathcal{H}_{0,2}$) and powers ($\mathcal{H}_{0,3}$) (as percentages) of all tests obtained under alternative A2 for the Dunnett constrasts with scaling function (D - distribution, $(\lambda_1,\lambda_2,\lambda_3,\lambda_4)$: (1,1,1,1) - homoscedastic case, (1,1.25,1.5,1.75) - heteroscedastic case (positive pairing), (1.75,1.5,1.25,1) - heteroscedastic case (negative pairing))}\\
\hline
D&$(n_1,n_2,n_3,n_4)$&$(\lambda_1,\lambda_2,\lambda_3,\lambda_4)$&$\mathcal{H}$&Fmax&GPF&L2b&Fb&CAFB&GPH&mGPH\\
\hline
\endfirsthead
\caption[]{Empirical sizes ($\mathcal{H}_{0,1}$, $\mathcal{H}_{0,2}$) and powers ($\mathcal{H}_{0,3}$) (as percentages) of all tests obtained under alternative A2 for the Dunnett constrasts with scaling function (D - distribution, $(\lambda_1,\lambda_2,\lambda_3,\lambda_4)$: (1,1,1,1) - homoscedastic case, (1,1.25,1.5,1.75) - heteroscedastic case (positive pairing), (1.75,1.5,1.25,1) - heteroscedastic case (negative pairing)) \textit{(continued)}}\\
\hline
D&$(n_1,n_2,n_3,n_4)$&$(\lambda_1,\lambda_2,\lambda_3,\lambda_4)$&$\mathcal{H}$&Fmax&GPF&L2b&Fb&CAFB&GPH&mGPH\\
\hline
\endhead
N&(15,20,25,30)&(1,1,1,1)&$\mathcal{H}_{0,1}$&1.80&1.25&3.30&2.30&1.45&0.85&1.50\\
&&&$\mathcal{H}_{0,2}$&1.55&1.00&2.60&1.85&1.30&1.20&1.45\\
&&&$\mathcal{H}_{0,3}$&81.05&98.55&82.00&74.55&86.15&98.10&98.45\\
N&(15,20,25,30)&(1,1.25,1.5,1.75)&$\mathcal{H}_{0,1}$&1.35&0.80&2.45&1.50&1.15&1.00&1.15\\
&&&$\mathcal{H}_{0,2}$&0.90&0.55&3.15&2.15&1.20&1.40&1.50\\
&&&$\mathcal{H}_{0,3}$&22.15&58.70&54.15&48.90&60.25&84.85&86.10\\
N&(15,20,25,30)&(1.75,1.5,1.25,1)&$\mathcal{H}_{0,1}$&2.00&1.65&2.65&1.50&1.20&1.20&1.70\\
&&&$\mathcal{H}_{0,2}$&4.10&3.25&2.95&1.50&1.70&1.55&2.20\\
&&&$\mathcal{H}_{0,3}$&69.55&87.45&38.75&29.70&36.35&58.15&65.10\\
t&(15,20,25,30)&(1,1,1,1)&$\mathcal{H}_{0,1}$&1.85&1.35&2.25&1.30&1.55&1.20&1.45\\
&&&$\mathcal{H}_{0,2}$&1.50&0.90&2.20&1.50&1.35&0.70&1.10\\
&&&$\mathcal{H}_{0,3}$&52.20&84.75&55.30&47.40&57.55&82.45&84.45\\
t&(15,20,25,30)&(1,1.25,1.5,1.75)&$\mathcal{H}_{0,1}$&0.95&0.75&2.35&1.40&1.40&1.05&1.20\\
&&&$\mathcal{H}_{0,2}$&0.50&0.45&2.00&1.20&0.95&1.10&1.25\\
&&&$\mathcal{H}_{0,3}$&9.35&26.35&33.75&29.00&34.80&55.75&57.80\\
t&(15,20,25,30)&(1.75,1.5,1.25,1)&$\mathcal{H}_{0,1}$&2.05&1.50&2.85&1.65&1.35&1.10&1.45\\
&&&$\mathcal{H}_{0,2}$&3.70&3.20&3.05&1.60&1.00&0.65&0.95\\
&&&$\mathcal{H}_{0,3}$&49.10&66.20&24.75&18.10&20.15&32.75&38.65\\
$\chi^2$&(15,20,25,30)&(1,1,1,1)&$\mathcal{H}_{0,1}$&2.00&1.60&2.75&1.95&1.55&1.40&1.85\\
&&&$\mathcal{H}_{0,2}$&1.55&1.70&2.50&1.65&1.55&1.25&1.60\\
&&&$\mathcal{H}_{0,3}$&82.25&98.75&80.15&73.55&86.50&97.60&98.30\\
$\chi^2$&(15,20,25,30)&(1,1.25,1.5,1.75)&$\mathcal{H}_{0,1}$&1.60&0.65&2.45&1.65&1.40&1.00&1.10\\
&&&$\mathcal{H}_{0,2}$&0.45&0.40&2.60&2.05&1.70&0.90&1.00\\
&&&$\mathcal{H}_{0,3}$&19.75&56.40&54.95&48.85&59.05&83.40&84.95\\
$\chi^2$&(15,20,25,30)&(1.75,1.5,1.25,1)&$\mathcal{H}_{0,1}$&2.50&1.15&3.20&1.80&1.65&0.70&1.15\\
&&&$\mathcal{H}_{0,2}$&3.55&2.35&3.40&2.15&1.75&0.75&1.10\\
&&&$\mathcal{H}_{0,3}$&71.15&85.55&41.05&32.10&40.90&57.15&63.35\\
\hline
\end{longtable}

\begin{figure}
\centering
\includegraphics[width= 0.95\textwidth,height=0.9\textheight]{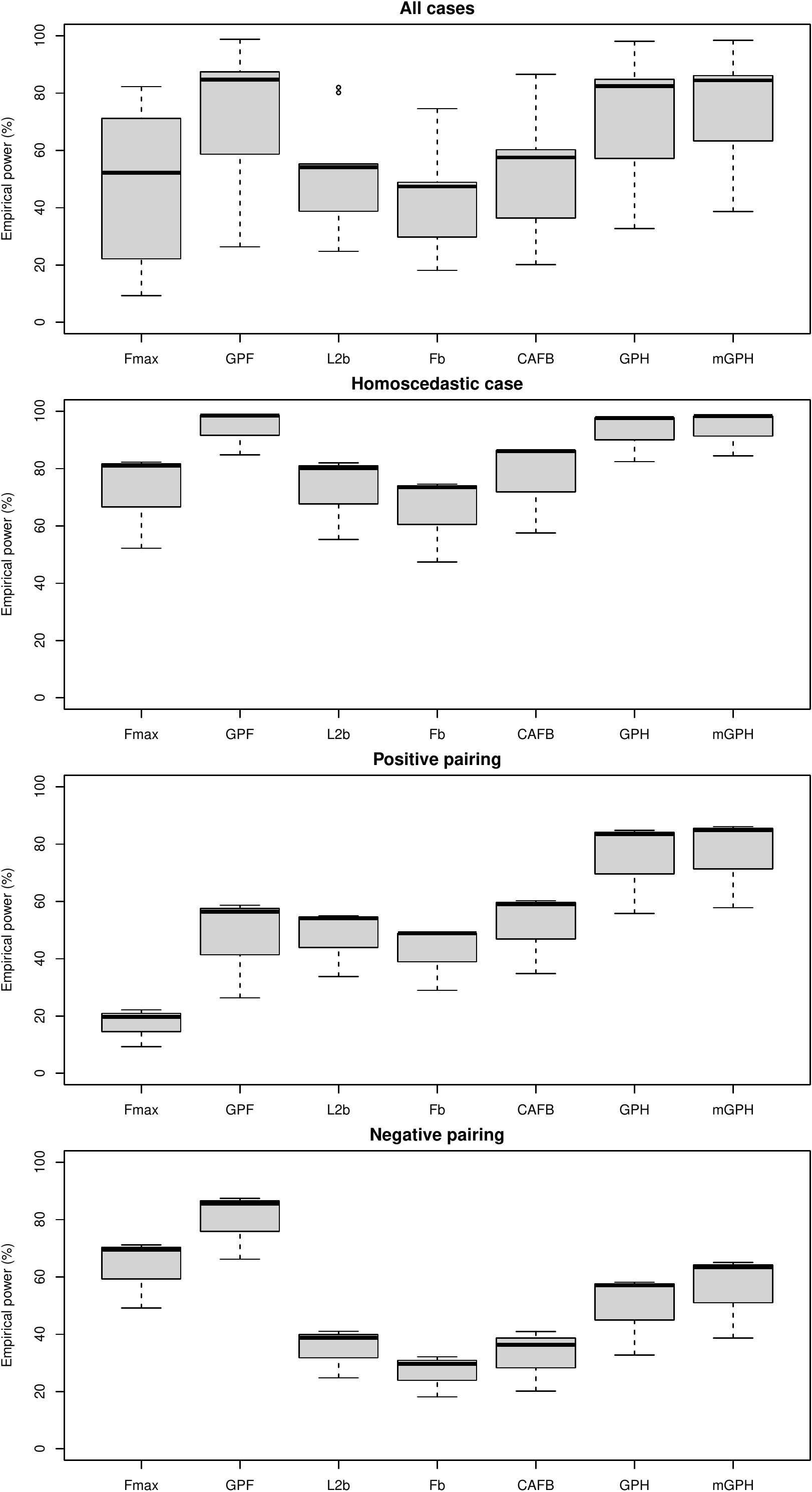}
\caption[Box-and-whisker plots for the empirical powers of all tests obtained under alternative A2 for the Dunnett constrasts with scaling function and homoscedastic and heteroscedastic cases]{Box-and-whisker plots for the empirical powers (as percentages) of all tests obtained under alternative A2 for the Dunnett constrasts with scaling function and homoscedastic and heteroscedastic cases}
\end{figure}

\begin{figure}
\centering
\includegraphics[width= 0.95\textwidth,height=0.9\textheight]{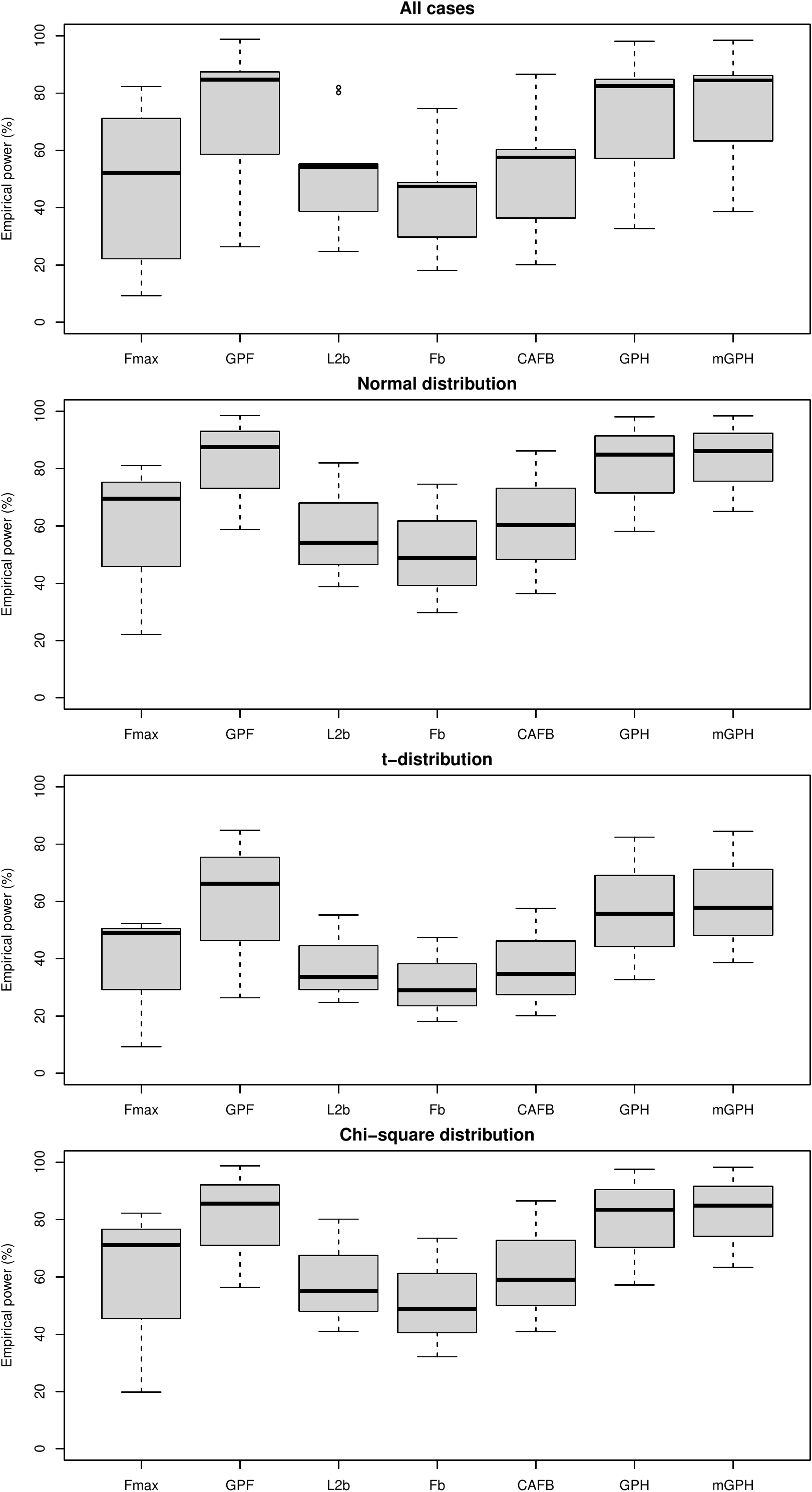}
\caption[Box-and-whisker plots for the empirical powers of all tests obtained under alternative A2 for the Dunnett constrasts with scaling function and different distributions]{Box-and-whisker plots for the empirical powers (as percentages) of all tests obtained under alternative A2 for the Dunnett constrasts with scaling function and different distributions}
\end{figure}

\newpage

\begin{longtable}[t]{rrrr|rrrrrrr}
\caption[Empirical powers of all tests obtained under alternative A3 for the Dunnett constrasts with scaling function]{\label{tab:unnamed-chunk-84}Empirical powers (as percentages) of all tests obtained under alternative A3 for the Dunnett constrasts with scaling function (D - distribution, $(\lambda_1,\lambda_2,\lambda_3,\lambda_4)$: (1,1,1,1) - homoscedastic case, (1,1.25,1.5,1.75) - heteroscedastic case (positive pairing), (1.75,1.5,1.25,1) - heteroscedastic case (negative pairing))}\\
\hline
D&$(n_1,n_2,n_3,n_4)$&$(\lambda_1,\lambda_2,\lambda_3,\lambda_4)$&$\mathcal{H}$&Fmax&GPF&L2b&Fb&CAFB&GPH&mGPH\\
\hline
\endfirsthead
\caption[]{Empirical powers (as percentages) of all tests obtained under alternative A3 for the Dunnett constrasts with scaling function (D - distribution, $(\lambda_1,\lambda_2,\lambda_3,\lambda_4)$: (1,1,1,1) - homoscedastic case, (1,1.25,1.5,1.75) - heteroscedastic case (positive pairing), (1.75,1.5,1.25,1) - heteroscedastic case (negative pairing)) \textit{(continued)}}\\
\hline
D&$(n_1,n_2,n_3,n_4)$&$(\lambda_1,\lambda_2,\lambda_3,\lambda_4)$&$\mathcal{H}$&Fmax&GPF&L2b&Fb&CAFB&GPH&mGPH\\
\hline
\endhead
N&(15,20,25,30)&(1,1,1,1)&$\mathcal{H}_{0,1}$&14.00&27.80&19.95&14.90&17.00&26.75&30.75\\
&&&$\mathcal{H}_{0,2}$&77.90&98.30&79.05&71.80&84.00&97.70&98.40\\
&&&$\mathcal{H}_{0,3}$&99.85&100.00&99.50&98.85&99.80&100.00&100.00\\
N&(15,20,25,30)&(1,1.25,1.5,1.75)&$\mathcal{H}_{0,1}$&11.15&19.40&16.30&12.20&13.40&22.60&23.90\\
&&&$\mathcal{H}_{0,2}$&34.75&76.00&56.15&49.60&62.25&87.15&87.80\\
&&&$\mathcal{H}_{0,3}$&68.75&98.90&92.35&89.40&95.90&99.85&99.85\\
N&(15,20,25,30)&(1.75,1.5,1.25,1)&$\mathcal{H}_{0,1}$&6.50&9.10&7.95&4.75&5.85&7.05&9.25\\
&&&$\mathcal{H}_{0,2}$&44.00&69.60&35.15&27.70&32.30&52.90&59.10\\
&&&$\mathcal{H}_{0,3}$&96.45&99.80&77.05&66.55&76.65&96.55&97.75\\
t&(15,20,25,30)&(1,1,1,1)&$\mathcal{H}_{0,1}$&8.45&14.00&12.95&8.85&8.85&12.75&15.55\\
&&&$\mathcal{H}_{0,2}$&48.40&81.10&52.05&44.20&53.90&79.30&82.90\\
&&&$\mathcal{H}_{0,3}$&92.80&99.75&91.00&86.45&94.60&99.40&99.60\\
t&(15,20,25,30)&(1,1.25,1.5,1.75)&$\mathcal{H}_{0,1}$&5.80&8.15&10.45&7.75&6.50&10.00&10.55\\
&&&$\mathcal{H}_{0,2}$&18.30&43.40&36.40&30.50&37.20&57.75&59.75\\
&&&$\mathcal{H}_{0,3}$&38.35&81.70&68.25&63.30&77.25&94.60&95.10\\
t&(15,20,25,30)&(1.75,1.5,1.25,1)&$\mathcal{H}_{0,1}$&5.10&4.55&4.70&3.40&3.00&3.40&4.75\\
&&&$\mathcal{H}_{0,2}$&28.65&44.85&20.75&14.35&18.10&27.70&33.25\\
&&&$\mathcal{H}_{0,3}$&81.50&93.75&53.70&42.45&51.50&75.65&81.10\\
$\chi^2$&(15,20,25,30)&(1,1,1,1)&$\mathcal{H}_{0,1}$&15.35&28.10&22.40&17.60&18.70&28.20&31.40\\
&&&$\mathcal{H}_{0,2}$&77.25&96.30&74.85&67.45&81.70&95.15&96.25\\
&&&$\mathcal{H}_{0,3}$&99.75&100.00&98.80&97.45&99.55&100.00&100.00\\
$\chi^2$&(15,20,25,30)&(1,1.25,1.5,1.75)&$\mathcal{H}_{0,1}$&9.10&16.10&14.40&11.35&10.85&19.55&21.00\\
&&&$\mathcal{H}_{0,2}$&35.45&77.00&57.65&51.00&62.10&86.05&87.40\\
&&&$\mathcal{H}_{0,3}$&69.55&98.85&91.15&87.85&96.00&99.85&99.85\\
$\chi^2$&(15,20,25,30)&(1.75,1.5,1.25,1)&$\mathcal{H}_{0,1}$&8.35&10.25&9.20&6.80&7.45&8.55&11.45\\
&&&$\mathcal{H}_{0,2}$&48.25&70.85&36.30&29.20&37.85&53.20&59.30\\
&&&$\mathcal{H}_{0,3}$&96.10&99.35&74.85&65.00&78.50&93.70&95.55\\
\hline
\end{longtable}

\begin{figure}
\centering
\includegraphics[width= 0.95\textwidth,height=0.9\textheight]{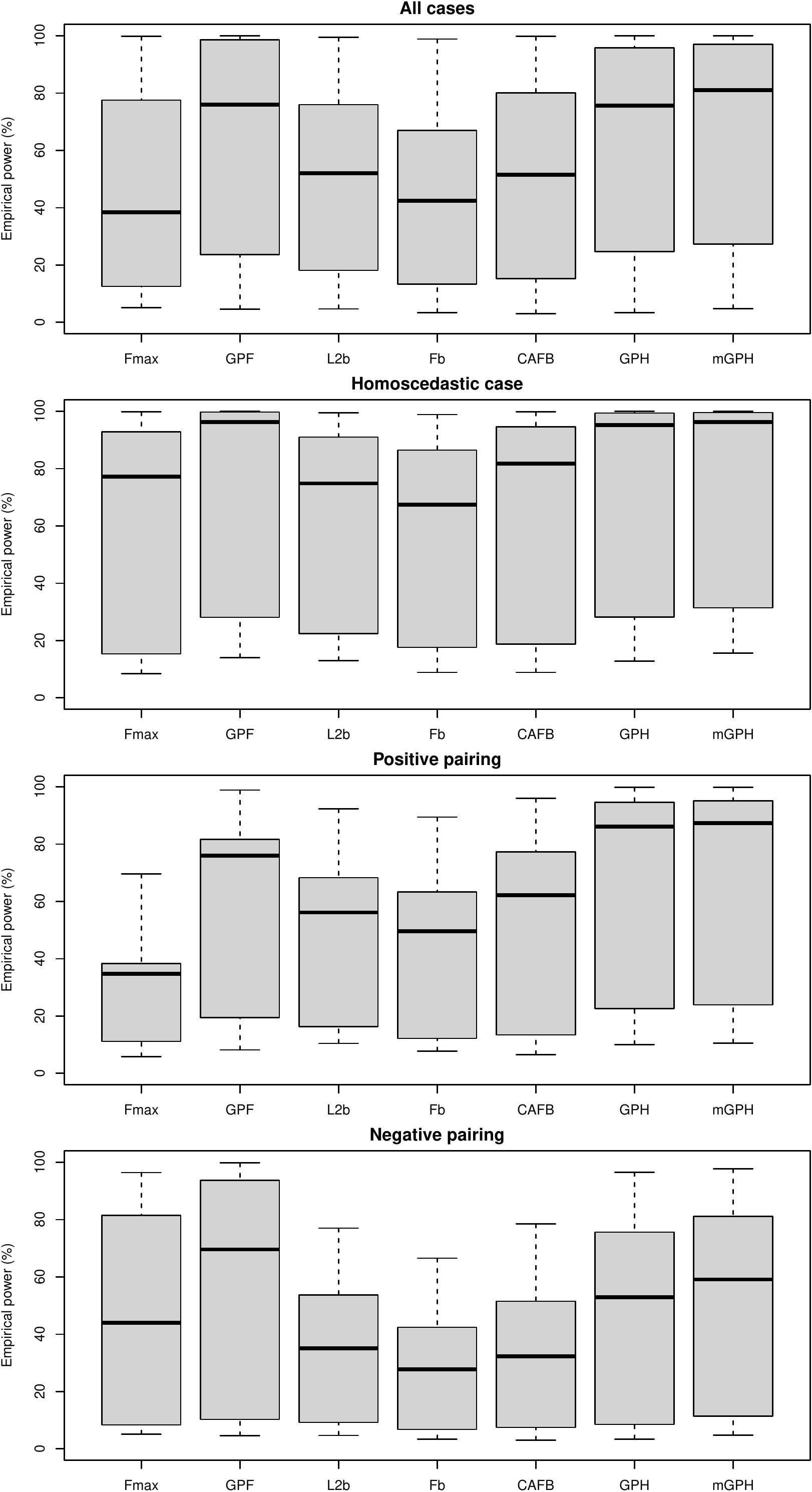}
\caption[Box-and-whisker plots for the empirical powers of all tests obtained under alternative A3 for the Dunnett constrasts with scaling function and homoscedastic and heteroscedastic cases]{Box-and-whisker plots for the empirical powers (as percentages) of all tests obtained under alternative A3 for the Dunnett constrasts with scaling function and homoscedastic and heteroscedastic cases}
\end{figure}

\begin{figure}
\centering
\includegraphics[width= 0.95\textwidth,height=0.9\textheight]{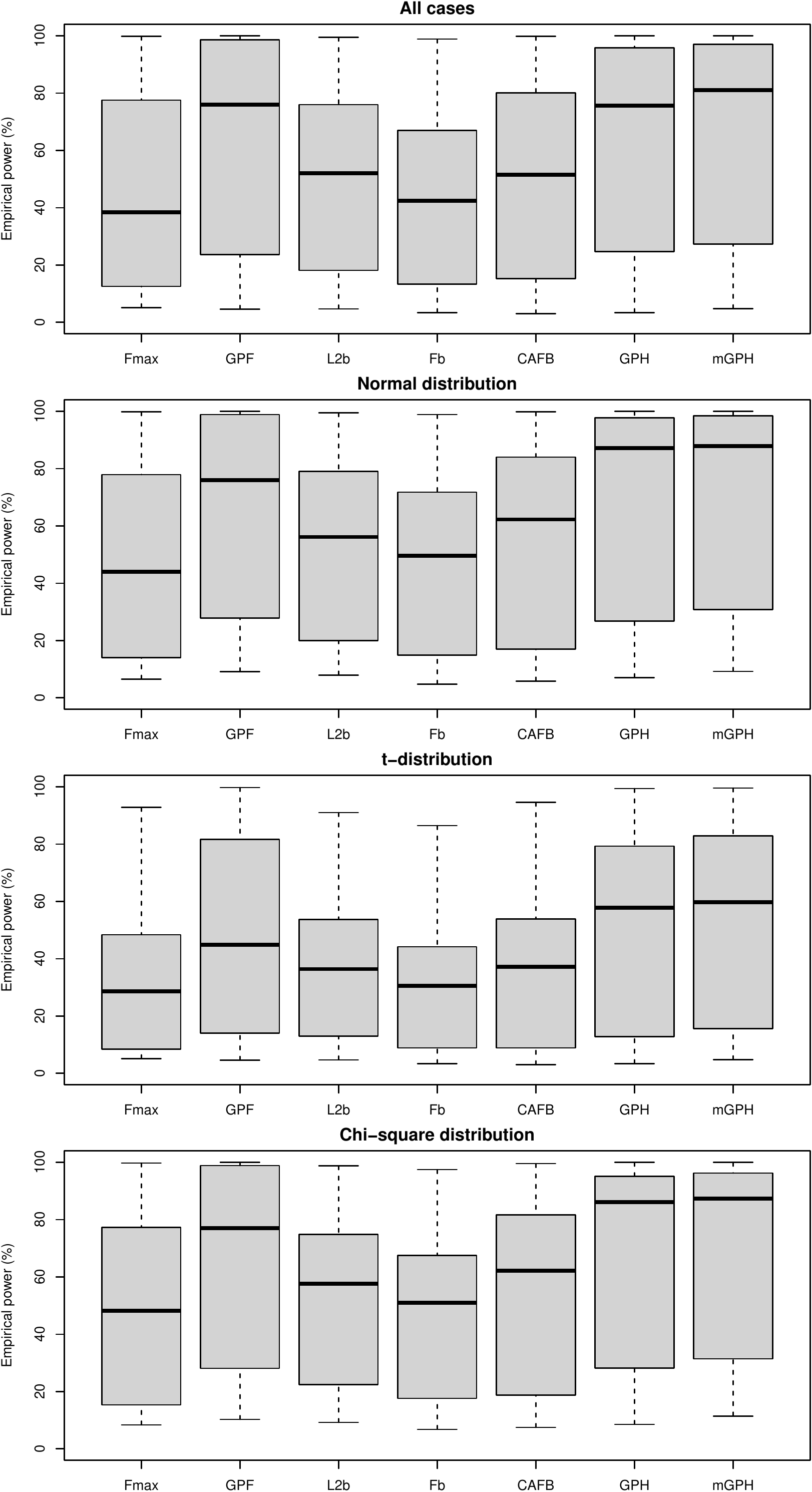}
\caption[Box-and-whisker plots for the empirical powers of all tests obtained under alternative A3 for the Dunnett constrasts with scaling function and different distributions]{Box-and-whisker plots for the empirical powers (as percentages) of all tests obtained under alternative A3 for the Dunnett constrasts with scaling function and different distributions}
\end{figure}

\newpage

\begin{longtable}[t]{rrrr|rrrrrrr}
\caption[Empirical powers of all tests obtained under alternative A4 for the Dunnett constrasts with scaling function]{\label{tab:unnamed-chunk-87}Empirical powers (as percentages) of all tests obtained under alternative A4 for the Dunnett constrasts with scaling function (D - distribution, $(\lambda_1,\lambda_2,\lambda_3,\lambda_4)$: (1,1,1,1) - homoscedastic case, (1,1.25,1.5,1.75) - heteroscedastic case (positive pairing), (1.75,1.5,1.25,1) - heteroscedastic case (negative pairing))}\\
\hline
D&$(n_1,n_2,n_3,n_4)$&$(\lambda_1,\lambda_2,\lambda_3,\lambda_4)$&$\mathcal{H}$&Fmax&GPF&L2b&Fb&CAFB&GPH&mGPH\\
\hline
\endfirsthead
\caption[]{Empirical powers (as percentages) of all tests obtained under alternative A4 for the Dunnett constrasts with scaling function (D - distribution, $(\lambda_1,\lambda_2,\lambda_3,\lambda_4)$: (1,1,1,1) - homoscedastic case, (1,1.25,1.5,1.75) - heteroscedastic case (positive pairing), (1.75,1.5,1.25,1) - heteroscedastic case (negative pairing)) \textit{(continued)}}\\
\hline
D&$(n_1,n_2,n_3,n_4)$&$(\lambda_1,\lambda_2,\lambda_3,\lambda_4)$&$\mathcal{H}$&Fmax&GPF&L2b&Fb&CAFB&GPH&mGPH\\
\hline
\endhead
N&(15,20,25,30)&(1,1,1,1)&$\mathcal{H}_{0,1}$&14.15&14.20&3.15&1.90&2.60&14.30&16.55\\
&&&$\mathcal{H}_{0,2}$&78.25&86.15&2.55&1.65&5.45&84.70&86.90\\
&&&$\mathcal{H}_{0,3}$&99.95&100.00&3.50&2.15&18.30&99.95&100.00\\
N&(15,20,25,30)&(1,1.25,1.5,1.75)&$\mathcal{H}_{0,1}$&8.55&7.55&2.20&1.45&1.35&9.80&10.45\\
&&&$\mathcal{H}_{0,2}$&38.85&43.55&2.60&1.45&4.20&62.50&63.85\\
&&&$\mathcal{H}_{0,3}$&74.55&84.55&2.50&1.50&7.80&96.95&97.60\\
N&(15,20,25,30)&(1.75,1.5,1.25,1)&$\mathcal{H}_{0,1}$&5.25&4.75&2.40&1.60&1.50&3.40&5.20\\
&&&$\mathcal{H}_{0,2}$&43.80&45.15&2.50&1.75&2.40&27.50&33.20\\
&&&$\mathcal{H}_{0,3}$&95.40&97.35&3.35&2.00&6.10&80.25&85.20\\
t&(15,20,25,30)&(1,1,1,1)&$\mathcal{H}_{0,1}$&7.35&7.05&2.90&1.80&1.80&6.15&7.65\\
&&&$\mathcal{H}_{0,2}$&47.95&54.80&2.20&1.30&3.80&52.10&57.15\\
&&&$\mathcal{H}_{0,3}$&94.15&97.35&3.30&1.95&10.05&95.20&96.80\\
t&(15,20,25,30)&(1,1.25,1.5,1.75)&$\mathcal{H}_{0,1}$&4.60&4.20&3.40&2.15&1.90&5.30&5.75\\
&&&$\mathcal{H}_{0,2}$&18.45&19.35&3.00&2.25&2.55&32.10&34.00\\
&&&$\mathcal{H}_{0,3}$&39.95&46.35&2.50&1.45&4.90&75.70&77.65\\
t&(15,20,25,30)&(1.75,1.5,1.25,1)&$\mathcal{H}_{0,1}$&3.20&2.65&2.30&1.35&1.35&1.95&2.90\\
&&&$\mathcal{H}_{0,2}$&24.60&25.80&2.65&1.35&1.55&13.50&17.80\\
&&&$\mathcal{H}_{0,3}$&79.85&82.35&2.55&1.60&3.70&48.05&56.10\\
$\chi^2$&(15,20,25,30)&(1,1,1,1)&$\mathcal{H}_{0,1}$&13.15&14.60&2.60&1.75&2.05&14.60&16.70\\
&&&$\mathcal{H}_{0,2}$&78.65&86.15&2.60&1.40&6.35&82.60&85.25\\
&&&$\mathcal{H}_{0,3}$&99.95&100.00&3.75&2.60&20.30&99.80&99.95\\
$\chi^2$&(15,20,25,30)&(1,1.25,1.5,1.75)&$\mathcal{H}_{0,1}$&9.15&7.40&2.50&1.65&1.55&10.05&10.60\\
&&&$\mathcal{H}_{0,2}$&37.70&42.30&2.50&1.75&4.10&61.85&64.00\\
&&&$\mathcal{H}_{0,3}$&79.70&86.80&3.10&2.35&8.90&97.15&97.80\\
$\chi^2$&(15,20,25,30)&(1.75,1.5,1.25,1)&$\mathcal{H}_{0,1}$&5.70&5.25&2.90&1.95&2.05&4.05&5.40\\
&&&$\mathcal{H}_{0,2}$&44.40&46.85&3.75&2.60&2.80&29.60&36.40\\
&&&$\mathcal{H}_{0,3}$&94.90&96.20&3.95&2.30&6.70&78.00&83.55\\
\hline
\end{longtable}

\begin{figure}
\centering
\includegraphics[width= 0.95\textwidth,height=0.9\textheight]{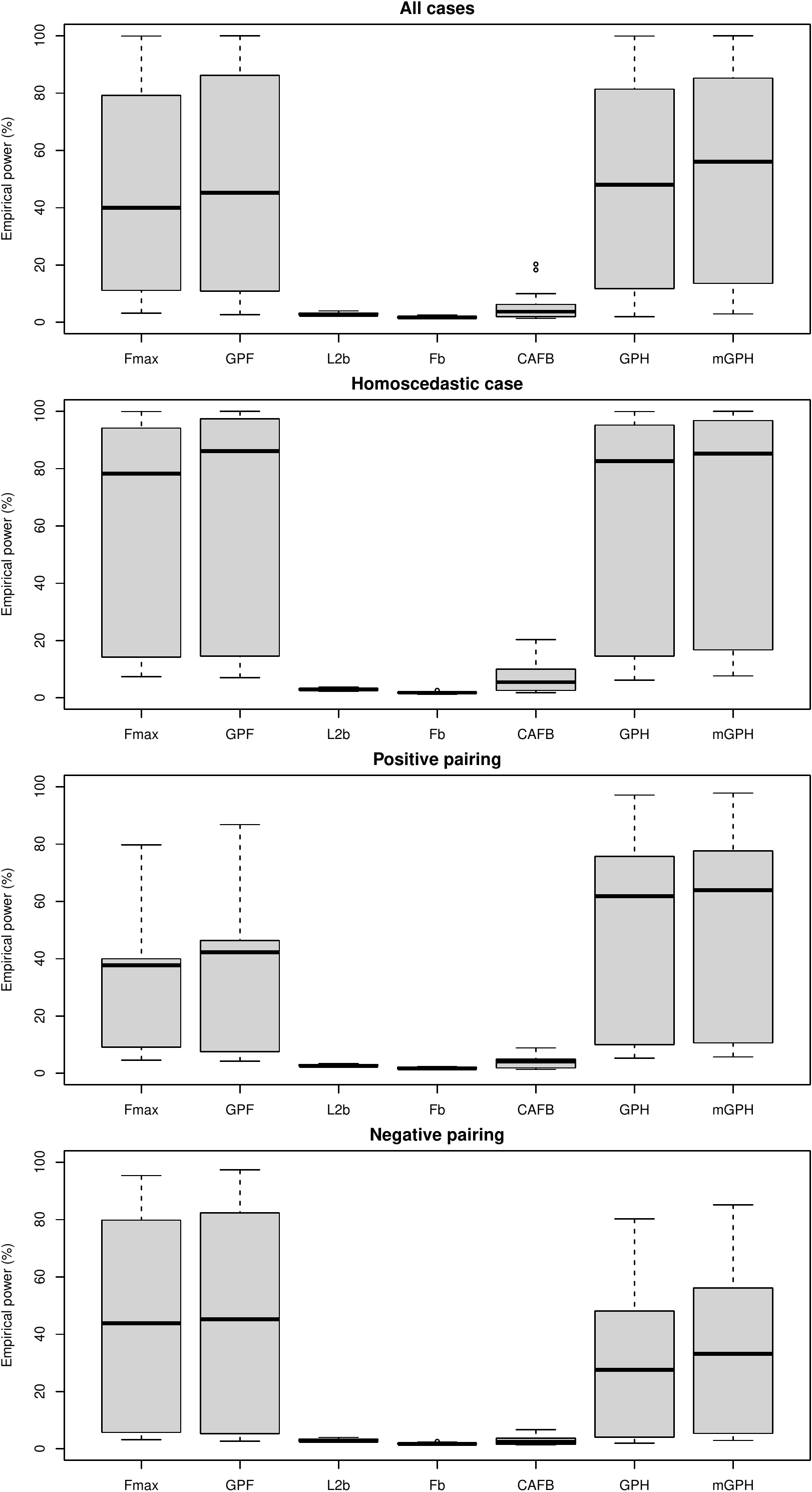}
\caption[Box-and-whisker plots for the empirical powers of all tests obtained under alternative A4 for the Dunnett constrasts with scaling function and homoscedastic and heteroscedastic cases]{Box-and-whisker plots for the empirical powers (as percentages) of all tests obtained under alternative A4 for the Dunnett constrasts with scaling function and homoscedastic and heteroscedastic cases}
\end{figure}

\begin{figure}
\centering
\includegraphics[width= 0.95\textwidth,height=0.9\textheight]{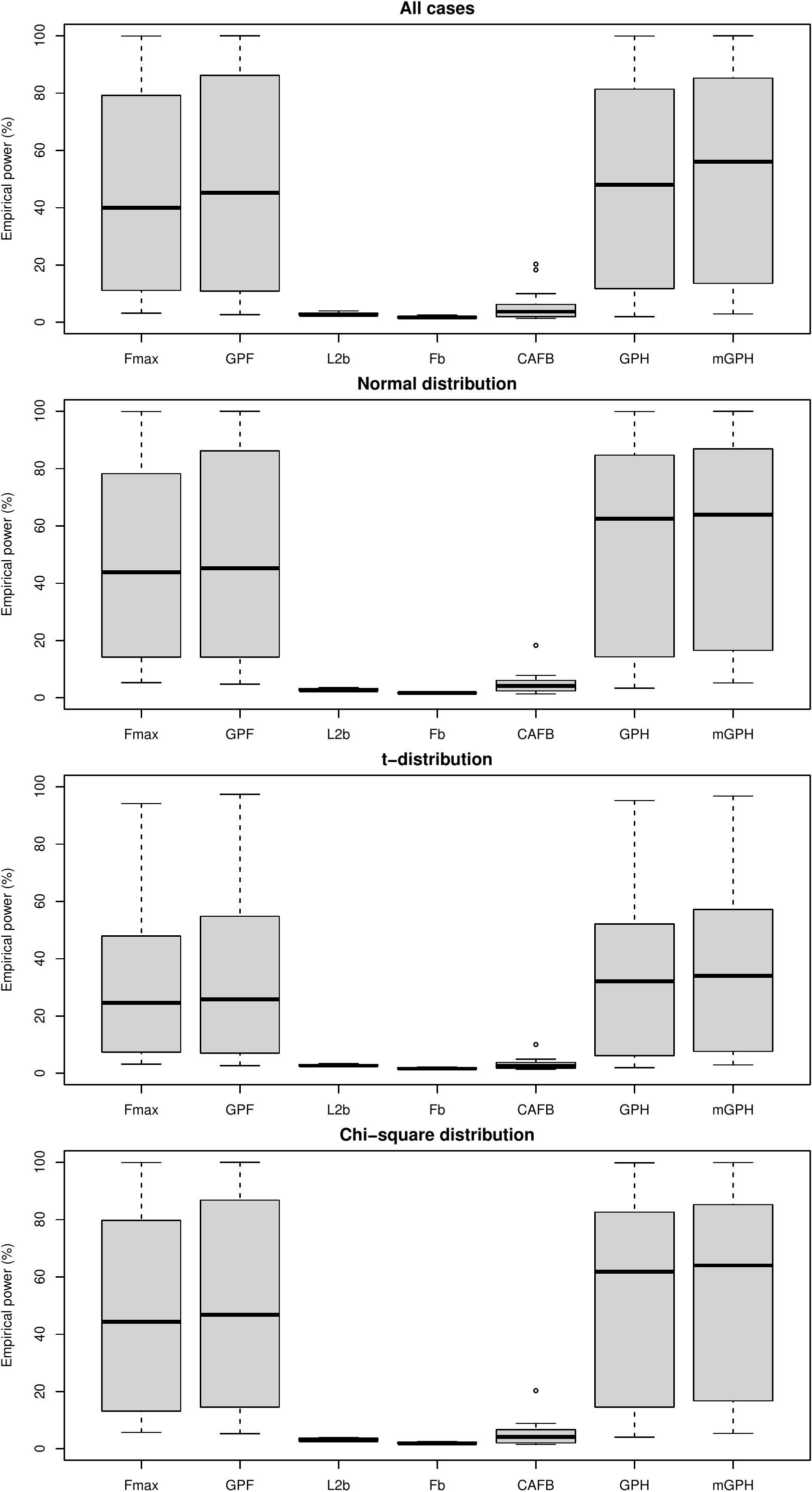}
\caption[Box-and-whisker plots for the empirical powers of all tests obtained under alternative A4 for the Dunnett constrasts with scaling function and different distributions]{Box-and-whisker plots for the empirical powers (as percentages) of all tests obtained under alternative A4 for the Dunnett constrasts with scaling function and different distributions}
\end{figure}

\newpage

\begin{longtable}[t]{rrrr|rrrrrrr}
\caption[Empirical sizes and powers of all tests obtained under alternative A5 for the Dunnett constrasts with scaling function]{\label{tab:unnamed-chunk-90}Empirical sizes ($\mathcal{H}_{0,1}$, $\mathcal{H}_{0,2}$) and powers ($\mathcal{H}_{0,3}$) (as percentages) of all tests obtained under alternative A5 for the Dunnett constrasts with scaling function (D - distribution, $(\lambda_1,\lambda_2,\lambda_3,\lambda_4)$: (1,1,1,1) - homoscedastic case, (1,1.25,1.5,1.75) - heteroscedastic case (positive pairing), (1.75,1.5,1.25,1) - heteroscedastic case (negative pairing))}\\
\hline
D&$(n_1,n_2,n_3,n_4)$&$(\lambda_1,\lambda_2,\lambda_3,\lambda_4)$&$\mathcal{H}$&Fmax&GPF&L2b&Fb&CAFB&GPH&mGPH\\
\hline
\endfirsthead
\caption[]{Empirical sizes ($\mathcal{H}_{0,1}$, $\mathcal{H}_{0,2}$) and powers ($\mathcal{H}_{0,3}$) (as percentages) of all tests obtained under alternative A5 for the Dunnett constrasts with scaling function (D - distribution, $(\lambda_1,\lambda_2,\lambda_3,\lambda_4)$: (1,1,1,1) - homoscedastic case, (1,1.25,1.5,1.75) - heteroscedastic case (positive pairing), (1.75,1.5,1.25,1) - heteroscedastic case (negative pairing)) \textit{(continued)}}\\
\hline
D&$(n_1,n_2,n_3,n_4)$&$(\lambda_1,\lambda_2,\lambda_3,\lambda_4)$&$\mathcal{H}$&Fmax&GPF&L2b&Fb&CAFB&GPH&mGPH\\
\hline
\endhead
N&(15,20,25,30)&(1,1,1,1)&$\mathcal{H}_{0,1}$&1.95&1.55&3.35&2.10&1.70&1.45&1.70\\
&&&$\mathcal{H}_{0,2}$&1.75&1.35&3.45&2.10&1.55&1.35&1.65\\
&&&$\mathcal{H}_{0,3}$&82.05&88.65&95.60&92.95&95.20&86.25&89.10\\
N&(15,20,25,30)&(1,1.25,1.5,1.75)&$\mathcal{H}_{0,1}$&1.80&0.90&2.65&1.70&1.70&1.05&1.15\\
&&&$\mathcal{H}_{0,2}$&0.80&0.25&2.25&1.35&1.10&1.00&1.10\\
&&&$\mathcal{H}_{0,3}$&21.20&25.85&76.60&71.70&72.75&54.85&56.95\\
N&(15,20,25,30)&(1.75,1.5,1.25,1)&$\mathcal{H}_{0,1}$&1.85&1.10&2.60&1.85&1.50&0.95&1.35\\
&&&$\mathcal{H}_{0,2}$&3.85&3.45&2.55&1.45&1.70&0.85&1.75\\
&&&$\mathcal{H}_{0,3}$&64.20&64.60&62.20&53.15&52.20&30.85&37.15\\
t&(15,20,25,30)&(1,1,1,1)&$\mathcal{H}_{0,1}$&1.25&0.75&2.80&1.50&1.20&0.80&1.00\\
&&&$\mathcal{H}_{0,2}$&1.60&0.85&2.40&1.65&1.15&0.70&0.80\\
&&&$\mathcal{H}_{0,3}$&48.65&56.05&76.80&69.65&73.10&52.15&56.90\\
t&(15,20,25,30)&(1,1.25,1.5,1.75)&$\mathcal{H}_{0,1}$&1.35&0.85&2.70&2.10&1.75&0.90&1.10\\
&&&$\mathcal{H}_{0,2}$&0.70&0.35&2.75&2.10&1.50&1.05&1.15\\
&&&$\mathcal{H}_{0,3}$&9.70&10.05&51.60&46.35&47.90&29.35&31.25\\
t&(15,20,25,30)&(1.75,1.5,1.25,1)&$\mathcal{H}_{0,1}$&1.90&1.40&2.60&1.45&1.10&1.10&1.40\\
&&&$\mathcal{H}_{0,2}$&3.60&3.00&2.85&1.85&1.75&1.15&1.45\\
&&&$\mathcal{H}_{0,3}$&42.75&44.30&37.65&28.65&27.55&13.60&18.05\\
$\chi^2$&(15,20,25,30)&(1,1,1,1)&$\mathcal{H}_{0,1}$&1.70&1.10&2.85&1.75&1.70&1.00&1.20\\
&&&$\mathcal{H}_{0,2}$&1.35&1.10&2.70&1.95&1.35&0.85&1.20\\
&&&$\mathcal{H}_{0,3}$&80.90&88.90&94.90&91.45&94.55&83.45&86.70\\
$\chi^2$&(15,20,25,30)&(1,1.25,1.5,1.75)&$\mathcal{H}_{0,1}$&1.15&1.00&2.40&1.55&0.95&1.40&1.50\\
&&&$\mathcal{H}_{0,2}$&0.90&0.30&2.35&1.75&1.30&1.45&1.60\\
&&&$\mathcal{H}_{0,3}$&19.90&21.75&78.15&72.70&74.60&54.35&56.25\\
$\chi^2$&(15,20,25,30)&(1.75,1.5,1.25,1)&$\mathcal{H}_{0,1}$&2.55&1.95&3.05&2.25&1.55&1.20&1.90\\
&&&$\mathcal{H}_{0,2}$&3.80&3.60&3.45&1.75&2.20&1.25&2.00\\
&&&$\mathcal{H}_{0,3}$&67.45&69.05&61.05&50.85&55.55&36.50&42.15\\
\hline
\end{longtable}

\begin{figure}
\centering
\includegraphics[width= 0.95\textwidth,height=0.9\textheight]{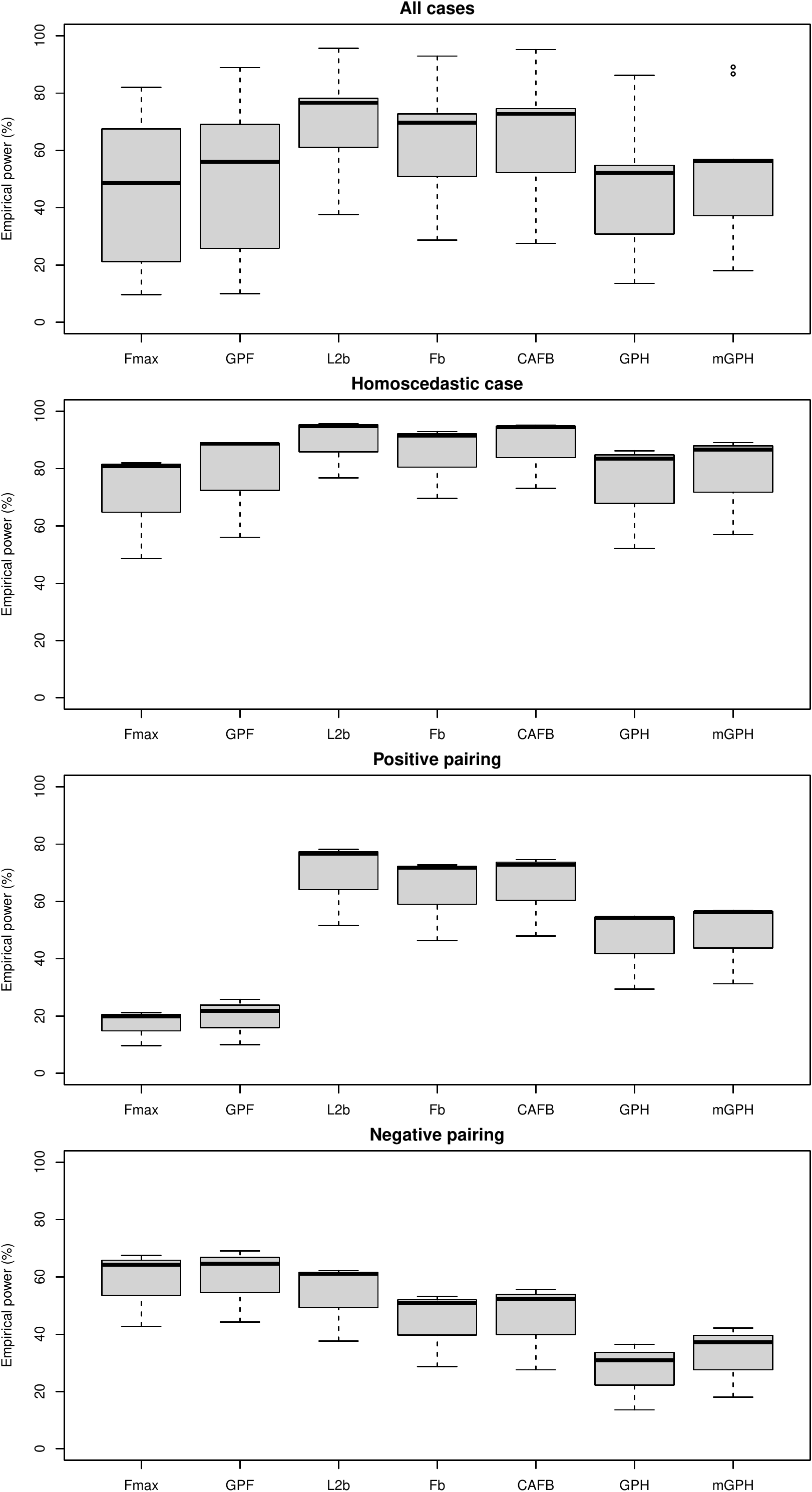}
\caption[Box-and-whisker plots for the empirical powers of all tests obtained under alternative A5 for the Dunnett constrasts with scaling function and homoscedastic and heteroscedastic cases]{Box-and-whisker plots for the empirical powers (as percentages) of all tests obtained under alternative A5 for the Dunnett constrasts with scaling function and homoscedastic and heteroscedastic cases}
\end{figure}

\begin{figure}
\centering
\includegraphics[width= 0.95\textwidth,height=0.9\textheight]{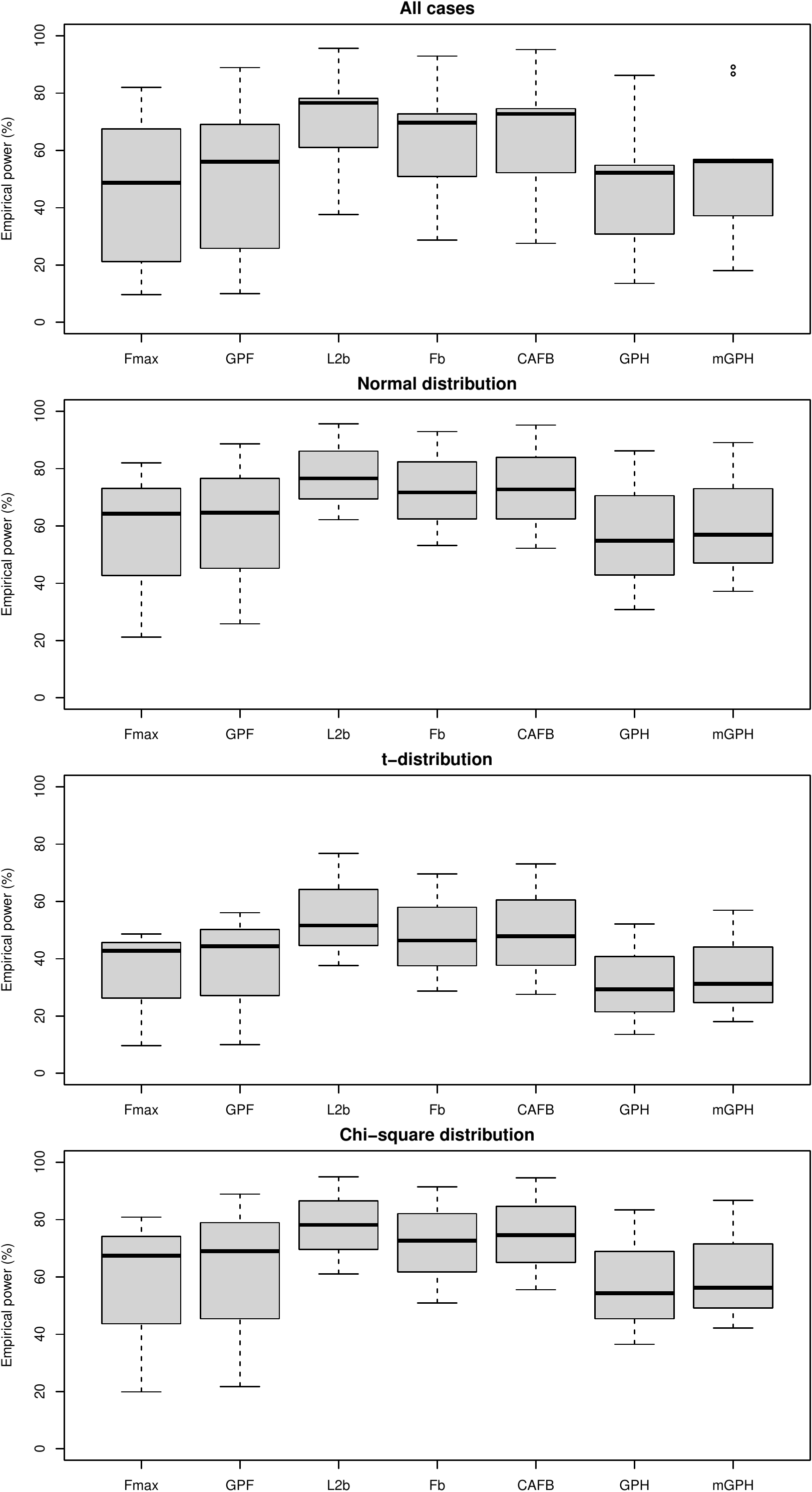}
\caption[Box-and-whisker plots for the empirical powers of all tests obtained under alternative A5 for the Dunnett constrasts with scaling function and different distributions]{Box-and-whisker plots for the empirical powers (as percentages) of all tests obtained under alternative A5 for the Dunnett constrasts with scaling function and different distributions}
\end{figure}

\newpage

\begin{longtable}[t]{rrrr|rrrrrrr}
\caption[Empirical powers of all tests obtained under alternative A6 for the Dunnett constrasts with scaling function]{\label{tab:unnamed-chunk-95}Empirical powers (as percentages) of all tests obtained under alternative A6 for the Dunnett constrasts with scaling function (D - distribution, $(\lambda_1,\lambda_2,\lambda_3,\lambda_4)$: (1,1,1,1) - homoscedastic case, (1,1.25,1.5,1.75) - heteroscedastic case (positive pairing), (1.75,1.5,1.25,1) - heteroscedastic case (negative pairing))}\\
\hline
D&$(n_1,n_2,n_3,n_4)$&$(\lambda_1,\lambda_2,\lambda_3,\lambda_4)$&$\mathcal{H}$&Fmax&GPF&L2b&Fb&CAFB&GPH&mGPH\\
\hline
\endfirsthead
\caption[]{Empirical powers (as percentages) of all tests obtained under alternative A6 for the Dunnett constrasts with scaling function (D - distribution, $(\lambda_1,\lambda_2,\lambda_3,\lambda_4)$: (1,1,1,1) - homoscedastic case, (1,1.25,1.5,1.75) - heteroscedastic case (positive pairing), (1.75,1.5,1.25,1) - heteroscedastic case (negative pairing)) \textit{(continued)}}\\
\hline
D&$(n_1,n_2,n_3,n_4)$&$(\lambda_1,\lambda_2,\lambda_3,\lambda_4)$&$\mathcal{H}$&Fmax&GPF&L2b&Fb&CAFB&GPH&mGPH\\
\hline
\endhead
N&(15,20,25,30)&(1,1,1,1)&$\mathcal{H}_{0,1}$&13.25&12.60&33.10&26.60&24.30&11.75&13.65\\
&&&$\mathcal{H}_{0,2}$&75.45&83.60&93.10&89.95&92.90&82.30&84.90\\
&&&$\mathcal{H}_{0,3}$&99.80&100.00&99.95&99.90&99.95&99.95&99.95\\
N&(15,20,25,30)&(1,1.25,1.5,1.75)&$\mathcal{H}_{0,1}$&7.20&6.50&25.85&19.65&18.20&9.15&9.80\\
&&&$\mathcal{H}_{0,2}$&35.10&39.15&80.75&76.60&77.75&58.10&60.10\\
&&&$\mathcal{H}_{0,3}$&72.75&81.15&98.90&98.40&99.00&96.20&96.55\\
N&(15,20,25,30)&(1.75,1.5,1.25,1)&$\mathcal{H}_{0,1}$&5.85&5.00&13.65&9.70&7.40&4.40&5.55\\
&&&$\mathcal{H}_{0,2}$&41.75&45.35&57.05&47.75&48.70&26.90&33.70\\
&&&$\mathcal{H}_{0,3}$&94.85&96.50&94.15&89.10&91.50&79.90&84.35\\
t&(15,20,25,30)&(1,1,1,1)&$\mathcal{H}_{0,1}$&6.95&6.85&18.90&14.50&12.40&6.00&7.55\\
&&&$\mathcal{H}_{0,2}$&45.55&53.10&77.05&70.35&72.80&51.45&56.30\\
&&&$\mathcal{H}_{0,3}$&93.60&96.20&98.65&97.85&98.90&94.40&95.60\\
t&(15,20,25,30)&(1,1.25,1.5,1.75)&$\mathcal{H}_{0,1}$&5.30&5.05&15.80&12.65&9.75&6.10&6.60\\
&&&$\mathcal{H}_{0,2}$&17.80&18.65&56.80&50.80&50.85&31.30&32.75\\
&&&$\mathcal{H}_{0,3}$&40.55&45.30&89.05&85.90&90.00&75.35&76.60\\
t&(15,20,25,30)&(1.75,1.5,1.25,1)&$\mathcal{H}_{0,1}$&3.40&3.50&7.10&4.90&5.00&2.20&3.10\\
&&&$\mathcal{H}_{0,2}$&23.70&24.35&33.25&25.95&24.70&12.25&15.35\\
&&&$\mathcal{H}_{0,3}$&77.85&80.45&74.10&64.45&67.85&45.70&53.55\\
$\chi^2$&(15,20,25,30)&(1,1,1,1)&$\mathcal{H}_{0,1}$&12.50&13.50&33.55&26.90&25.35&13.90&16.15\\
&&&$\mathcal{H}_{0,2}$&76.45&81.30&91.95&88.80&92.40&77.85&81.50\\
&&&$\mathcal{H}_{0,3}$&99.90&99.95&99.95&99.70&100.00&99.70&99.85\\
$\chi^2$&(15,20,25,30)&(1,1.25,1.5,1.75)&$\mathcal{H}_{0,1}$&7.00&6.10&24.65&19.20&16.95&8.60&9.20\\
&&&$\mathcal{H}_{0,2}$&32.40&38.70&80.45&74.95&77.00&57.40&59.55\\
&&&$\mathcal{H}_{0,3}$&73.90&85.20&99.15&98.70&99.30&96.65&96.95\\
$\chi^2$&(15,20,25,30)&(1.75,1.5,1.25,1)&$\mathcal{H}_{0,1}$&7.30&5.75&14.65&11.05&9.65&4.95&6.80\\
&&&$\mathcal{H}_{0,2}$&42.60&45.65&55.10&45.90&49.40&29.45&35.25\\
&&&$\mathcal{H}_{0,3}$&94.80&95.65&90.60&85.10&89.55&76.35&82.35\\
\hline
\end{longtable}

\begin{figure}
\centering
\includegraphics[width= 0.95\textwidth,height=0.9\textheight]{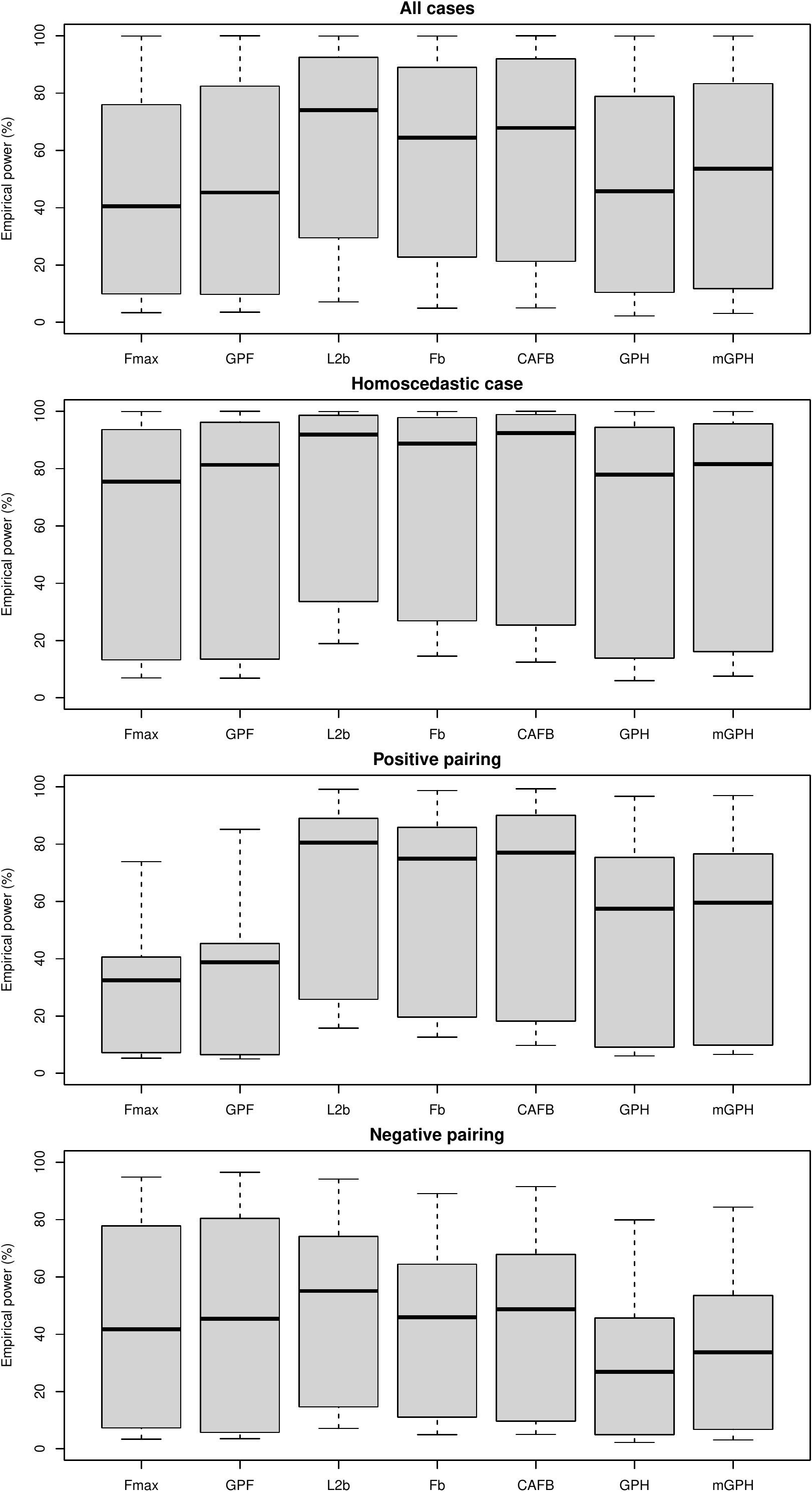}
\caption[Box-and-whisker plots for the empirical powers of all tests obtained under alternative A6 for the Dunnett constrasts with scaling function and homoscedastic and heteroscedastic cases]{Box-and-whisker plots for the empirical powers (as percentages) of all tests obtained under alternative A6 for the Dunnett constrasts with scaling function and homoscedastic and heteroscedastic cases}
\end{figure}

\begin{figure}
\centering
\includegraphics[width= 0.95\textwidth,height=0.9\textheight]{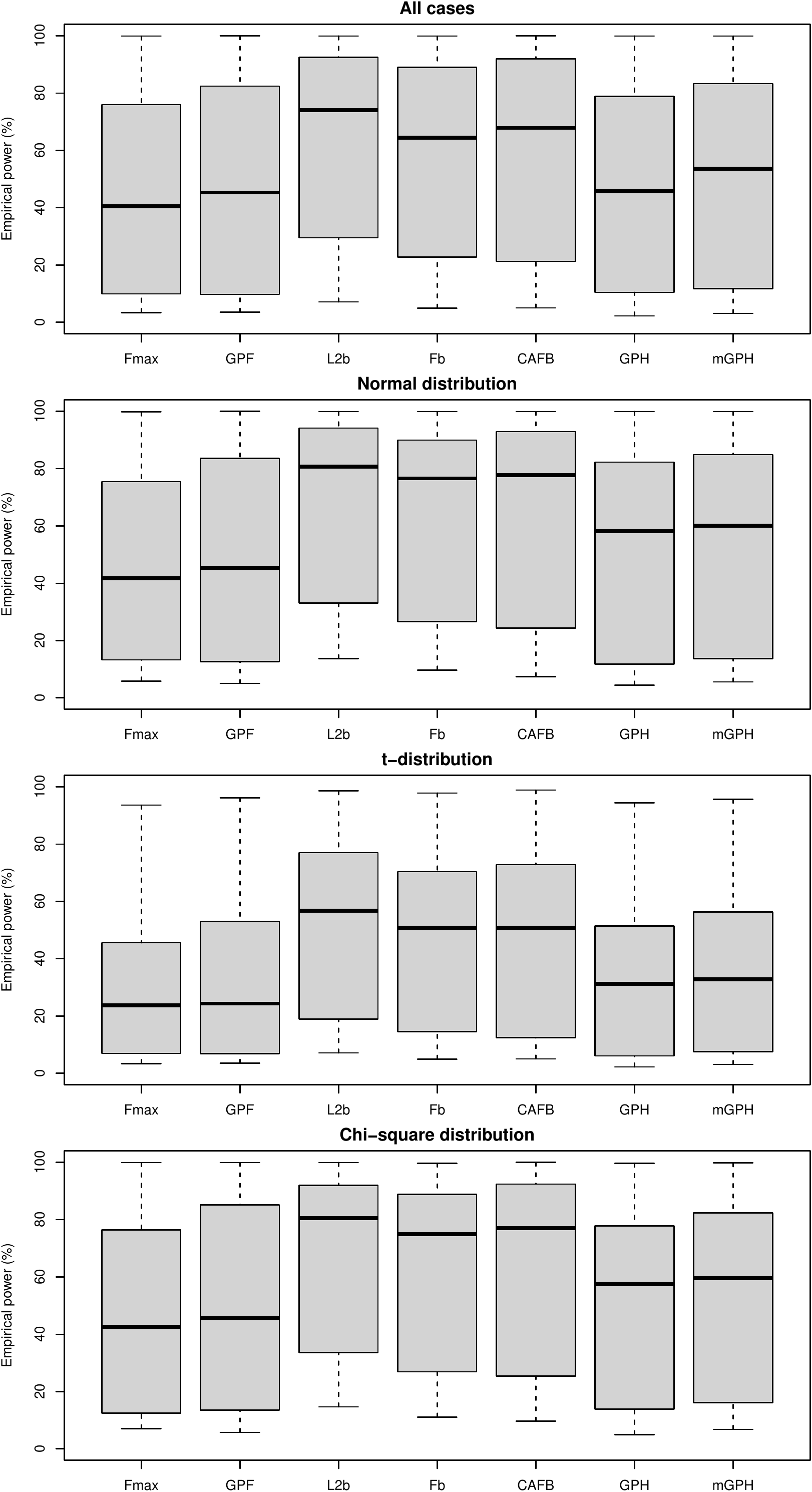}
\caption[Box-and-whisker plots for the empirical powers of all tests obtained under alternative A6 for the Dunnett constrasts with scaling function and different distributions]{Box-and-whisker plots for the empirical powers (as percentages) of all tests obtained under alternative A6 for the Dunnett constrasts with scaling function and different distributions}
\end{figure}

\newpage

\begin{longtable}[t]{rrrr|rrrrrrr}
\caption[Empirical FWER and sizes of all tests obtained for the Tukey constrasts with scaling function]{\label{tab:unnamed-chunk-98}Empirical FWER and sizes (as percentages) of all tests obtained for the Tukey constrasts with scaling function (D - distribution, $(\lambda_1,\lambda_2,\lambda_3,\lambda_4)$: (1,1,1,1) - homoscedastic case, (1,1.25,1.5,1.75) - heteroscedastic case (positive pairing), (1.75,1.5,1.25,1) - heteroscedastic case (negative pairing))}\\
\hline
D&$(n_1,n_2,n_3,n_4)$&$(\lambda_1,\lambda_2,\lambda_3,\lambda_4)$&$\mathcal{H}$&Fmax&GPF&L2b&Fb&CAFB&GPH&mGPH\\
\hline
\endfirsthead
\caption[]{Empirical FWER and sizes (as percentages) of all tests obtained for the Tukey constrasts with scaling function (D - distribution, $(\lambda_1,\lambda_2,\lambda_3,\lambda_4)$: (1,1,1,1) - homoscedastic case, (1,1.25,1.5,1.75) - heteroscedastic case (positive pairing), (1.75,1.5,1.25,1) - heteroscedastic case (negative pairing)) \textit{(continued)}}\\
\hline
D&$(n_1,n_2,n_3,n_4)$&$(\lambda_1,\lambda_2,\lambda_3,\lambda_4)$&$\mathcal{H}$&Fmax&GPF&L2b&Fb&CAFB&GPH&mGPH\\
\hline
\endhead
N&(15,20,25,30)&(1,1,1,1)&FWER&4.05&3.35&7.80&5.10&4.00&3.35&3.85\\
&&&$\mathcal{H}_{0,1}$&0.75&0.75&2.10&1.35&0.95&0.65&0.70\\
&&&$\mathcal{H}_{0,2}$&0.80&0.70&1.35&0.80&0.80&0.90&0.95\\
&&&$\mathcal{H}_{0,3}$&0.50&0.55&1.80&1.00&0.95&0.80&0.95\\
&&&$\mathcal{H}_{0,4}$&0.80&0.65&1.35&0.85&1.05&0.60&0.65\\
&&&$\mathcal{H}_{0,5}$&1.05&0.75&1.95&1.20&0.80&0.70&0.75\\
&&&$\mathcal{H}_{0,6}$&0.70&0.65&1.40&1.10&0.35&0.60&0.90\\
N&(15,20,25,30)&(1,1.25,1.5,1.75)&FWER&3.00&2.00&5.60&3.85&2.90&3.35&3.90\\
&&&$\mathcal{H}_{0,1}$&0.55&0.50&1.00&0.65&0.30&0.65&0.70\\
&&&$\mathcal{H}_{0,2}$&0.30&0.15&0.90&0.65&0.45&0.75&0.95\\
&&&$\mathcal{H}_{0,3}$&0.35&0.00&1.40&0.65&0.40&0.70&0.75\\
&&&$\mathcal{H}_{0,4}$&1.05&0.70&1.15&0.70&0.60&0.75&0.90\\
&&&$\mathcal{H}_{0,5}$&0.60&0.60&1.60&1.10&0.80&0.65&0.75\\
&&&$\mathcal{H}_{0,6}$&0.65&0.30&1.00&0.75&0.70&0.60&0.65\\
N&(15,20,25,30)&(1.75,1.5,1.25,1)&FWER&13.30&9.15&6.80&3.70&4.10&3.85&4.45\\
&&&$\mathcal{H}_{0,1}$&1.05&0.80&1.80&0.75&0.85&0.65&1.00\\
&&&$\mathcal{H}_{0,2}$&2.60&1.55&2.05&1.30&0.95&0.55&0.70\\
&&&$\mathcal{H}_{0,3}$&7.00&5.10&1.70&0.70&0.85&0.80&0.95\\
&&&$\mathcal{H}_{0,4}$&1.45&0.90&1.25&0.90&0.75&0.65&0.65\\
&&&$\mathcal{H}_{0,5}$&2.50&1.80&1.30&0.60&0.75&0.75&0.95\\
&&&$\mathcal{H}_{0,6}$&1.40&1.30&1.20&0.70&0.70&1.00&1.15\\
N&(30,40,50,60)&(1,1,1,1)&FWER&5.15&4.15&5.20&4.15&3.40&3.90&5.00\\
&&&$\mathcal{H}_{0,1}$&1.20&0.75&1.20&1.00&0.75&0.75&0.85\\
&&&$\mathcal{H}_{0,2}$&1.25&0.75&1.45&1.25&0.95&0.65&0.85\\
&&&$\mathcal{H}_{0,3}$&1.20&1.15&1.40&1.25&0.85&1.30&1.55\\
&&&$\mathcal{H}_{0,4}$&0.90&0.95&1.00&0.70&0.20&0.85&1.05\\
&&&$\mathcal{H}_{0,5}$&0.80&0.50&0.85&0.65&0.65&0.50&0.70\\
&&&$\mathcal{H}_{0,6}$&0.55&0.70&0.65&0.50&0.55&0.70&0.90\\
N&(30,40,50,60)&(1,1.25,1.5,1.75)&FWER&3.05&1.25&5.80&5.25&3.55&3.15&3.65\\
&&&$\mathcal{H}_{0,1}$&0.65&0.30&1.25&1.10&0.65&0.55&0.65\\
&&&$\mathcal{H}_{0,2}$&0.40&0.15&1.30&1.05&0.90&0.85&0.95\\
&&&$\mathcal{H}_{0,3}$&0.00&0.00&1.05&0.85&0.60&0.70&0.75\\
&&&$\mathcal{H}_{0,4}$&0.65&0.25&1.50&1.25&0.60&0.35&0.45\\
&&&$\mathcal{H}_{0,5}$&0.65&0.10&1.05&1.00&0.60&0.50&0.55\\
&&&$\mathcal{H}_{0,6}$&0.85&0.50&1.15&1.05&0.65&0.60&0.75\\
N&(30,40,50,60)&(1.75,1.5,1.25,1)&FWER&11.35&9.55&5.35&4.20&3.55&3.60&4.45\\
&&&$\mathcal{H}_{0,1}$&0.90&1.10&1.25&0.90&0.70&0.95&1.20\\
&&&$\mathcal{H}_{0,2}$&2.35&1.75&1.50&1.15&0.75&0.80&0.90\\
&&&$\mathcal{H}_{0,3}$&6.35&4.95&1.45&1.15&0.70&0.75&0.80\\
&&&$\mathcal{H}_{0,4}$&1.10&0.80&0.70&0.60&0.60&0.60&0.80\\
&&&$\mathcal{H}_{0,5}$&2.70&2.15&1.15&0.80&0.80&0.80&1.00\\
&&&$\mathcal{H}_{0,6}$&0.80&1.00&0.90&0.60&0.35&0.60&0.90\\
N&(60,80,100,120)&(1,1,1,1)&FWER&3.80&3.70&5.35&4.70&2.90&3.80&4.25\\
&&&$\mathcal{H}_{0,1}$&0.50&0.65&1.30&1.20&0.45&0.75&0.80\\
&&&$\mathcal{H}_{0,2}$&0.55&0.70&0.75&0.60&0.45&0.75&0.90\\
&&&$\mathcal{H}_{0,3}$&0.75&0.85&1.20&1.05&0.70&0.85&0.95\\
&&&$\mathcal{H}_{0,4}$&0.70&1.00&1.15&1.00&0.55&0.90&0.90\\
&&&$\mathcal{H}_{0,5}$&0.55&0.40&0.95&0.90&0.40&0.50&0.60\\
&&&$\mathcal{H}_{0,6}$&1.05&0.70&1.10&1.00&0.65&0.55&0.80\\
N&(60,80,100,120)&(1,1.25,1.5,1.75)&FWER&2.50&2.40&4.65&4.50&3.05&4.15&5.15\\
&&&$\mathcal{H}_{0,1}$&0.50&0.50&0.80&0.70&0.45&0.70&0.95\\
&&&$\mathcal{H}_{0,2}$&0.30&0.25&0.80&0.85&0.50&0.85&1.15\\
&&&$\mathcal{H}_{0,3}$&0.05&0.15&1.05&1.00&0.30&0.75&1.00\\
&&&$\mathcal{H}_{0,4}$&0.85&0.65&0.90&0.95&0.65&0.65&0.80\\
&&&$\mathcal{H}_{0,5}$&0.35&0.55&1.00&0.85&0.70&1.20&1.40\\
&&&$\mathcal{H}_{0,6}$&0.70&0.55&0.90&0.90&0.75&0.60&0.75\\
N&(60,80,100,120)&(1.75,1.5,1.25,1)&FWER&11.30&10.60&5.10&4.60&2.90&4.05&4.90\\
&&&$\mathcal{H}_{0,1}$&1.25&0.85&1.35&1.40&0.75&0.80&1.05\\
&&&$\mathcal{H}_{0,2}$&2.15&1.90&0.95&0.85&0.60&0.70&0.85\\
&&&$\mathcal{H}_{0,3}$&6.25&5.60&1.15&0.90&0.70&0.80&0.90\\
&&&$\mathcal{H}_{0,4}$&1.05&1.30&0.90&0.70&0.35&0.80&0.90\\
&&&$\mathcal{H}_{0,5}$&2.55&2.75&0.90&0.85&0.55&1.05&1.40\\
&&&$\mathcal{H}_{0,6}$&0.95&1.00&1.20&1.05&0.50&0.65&0.85\\
t&(15,20,25,30)&(1,1,1,1)&FWER&4.20&3.65&5.95&4.00&3.30&3.20&3.90\\
&&&$\mathcal{H}_{0,1}$&1.10&0.60&0.65&0.30&0.40&0.55&0.70\\
&&&$\mathcal{H}_{0,2}$&0.55&0.35&1.30&0.80&0.70&0.45&0.60\\
&&&$\mathcal{H}_{0,3}$&0.80&0.95&1.65&1.00&0.65&0.85&1.00\\
&&&$\mathcal{H}_{0,4}$&0.50&0.60&1.25&0.75&0.60&0.45&0.60\\
&&&$\mathcal{H}_{0,5}$&0.65&0.65&1.10&0.75&0.55&0.55&0.60\\
&&&$\mathcal{H}_{0,6}$&0.90&0.65&1.40&1.10&0.70&0.60&0.75\\
t&(15,20,25,30)&(1,1.25,1.5,1.75)&FWER&3.25&1.95&5.70&3.45&3.10&3.50&4.05\\
&&&$\mathcal{H}_{0,1}$&0.50&0.25&1.40&0.70&0.50&0.50&0.65\\
&&&$\mathcal{H}_{0,2}$&0.60&0.30&1.10&0.50&0.65&0.65&0.70\\
&&&$\mathcal{H}_{0,3}$&0.20&0.10&1.55&0.75&0.50&0.75&0.75\\
&&&$\mathcal{H}_{0,4}$&0.65&0.65&1.05&0.85&0.75&0.80&0.90\\
&&&$\mathcal{H}_{0,5}$&0.20&0.10&0.80&0.65&0.40&0.45&0.50\\
&&&$\mathcal{H}_{0,6}$&1.25&0.75&1.00&0.60&0.60&0.70&0.90\\
t&(15,20,25,30)&(1.75,1.5,1.25,1)&FWER&11.10&9.35&6.65&3.90&3.70&2.40&3.20\\
&&&$\mathcal{H}_{0,1}$&1.00&0.65&1.15&0.60&0.70&0.25&0.35\\
&&&$\mathcal{H}_{0,2}$&2.35&1.70&1.80&1.10&0.70&0.50&0.55\\
&&&$\mathcal{H}_{0,3}$&6.35&5.60&2.15&1.05&0.90&0.45&0.55\\
&&&$\mathcal{H}_{0,4}$&1.10&0.55&1.15&0.75&0.65&0.35&0.60\\
&&&$\mathcal{H}_{0,5}$&1.70&1.85&1.15&0.65&0.65&0.50&0.80\\
&&&$\mathcal{H}_{0,6}$&1.10&1.05&1.25&0.95&0.60&0.70&0.85\\
t&(30,40,50,60)&(1,1,1,1)&FWER&5.00&4.00&5.35&4.50&3.25&4.00&4.50\\
&&&$\mathcal{H}_{0,1}$&0.90&0.55&1.05&0.90&0.45&0.50&0.55\\
&&&$\mathcal{H}_{0,2}$&0.80&0.90&0.95&0.80&0.65&0.95&1.05\\
&&&$\mathcal{H}_{0,3}$&0.85&0.90&1.00&0.90&0.70&0.90&0.95\\
&&&$\mathcal{H}_{0,4}$&0.90&0.80&1.00&0.90&0.60&0.80&1.00\\
&&&$\mathcal{H}_{0,5}$&1.10&0.95&1.60&1.40&0.75&0.95&1.00\\
&&&$\mathcal{H}_{0,6}$&1.10&0.60&1.05&0.70&0.40&0.60&0.65\\
t&(30,40,50,60)&(1,1.25,1.5,1.75)&FWER&2.95&2.05&6.20&5.30&3.80&3.90&4.55\\
&&&$\mathcal{H}_{0,1}$&0.90&0.50&1.55&1.15&1.05&0.75&0.85\\
&&&$\mathcal{H}_{0,2}$&0.35&0.10&0.85&0.75&0.75&0.65&0.65\\
&&&$\mathcal{H}_{0,3}$&0.05&0.05&1.40&1.25&0.60&0.60&0.75\\
&&&$\mathcal{H}_{0,4}$&0.65&0.60&0.90&0.75&0.40&0.65&0.85\\
&&&$\mathcal{H}_{0,5}$&0.45&0.35&1.60&1.40&0.90&1.05&1.40\\
&&&$\mathcal{H}_{0,6}$&0.55&0.70&0.90&0.80&0.45&0.70&0.75\\
t&(30,40,50,60)&(1.75,1.5,1.25,1)&FWER&11.55&10.55&5.40&3.80&3.30&3.70&4.85\\
&&&$\mathcal{H}_{0,1}$&1.15&1.10&0.95&0.70&0.55&0.55&0.75\\
&&&$\mathcal{H}_{0,2}$&2.25&2.15&1.15&0.80&0.70&0.60&0.95\\
&&&$\mathcal{H}_{0,3}$&5.65&5.55&1.15&0.55&0.75&0.90&1.10\\
&&&$\mathcal{H}_{0,4}$&1.40&0.70&0.75&0.65&0.30&0.65&0.70\\
&&&$\mathcal{H}_{0,5}$&2.40&2.50&1.40&0.95&0.65&1.00&1.20\\
&&&$\mathcal{H}_{0,6}$&1.15&1.05&1.30&0.95&0.75&0.70&1.05\\
t&(60,80,100,120)&(1,1,1,1)&FWER&4.40&4.35&4.80&4.40&3.35&4.25&4.95\\
&&&$\mathcal{H}_{0,1}$&0.80&0.75&0.85&0.80&0.65&0.75&0.90\\
&&&$\mathcal{H}_{0,2}$&1.15&1.20&0.70&0.65&0.80&1.10&1.25\\
&&&$\mathcal{H}_{0,3}$&0.70&0.95&1.20&1.05&0.40&1.10&1.15\\
&&&$\mathcal{H}_{0,4}$&0.95&0.80&1.20&1.00&0.50&0.85&0.95\\
&&&$\mathcal{H}_{0,5}$&0.75&0.85&1.00&0.95&0.65&0.60&0.80\\
&&&$\mathcal{H}_{0,6}$&0.60&0.70&0.90&0.90&0.50&0.60&0.70\\
t&(60,80,100,120)&(1,1.25,1.5,1.75)&FWER&1.70&2.35&5.15&4.85&3.05&3.80&4.05\\
&&&$\mathcal{H}_{0,1}$&0.35&0.55&1.25&1.00&0.70&0.75&0.85\\
&&&$\mathcal{H}_{0,2}$&0.15&0.30&1.20&1.15&0.50&0.95&0.95\\
&&&$\mathcal{H}_{0,3}$&0.10&0.10&1.45&1.45&0.70&0.75&0.85\\
&&&$\mathcal{H}_{0,4}$&0.55&0.50&0.85&0.65&0.25&0.75&0.75\\
&&&$\mathcal{H}_{0,5}$&0.25&0.65&1.20&1.05&0.75&1.00&1.00\\
&&&$\mathcal{H}_{0,6}$&0.50&0.60&1.20&1.25&0.55&0.45&0.75\\
t&(60,80,100,120)&(1.75,1.5,1.25,1)&FWER&11.50&9.25&5.10&4.50&3.05&4.25&5.15\\
&&&$\mathcal{H}_{0,1}$&1.70&1.40&0.85&0.75&0.55&1.30&1.55\\
&&&$\mathcal{H}_{0,2}$&2.75&2.55&1.25&1.00&0.65&1.10&1.45\\
&&&$\mathcal{H}_{0,3}$&5.90&5.05&1.40&1.20&0.55&0.90&1.05\\
&&&$\mathcal{H}_{0,4}$&0.50&0.60&1.05&0.95&0.50&0.50&0.60\\
&&&$\mathcal{H}_{0,5}$&2.10&1.45&1.00&0.90&0.60&0.70&0.80\\
&&&$\mathcal{H}_{0,6}$&1.20&1.20&1.20&1.10&0.85&0.85&1.05\\
$\chi^2$&(15,20,25,30)&(1,1,1,1)&FWER&4.95&3.25&7.35&4.25&3.00&2.75&3.80\\
&&&$\mathcal{H}_{0,1}$&0.55&0.75&1.65&0.60&0.65&0.60&0.70\\
&&&$\mathcal{H}_{0,2}$&0.85&0.60&1.55&0.80&0.90&0.55&0.70\\
&&&$\mathcal{H}_{0,3}$&1.15&0.65&1.45&0.80&0.45&0.40&0.55\\
&&&$\mathcal{H}_{0,4}$&0.80&0.30&1.40&0.90&0.60&0.25&0.45\\
&&&$\mathcal{H}_{0,5}$&1.05&0.70&1.65&1.15&0.55&0.50&0.85\\
&&&$\mathcal{H}_{0,6}$&1.05&0.65&1.10&0.80&0.10&0.75&1.00\\
$\chi^2$&(15,20,25,30)&(1,1.25,1.5,1.75)&FWER&2.85&2.15&6.15&4.10&3.65&3.35&4.00\\
&&&$\mathcal{H}_{0,1}$&0.55&0.45&1.70&0.80&0.85&0.65&0.75\\
&&&$\mathcal{H}_{0,2}$&0.50&0.05&1.15&0.85&0.55&0.35&0.55\\
&&&$\mathcal{H}_{0,3}$&0.05&0.05&0.90&0.65&0.70&0.70&0.80\\
&&&$\mathcal{H}_{0,4}$&0.75&0.70&1.45&0.90&0.55&0.85&0.85\\
&&&$\mathcal{H}_{0,5}$&0.55&0.35&1.45&0.85&0.70&0.65&0.90\\
&&&$\mathcal{H}_{0,6}$&0.60&0.80&0.85&0.60&0.60&0.75&0.80\\
$\chi^2$&(15,20,25,30)&(1.75,1.5,1.25,1)&FWER&12.05&8.30&6.10&4.25&4.25&2.15&2.80\\
&&&$\mathcal{H}_{0,1}$&0.85&0.90&1.15&0.65&0.40&0.45&0.60\\
&&&$\mathcal{H}_{0,2}$&2.50&2.10&1.50&1.15&1.00&0.50&0.65\\
&&&$\mathcal{H}_{0,3}$&5.75&4.45&1.65&0.95&1.10&0.50&0.70\\
&&&$\mathcal{H}_{0,4}$&1.35&0.75&1.95&1.05&0.95&0.55&0.70\\
&&&$\mathcal{H}_{0,5}$&2.65&1.55&1.45&0.95&0.65&0.25&0.30\\
&&&$\mathcal{H}_{0,6}$&1.25&0.75&0.80&0.65&0.60&0.45&0.55\\
$\chi^2$&(30,40,50,60)&(1,1,1,1)&FWER&4.55&3.25&4.35&3.60&3.20&3.25&3.95\\
&&&$\mathcal{H}_{0,1}$&1.15&0.65&0.80&0.65&0.40&0.80&0.85\\
&&&$\mathcal{H}_{0,2}$&0.65&0.65&1.15&1.00&0.45&0.55&0.70\\
&&&$\mathcal{H}_{0,3}$&0.75&0.70&1.15&0.85&0.70&0.65&0.65\\
&&&$\mathcal{H}_{0,4}$&0.50&0.60&0.75&0.60&0.55&0.50&0.65\\
&&&$\mathcal{H}_{0,5}$&0.95&0.65&0.90&0.70&0.75&0.60&0.85\\
&&&$\mathcal{H}_{0,6}$&1.00&0.60&0.60&0.50&0.65&0.65&0.85\\
$\chi^2$&(30,40,50,60)&(1,1.25,1.5,1.75)&FWER&2.55&1.95&5.30&4.75&3.40&3.25&3.90\\
&&&$\mathcal{H}_{0,1}$&0.40&0.45&1.10&0.85&0.50&0.55&0.60\\
&&&$\mathcal{H}_{0,2}$&0.05&0.20&1.00&0.90&0.30&0.55&0.65\\
&&&$\mathcal{H}_{0,3}$&0.10&0.00&0.85&0.70&0.40&0.75&0.95\\
&&&$\mathcal{H}_{0,4}$&0.60&0.30&1.30&1.30&0.95&0.20&0.40\\
&&&$\mathcal{H}_{0,5}$&0.50&0.35&1.05&0.90&0.75&0.65&0.70\\
&&&$\mathcal{H}_{0,6}$&1.15&0.90&1.20&1.00&0.80&1.05&1.30\\
$\chi^2$&(30,40,50,60)&(1.75,1.5,1.25,1)&FWER&12.30&10.25&5.90&4.75&4.25&4.05&4.85\\
&&&$\mathcal{H}_{0,1}$&1.35&1.10&1.40&1.10&1.05&0.75&0.90\\
&&&$\mathcal{H}_{0,2}$&2.15&1.75&1.35&1.00&0.80&0.75&0.90\\
&&&$\mathcal{H}_{0,3}$&6.25&5.90&1.90&1.25&1.00&0.80&0.95\\
&&&$\mathcal{H}_{0,4}$&1.70&1.20&1.15&1.10&0.90&0.75&1.05\\
&&&$\mathcal{H}_{0,5}$&2.30&2.00&1.10&0.80&0.55&0.70&0.95\\
&&&$\mathcal{H}_{0,6}$&1.50&1.05&1.15&1.00&0.85&0.75&0.95\\
$\chi^2$&(60,80,100,120)&(1,1,1,1)&FWER&4.25&3.80&5.20&4.70&3.05&3.95&4.75\\
&&&$\mathcal{H}_{0,1}$&0.55&1.00&1.00&0.95&0.65&1.00&1.15\\
&&&$\mathcal{H}_{0,2}$&1.05&0.65&0.85&0.65&0.35&0.65&0.70\\
&&&$\mathcal{H}_{0,3}$&0.85&0.90&0.90&0.75&0.60&0.95&1.10\\
&&&$\mathcal{H}_{0,4}$&1.00&0.90&1.40&1.15&0.70&0.75&1.00\\
&&&$\mathcal{H}_{0,5}$&0.75&0.50&1.20&1.25&0.85&0.45&0.60\\
&&&$\mathcal{H}_{0,6}$&0.80&0.60&1.05&0.75&0.40&0.80&0.95\\
$\chi^2$&(60,80,100,120)&(1,1.25,1.5,1.75)&FWER&2.55&2.10&4.60&4.15&3.45&4.00&4.35\\
&&&$\mathcal{H}_{0,1}$&0.85&0.45&1.10&0.95&0.80&0.60&0.75\\
&&&$\mathcal{H}_{0,2}$&0.35&0.20&0.65&0.60&0.75&0.95&1.05\\
&&&$\mathcal{H}_{0,3}$&0.25&0.10&1.30&1.20&0.65&0.60&0.65\\
&&&$\mathcal{H}_{0,4}$&0.65&0.50&1.05&0.90&0.40&0.65&0.75\\
&&&$\mathcal{H}_{0,5}$&0.35&0.35&0.90&0.85&0.55&1.05&1.05\\
&&&$\mathcal{H}_{0,6}$&0.30&0.60&0.70&0.50&0.55&0.75&0.75\\
$\chi^2$&(60,80,100,120)&(1.75,1.5,1.25,1)&FWER&11.55&10.10&5.00&4.50&2.50&4.20&4.85\\
&&&$\mathcal{H}_{0,1}$&1.10&1.05&1.20&1.05&0.60&0.90&0.95\\
&&&$\mathcal{H}_{0,2}$&2.85&2.40&1.30&1.10&0.55&1.25&1.30\\
&&&$\mathcal{H}_{0,3}$&6.00&6.35&1.25&1.30&0.55&1.00&1.30\\
&&&$\mathcal{H}_{0,4}$&0.95&1.00&0.70&0.55&0.35&0.90&1.05\\
&&&$\mathcal{H}_{0,5}$&2.95&1.90&1.25&1.10&0.65&0.75&0.90\\
&&&$\mathcal{H}_{0,6}$&0.90&0.70&0.60&0.40&0.35&0.55&0.80\\
\hline
\end{longtable}

\begin{figure}
\centering
\includegraphics[width= 0.95\textwidth,height=0.9\textheight]{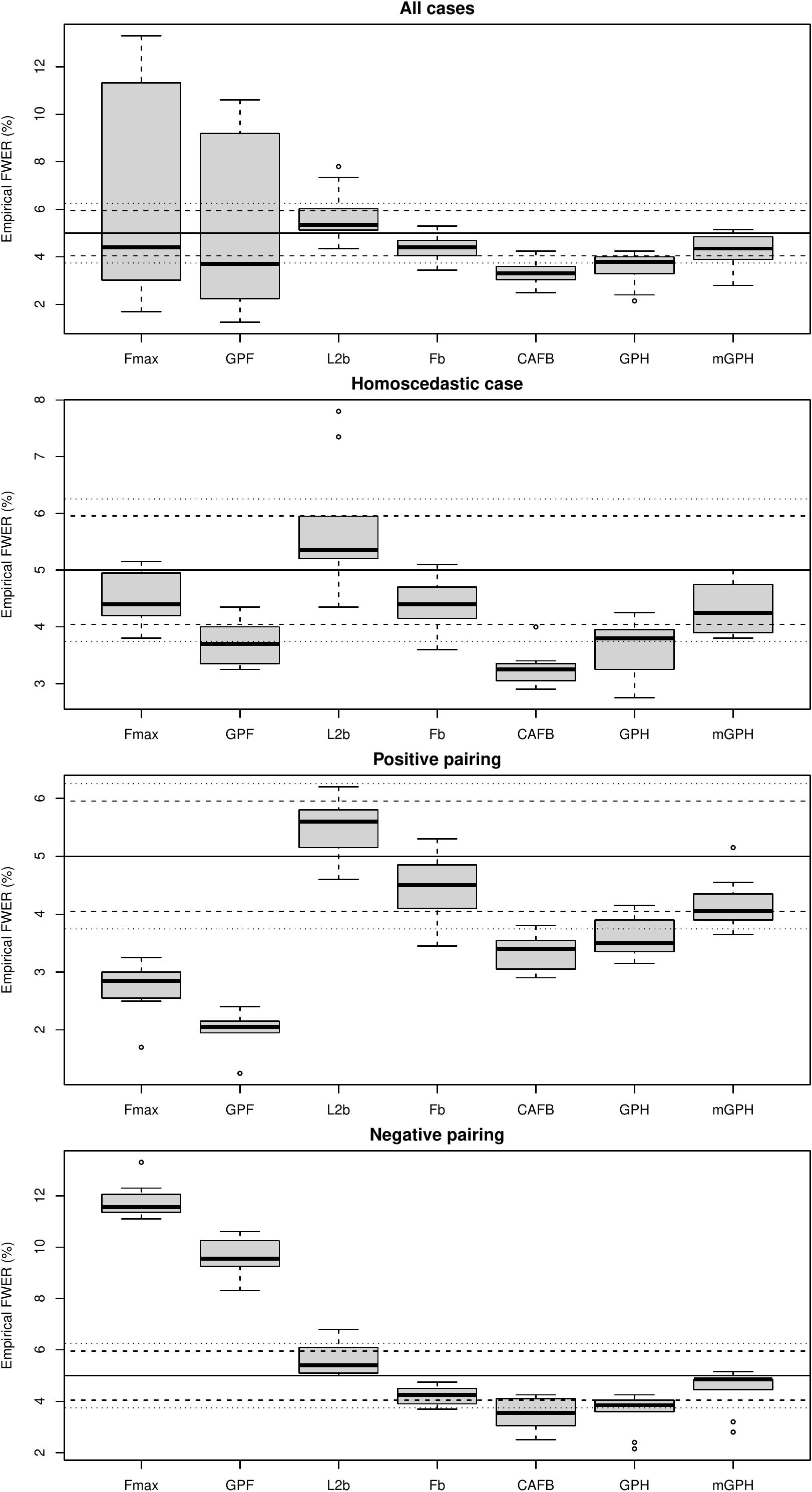}
\caption[Box-and-whisker plots for the empirical FWER of all tests obtained for the Tukey constrasts with scaling function and homoscedastic and heteroscedastic cases]{Box-and-whisker plots for the empirical FWER (as percentages) of all tests obtained for the Tukey constrasts with scaling function and homoscedastic and heteroscedastic cases}
\end{figure}

\begin{figure}
\centering
\includegraphics[width= 0.95\textwidth,height=0.9\textheight]{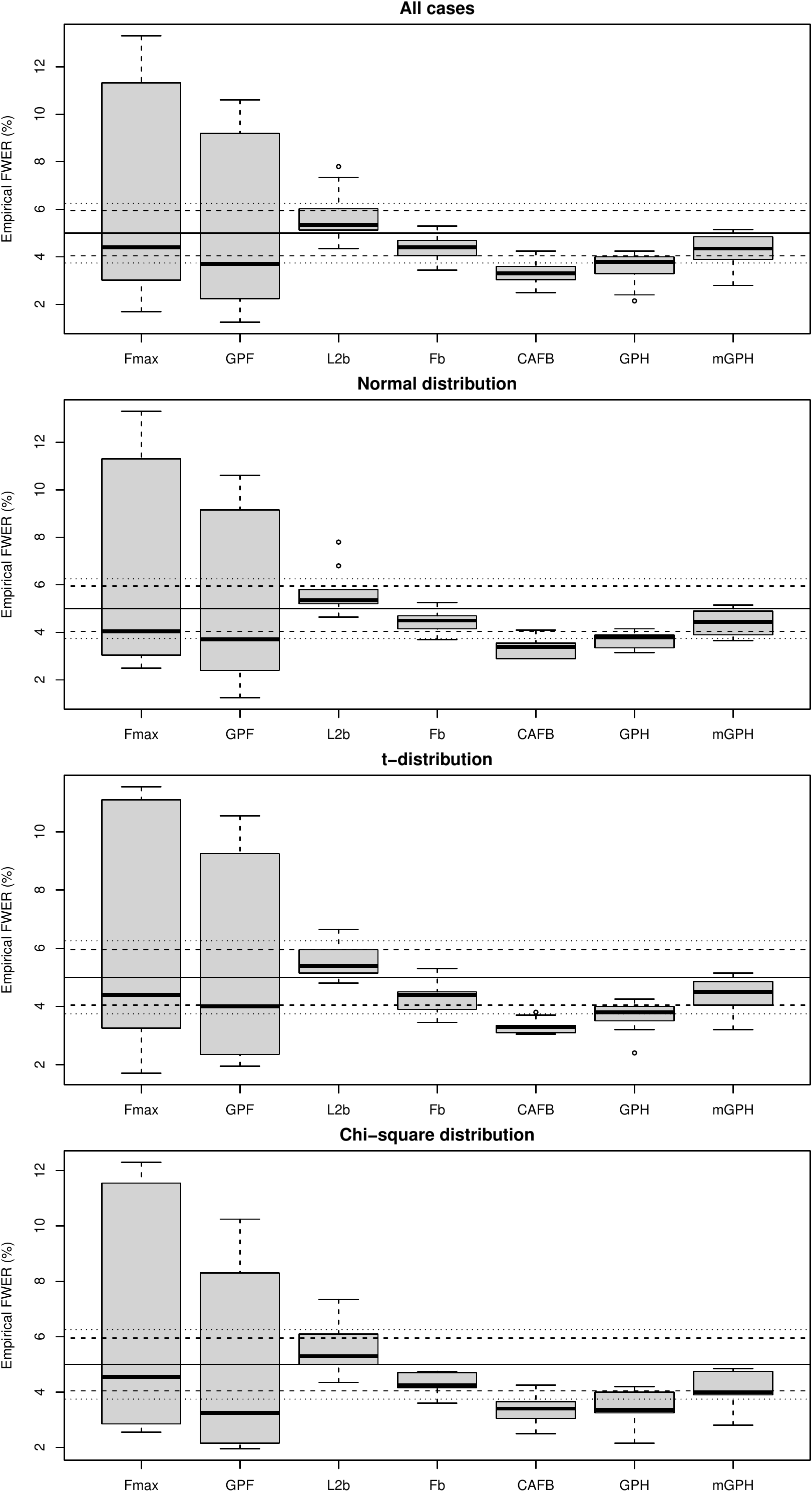}
\caption[Box-and-whisker plots for the empirical FWER of all tests obtained for the Tukey constrasts with scaling function and different distributions]{Box-and-whisker plots for the empirical FWER (as percentages) of all tests obtained for the Tukey constrasts with scaling function and different distributions}
\end{figure}

\newpage

\begin{longtable}[t]{rrrr|rrrrrrr}
\caption[Empirical sizes and powers of all tests obtained under alternative A1 for the Tukey constrasts with scaling function]{\label{tab:unnamed-chunk-108}Empirical sizes ($\mathcal{H}_{0,1}$, $\mathcal{H}_{0,2}$, $\mathcal{H}_{0,4}$) and powers ($\mathcal{H}_{0,3}$, $\mathcal{H}_{0,5}$, $\mathcal{H}_{0,6}$) (as percentages) of all tests obtained under alternative A1 for the Tukey constrasts with scaling function (D - distribution, $(\lambda_1,\lambda_2,\lambda_3,\lambda_4)$: (1,1,1,1) - homoscedastic case, (1,1.25,1.5,1.75) - heteroscedastic case (positive pairing), (1.75,1.5,1.25,1) - heteroscedastic case (negative pairing))}\\
\hline
D&$(n_1,n_2,n_3,n_4)$&$(\lambda_1,\lambda_2,\lambda_3,\lambda_4)$&$\mathcal{H}$&Fmax&GPF&L2b&Fb&CAFB&GPH&mGPH\\
\hline
\endfirsthead
\caption[]{Empirical sizes ($\mathcal{H}_{0,1}$, $\mathcal{H}_{0,2}$, $\mathcal{H}_{0,4}$) and powers ($\mathcal{H}_{0,3}$, $\mathcal{H}_{0,5}$, $\mathcal{H}_{0,6}$) (as percentages) of all tests obtained under alternative A1 for the Tukey constrasts with scaling function (D - distribution, $(\lambda_1,\lambda_2,\lambda_3,\lambda_4)$: (1,1,1,1) - homoscedastic case, (1,1.25,1.5,1.75) - heteroscedastic case (positive pairing), (1.75,1.5,1.25,1) - heteroscedastic case (negative pairing)) \textit{(continued)}}\\
\hline
D&$(n_1,n_2,n_3,n_4)$&$(\lambda_1,\lambda_2,\lambda_3,\lambda_4)$&$\mathcal{H}$&Fmax&GPF&L2b&Fb&CAFB&GPH&mGPH\\
\hline
\endhead
N&(15,20,25,30)&(1,1,1,1)&$\mathcal{H}_{0,1}$&0.85&0.80&1.40&0.85&0.70&0.70&0.75\\
&&&$\mathcal{H}_{0,2}$&0.80&0.80&1.95&1.40&0.65&0.85&0.95\\
&&&$\mathcal{H}_{0,3}$&73.10&82.65&1.95&1.30&3.00&79.15&81.70\\
&&&$\mathcal{H}_{0,4}$&1.15&0.65&1.40&0.90&0.90&0.75&0.85\\
&&&$\mathcal{H}_{0,5}$&83.40&92.00&1.70&1.20&4.90&91.50&92.70\\
&&&$\mathcal{H}_{0,6}$&91.00&96.25&1.90&1.50&7.05&96.00&96.90\\
N&(15,20,25,30)&(1,1.25,1.5,1.75)&$\mathcal{H}_{0,1}$&0.50&0.40&1.55&1.05&0.60&0.55&0.65\\
&&&$\mathcal{H}_{0,2}$&0.20&0.25&2.00&1.05&0.80&0.65&0.75\\
&&&$\mathcal{H}_{0,3}$&14.95&16.90&1.00&0.50&2.25&44.30&47.75\\
&&&$\mathcal{H}_{0,4}$&0.85&0.70&1.10&0.85&0.70&1.05&1.15\\
&&&$\mathcal{H}_{0,5}$&26.45&31.10&1.45&1.00&2.20&43.40&45.90\\
&&&$\mathcal{H}_{0,6}$&31.35&37.75&1.30&0.85&2.15&40.95&43.40\\
N&(15,20,25,30)&(1.75,1.5,1.25,1)&$\mathcal{H}_{0,1}$&0.90&0.65&1.70&0.95&0.95&0.55&0.65\\
&&&$\mathcal{H}_{0,2}$&2.15&1.45&1.80&1.00&0.80&0.40&0.50\\
&&&$\mathcal{H}_{0,3}$&55.05&59.35&2.05&1.30&1.95&22.55&25.25\\
&&&$\mathcal{H}_{0,4}$&1.35&0.75&1.80&1.15&0.90&0.45&0.70\\
&&&$\mathcal{H}_{0,5}$&65.30&70.85&1.50&0.90&2.30&52.45&56.85\\
&&&$\mathcal{H}_{0,6}$&80.30&88.85&1.50&1.05&4.05&85.35&87.80\\
t&(15,20,25,30)&(1,1,1,1)&$\mathcal{H}_{0,1}$&0.85&0.95&1.95&0.65&0.75&0.80&0.85\\
&&&$\mathcal{H}_{0,2}$&1.40&0.50&1.40&1.00&0.65&0.60&0.70\\
&&&$\mathcal{H}_{0,3}$&43.20&50.95&1.95&1.10&2.90&46.75&50.05\\
&&&$\mathcal{H}_{0,4}$&0.80&0.65&1.05&0.60&0.65&0.45&0.60\\
&&&$\mathcal{H}_{0,5}$&52.80&63.60&1.60&1.00&2.70&61.25&64.45\\
&&&$\mathcal{H}_{0,6}$&63.35&71.35&1.60&1.05&3.05&69.95&72.30\\
t&(15,20,25,30)&(1,1.25,1.5,1.75)&$\mathcal{H}_{0,1}$&0.85&0.45&1.30&0.75&0.60&0.60&0.70\\
&&&$\mathcal{H}_{0,2}$&0.45&0.15&1.65&1.15&0.55&0.55&0.60\\
&&&$\mathcal{H}_{0,3}$&6.55&5.20&1.20&0.80&1.65&21.85&24.30\\
&&&$\mathcal{H}_{0,4}$&0.80&0.40&1.35&0.65&0.75&0.45&0.60\\
&&&$\mathcal{H}_{0,5}$&12.50&13.15&1.35&1.00&1.45&20.95&22.75\\
&&&$\mathcal{H}_{0,6}$&15.15&16.75&1.10&0.70&1.05&17.70&20.15\\
t&(15,20,25,30)&(1.75,1.5,1.25,1)&$\mathcal{H}_{0,1}$&1.00&0.50&1.35&0.75&0.80&0.40&0.40\\
&&&$\mathcal{H}_{0,2}$&1.70&1.65&1.70&1.35&0.85&0.40&0.50\\
&&&$\mathcal{H}_{0,3}$&34.20&35.75&2.40&1.20&1.55&10.05&11.50\\
&&&$\mathcal{H}_{0,4}$&1.15&0.70&1.40&0.85&0.90&0.30&0.50\\
&&&$\mathcal{H}_{0,5}$&38.50&43.40&2.00&1.35&1.80&25.55&29.15\\
&&&$\mathcal{H}_{0,6}$&49.20&57.70&1.55&1.05&2.25&51.00&55.60\\
$\chi^2$&(15,20,25,30)&(1,1,1,1)&$\mathcal{H}_{0,1}$&0.60&0.45&1.40&0.85&0.50&0.45&0.50\\
&&&$\mathcal{H}_{0,2}$&0.80&0.70&1.25&1.05&0.70&0.60&0.80\\
&&&$\mathcal{H}_{0,3}$&73.80&82.15&1.40&0.90&4.40&77.05&79.10\\
&&&$\mathcal{H}_{0,4}$&0.90&0.80&1.10&0.75&0.75&0.85&0.95\\
&&&$\mathcal{H}_{0,5}$&85.40&91.85&1.65&0.90&4.95&89.80&91.00\\
&&&$\mathcal{H}_{0,6}$&91.25&95.25&1.55&1.10&6.25&95.10&95.80\\
$\chi^2$&(15,20,25,30)&(1,1.25,1.5,1.75)&$\mathcal{H}_{0,1}$&0.90&0.15&1.45&0.95&0.95&0.25&0.30\\
&&&$\mathcal{H}_{0,2}$&0.60&0.25&1.50&0.95&0.55&0.35&0.40\\
&&&$\mathcal{H}_{0,3}$&14.60&13.80&1.15&0.65&1.75&45.70&48.40\\
&&&$\mathcal{H}_{0,4}$&0.70&0.40&1.40&0.85&0.65&0.50&0.55\\
&&&$\mathcal{H}_{0,5}$&25.40&30.90&1.25&0.95&1.75&42.85&45.60\\
&&&$\mathcal{H}_{0,6}$&31.15&37.50&1.35&1.10&1.45&39.70&42.20\\
$\chi^2$&(15,20,25,30)&(1.75,1.5,1.25,1)&$\mathcal{H}_{0,1}$&1.05&0.65&1.35&0.90&0.80&0.60&0.65\\
&&&$\mathcal{H}_{0,2}$&2.00&1.40&1.65&0.95&0.65&0.35&0.50\\
&&&$\mathcal{H}_{0,3}$&56.80&58.50&2.05&1.35&2.35&24.75&27.90\\
&&&$\mathcal{H}_{0,4}$&1.35&0.55&1.35&1.00&0.75&0.40&0.60\\
&&&$\mathcal{H}_{0,5}$&66.20&69.50&1.70&1.15&2.90&52.55&55.45\\
&&&$\mathcal{H}_{0,6}$&83.10&87.30&1.45&1.05&4.60&83.60&86.15\\
\hline
\end{longtable}

\begin{figure}
\centering
\includegraphics[width= 0.95\textwidth,height=0.9\textheight]{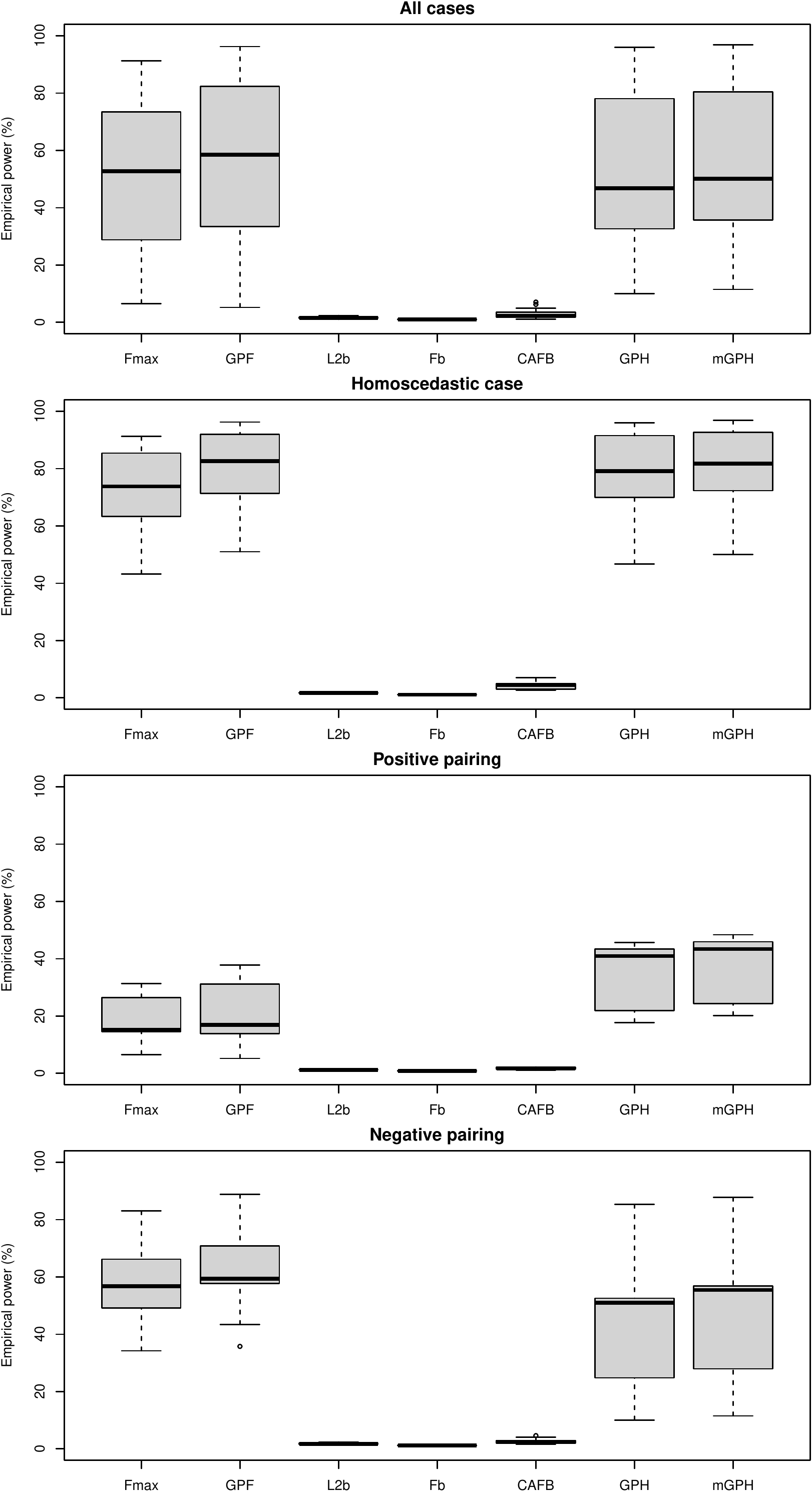}
\caption[Box-and-whisker plots for the empirical powers of all tests obtained under alternative A1 for the Tukey constrasts with scaling function and homoscedastic and heteroscedastic cases]{Box-and-whisker plots for the empirical powers (as percentages) of all tests obtained under alternative A1 for the Tukey constrasts with scaling function and homoscedastic and heteroscedastic cases}
\end{figure}

\begin{figure}
\centering
\includegraphics[width= 0.95\textwidth,height=0.9\textheight]{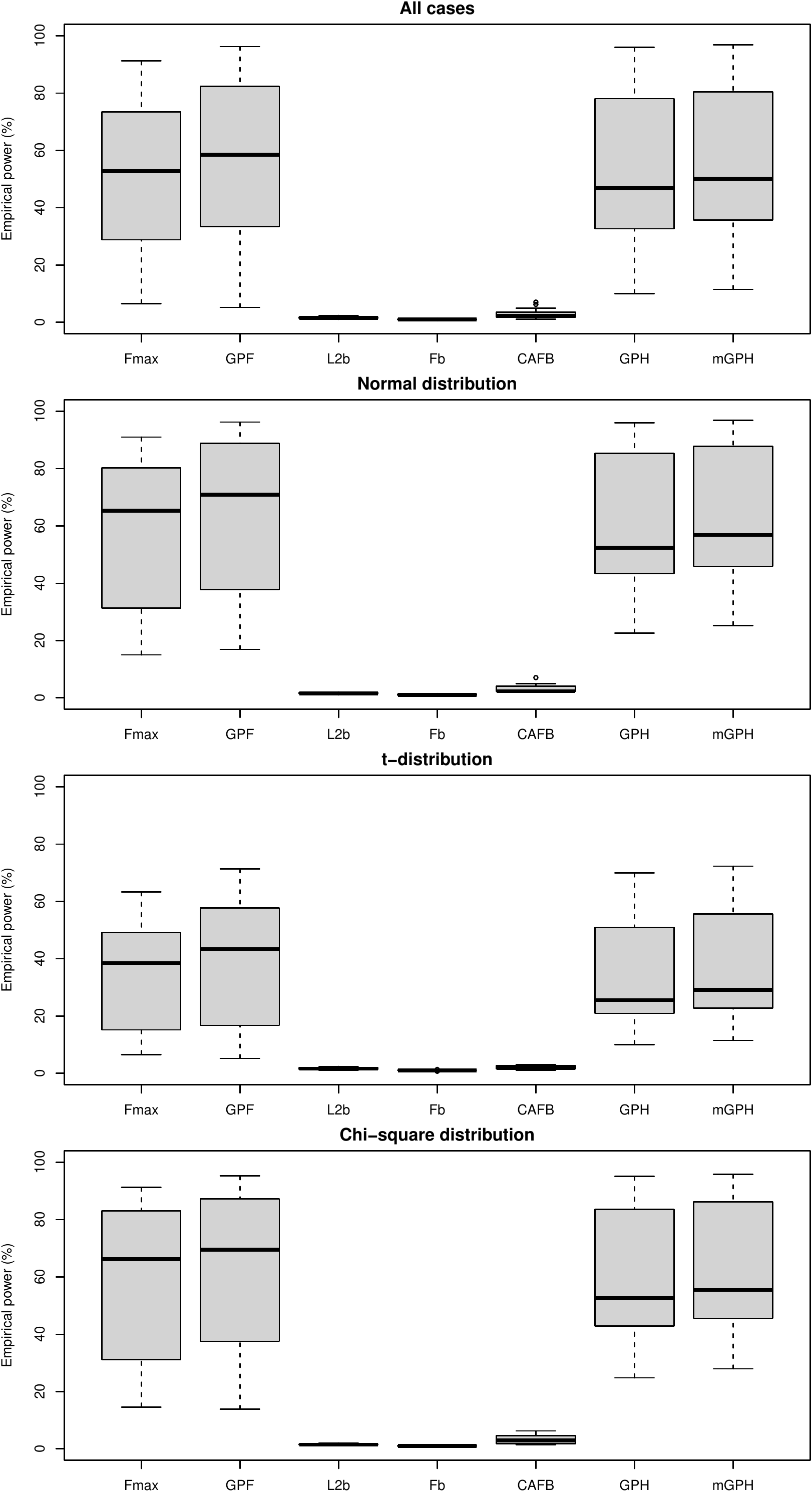}
\caption[Box-and-whisker plots for the empirical powers all tests obtained under alternative A1 for the Tukey constrasts with scaling function and different distributions]{Box-and-whisker plots for the empirical powers (as percentages) of all tests obtained under alternative A1 for the Tukey constrasts with scaling function and different distributions}
\end{figure}

\newpage

\begin{longtable}[t]{rrrr|rrrrrrr}
\caption[Empirical sizes and powers of all tests obtained under alternative A2 for the Tukey constrasts with scaling function]{\label{tab:unnamed-chunk-103}Empirical sizes ($\mathcal{H}_{0,1}$, $\mathcal{H}_{0,2}$, $\mathcal{H}_{0,4}$) and powers ($\mathcal{H}_{0,3}$, $\mathcal{H}_{0,5}$, $\mathcal{H}_{0,6}$) (as percentages) of all tests obtained under alternative A2 for the Tukey constrasts with scaling function (D - distribution, $(\lambda_1,\lambda_2,\lambda_3,\lambda_4)$: (1,1,1,1) - homoscedastic case, (1,1.25,1.5,1.75) - heteroscedastic case (positive pairing), (1.75,1.5,1.25,1) - heteroscedastic case (negative pairing))}\\
\hline
D&$(n_1,n_2,n_3,n_4)$&$(\lambda_1,\lambda_2,\lambda_3,\lambda_4)$&$\mathcal{H}$&Fmax&GPF&L2b&Fb&CAFB&GPH&mGPH\\
\hline
\endfirsthead
\caption[]{Empirical sizes ($\mathcal{H}_{0,1}$, $\mathcal{H}_{0,2}$, $\mathcal{H}_{0,4}$) and powers ($\mathcal{H}_{0,3}$, $\mathcal{H}_{0,5}$, $\mathcal{H}_{0,6}$) (as percentages) of all tests obtained under alternative A2 for the Tukey constrasts with scaling function (D - distribution, $(\lambda_1,\lambda_2,\lambda_3,\lambda_4)$: (1,1,1,1) - homoscedastic case, (1,1.25,1.5,1.75) - heteroscedastic case (positive pairing), (1.75,1.5,1.25,1) - heteroscedastic case (negative pairing)) \textit{(continued)}}\\
\hline
D&$(n_1,n_2,n_3,n_4)$&$(\lambda_1,\lambda_2,\lambda_3,\lambda_4)$&$\mathcal{H}$&Fmax&GPF&L2b&Fb&CAFB&GPH&mGPH\\
\hline
\endhead
N&(15,20,25,30)&(1,1,1,1)&$\mathcal{H}_{0,1}$&0.85&0.40&2.00&1.25&1.00&0.30&0.35\\
&&&$\mathcal{H}_{0,2}$&0.65&0.55&1.70&0.85&0.85&0.60&0.75\\
&&&$\mathcal{H}_{0,3}$&71.30&97.15&74.15&64.60&79.50&96.80&97.20\\
&&&$\mathcal{H}_{0,4}$&0.50&0.65&1.60&0.90&0.50&0.75&1.10\\
&&&$\mathcal{H}_{0,5}$&81.00&99.40&81.05&75.85&87.60&99.20&99.40\\
&&&$\mathcal{H}_{0,6}$&88.00&99.90&87.00&84.20&93.40&99.90&99.90\\
N&(15,20,25,30)&(1,1.25,1.5,1.75)&$\mathcal{H}_{0,1}$&0.40&0.35&1.45&0.80&0.50&0.35&0.55\\
&&&$\mathcal{H}_{0,2}$&0.40&0.05&1.90&1.05&0.75&0.80&0.90\\
&&&$\mathcal{H}_{0,3}$&13.60&47.00&46.30&38.75&49.90&77.55&79.25\\
&&&$\mathcal{H}_{0,4}$&0.70&0.60&1.40&1.05&0.65&0.60&0.65\\
&&&$\mathcal{H}_{0,5}$&24.15&63.20&40.75&35.25&43.90&72.50&74.80\\
&&&$\mathcal{H}_{0,6}$&31.40&68.70&37.75&32.30&40.80&70.45&72.85\\
N&(15,20,25,30)&(1.75,1.5,1.25,1)&$\mathcal{H}_{0,1}$&1.15&0.85&1.25&0.80&0.70&0.70&0.95\\
&&&$\mathcal{H}_{0,2}$&2.70&2.30&1.45&0.95&1.15&0.80&1.15\\
&&&$\mathcal{H}_{0,3}$&58.95&82.25&31.40&21.95&27.35&46.00&50.25\\
&&&$\mathcal{H}_{0,4}$&1.30&0.75&1.25&0.95&0.80&0.50&0.75\\
&&&$\mathcal{H}_{0,5}$&63.45&91.70&50.65&42.55&54.90&82.70&85.40\\
&&&$\mathcal{H}_{0,6}$&78.60&98.40&75.45&70.00&82.85&97.80&98.15\\
t&(15,20,25,30)&(1,1,1,1)&$\mathcal{H}_{0,1}$&1.30&0.60&1.30&0.65&0.85&0.65&0.80\\
&&&$\mathcal{H}_{0,2}$&0.95&0.30&1.40&0.75&0.65&0.30&0.40\\
&&&$\mathcal{H}_{0,3}$&39.50&78.20&46.25&37.80&48.00&75.60&78.05\\
&&&$\mathcal{H}_{0,4}$&0.50&0.40&1.30&0.90&0.45&0.25&0.30\\
&&&$\mathcal{H}_{0,5}$&50.95&88.00&54.60&48.55&60.30&87.00&88.55\\
&&&$\mathcal{H}_{0,6}$&58.35&93.55&61.95&57.15&70.60&92.80&93.90\\
t&(15,20,25,30)&(1,1.25,1.5,1.75)&$\mathcal{H}_{0,1}$&0.45&0.40&1.40&0.95&0.60&0.45&0.60\\
&&&$\mathcal{H}_{0,2}$&0.30&0.15&0.95&0.50&0.45&0.60&0.65\\
&&&$\mathcal{H}_{0,3}$&5.70&17.70&26.55&22.05&26.00&46.25&48.85\\
&&&$\mathcal{H}_{0,4}$&0.75&0.65&1.50&0.75&0.60&0.80&0.90\\
&&&$\mathcal{H}_{0,5}$&12.25&31.90&24.45&20.35&22.60&41.45&44.20\\
&&&$\mathcal{H}_{0,6}$&14.95&37.55&21.95&18.40&20.70&38.45&41.50\\
t&(15,20,25,30)&(1.75,1.5,1.25,1)&$\mathcal{H}_{0,1}$&0.85&0.75&1.60&0.85&0.65&0.55&0.60\\
&&&$\mathcal{H}_{0,2}$&1.80&1.35&1.60&0.70&0.60&0.15&0.20\\
&&&$\mathcal{H}_{0,3}$&39.50&57.10&18.90&12.60&15.05&22.50&26.05\\
&&&$\mathcal{H}_{0,4}$&0.90&0.50&1.30&0.70&0.95&0.45&0.45\\
&&&$\mathcal{H}_{0,5}$&39.70&68.30&29.65&22.70&30.80&52.50&56.20\\
&&&$\mathcal{H}_{0,6}$&48.50&84.55&47.60&41.65&53.20&80.35&83.15\\
$\chi^2$&(15,20,25,30)&(1,1,1,1)&$\mathcal{H}_{0,1}$&1.05&0.90&1.80&1.20&0.70&0.90&0.90\\
&&&$\mathcal{H}_{0,2}$&0.95&0.90&1.40&0.95&0.75&0.80&0.95\\
&&&$\mathcal{H}_{0,3}$&70.95&97.80&72.95&65.55&80.15&95.55&96.35\\
&&&$\mathcal{H}_{0,4}$&0.90&0.60&1.70&1.00&0.85&0.65&0.80\\
&&&$\mathcal{H}_{0,5}$&83.25&99.30&80.80&75.95&89.00&98.95&99.20\\
&&&$\mathcal{H}_{0,6}$&89.05&99.55&87.65&84.55&94.10&99.50&99.70\\
$\chi^2$&(15,20,25,30)&(1,1.25,1.5,1.75)&$\mathcal{H}_{0,1}$&0.75&0.50&1.45&0.90&0.85&0.50&0.55\\
&&&$\mathcal{H}_{0,2}$&0.20&0.25&1.70&1.10&0.65&0.40&0.50\\
&&&$\mathcal{H}_{0,3}$&12.40&44.95&45.85&38.70&49.15&75.20&77.45\\
&&&$\mathcal{H}_{0,4}$&0.70&0.65&1.20&0.60&0.50&0.50&0.80\\
&&&$\mathcal{H}_{0,5}$&25.60&64.70&42.25&36.70&45.00&74.65&76.90\\
&&&$\mathcal{H}_{0,6}$&31.85&69.20&39.85&33.35&43.55&70.90&73.40\\
$\chi^2$&(15,20,25,30)&(1.75,1.5,1.25,1)&$\mathcal{H}_{0,1}$&1.20&0.60&1.70&1.20&0.80&0.30&0.40\\
&&&$\mathcal{H}_{0,2}$&1.85&1.40&2.30&1.20&1.00&0.30&0.45\\
&&&$\mathcal{H}_{0,3}$&60.60&80.10&34.10&24.50&33.00&47.85&51.05\\
&&&$\mathcal{H}_{0,4}$&1.30&0.90&1.85&1.25&0.85&0.50&0.75\\
&&&$\mathcal{H}_{0,5}$&66.45&90.45&53.35&46.40&59.45&80.00&82.60\\
&&&$\mathcal{H}_{0,6}$&81.00&97.85&75.00&69.80&83.60&97.05&97.30\\
\hline
\end{longtable}

\begin{figure}
\centering
\includegraphics[width= 0.95\textwidth,height=0.9\textheight]{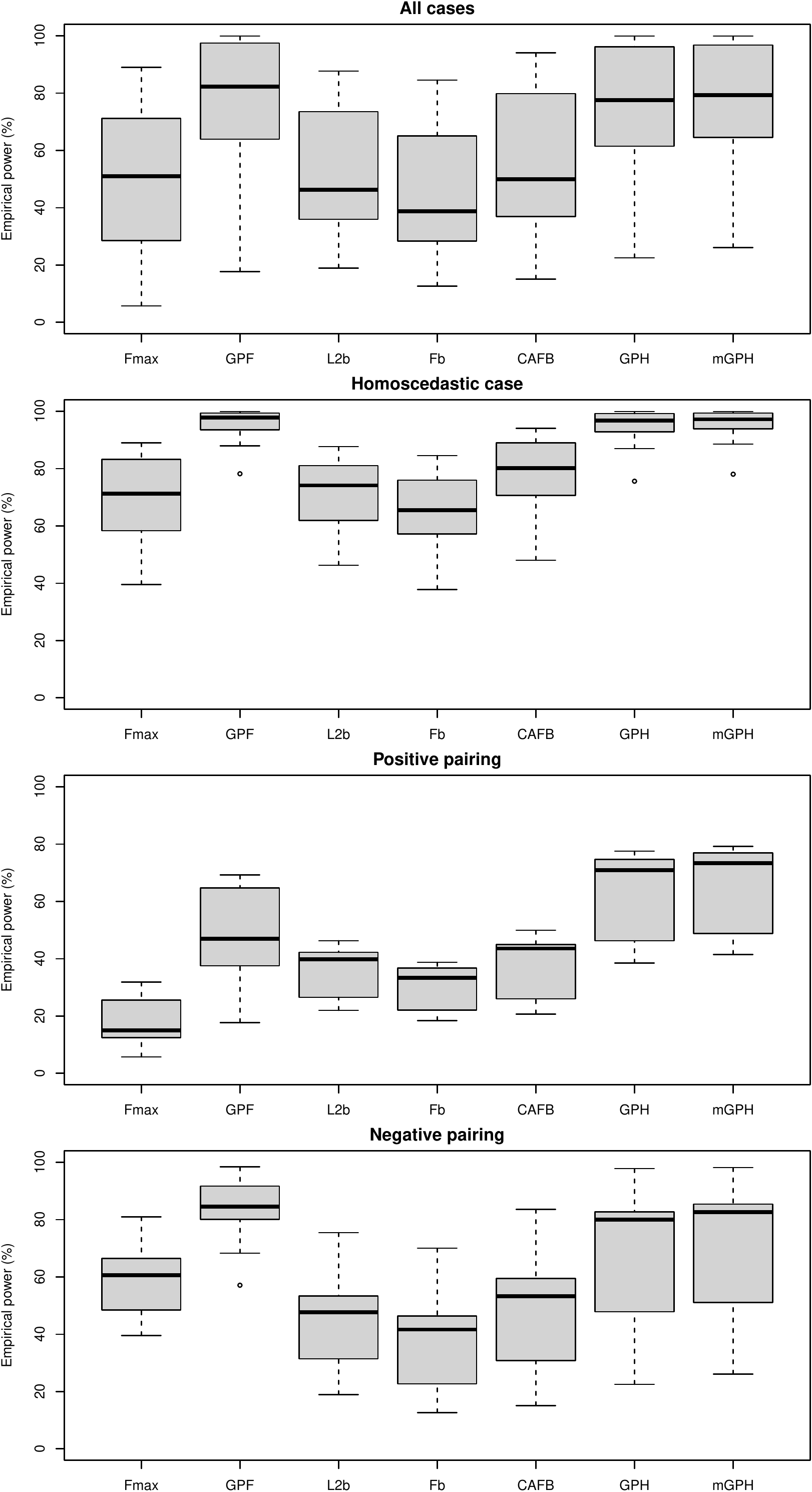}
\caption[Box-and-whisker plots for the empirical powers of all tests obtained under alternative A2 for the Tukey constrasts with scaling function and homoscedastic and heteroscedastic cases]{Box-and-whisker plots for the empirical powers (as percentages) of all tests obtained under alternative A2 for the Tukey constrasts with scaling function and homoscedastic and heteroscedastic cases}
\end{figure}

\begin{figure}
\centering
\includegraphics[width= 0.95\textwidth,height=0.9\textheight]{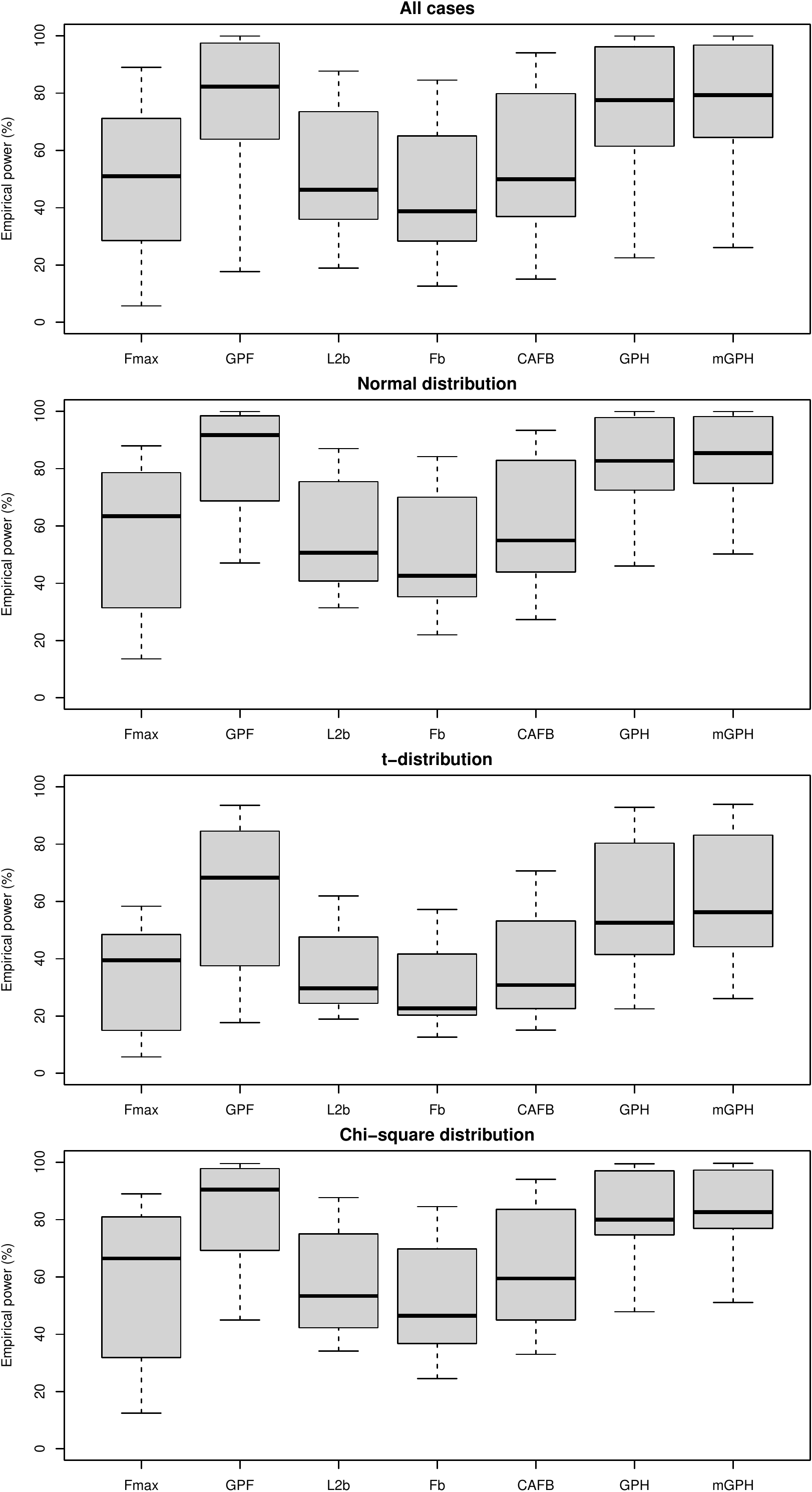}
\caption[Box-and-whisker plots for the empirical powers of all tests obtained under alternative A2 for the Tukey constrasts with scaling function and different distributions]{Box-and-whisker plots for the empirical powers (as percentages) of all tests obtained under alternative A2 for the Tukey constrasts with scaling function and different distributions}
\end{figure}

\newpage

\begin{longtable}[t]{rrrr|rrrrrrr}
\caption[Empirical powers of all tests obtained under alternative A3 for the Tukey constrasts with scaling function]{\label{tab:unnamed-chunk-113}Empirical powers (as percentages) of all tests obtained under alternative A3 for the Tukey constrasts with scaling function (D - distribution, $(\lambda_1,\lambda_2,\lambda_3,\lambda_4)$: (1,1,1,1) - homoscedastic case, (1,1.25,1.5,1.75) - heteroscedastic case (positive pairing), (1.75,1.5,1.25,1) - heteroscedastic case (negative pairing))}\\
\hline
D&$(n_1,n_2,n_3,n_4)$&$(\lambda_1,\lambda_2,\lambda_3,\lambda_4)$&$\mathcal{H}$&Fmax&GPF&L2b&Fb&CAFB&GPH&mGPH\\
\hline
\endfirsthead
\caption[]{Empirical powers (as percentages) of all tests obtained under alternative A3 for the Tukey constrasts with scaling function (D - distribution, $(\lambda_1,\lambda_2,\lambda_3,\lambda_4)$: (1,1,1,1) - homoscedastic case, (1,1.25,1.5,1.75) - heteroscedastic case (positive pairing), (1.75,1.5,1.25,1) - heteroscedastic case (negative pairing)) \textit{(continued)}}\\
\hline
D&$(n_1,n_2,n_3,n_4)$&$(\lambda_1,\lambda_2,\lambda_3,\lambda_4)$&$\mathcal{H}$&Fmax&GPF&L2b&Fb&CAFB&GPH&mGPH\\
\hline
\endhead
N&(15,20,25,30)&(1,1,1,1)&$\mathcal{H}_{0,1}$&8.75&19.05&14.10&9.85&10.75&19.10&20.85\\
&&&$\mathcal{H}_{0,2}$&65.40&96.20&71.05&62.15&76.55&95.30&96.05\\
&&&$\mathcal{H}_{0,3}$&99.30&100.00&98.75&97.65&99.65&100.00&100.00\\
&&&$\mathcal{H}_{0,4}$&12.95&30.25&18.65&14.55&16.65&30.45&33.40\\
&&&$\mathcal{H}_{0,5}$&81.45&99.20&81.70&75.50&88.75&99.10&99.20\\
&&&$\mathcal{H}_{0,6}$&19.00&44.90&22.80&18.85&23.10&44.50&47.50\\
N&(15,20,25,30)&(1,1.25,1.5,1.75)&$\mathcal{H}_{0,1}$&6.65&12.75&11.50&8.45&8.60&15.80&17.65\\
&&&$\mathcal{H}_{0,2}$&24.40&66.55&47.35&40.40&53.70&80.65&82.50\\
&&&$\mathcal{H}_{0,3}$&54.95&97.20&87.75&82.95&93.35&99.70&99.75\\
&&&$\mathcal{H}_{0,4}$&5.80&11.05&8.60&6.40&6.60&12.45&14.25\\
&&&$\mathcal{H}_{0,5}$&24.25&64.25&41.25&35.85&44.45&73.20&76.05\\
&&&$\mathcal{H}_{0,6}$&4.90&9.90&8.25&6.20&5.85&10.80&11.80\\
N&(15,20,25,30)&(1.75,1.5,1.25,1)&$\mathcal{H}_{0,1}$&3.60&5.00&5.00&3.35&3.60&4.35&5.35\\
&&&$\mathcal{H}_{0,2}$&33.05&60.40&28.10&20.45&23.70&41.85&46.15\\
&&&$\mathcal{H}_{0,3}$&93.45&99.40&69.00&55.55&67.75&93.40&94.75\\
&&&$\mathcal{H}_{0,4}$&7.10&12.50&9.15&7.40&7.25&9.65&11.90\\
&&&$\mathcal{H}_{0,5}$&63.90&92.25&52.35&44.40&55.85&82.80&85.80\\
&&&$\mathcal{H}_{0,6}$&15.30&33.05&17.00&13.90&16.05&28.85&32.05\\
t&(15,20,25,30)&(1,1,1,1)&$\mathcal{H}_{0,1}$&5.20&9.20&8.55&5.95&5.40&8.40&9.25\\
&&&$\mathcal{H}_{0,2}$&36.45&73.25&42.70&34.70&43.60&71.05&74.15\\
&&&$\mathcal{H}_{0,3}$&86.80&99.50&86.35&79.85&90.80&99.05&99.20\\
&&&$\mathcal{H}_{0,4}$&7.10&14.75&10.40&7.75&7.30&14.30&16.40\\
&&&$\mathcal{H}_{0,5}$&49.65&87.75&53.85&48.55&60.60&86.65&88.55\\
&&&$\mathcal{H}_{0,6}$&9.55&19.10&12.95&10.40&10.75&18.75&20.45\\
t&(15,20,25,30)&(1,1.25,1.5,1.75)&$\mathcal{H}_{0,1}$&3.60&4.80&7.20&4.90&3.90&5.65&6.85\\
&&&$\mathcal{H}_{0,2}$&11.60&33.80&28.35&22.00&27.40&47.95&51.35\\
&&&$\mathcal{H}_{0,3}$&25.80&72.40&59.85&52.85&68.70&91.00&92.05\\
&&&$\mathcal{H}_{0,4}$&3.30&5.00&5.30&3.70&3.45&5.05&6.00\\
&&&$\mathcal{H}_{0,5}$&11.60&32.65&22.35&17.90&22.05&43.10&45.65\\
&&&$\mathcal{H}_{0,6}$&2.95&4.95&4.60&3.15&3.55&5.60&6.15\\
t&(15,20,25,30)&(1.75,1.5,1.25,1)&$\mathcal{H}_{0,1}$&2.40&2.65&3.20&2.05&1.70&1.65&2.30\\
&&&$\mathcal{H}_{0,2}$&20.00&35.70&14.30&9.80&11.80&18.90&21.55\\
&&&$\mathcal{H}_{0,3}$&72.20&90.15&45.20&33.55&41.30&67.00&70.30\\
&&&$\mathcal{H}_{0,4}$&4.40&5.40&5.85&4.00&4.00&4.40&5.05\\
&&&$\mathcal{H}_{0,5}$&38.90&68.95&30.10&23.80&29.30&51.25&55.85\\
&&&$\mathcal{H}_{0,6}$&9.30&17.45&10.05&8.25&8.35&14.20&16.40\\
$\chi^2$&(15,20,25,30)&(1,1,1,1)&$\mathcal{H}_{0,1}$&9.80&20.75&17.00&11.60&13.65&20.05&22.00\\
&&&$\mathcal{H}_{0,2}$&66.65&94.95&67.20&59.50&74.70&92.70&93.65\\
&&&$\mathcal{H}_{0,3}$&99.00&100.00&97.85&95.45&99.10&100.00&100.00\\
&&&$\mathcal{H}_{0,4}$&14.35&33.50&19.35&15.05&18.55&33.65&36.45\\
&&&$\mathcal{H}_{0,5}$&84.20&98.90&80.15&76.40&89.65&98.50&98.60\\
&&&$\mathcal{H}_{0,6}$&19.65&43.75&24.75&20.35&23.65&43.65&46.70\\
$\chi^2$&(15,20,25,30)&(1,1.25,1.5,1.75)&$\mathcal{H}_{0,1}$&5.25&10.10&10.25&7.10&6.80&12.35&14.10\\
&&&$\mathcal{H}_{0,2}$&24.60&67.00&48.45&40.15&52.35&80.20&82.60\\
&&&$\mathcal{H}_{0,3}$&54.80&97.65&86.20&81.80&92.85&99.60&99.60\\
&&&$\mathcal{H}_{0,4}$&5.20&10.60&10.05&7.00&6.00&13.10&14.75\\
&&&$\mathcal{H}_{0,5}$&25.25&63.75&42.10&36.10&45.20&73.60&76.00\\
&&&$\mathcal{H}_{0,6}$&5.45&10.05&7.40&6.00&5.90&10.85&11.75\\
$\chi^2$&(15,20,25,30)&(1.75,1.5,1.25,1)&$\mathcal{H}_{0,1}$&4.50&6.30&6.50&4.85&4.35&5.35&6.60\\
&&&$\mathcal{H}_{0,2}$&35.55&61.70&29.65&21.65&30.20&44.50&47.90\\
&&&$\mathcal{H}_{0,3}$&92.05&98.55&68.10&56.05&72.00&90.15&92.00\\
&&&$\mathcal{H}_{0,4}$&7.40&14.85&9.65&7.00&7.60&12.60&14.60\\
&&&$\mathcal{H}_{0,5}$&67.20&91.55&52.90&44.65&58.85&82.15&84.55\\
&&&$\mathcal{H}_{0,6}$&16.25&34.55&17.50&14.35&17.85&31.20&33.90\\
\hline
\end{longtable}

\begin{figure}
\centering
\includegraphics[width= 0.95\textwidth,height=0.9\textheight]{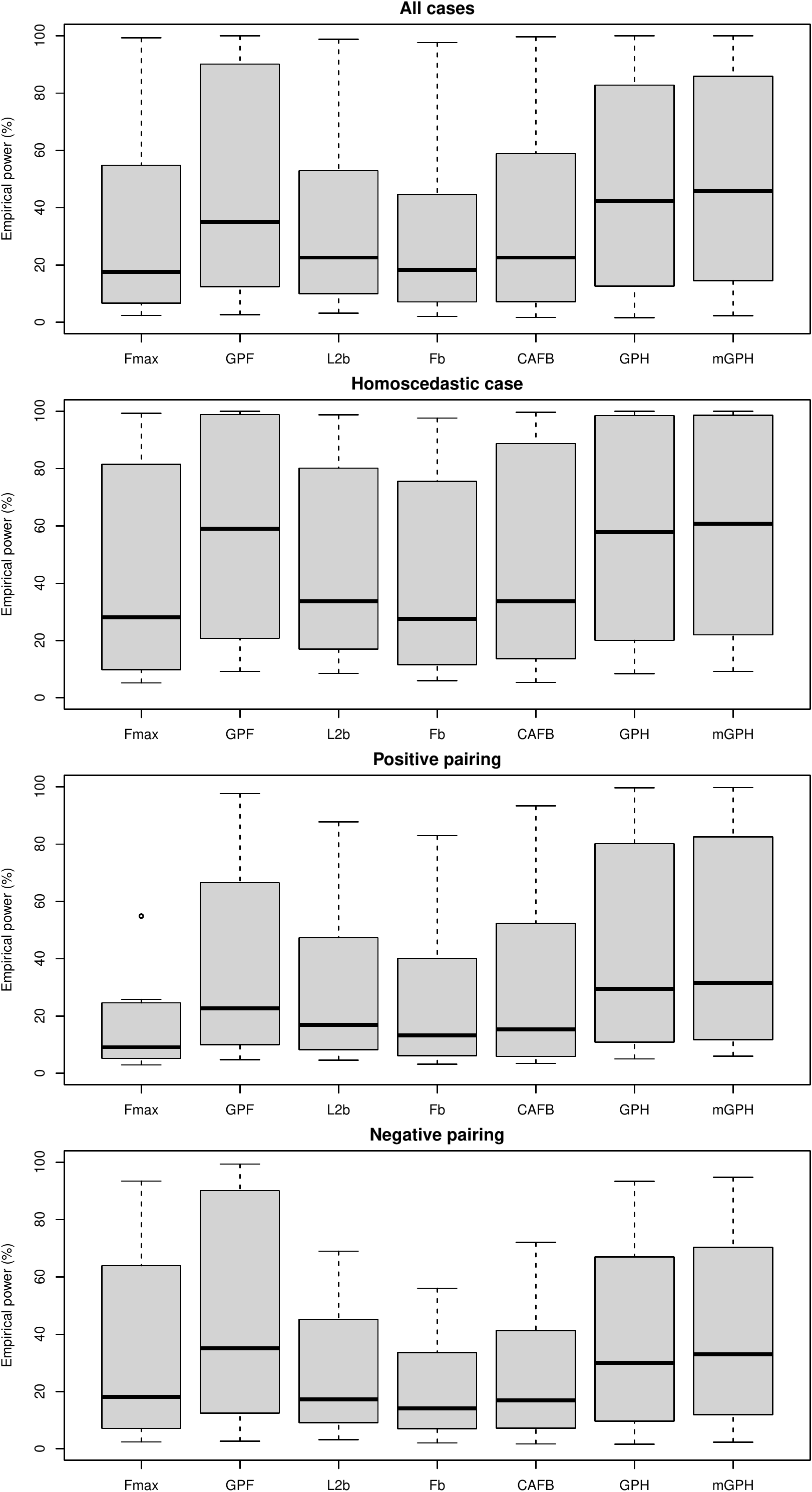}
\caption[Box-and-whisker plots for the empirical powers of all tests obtained under alternative A3 for the Tukey constrasts with scaling function and homoscedastic and heteroscedastic cases]{Box-and-whisker plots for the empirical powers (as percentages) of all tests obtained under alternative A3 for the Tukey constrasts with scaling function and homoscedastic and heteroscedastic cases}
\end{figure}

\begin{figure}
\centering
\includegraphics[width= 0.95\textwidth,height=0.9\textheight]{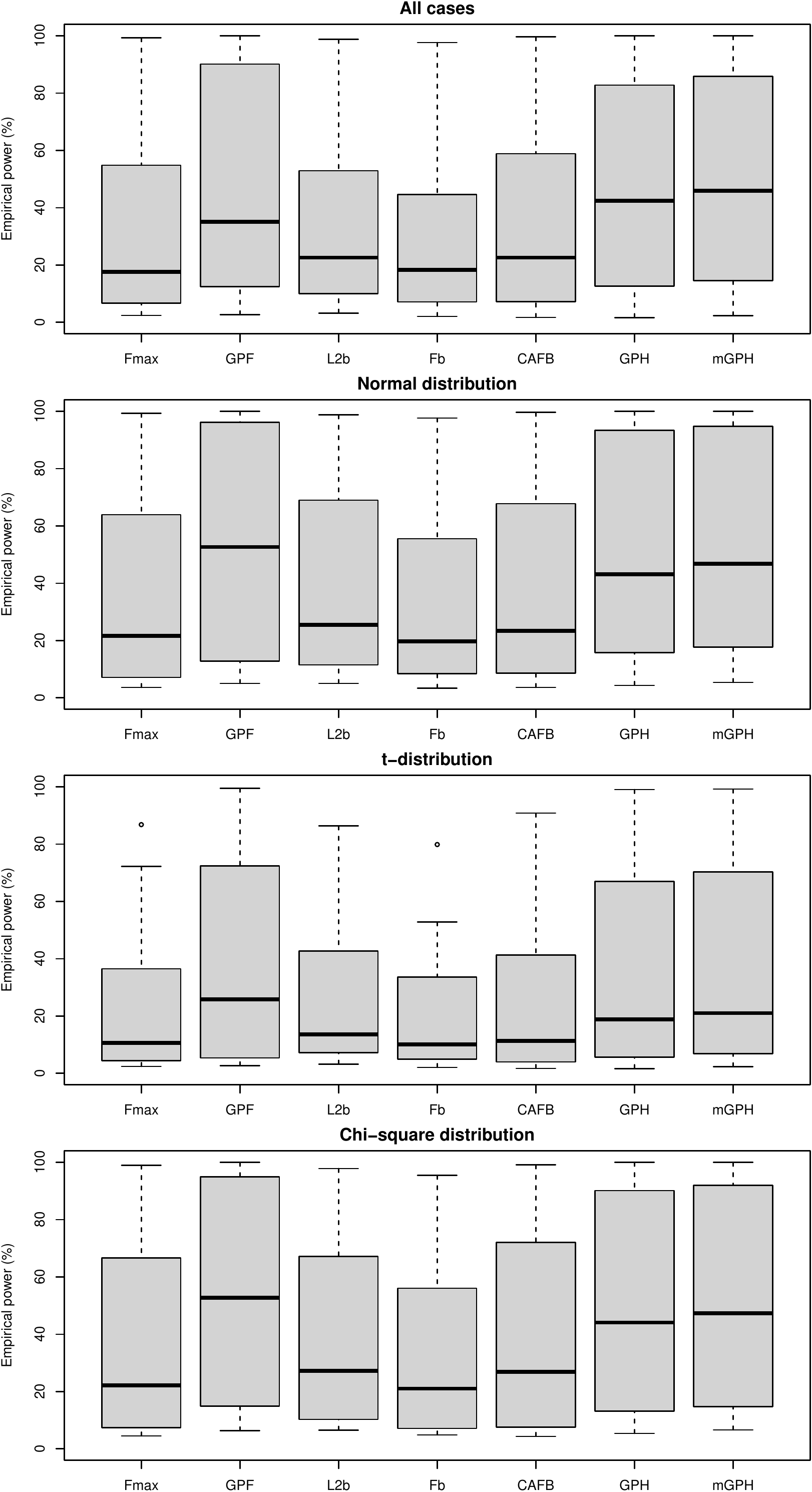}
\caption[Box-and-whisker plots for the empirical powers of all tests obtained under alternative A3 for the Tukey constrasts with scaling function and different distributions]{Box-and-whisker plots for the empirical powers (as percentages) of all tests obtained under alternative A3 for the Tukey constrasts with scaling function and different distributions}
\end{figure}

\newpage

\begin{longtable}[t]{rrrr|rrrrrrr}
\caption[Empirical powers of all tests obtained under alternative A4 for the Tukey constrasts with scaling function]{\label{tab:unnamed-chunk-116}Empirical powers (as percentages) of all tests obtained under alternative A4 for the Tukey constrasts with scaling function (D - distribution, $(\lambda_1,\lambda_2,\lambda_3,\lambda_4)$: (1,1,1,1) - homoscedastic case, (1,1.25,1.5,1.75) - heteroscedastic case (positive pairing), (1.75,1.5,1.25,1) - heteroscedastic case (negative pairing))}\\
\hline
D&$(n_1,n_2,n_3,n_4)$&$(\lambda_1,\lambda_2,\lambda_3,\lambda_4)$&$\mathcal{H}$&Fmax&GPF&L2b&Fb&CAFB&GPH&mGPH\\
\hline
\endfirsthead
\caption[]{Empirical powers (as percentages) of all tests obtained under alternative A4 for the Tukey constrasts with scaling function (D - distribution, $(\lambda_1,\lambda_2,\lambda_3,\lambda_4)$: (1,1,1,1) - homoscedastic case, (1,1.25,1.5,1.75) - heteroscedastic case (positive pairing), (1.75,1.5,1.25,1) - heteroscedastic case (negative pairing)) \textit{(continued)}}\\
\hline
D&$(n_1,n_2,n_3,n_4)$&$(\lambda_1,\lambda_2,\lambda_3,\lambda_4)$&$\mathcal{H}$&Fmax&GPF&L2b&Fb&CAFB&GPH&mGPH\\
\hline
\endhead
N&(15,20,25,30)&(1,1,1,1)&$\mathcal{H}_{0,1}$&9.00&9.35&2.00&1.35&1.65&9.15&10.20\\
&&&$\mathcal{H}_{0,2}$&67.25&78.45&1.55&0.80&3.20&76.10&78.55\\
&&&$\mathcal{H}_{0,3}$&99.75&100.00&1.95&1.15&13.40&99.95&99.95\\
&&&$\mathcal{H}_{0,4}$&12.65&13.80&1.45&1.00&1.15&14.00&15.25\\
&&&$\mathcal{H}_{0,5}$&84.75&91.20&2.00&1.35&5.90&91.45&92.65\\
&&&$\mathcal{H}_{0,6}$&17.45&19.70&1.55&1.15&1.50&19.70&21.80\\
N&(15,20,25,30)&(1,1.25,1.5,1.75)&$\mathcal{H}_{0,1}$&5.05&4.60&1.25&0.60&0.60&5.70&6.55\\
&&&$\mathcal{H}_{0,2}$&28.15&31.40&1.20&0.80&2.25&50.75&53.40\\
&&&$\mathcal{H}_{0,3}$&64.45&76.25&1.40&0.80&4.85&94.25&95.30\\
&&&$\mathcal{H}_{0,4}$&3.75&4.60&1.10&0.80&1.25&5.35&6.25\\
&&&$\mathcal{H}_{0,5}$&25.95&30.40&1.25&0.95&1.90&41.25&43.60\\
&&&$\mathcal{H}_{0,6}$&4.25&3.95&1.20&0.95&0.90&4.55&5.15\\
N&(15,20,25,30)&(1.75,1.5,1.25,1)&$\mathcal{H}_{0,1}$&3.60&2.70&1.50&0.95&0.95&1.90&2.40\\
&&&$\mathcal{H}_{0,2}$&33.65&35.90&1.65&1.05&1.45&18.95&21.95\\
&&&$\mathcal{H}_{0,3}$&92.35&95.20&2.00&1.05&3.65&69.70&73.55\\
&&&$\mathcal{H}_{0,4}$&6.40&6.80&1.30&0.95&0.95&5.25&6.50\\
&&&$\mathcal{H}_{0,5}$&64.15&70.25&1.65&0.95&2.25&51.90&56.70\\
&&&$\mathcal{H}_{0,6}$&14.00&16.25&1.15&0.75&0.90&13.10&15.20\\
t&(15,20,25,30)&(1,1,1,1)&$\mathcal{H}_{0,1}$&4.80&4.05&1.75&1.15&0.75&3.80&4.50\\
&&&$\mathcal{H}_{0,2}$&37.55&44.35&1.10&0.65&2.25&41.90&45.15\\
&&&$\mathcal{H}_{0,3}$&89.95&94.60&2.10&0.90&6.50&91.90&93.20\\
&&&$\mathcal{H}_{0,4}$&6.25&7.20&1.25&0.95&0.95&6.00&7.40\\
&&&$\mathcal{H}_{0,5}$&54.90&63.35&1.40&0.85&2.55&62.05&64.85\\
&&&$\mathcal{H}_{0,6}$&8.30&9.70&1.00&0.75&1.10&9.35&10.55\\
t&(15,20,25,30)&(1,1.25,1.5,1.75)&$\mathcal{H}_{0,1}$&2.15&2.25&1.85&1.25&0.85&2.75&3.15\\
&&&$\mathcal{H}_{0,2}$&11.95&12.20&1.85&1.05&1.55&23.35&26.25\\
&&&$\mathcal{H}_{0,3}$&29.50&35.15&1.40&0.90&2.75&65.95&69.10\\
&&&$\mathcal{H}_{0,4}$&2.85&2.60&1.50&1.15&0.80&2.90&3.25\\
&&&$\mathcal{H}_{0,5}$&11.35&12.55&1.15&0.75&1.20&18.40&20.30\\
&&&$\mathcal{H}_{0,6}$&2.10&1.95&1.50&1.10&1.10&2.10&2.35\\
t&(15,20,25,30)&(1.75,1.5,1.25,1)&$\mathcal{H}_{0,1}$&1.75&1.45&1.30&0.70&0.75&1.10&1.25\\
&&&$\mathcal{H}_{0,2}$&17.45&18.55&1.25&0.80&0.95&8.45&10.45\\
&&&$\mathcal{H}_{0,3}$&72.50&74.90&1.75&1.00&2.00&37.10&41.05\\
&&&$\mathcal{H}_{0,4}$&3.45&3.10&1.25&0.80&0.90&2.40&2.90\\
&&&$\mathcal{H}_{0,5}$&39.05&42.95&1.55&1.10&1.50&25.65&29.05\\
&&&$\mathcal{H}_{0,6}$&8.20&8.60&1.00&0.65&0.85&6.40&7.85\\
$\chi^2$&(15,20,25,30)&(1,1,1,1)&$\mathcal{H}_{0,1}$&8.60&9.40&1.50&0.75&1.00&9.50&10.95\\
&&&$\mathcal{H}_{0,2}$&69.05&78.00&1.15&0.90&3.75&74.60&78.00\\
&&&$\mathcal{H}_{0,3}$&99.80&99.95&2.35&1.30&14.70&99.65&99.75\\
&&&$\mathcal{H}_{0,4}$&12.00&13.10&1.45&0.85&1.00&13.40&15.10\\
&&&$\mathcal{H}_{0,5}$&84.30&90.80&1.25&0.95&5.45&88.20&89.95\\
&&&$\mathcal{H}_{0,6}$&19.20&21.65&1.20&0.65&0.95&20.75&22.95\\
$\chi^2$&(15,20,25,30)&(1,1.25,1.5,1.75)&$\mathcal{H}_{0,1}$&5.85&4.60&1.70&1.15&0.95&6.40&7.50\\
&&&$\mathcal{H}_{0,2}$&27.50&31.55&1.20&0.80&2.15&48.25&52.00\\
&&&$\mathcal{H}_{0,3}$&67.60&76.20&1.80&1.25&5.40&94.30&95.15\\
&&&$\mathcal{H}_{0,4}$&4.45&4.15&1.95&1.20&0.70&5.25&6.35\\
&&&$\mathcal{H}_{0,5}$&29.20&31.85&1.70&1.15&2.45&44.15&46.65\\
&&&$\mathcal{H}_{0,6}$&4.80&4.70&1.20&0.95&1.05&5.25&6.15\\
$\chi^2$&(15,20,25,30)&(1.75,1.5,1.25,1)&$\mathcal{H}_{0,1}$&3.60&2.65&2.00&1.05&1.35&1.90&2.70\\
&&&$\mathcal{H}_{0,2}$&34.80&37.35&2.55&1.40&1.45&21.60&24.85\\
&&&$\mathcal{H}_{0,3}$&91.30&93.45&2.45&1.60&4.70&70.05&73.35\\
&&&$\mathcal{H}_{0,4}$&7.10&6.55&1.30&1.05&1.00&4.75&6.05\\
&&&$\mathcal{H}_{0,5}$&67.85&70.80&1.60&1.00&2.85&53.50&58.10\\
&&&$\mathcal{H}_{0,6}$&15.15&15.30&1.40&0.95&1.30&13.40&15.75\\
\hline
\end{longtable}

\begin{figure}
\centering
\includegraphics[width= 0.95\textwidth,height=0.9\textheight]{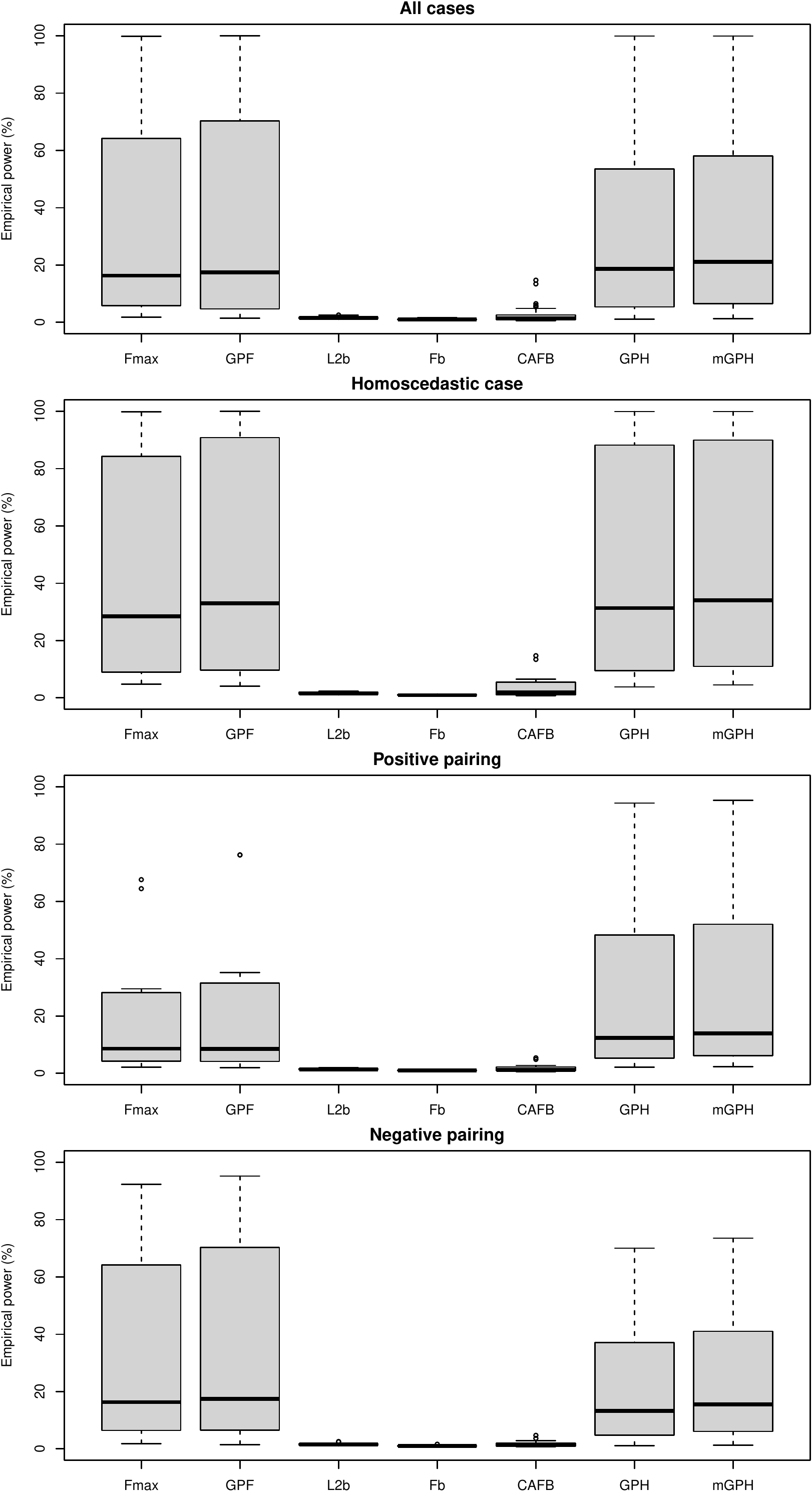}
\caption[Box-and-whisker plots for the empirical powers of all tests obtained under alternative A4 for the Tukey constrasts with scaling function and homoscedastic and heteroscedastic cases]{Box-and-whisker plots for the empirical powers (as percentages) of all tests obtained under alternative A4 for the Tukey constrasts with scaling function and homoscedastic and heteroscedastic cases}
\end{figure}

\begin{figure}
\centering
\includegraphics[width= 0.95\textwidth,height=0.9\textheight]{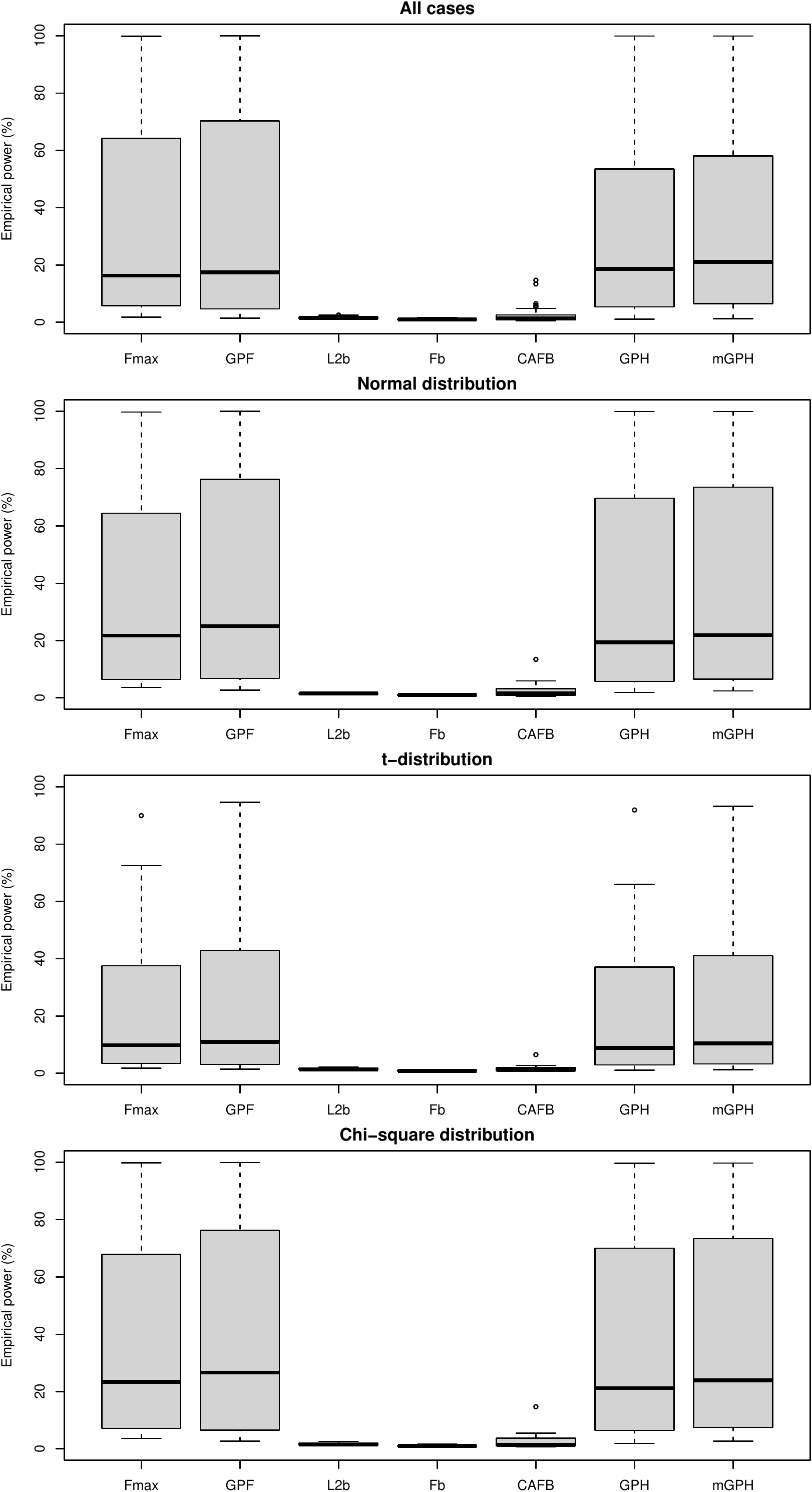}
\caption[Box-and-whisker plots for the empirical powers of all tests obtained under alternative A4 for the Tukey constrasts with scaling function and different distributions]{Box-and-whisker plots for the empirical powers (as percentages) of all tests obtained under alternative A4 for the Tukey constrasts with scaling function and different distributions}
\end{figure}

\newpage

\begin{longtable}[t]{rrrr|rrrrrrr}
\caption[Empirical sizes and powers of all tests obtained under alternative A5 for the Tukey constrasts with scaling function]{\label{tab:unnamed-chunk-119}Empirical sizes ($\mathcal{H}_{0,1}$, $\mathcal{H}_{0,2}$, $\mathcal{H}_{0,4}$) and powers ($\mathcal{H}_{0,3}$, $\mathcal{H}_{0,5}$, $\mathcal{H}_{0,6}$) (as percentages) of all tests obtained under alternative A5 for the Tukey constrasts with scaling function (D - distribution, $(\lambda_1,\lambda_2,\lambda_3,\lambda_4)$: (1,1,1,1) - homoscedastic case, (1,1.25,1.5,1.75) - heteroscedastic case (positive pairing), (1.75,1.5,1.25,1) - heteroscedastic case (negative pairing))}\\
\hline
D&$(n_1,n_2,n_3,n_4)$&$(\lambda_1,\lambda_2,\lambda_3,\lambda_4)$&$\mathcal{H}$&Fmax&GPF&L2b&Fb&CAFB&GPH&mGPH\\
\hline
\endfirsthead
\caption[]{Empirical sizes ($\mathcal{H}_{0,1}$, $\mathcal{H}_{0,2}$, $\mathcal{H}_{0,4}$) and powers ($\mathcal{H}_{0,3}$, $\mathcal{H}_{0,5}$, $\mathcal{H}_{0,6}$) (as percentages) of all tests obtained under alternative A5 for the Tukey constrasts with scaling function (D - distribution, $(\lambda_1,\lambda_2,\lambda_3,\lambda_4)$: (1,1,1,1) - homoscedastic case, (1,1.25,1.5,1.75) - heteroscedastic case (positive pairing), (1.75,1.5,1.25,1) - heteroscedastic case (negative pairing)) \textit{(continued)}}\\
\hline
D&$(n_1,n_2,n_3,n_4)$&$(\lambda_1,\lambda_2,\lambda_3,\lambda_4)$&$\mathcal{H}$&Fmax&GPF&L2b&Fb&CAFB&GPH&mGPH\\
\hline
\endhead
N&(15,20,25,30)&(1,1,1,1)&$\mathcal{H}_{0,1}$&0.70&0.75&2.10&1.20&0.80&0.75&0.95\\
&&&$\mathcal{H}_{0,2}$&0.95&0.65&1.75&0.70&0.80&0.75&0.85\\
&&&$\mathcal{H}_{0,3}$&72.10&82.30&92.70&88.10&91.65&77.90&80.50\\
&&&$\mathcal{H}_{0,4}$&0.50&0.50&1.50&0.95&0.50&0.55&0.85\\
&&&$\mathcal{H}_{0,5}$&83.10&90.80&95.95&94.70&96.80&90.25&91.50\\
&&&$\mathcal{H}_{0,6}$&90.30&95.85&98.50&97.65&99.15&95.90&96.60\\
N&(15,20,25,30)&(1,1.25,1.5,1.75)&$\mathcal{H}_{0,1}$&0.80&0.45&1.55&0.85&0.90&0.60&0.60\\
&&&$\mathcal{H}_{0,2}$&0.20&0.00&1.15&0.65&0.50&0.55&0.65\\
&&&$\mathcal{H}_{0,3}$&14.75&17.05&69.25&60.95&64.85&44.05&46.45\\
&&&$\mathcal{H}_{0,4}$&0.75&0.50&1.45&0.90&0.55&0.55&0.60\\
&&&$\mathcal{H}_{0,5}$&24.00&29.60&65.90&59.75&62.65&40.65&43.90\\
&&&$\mathcal{H}_{0,6}$&29.05&34.75&61.30&55.50&57.50&36.90&39.80\\
N&(15,20,25,30)&(1.75,1.5,1.25,1)&$\mathcal{H}_{0,1}$&1.10&0.60&1.55&0.60&0.65&0.45&0.55\\
&&&$\mathcal{H}_{0,2}$&2.25&1.65&1.45&0.90&1.10&0.45&0.55\\
&&&$\mathcal{H}_{0,3}$&54.55&56.90&54.60&42.75&42.30&22.20&24.75\\
&&&$\mathcal{H}_{0,4}$&1.25&0.90&2.00&1.05&0.55&0.60&0.70\\
&&&$\mathcal{H}_{0,5}$&61.45&69.10&76.80&69.20&71.65&49.40&54.35\\
&&&$\mathcal{H}_{0,6}$&78.75&85.75&92.85&90.95&93.70&82.80&85.40\\
t&(15,20,25,30)&(1,1,1,1)&$\mathcal{H}_{0,1}$&0.70&0.25&1.45&0.85&0.75&0.20&0.35\\
&&&$\mathcal{H}_{0,2}$&0.85&0.40&1.70&0.90&0.60&0.30&0.45\\
&&&$\mathcal{H}_{0,3}$&37.15&43.85&68.95&60.55&64.20&39.45&43.90\\
&&&$\mathcal{H}_{0,4}$&1.20&1.15&1.40&0.75&0.60&0.90&1.05\\
&&&$\mathcal{H}_{0,5}$&50.15&58.05&79.15&73.60&77.30&56.20&59.60\\
&&&$\mathcal{H}_{0,6}$&61.05&68.05&83.85&80.75&84.80&66.55&69.50\\
t&(15,20,25,30)&(1,1.25,1.5,1.75)&$\mathcal{H}_{0,1}$&0.60&0.45&1.65&1.05&1.05&0.50&0.80\\
&&&$\mathcal{H}_{0,2}$&0.20&0.20&1.80&1.00&0.95&0.50&0.75\\
&&&$\mathcal{H}_{0,3}$&6.45&6.05&43.80&38.40&38.40&20.40&23.10\\
&&&$\mathcal{H}_{0,4}$&0.65&0.55&1.50&1.10&0.65&0.65&0.75\\
&&&$\mathcal{H}_{0,5}$&12.20&13.70&39.60&34.45&35.85&19.65&21.45\\
&&&$\mathcal{H}_{0,6}$&14.65&17.65&36.15&32.50&31.90&18.40&20.55\\
t&(15,20,25,30)&(1.75,1.5,1.25,1)&$\mathcal{H}_{0,1}$&1.00&0.75&1.35&0.45&0.60&0.55&0.75\\
&&&$\mathcal{H}_{0,2}$&1.85&1.80&1.80&1.20&1.05&0.60&0.75\\
&&&$\mathcal{H}_{0,3}$&34.30&34.65&30.65&20.95&21.10&8.70&10.70\\
&&&$\mathcal{H}_{0,4}$&1.10&0.65&1.25&0.80&0.60&0.50&0.65\\
&&&$\mathcal{H}_{0,5}$&36.45&41.50&52.95&45.10&45.25&25.10&28.75\\
&&&$\mathcal{H}_{0,6}$&46.50&56.45&72.35&67.45&71.10&50.70&54.50\\
$\chi^2$&(15,20,25,30)&(1,1,1,1)&$\mathcal{H}_{0,1}$&0.75&0.45&1.55&1.05&0.95&0.50&0.55\\
&&&$\mathcal{H}_{0,2}$&0.70&0.50&1.65&0.70&0.75&0.35&0.40\\
&&&$\mathcal{H}_{0,3}$&71.35&80.30&92.00&85.30&91.10&75.25&78.10\\
&&&$\mathcal{H}_{0,4}$&0.95&0.85&1.25&0.95&0.90&0.80&0.95\\
&&&$\mathcal{H}_{0,5}$&83.80&90.45&94.95&92.55&96.50&88.05&89.65\\
&&&$\mathcal{H}_{0,6}$&89.60&94.55&97.60&96.60&98.45&93.50&94.55\\
$\chi^2$&(15,20,25,30)&(1,1.25,1.5,1.75)&$\mathcal{H}_{0,1}$&0.65&0.50&1.45&0.85&0.45&0.60&0.75\\
&&&$\mathcal{H}_{0,2}$&0.45&0.05&1.25&0.65&0.70&0.60&0.75\\
&&&$\mathcal{H}_{0,3}$&12.30&12.85&70.00&64.35&66.00&43.35&46.40\\
&&&$\mathcal{H}_{0,4}$&0.75&0.70&1.05&0.85&0.70&0.75&0.80\\
&&&$\mathcal{H}_{0,5}$&25.60&29.45&65.30&60.35&62.25&41.60&44.75\\
&&&$\mathcal{H}_{0,6}$&30.85&35.20&63.25&57.75&59.30&37.50&40.05\\
$\chi^2$&(15,20,25,30)&(1.75,1.5,1.25,1)&$\mathcal{H}_{0,1}$&1.15&0.90&2.25&1.20&0.80&0.50&0.90\\
&&&$\mathcal{H}_{0,2}$&2.25&2.00&1.75&1.00&1.45&0.60&0.70\\
&&&$\mathcal{H}_{0,3}$&57.15&60.25&53.25&42.35&46.75&27.75&30.20\\
&&&$\mathcal{H}_{0,4}$&0.70&0.35&1.45&0.90&0.60&0.35&0.60\\
&&&$\mathcal{H}_{0,5}$&63.60&70.25&74.80&66.95&73.35&52.80&56.45\\
&&&$\mathcal{H}_{0,6}$&80.10&85.30&92.50&90.05&93.90&81.30&83.80\\
\hline
\end{longtable}

\begin{figure}
\centering
\includegraphics[width= 0.95\textwidth,height=0.9\textheight]{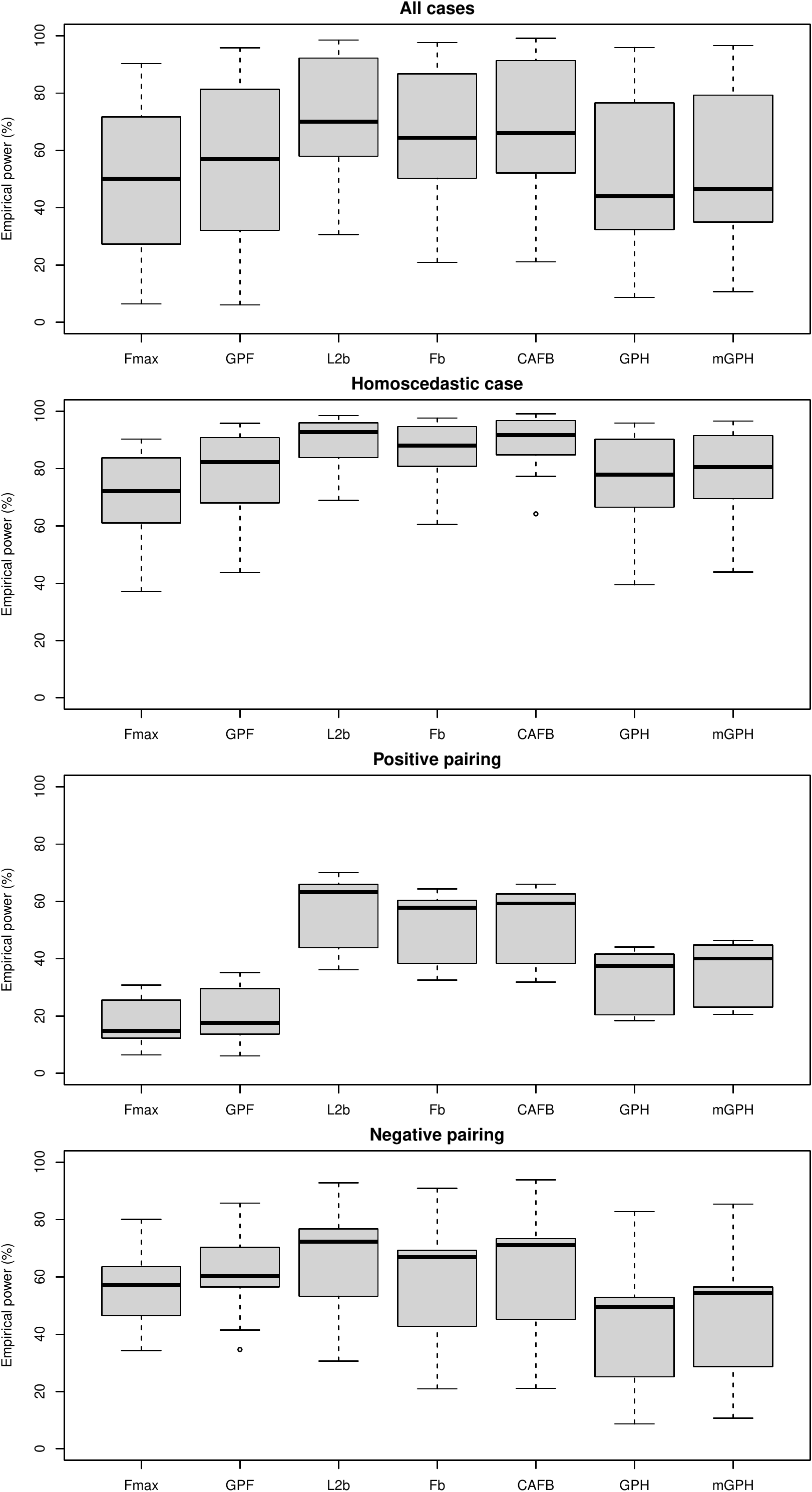}
\caption[Box-and-whisker plots for the empirical powers of all tests obtained under alternative A5 for the Tukey constrasts with scaling function and homoscedastic and heteroscedastic cases]{Box-and-whisker plots for the empirical powers (as percentages) of all tests obtained under alternative A5 for the Tukey constrasts with scaling function and homoscedastic and heteroscedastic cases}
\end{figure}

\begin{figure}
\centering
\includegraphics[width= 0.95\textwidth,height=0.9\textheight]{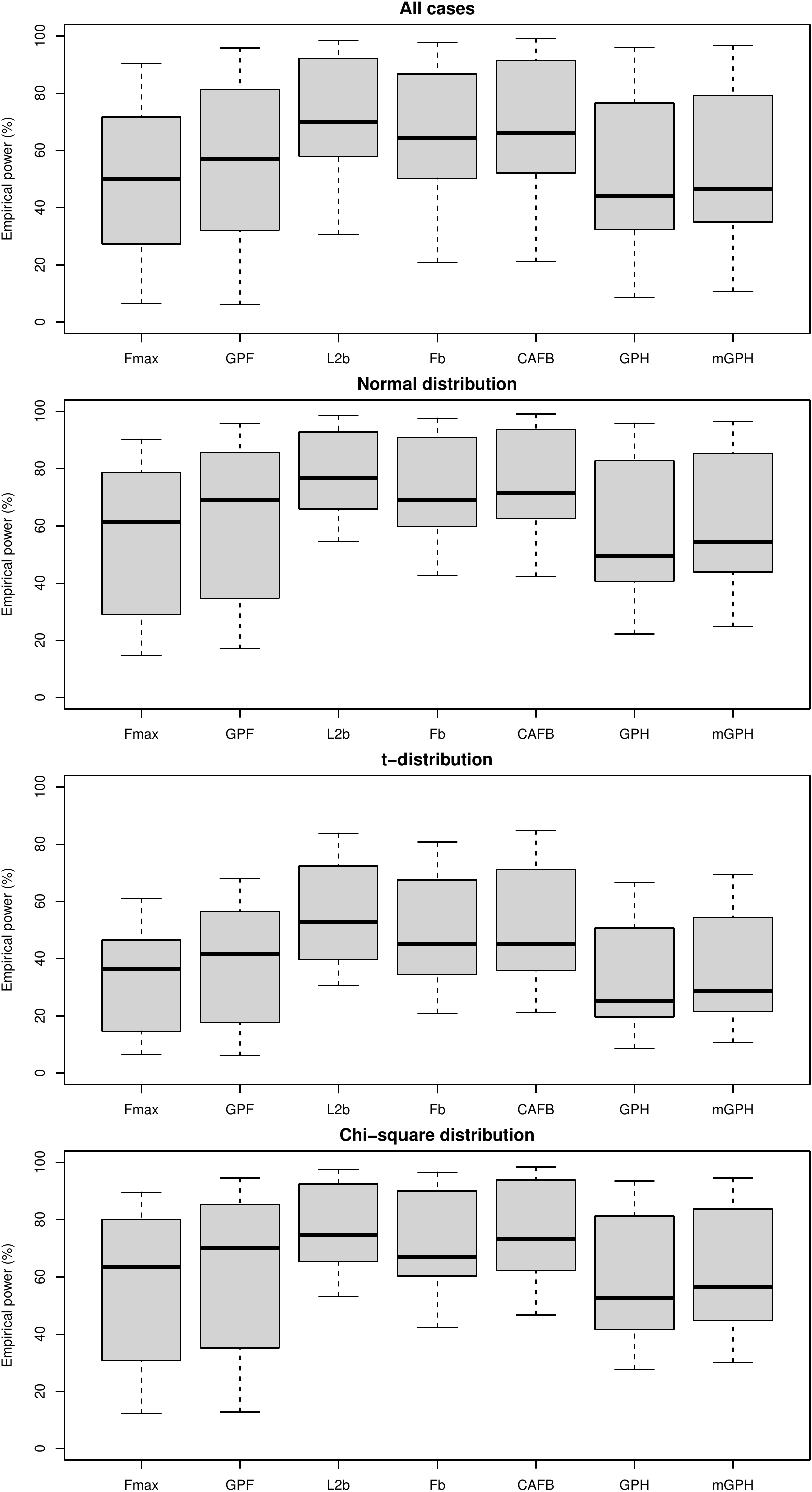}
\caption[Box-and-whisker plots for the empirical powers of all tests obtained under alternative A5 for the Tukey constrasts with scaling function and different distributions]{Box-and-whisker plots for the empirical powers (as percentages) of all tests obtained under alternative A5 for the Tukey constrasts with scaling function and different distributions}
\end{figure}

\newpage

\begin{longtable}[t]{rrrr|rrrrrrr}
\caption[Empirical powers of all tests obtained under alternative A6 for the Tukey constrasts with scaling function]{\label{tab:unnamed-chunk-124}Empirical powers (as percentages) of all tests obtained under alternative A6 for the Tukey constrasts with scaling function (D - distribution, $(\lambda_1,\lambda_2,\lambda_3,\lambda_4)$: (1,1,1,1) - homoscedastic case, (1,1.25,1.5,1.75) - heteroscedastic case (positive pairing), (1.75,1.5,1.25,1) - heteroscedastic case (negative pairing))}\\
\hline
D&$(n_1,n_2,n_3,n_4)$&$(\lambda_1,\lambda_2,\lambda_3,\lambda_4)$&$\mathcal{H}$&Fmax&GPF&L2b&Fb&CAFB&GPH&mGPH\\
\hline
\endfirsthead
\caption[]{Empirical powers (as percentages) of all tests obtained under alternative A6 for the Tukey constrasts with scaling function (D - distribution, $(\lambda_1,\lambda_2,\lambda_3,\lambda_4)$: (1,1,1,1) - homoscedastic case, (1,1.25,1.5,1.75) - heteroscedastic case (positive pairing), (1.75,1.5,1.25,1) - heteroscedastic case (negative pairing)) \textit{(continued)}}\\
\hline
D&$(n_1,n_2,n_3,n_4)$&$(\lambda_1,\lambda_2,\lambda_3,\lambda_4)$&$\mathcal{H}$&Fmax&GPF&L2b&Fb&CAFB&GPH&mGPH\\
\hline
\endhead
N&(15,20,25,30)&(1,1,1,1)&$\mathcal{H}_{0,1}$&8.20&7.95&26.15&19.45&18.30&7.70&9.05\\
&&&$\mathcal{H}_{0,2}$&64.50&74.75&89.40&83.35&88.80&72.20&74.75\\
&&&$\mathcal{H}_{0,3}$&99.30&99.85&99.90&99.80&99.85&99.85&99.90\\
&&&$\mathcal{H}_{0,4}$&11.70&14.10&31.45&25.85&24.10&13.10&14.85\\
&&&$\mathcal{H}_{0,5}$&81.25&89.85&96.50&94.60&96.30&89.10&90.55\\
&&&$\mathcal{H}_{0,6}$&16.95&18.35&39.20&34.10&34.75&17.20&19.30\\
N&(15,20,25,30)&(1,1.25,1.5,1.75)&$\mathcal{H}_{0,1}$&4.20&3.40&18.95&13.25&12.75&5.25&6.25\\
&&&$\mathcal{H}_{0,2}$&23.95&29.35&74.55&66.65&68.90&46.20&48.80\\
&&&$\mathcal{H}_{0,3}$&60.95&70.65&97.70&96.40&98.15&92.70&93.65\\
&&&$\mathcal{H}_{0,4}$&4.65&4.15&15.95&12.15&9.70&5.35&6.20\\
&&&$\mathcal{H}_{0,5}$&24.80&30.00&65.05&59.40&63.60&42.40&44.50\\
&&&$\mathcal{H}_{0,6}$&4.85&3.80&13.80&11.25&8.95&4.60&5.50\\
N&(15,20,25,30)&(1.75,1.5,1.25,1)&$\mathcal{H}_{0,1}$&3.60&3.25&9.60&6.10&4.65&3.30&3.55\\
&&&$\mathcal{H}_{0,2}$&31.05&35.20&48.20&37.30&39.15&19.05&21.70\\
&&&$\mathcal{H}_{0,3}$&91.05&94.00&90.70&83.15&86.05&69.40&73.65\\
&&&$\mathcal{H}_{0,4}$&6.40&6.20&14.65&11.30&9.25&5.05&6.30\\
&&&$\mathcal{H}_{0,5}$&63.20&67.45&76.70&69.95&73.60&49.45&53.60\\
&&&$\mathcal{H}_{0,6}$&13.05&16.05&31.15&25.45&22.90&14.10&15.85\\
t&(15,20,25,30)&(1,1,1,1)&$\mathcal{H}_{0,1}$&4.00&4.00&13.90&9.80&7.75&3.60&4.50\\
&&&$\mathcal{H}_{0,2}$&35.50&43.50&69.90&59.70&62.45&39.10&42.60\\
&&&$\mathcal{H}_{0,3}$&88.20&93.25&97.70&95.70&97.70&90.20&92.00\\
&&&$\mathcal{H}_{0,4}$&6.50&6.60&18.65&14.95&12.35&6.25&6.75\\
&&&$\mathcal{H}_{0,5}$&50.25&57.70&80.20&75.05&79.45&55.45&59.10\\
&&&$\mathcal{H}_{0,6}$&8.15&8.45&22.00&18.75&15.15&8.15&8.90\\
t&(15,20,25,30)&(1,1.25,1.5,1.75)&$\mathcal{H}_{0,1}$&3.30&2.50&11.35&7.65&6.45&3.45&4.20\\
&&&$\mathcal{H}_{0,2}$&12.05&12.95&48.20&40.50&41.60&22.65&25.00\\
&&&$\mathcal{H}_{0,3}$&29.60&33.85&83.85&79.65&84.35&64.75&68.10\\
&&&$\mathcal{H}_{0,4}$&3.10&2.40&7.70&5.85&5.45&3.10&3.80\\
&&&$\mathcal{H}_{0,5}$&10.20&11.80&39.60&34.85&33.70&19.05&21.00\\
&&&$\mathcal{H}_{0,6}$&2.55&2.35&7.20&5.35&4.00&2.55&3.25\\
t&(15,20,25,30)&(1.75,1.5,1.25,1)&$\mathcal{H}_{0,1}$&2.10&1.45&5.00&3.55&3.05&0.95&1.55\\
&&&$\mathcal{H}_{0,2}$&16.45&17.25&26.45&19.60&17.90&7.65&9.35\\
&&&$\mathcal{H}_{0,3}$&68.75&73.00&66.45&54.85&57.30&34.80&38.75\\
&&&$\mathcal{H}_{0,4}$&3.60&3.50&9.10&6.45&4.75&2.90&3.35\\
&&&$\mathcal{H}_{0,5}$&37.90&41.00&51.65&43.95&44.20&24.25&28.40\\
&&&$\mathcal{H}_{0,6}$&7.20&7.50&17.20&13.85&12.30&5.75&6.85\\
$\chi^2$&(15,20,25,30)&(1,1,1,1)&$\mathcal{H}_{0,1}$&7.60&8.50&26.35&19.50&19.00&9.10&10.65\\
&&&$\mathcal{H}_{0,2}$&65.60&72.85&88.70&83.55&88.40&70.25&72.75\\
&&&$\mathcal{H}_{0,3}$&99.65&99.85&99.80&99.25&99.95&99.25&99.50\\
&&&$\mathcal{H}_{0,4}$&12.20&13.20&32.40&27.15&27.30&13.00&15.20\\
&&&$\mathcal{H}_{0,5}$&84.25&91.20&95.50&92.90&96.45&88.85&90.55\\
&&&$\mathcal{H}_{0,6}$&17.90&21.40&41.80&36.65&35.50&21.50&23.85\\
$\chi^2$&(15,20,25,30)&(1,1.25,1.5,1.75)&$\mathcal{H}_{0,1}$&3.70&3.20&17.90&12.85&11.45&4.60&5.70\\
&&&$\mathcal{H}_{0,2}$&22.20&28.20&72.90&66.00&70.20&45.40&48.85\\
&&&$\mathcal{H}_{0,3}$&62.25&75.20&98.20&96.95&98.60&93.40&94.25\\
&&&$\mathcal{H}_{0,4}$&4.40&5.00&16.10&12.95&10.40&5.75&6.90\\
&&&$\mathcal{H}_{0,5}$&25.75&30.35&67.20&62.20&64.80&41.55&44.35\\
&&&$\mathcal{H}_{0,6}$&3.85&3.90&14.40&11.70&10.45&4.40&4.85\\
$\chi^2$&(15,20,25,30)&(1.75,1.5,1.25,1)&$\mathcal{H}_{0,1}$&4.80&3.55&10.75&7.15&6.70&2.75&3.35\\
&&&$\mathcal{H}_{0,2}$&32.70&35.35&46.60&37.55&42.30&22.00&24.70\\
&&&$\mathcal{H}_{0,3}$&91.65&92.85&87.05&78.75&85.00&67.40&70.50\\
&&&$\mathcal{H}_{0,4}$&6.70&6.30&16.05&12.85&12.00&5.20&6.40\\
&&&$\mathcal{H}_{0,5}$&64.05&69.35&75.40&67.65&72.85&51.55&55.45\\
&&&$\mathcal{H}_{0,6}$&14.25&16.00&31.80&26.65&25.80&13.30&15.00\\
\hline
\end{longtable}

\begin{figure}
\centering
\includegraphics[width= 0.95\textwidth,height=0.9\textheight]{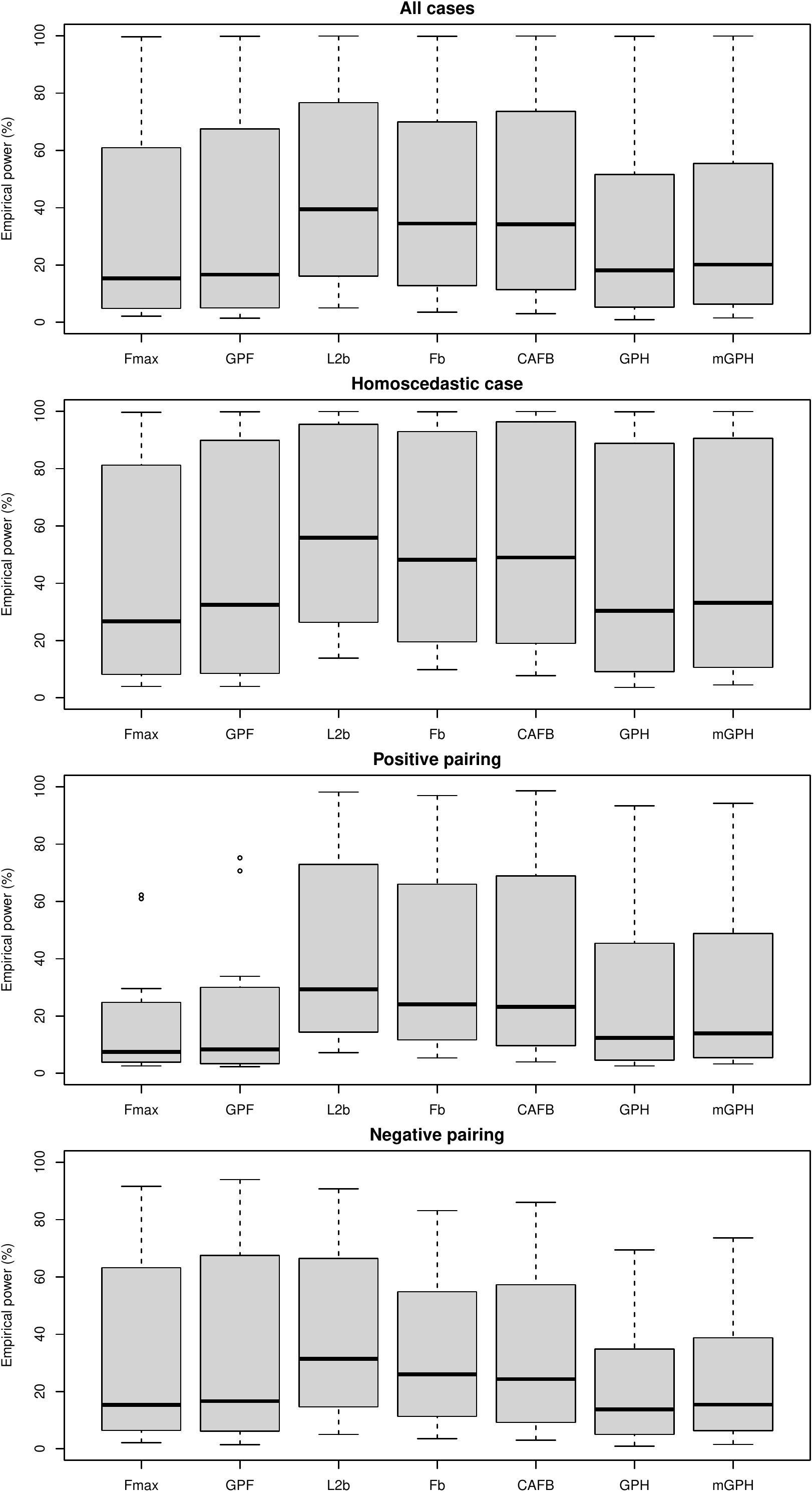}
\caption[Box-and-whisker plots for the empirical powers of all tests obtained under alternative A6 for the Tukey constrasts with scaling function and homoscedastic and heteroscedastic cases]{Box-and-whisker plots for the empirical powers (as percentages) of all tests obtained under alternative A6 for the Tukey constrasts with scaling function and homoscedastic and heteroscedastic cases}
\end{figure}

\begin{figure}
\centering
\includegraphics[width= 0.95\textwidth,height=0.9\textheight]{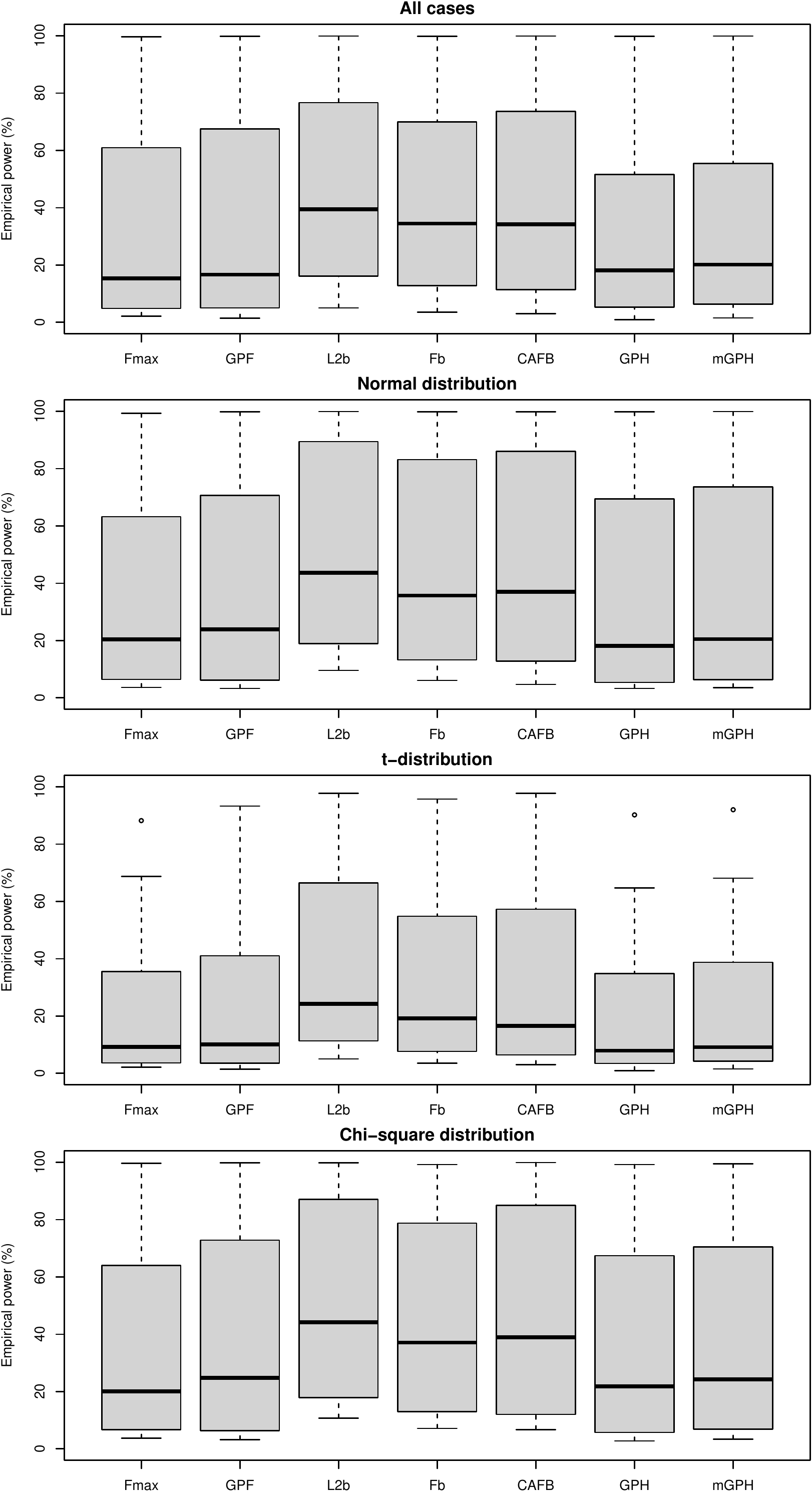}
\caption[Box-and-whisker plots for the empirical powers of all tests obtained under alternative A6 for the Tukey constrasts with scaling function and different distributions]{Box-and-whisker plots for the empirical powers (as percentages) of all tests obtained under alternative A6 for the Tukey constrasts with scaling function and different distributions}
\end{figure}

\newpage

\section{Details on the Competitors for the Simulation Studies}
In this section, more details on the competitors for the simulation studies are provided. Here, we use the notation of the main paper.

The Fmax- and GPF-test by \citeauthor{SmaZha} \cite{SmaZha} are using the test statistics
\begin{align*}
   \text{Fmax}_n := \sup\limits_{t\in\T} \left\{ \dfrac{\text{SSH}_n(t) / r}{\sum_{i=1}^k (n_i -1) \widehat{\gamma}_i (t,t) / (n-k)} \right\}\qquad
    \text{and}\qquad
     \text{GPF}_n := \int\limits_{\T} \dfrac{\text{SSH}_n(t) / r}{\sum_{i=1}^k (n_i -1) \widehat{\gamma}_i (t,t) / (n-k)} \;\mathrm{d}t,
\end{align*}
where $\text{SSH}_n(t) :=  ( \mathbf{H} \widehat{\boldsymbol\eta}(t) - \mathbf{c}(t) )^{\top} (\mathbf{H}\mathbf{D}_n \mathbf{H}^{\top})^{-1} ( \mathbf{H} \widehat{\boldsymbol\eta}(t) - \mathbf{c}(t) )$ for all $t\in\T$ and $\mathbf{D}_n := \mathrm{diag}(1/n_1,...,1/n_k).$
The critical values are approximated with a nonparametric pooled bootstrap approach as described in \cite[p.~159]{Zhang2013}. Here, the bootstrap counterparts $v_{ij}^*, j\in\{1,...,n_i\}, i\in\{1,...,k\},$ of the subject-effect functions are randomly generated from $x_{ij}-\widehat{\eta}_i, j\in\{1,...,n_i\}, i\in\{1,...,k\}.$ Based on the resulting bootstrap samples
$x_{ij}^* = v_{ij}^* + \widehat{\eta}_i, j\in\{1,...,n_i\}, i\in\{1,...,k\},$ the bootstrap counterparts of the Fmax- and GPF-test statistic can be calculated.

The L2b- and Fb-test statistics by \citeauthor{Zhang2013}~\cite{Zhang2013} are defined by
\begin{align*}
    \text{L2b}_n := \int\limits_{\T} \text{SSH}_n(t) \;\mathrm{d}t\qquad
    \text{and \qquad Fb}_n := \dfrac{\text{L2b}_n}{\sum_{i=1}^k a_{n,ii} \int_{\T}\widehat{\gamma}_i (t,t)\;\mathrm{d}t }
\end{align*}
with $a_{n,ii}$ being the $i$th diagonal element of $\mathbf{A}_n = \mathbf{D}_n^{1/2} \mathbf{H}^{\top} (\mathbf{H}\mathbf{D}_n\mathbf{H}^{\top})^{-1} \mathbf{H}\mathbf{D}_n^{1/2}$ for all $i\in\{1,...,k\}$.
The critical values are approximated by using a nonparametric groupwise bootstrap approach as described in \cite[p.~329]{Zhang2013}.
In contrast to the nonparametric pooled bootstrap, the $i$th bootstrap sample $x_{ij}^*, j\in\{1,...,n_i\},$ is randomly generated from the $i$th sample $x_{ij}, j\in\{1,...,n_i\},$ for each $i\in\{1,...,k\}$.

For the CAFB-test by \citeauthor{febrero}~\cite{febrero}, the null hypothesis 
\begin{align}\label{eq:local}
    \mathcal{H}_0^{\nu}: \mathbf{H}\int_{\T} \nu(t) {\boldsymbol\eta}(t)\;\mathrm{d}t = \int_{\T} \nu(t) \mathbf{c}(t)\;\mathrm{d}t
\end{align}
is examined with the robust ANOVA test of \citeauthor{brunner:dette:munk:1997} \cite{brunner:dette:munk:1997} based on the projected data $\int_{\T} \nu(t) x_{ij}(t)\;\mathrm{d}t, j\in\{1,...,n_i\}, i\in\{1,...,k\}$, where $\nu$ is randomly chosen from a Gaussian distribution 
such that each of its one-dimensional projections is non-degenerate.
This leads to the CAFB-test statistic
\begin{align*}
    \text{CAFB}_n(\nu) := \frac{\left(  \int_{\T} \nu(t) (\mathbf{H}\widehat{\boldsymbol\eta}(t)- \mathbf{c}(t))\;\mathrm{d}t \right)^{\top}(\mathbf{H}\mathbf{H}^{\top})^{-1}\left(  \int_{\T} \nu(t) (\mathbf{H}\widehat{\boldsymbol\eta}(t)- \mathbf{c}(t))\;\mathrm{d}t \right) }{\sum_{i=1}^k m_{ii} \widehat{\sigma}_i^2(\nu) / n_i},
\end{align*}
where $m_{ii}$ denotes the $i$th diagonal element of $\mathbf{M} := \mathbf{H}^{\top}(\mathbf{H}\mathbf{H}^{\top})^{-1}\mathbf{H}$ and $\widehat{\sigma}_i^2(\nu)$ denotes the empirical variance of $\int_{\T} \nu(t) x_{ij}(t)\;\mathrm{d}t, j\in\{1,...,n_i\}, $ for all $i\in\{1,...,k\}.$ The p-value can be determined from the F-distribution, see \cite{brunner:dette:munk:1997}.
For stabilization, this is done for 30 independent Gaussian white noise processes $\nu_1,...,\nu_{30}$ and, then, the resulting p-values are adjusted with the Bonferroni correction. The global null hypothesis (2) is rejected whenever at least one of the local hypotheses (\ref{eq:local}) is.  

\newpage
\bibliographystyle{myjmva}
\bibliography{sample}